%% file: thesis.tex
\documentclass[12pt]{report}

\usepackage[T1]{fontenc}
\usepackage{blindtext, color}
\usepackage[raggedright]{titlesec}
\definecolor{gray75}{gray}{0.75}
\newcommand{\hsp}{\hspace{20pt}}
\titleformat{\chapter}[hang]{\Huge\bfseries}{\thechapter\hsp\textcolor{gray75}{|}\hsp}{0pt}{\Huge\bfseries}

\usepackage{etoolbox} 
\usepackage{titling}
\usepackage{setspace} 
\usepackage[british]{babel}
\usepackage[nodayofweek]{datetime}

\newtoggle{debugging}
\input{./user_settings.tex}

\usepackage[a4paper,margin=20mm, bindingoffset=20mm]{geometry}

\usepackage{graphicx} 

\iftoggle{debugging}{%
\usepackage{layout}
}

\newdateformat{monthyear}{\monthname[\THEMONTH] \THEYEAR} 
\input{./user_preamble.tex}

\begin{document}

\pagestyle{empty} 

\begin{center}
\vspace*{2cm}
{\Huge\textbf{\thetitle}}\\ 
\vspace*{6cm}
{\Large \theauthor}\\ 

\vfill

\vspace*{0.2cm}
\begin{figure}[h!]
 \centering
  \includegraphics[width=50mm]{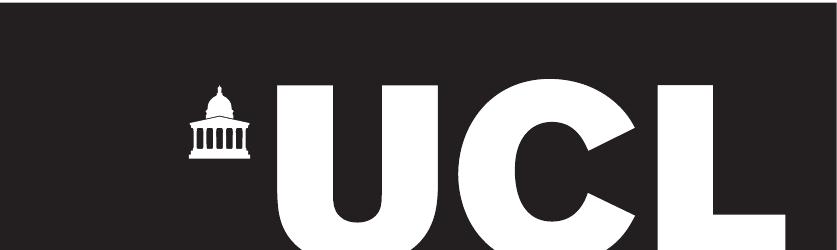}
\end{figure}
{\large\sc University College London}

\vspace*{1.2cm}
\large A Thesis Submitted for the Degree of

{\Large Doctor of Philosophy}

\monthyear\displaydate{thesis_date}
\end{center}

\newpage

\pagestyle{plain} 

\clearpage
\phantomsection
\addcontentsline{toc}{chapter}{Declaration}

\chapter*{Declaration}

I, \theauthor{}, confirm that the work presented in this thesis is my own. Where information has been derived from other sources, I confirm that this has been indicated in the thesis.  The work contains nothing which is the outcome of work done in collaboration except where specifically indicated in the text.

\IfFileExists{./user_publications.tex}{%
Parts of this thesis have been published, or submitted for publication, as follows.
\input{./user_publications.tex}
}{}
\vfill

\hfill \theauthor

\hfill \monthyear\displaydate{thesis_date}

\clearpage
\phantomsection
\addcontentsline{toc}{chapter}{Abstract}

\chapter*{Abstract}
\input{./user_abstract.tex}

\clearpage

\vspace*{\fill}
\begingroup
\begin{center}
{\em In memory of Siroun \& Alidz}
\end{center}
\endgroup
\vspace*{\fill}

%

\clearpage
\phantomsection
\addcontentsline{toc}{chapter}{Acknowledgements}

\IfFileExists{./user_acknowledgements.tex}{%
\chapter*{Acknowledgements}
\input{./user_acknowledgements.tex}
}{}

\clearpage

\vspace*{\fill}
\begingroup
\begin{center}
Ask not what your bath can do for your qubit,\\
ask what your qubit can do for your bath.
\end{center}
\endgroup
\vspace*{\fill}

\clearpage

\pagestyle{fancy}
\fancyhf{}
\fancyhead[L]{\footnotesize \leftmark}
\fancyfoot[C]{\thepage}

\dominitoc

\makenomenclature
\renewcommand{\nomlabel}[1]{\hfil #1\dotfill\hfil}
\input{./user_abbreviations.tex}

\singlespacing
\tableofcontents
\small
\listoffigures
\listoftables
\clearpage
\phantomsection
\markboth{NOMENCLATURE}{NOMENCLATURE}
\cleardoublepage
\printnomenclature[15em]
\normalsize
\doublespacing

\iftoggle{debugging}{%
  \chapter*{LAYOUT DEBUGGING}
  \layout*{}
}


%

\chapter{Introduction}
\label{chap:intro}

Decoherence is the loss of phase information encoded in a quantum system
as the system interacts with a far larger environment \citep{Zurek2003}.
A certain degree of immunity from the destructive effects of decoherence,
sometimes even achievable by directly suppressing the process, is an essential requirement
for the successful realisation of technological
devices that actively exploit quantum phenomena.
These include fault-tolerant quantum processors \citep{Shor1996} and
quantum memory \citep{Simon2010}.
Thus, it is of great practical importance to accurately predict the
timescale of decoherence -- characterised by the coherence time $T_2$ --
and also to develop methods of extending $T_2$ times.

It is also of fundamental interest to understand how decoherence due to quantum environments
differs from decoherence driven by classical noise sources. By quantum environment, we mean that
the system encoding the quantum information is situated in an environment with which it is highly correlated
or entangled, leading to significant system-environment `back-action' and environment-memory effects \citep{Breuer2002,Maniscalco2006,Mazzola2012}.
More specifically, the environment dynamics is sensitive to the state of the central spin system \citep{Yao2006,Liu2007}.
A quantum {\em spin bath} is an example of such an environment; in general, decoherence of a central spin system
coupled to a spin bath arises from many-body spin interactions inside the bath \citep{Witzel2005,Yang2008a}.
The extent to which many-body correlations play a role in quantum dynamics
is of broad interest in condensed matter physics \citep{Ma2014}.

Thus, in this thesis, we address problems of both practical and fundamental physical importance.
On one hand, understanding and reliably predicting decoherence provides a useful guide to experimentalists
working on implementing quantum technologies; also of importance is developing methods of mitigating decoherence.
On the other hand, the study of decoherence serves as a valuable tool to probe the
rich physics of many-body quantum systems and the extent to which these can be
approximated using classical models.

\section{Motivation}

Individual electronic and nuclear spins in silicon are among the prime contenders for realising
scalable quantum technologies \citep{Zwanenburg2013}.
In particular, due to its long-lived coherence and fast manipulation time, the electronic spin
of a shallow donor in silicon is a promising candidate for implementing the quantum analogue of
the classical bit -- the qubit; in a solid state system \citep{Morley2015}.

Decoherence for silicon donor qubits is often limited by the naturally-occurring
\sitn\ nuclear spin bath \citep{DeSousa2003a,DeSousa2003b}. The phosphorus donor
has been widely studied \citep{Kane1998}, but more recently, there has been growing interest in bismuth,
the deepest of the Group V donors in silicon \citep{Morley2010,George2010,Mohammady2010}.
It was proposed that decoherence would be strongly suppressed and $T_2$ significantly enhanced for the bismuth system
at particular magnetic field values termed `optimal working points' (OWPs) \citep{Mohammady2010}.
The presence of OWPs at experimentally accessible magnetic fields is due to
the strong quantum state-mixing of the donor electronic spin with the host nuclear spin,
hence the term `hybrid electron-nuclear qubit'.

The scenario of decoherence driven by a spin bath is not only limited to silicon donor qubits \citep{DeSousa2003b,Witzel2005,Abe2010,Witzel2010},
but is of considerable significance for a range of other physical implementations of quantum
information processing, including quantum dots in environments with a variety of nuclear spin impurities
\citep{DeSousa2003b,Witzel2005,Yao2006,Liu2007,Witzel2008,Weiss2012,Weiss2013,Webster2014},
and nitrogen-vacancy (NV) centres in the $^{13}$C spin bath of
diamond.\footnote{See \citep{Takahashi2008,Maze2008,Bar-Gill2012,Zhao2011b,Zhao2012a,Friedemann2012,DeLange2012}.}

At first glance, it seems impossible to accurately solve for
the closed system-bath dynamics for a bath of spins, due to the large number of spin degrees of freedom
involved. Nevertheless, cluster expansion techniques, the most general of which is the cluster correlation expansion
(CCE) \citep{Yang2008a,Yang2008b,Yang2009} have provided a solution.
In the CCE and analogous formalisms, accurate simulation of experimental
coherence decays becomes computationally tractable since
the bath is decomposed into independent contributions from many small sets or
clusters of spins \citep{DeSousa2003a,DeSousa2003b,Witzel2005,Witzel2006,Yao2006}.
Fortunately, it turns out that for most problems of practical interest
in quantum information the expansions converge for clusters containing at most
half a dozen or so spins.

The CCE has been used with considerable success to model central spin decoherence in
a variety of systems, including the the silicon spin bath, despite
the large number of bath spins involved \citep{Abe2004,Witzel2006,Abe2010,Witzel2010}.
However, in all cases prior to the work presented herein,
the CCE was implemented and applied for the central system limited to the case of a simple electronic or
nuclear
spin.\footnote{At the time of writing, and after correspondence with S.J.B. and Professor Tania Monteiro, Dr. Wen-Long Ma and Professor Ren-Bao Liu applied the CCE to
the hybrid qubit for the purpose of investigating the semi-classical nature of a nuclear spin bath
near OWPs \citep{Ma2015}. The code used in \citep{Ma2015} was checked against our code.}
Moreover, previous calculations of $T_2$ for the hybrid qubit relied on
analyses involving classical noise models \citep{George2010}. As we shall see,
these models do not give reliable $T_2$ times in all regimes.
In \citet{George2010}, {\em weak} state-mixing of the central spin in a nuclear spin bath
was investigated by simply allowing for the variation of an effective electronic gyromagnetic ratio which quantifies the response
to external classical magnetic fields.
Although this classical treatment and analogous ones are valid in some regimes,
they do not reliably describe the crucial OWP regions,
and also, cannot account for certain `forbidden transitions' which allow
fast quantum control of the hybrid qubits.
Our primary aim was to solve this problem by considering the full quantum state-mixing of
the hybrid qubit in many-body calculations of decoherence driven by a nuclear spin bath.

It is also of experimental interest to investigate how and when
the commonly applied method of dynamical decoupling \citep{Viola1998} can be combined
with operating near OWPs in order to further extend coherence times.
In dynamical decoupling, the central qubit is subjected to a
sequence of electromagnetic pulses separated in time; environmental noise
is suppressed when the frequency of the noise spectrum is less than or equal to the inverse of the pulse
spacing in the sequence.

However, interest in the silicon spin bath has recently shifted beyond
its destructive decohering role.
For example, the need remains to establish the feasibility of using nuclear
spin impurities for quantum information applications \citep{Cappellaro2009,Akhtar2012,Pla2014},
especially when the nuclear spins are in proximity to a donor.
For example, nuclear spins in the bath can act as registers storing quantum information \citep{Cappellaro2009,Waldherr2014,Taminiau2014}.

As mentioned in the opening paragraphs, understanding decoherence is not only motivated by practical reasons.
It is of fundamental importance in physics to determine the differences
between decoherence caused by classical magnetic field fluctuations
and decoherence driven by quantum baths. Also, it is interesting
to elucidate the many-body nature of a spin bath \citep{Witzel2005,Yang2008a,Witzel2010,Zhao2012a,Witzel2012,Ma2014};
are experiments fully described by only considering sets
of independent pairs of bath spins? Or are sets containing $n > 2$
bath spins required? In other words, we wish to determine
to what degree many-body system-bath correlations are important.

In many cases of central spin decoherence problems,
the dominant contribution to the combined dynamics arises from
{\em pairs} of bath spins (the so-called pair correlation) \citep{Yao2006};
in effect, from the magnetic noise due to the independent
`flip-flopping' of spin pairs. Contributions from larger clusters are
usually only needed for high accuracy \citep{Witzel2010,Witzel2012,Zhao2012a}.

\section{Outcomes}
\label{sec:outcomes}

In the work presented herein, the numerical CCE method was adapted and implemented
to include the state-mixing of the hybrid qubit (details of the code are given in \App{spindec}).
In fact, any complex multi-spin {\em system} coupled to other spin systems in the interacting
many-body bath can be simulated with our implementation.
It provided the first theoretical demonstration of suppression of spin bath decoherence near OWPs \citep{Balian2012}, subsequently
verified in experiments \citep{Wolfowicz2013,Balian2014}.
Coherence decays and $T_2$ times were obtained in
perfect agreement with experiments for forbidden transitions \citep{Morley2013}, near OWPs \citep{Balian2014},
and in the usual regimes far from OWPs \citep{Balian2014}.
As for dynamical decoupling, in order to extend the already long coherence
times near OWPs, a large number of dynamical decoupling pulses must be applied, in contrast
to the usual regimes away from OWPs \citep{Balian2015}.

As a parallel study to complement our understanding of nuclear spin bath decoherence, we
analysed the system-bath interaction for the case of the hybrid qubit,
and found clear spectroscopic signatures of the central state-mixing
and of OWPs by comparing our theory with pulsed magnetic resonance experiments \citep{Balian2012}.
These experiments resolved groups of nuclear bath spins at equivalent crystal sites and thus motivated us to
investigate the feasibility of using nuclear spins for quantum memory.
We studied the decoherence of such nuclear impurities in proximity
to a donor and found that the nuclear $T_2$ time far exceeds
that for the case of an impurity in the absence of a donor \citep{Guichard2014}.

It was already mentioned that some effects of the state-mixing
of the hybrid qubit can be adequately described using
classical noise models. In some cases, this description
is sufficient, however, we find that near the important OWP regions,
a full quantum treatment of the system-bath dynamics including the central system mixing
is necessary for obtaining the experimentally observed coherence decays \citep{Balian2014}.

We also identify qualitative differences between
the classical and quantum models, using a closed-form
analytical $T_2$ formula for the decoherence of donors in silicon
which we derive \citep{Balian2014}. The formula also
predicts $T_2$ values in excellent agreement with experiment
and numerical CCE calculations. Also, we find that classical noise models sometimes
give `false positives' for the existence of sweet-spots for decoherence
in quantum baths.

Finally, we present the only case where there is almost complete suppression of the usual pair correlations
provided that one is operating near OWPs and with sufficiently low orders of dynamical decoupling \citep{Balian2015}.
We find that clusters containing at least three bath spins (3-body clusters) are required
to recover the experimentally measured decays.\\

For the rest of this chapter, before providing an outline of the thesis,
we review the field of quantum information processing with donor qubits
in silicon, the various methods of mitigating decoherence, and the quantum
theory of spin decoherence. We start with briefly defining the general problem of decoherence and give a 
more comprehensive account for the case of spin baths in \Chap{background}.

\section{The Problem of Decoherence}

All known quantum algorithms offering speed-up over their classical counterparts rely either on quantum superposition,
quantum entanglement or both \citep{Nielsen2010}. We begin by introducing these two concepts.
In quantum computing, the classical two level system known as the `bit' is replaced by
its quantum version -- the qubit \citep{Audretsch2007,Nielsen2010}:
\begin{equation}
\ket{\psi} = \alpha \ket{0} + \beta \ket{1}.
\label{eq:qubit}
\end{equation}
\begin{figure}[h]
\centering\includegraphics[width=2.8in]{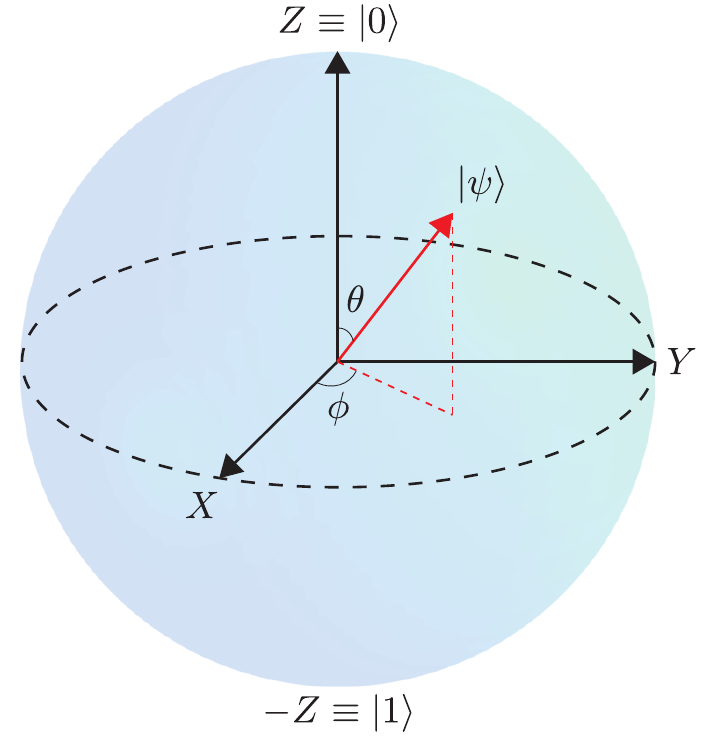}
\caption{
The states of any two-level quantum system (a qubit) can be represented as points on the surface of the Bloch sphere \citep{Nielsen2010}.
}
\label{fig:Other_bloch}
\end{figure}
The classical bit is always either in state $\ket{0}$ or $\ket{1}$, whereas the qubit can be in any general superposition $\ket{\psi}$
of the two states forming the complete orthonormal basis $\{\ket{0},\ket{1}\}$, as shown in \Eq{qubit}, where  $\alpha$ and $\beta$ are complex numbers.
As illustrated in \Fig{Other_bloch},
all normalised single qubit states ($|\alpha|^2 + |\beta|^2 = 1$) can be represented as points on the surface of a
unit sphere known as the Bloch sphere,
with the polar ($\theta$) and azimuthal ($\phi$) angles related to the amplitudes $\alpha$ and $\beta$
according to
\begin{align}
\alpha &= \cos{\left(\frac{\theta}{2} \right)},&\beta = e^{i\phi} \sin{\left( \frac{\theta}{2} \right)}.
\label{eq:blochangles}
\end{align}

Quantum entanglement is a property of multipartite quantum systems. Two qubits ($A$ and $B$)
are said to be entangled if their combined state is not separable, or equivalently, not a product state such as
$\ket{0}_A \otimes \ket{1}_B$.
For example,
\begin{equation}
\ket{\psi}_{AB} = \frac{1}{ \sqrt{2} } \left(  \ket{0}_A \otimes \ket{1}_B + \ket{1}_A \otimes \ket{0}_B  \right) 
\label{eq:entanglement}
\end{equation}
is a (maximally) entangled state and has no classical analogue.

\begin{figure}[h]
\centering\includegraphics[width=4.0in]{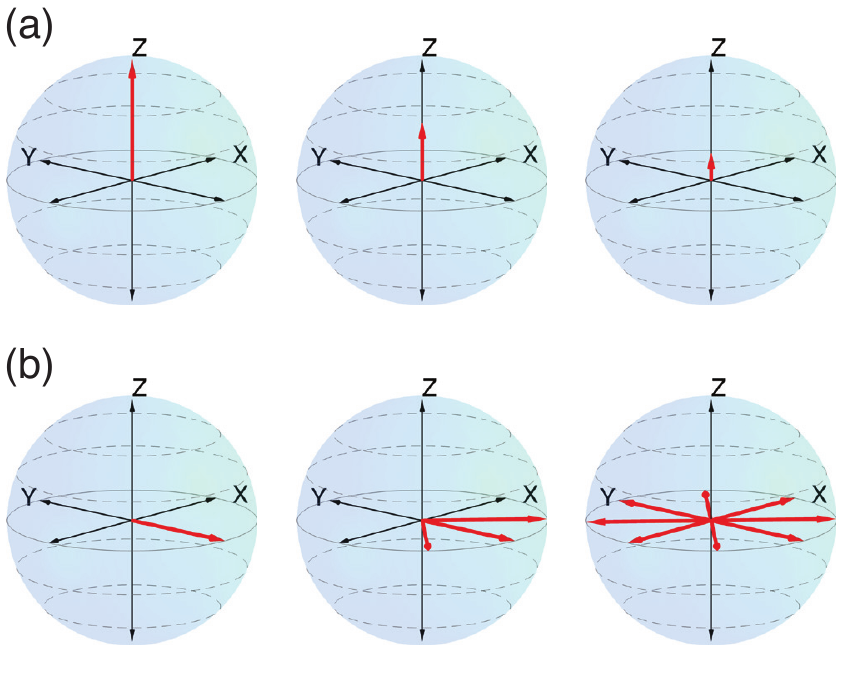}
\caption{
Loss of information encoded in a qubit can be (a) a classical or (b) a purely quantum process.
(a) Classical loss of information is decay along the $z$-axis of the Bloch sphere (\Fig{Other_bloch}) and
is characterised by $T_1$.
(b) Decoherence is the process by which the phase information is lost (on a timescale defined by $T_2$)
and can be visualized as `spreading' of
the qubit state on the equator of the Bloch sphere.
}
\label{fig:Other_t1t2}
\end{figure}

The loss of information contained in a qubit state due to its interaction with a far larger environment
is quantified by two characteristic timescales: $T_1$ and $T_2$ \citep{Schweiger2001}. These are illustrated in \Fig{Other_t1t2}.
Decoherence is the mechanism by which the quantum phase information 
is lost and is represented by $T_2$ \citep{Breuer2002}.
The single qubit state can be expressed using a density matrix which acts on the 2-dimensional Hilbert space (spanned by the basis $\{\ket{0},\ket{1}\}$):
\begin{equation}
\boldsymbol{\rho} = 
\left(
\begin{array}{cc}
|\alpha|^2 & \alpha \beta^* \\
\alpha^* \beta & |\beta|^2
\end{array}
\right).
\end{equation}
The phase information is contained in the off-diagonals of the density matrix and as the system evolves in its environment,
the decay rate of the off-diagonals is given by $1/T_2$.

Classical information is lost by `$T_1$' or `relaxation' processes which involve direct bit flips (depolarisation);
i.e.\ $\ket{0} \leftrightarrow \ket{1}$.
Unlike a typical `$T_2$ process', relaxation involves the exchange of some form of energy,
usually mediated by phonons in the bath and is manifested as time decay in the diagonals of the density matrix.
Relaxation is also a source of errors in quantum computing,
however, for our systems of interest, temperatures are low enough ($ < 15$~K) to completely
ignore $T_1$ processes, and the $T_1$ time far exceeds the coherence time $T_2$.

Given a qubit system prepared in some superposition, or two qubits in an entangled state,
it is desirable to preserve these initial states for as long as possible as the system interacts with
its often uncontrollable environment.
Ignoring relaxation, this translates to extending the coherence time $T_2$.
A wide range of quantum technologies, including fault-tolerant quantum computation, rely on
coherence times longer than the times required to
navigate the state on the Bloch sphere, and preferably as long as possible.

\section{Quantum Information Processing in Silicon}

There are two primary advantages of choosing the silicon platform for quantum information applications.
First, silicon is a good `semiconductor vacuum'; in other words, coherence times in silicon
are long compared to those in other solids.
Second, there has been decades of unprecedented technological progress in conventional
silicon electronics since the invention of the transistor around 1950; silicon is also cheap and easily available, and has good potential for scalability.
In this section, we review the recent progress in silicon quantum electronics with a particular
focus on the hybrid donor qubit.
\citet{Zwanenburg2013} provide a recent and comprehensive review of the field.

A novel proposal for quantum computing in silicon was put forward by Kane \citep{Kane1998},
in which the nuclear spins of phosphorus donors would be used as qubits,
with the donor electrons mediating qubit interactions.
The nuclei were chosen as qubits since nuclear spin coherence times
typically far exceed electronic spin coherence times.
However, the price to pay is the much longer manipulation time of nuclear spins
compared to that of electrons.
More recently, coherence times of electronic spins have caught up
and interest has shifted towards using electronic spins as long-lived
qubits with fast quantum control.

In our case, the qubit is formed out of a pair of eigenstates of the mixed system comprised of
a host nuclear spin interacting with the donor electron spin.
Hence, it only makes strict sense to talk of separate electronic and nuclear spins
of the mixed system in the high-field limit, where the interaction Hamiltonian
becomes negligible. As we shall see, the advantages of the hybrid qubit
for quantum computing arise when operating
in regimes where the electronic and nuclear spin states are strongly mixed.

Most experiments measuring coherence times are performed on
ensembles of spins. However, for quantum information applications
it is essential to build to single-atom devices.
Fortunately, in some cases, ensemble $T_2$ measurements
are in good agreement with the corresponding $T_2$
for single-atom devices.
It is important to note that there has been much progress in
recent years in single-spin detection and read-out for both quantum dots \citep{Kawakami2014,Veldhorst2014} and
donor qubits in silicon \citep{Morello2010,Pla2012,Pla2013,Muhonen2014,Pla2014}.

\subsection{Silicon Spin Bath}

In {\em natural silicon}, 4.67\% of crystal sites are occupied by the nuclear spin-1/2
$^{29}\text{Si}$ isotope, rather than the spin-0 $^{28}\text{Si}$.
It is this spin bath that provides the leading
source of decoherence in silicon at low temperatures ($T \lesssim 15$~K).\footnote{See \citep{DeSousa2003a,DeSousa2003b,Tyryshkin2003,Witzel2005,George2010,Morley2013,Balian2014,Balian2015}.}
For a donor electron spin (without OWP or dynamical decoupling enhancement),
$T_2$ is limited to a few hundred microseconds \citep{Tyryshkin2003,George2010,Morley2010}.
Similarly, a spin bath highly rich in nuclear spins exists for III-V semiconductor quantum dots,
limiting $T_2$ to less than 1~$\mu$s \citep{Koppens2008}.
This is also the case in diamond, where decoherence is instead driven by
1\% $^{13}$C spin-1/2 isotopes, resulting in $T_2 \simeq 200$~$\mu$s of an NV centre \citep{Gaebel2006}.

A successful means of controlling decoherence is to employ isotopically enriched samples,\footnote{See \citep{Abe2004,Abe2010,Tyryshkin2003,Tyryshkin2006,Steger2011,Tyryshkin2012a,Simmons2011,Steger2012,Weis2012,Wolfowicz2012,Saeedi2013,Muhonen2014}.}
whereby the percentage of $^{29}$Si impurities is reduced.
The donor electron spin in such samples can exhibit long $T_2$ times up to 20~ms \citep{Tyryshkin2012a}.
However, isotopic enrichment is a difficult process and some nuclear spins remain.
Even in isotopically enriched silicon, $T_2$ of an ensemble of donors is limited by an all-dipolar many-body spin system \citep{Witzel2010,Tyryshkin2012a,Wolfowicz2012,Witzel2012}.
Therefore, studying the nuclear bath is useful even in the case of its absence in enriched samples
as the decoherence mechanisms in that case can be analogous to the case of the nuclear bath.
Moreover, as discussed below, the nuclear impurity spins can be potentially useful for quantum memory \citep{Akhtar2012,Pla2014,Guichard2014}.

\subsection{Donors in Silicon}

A promising approach for silicon-based quantum information processing and memory
involves electronic or nuclear spins of donor atoms in silicon, which are
amenable to high fidelity manipulation by means of electron spin resonance (ESR)
and nuclear magnetic resonance (NMR), respectively.
Most studies have considered phosphorus ($^{31}\text{P}$) donors in silicon.\footnote{See \citep{Kane1998,Schofield2003,Stoneham2003,Tyryshkin2003,Fu2004,Morley2008,McCamey2010,Morello2010,Greenland2010,Simmons2011,Steger2011,Dreher2012,Fuechsle2012,Pla2012,Tyryshkin2012a,Pla2013,Saeedi2013,Muhonen2014}.}
More recently, several different groups have investigated another Group V donor, $^{209}\text{Bi}$.\footnote{See \citep{Morley2010,George2010,Mohammady2010,Sekiguchi2010,Belli2011,Weis2012,Mohammady2012,Wolfowicz2012,Balian2012,Morley2013,Wolfowicz2013,Balian2014,Balian2015}.}
The bismuth system offers new possibilities for quantum information processing. For example,
strong optical hyperpolarisation was demonstrated \citep{Morley2010,Sekiguchi2010}, allowing
for efficient initialization of the host nuclear spin. Transitions which are forbidden at
high magnetic fields and which allow for fast control of the hybrid bismuth system were predicted
\citep{Mohammady2010,Mohammady2012} and observed later in \citet{Morley2013}.
Most importantly, the bismuth donor has OWPs, where both spin bath decoherence is suppressed \citep{Balian2012,Wolfowicz2013,Balian2014}
and the donor becomes insensitive to classical field fluctuations (e.g.\ instrument noise) \citep{Mohammady2012,Wolfowicz2013}.
Near OWPs in natural silicon, the electronic spin coherence time is increased by over two orders of magnitude \citep{Wolfowicz2013,Balian2014}
from $0.5$~ms \citep{Morley2010,George2010}.
A review of donors in silicon for quantum information processing was recently conducted by \citet{Morley2015}.

\subsection{Nuclear Spin Impurities}

Interest in nuclear spin impurities has now moved far beyond their role as a destructive source of decoherence.
One application is sensing of a few nuclear $^{29}$Si spins in silicon \citep{Muller2014,Lang2015} and $^{13}$C
spins in diamond \citep{Zhao2011a,Kolkowitz2012,Zhao2012b,Kolkowitz2012,Taminiau2012,Muller2014}.\footnote{We note that another solid-state system with great potential for quantum technologies is that of nitrogen vacancy (NV) colour centres in diamond
\citep{Gaebel2006,Robledo2011,Zhao2012a,Zhao2012b,Kolkowitz2012,Bernien2013,Bar-Gill2013}.
Spin-dependent optical read-out and polarisation are possible,
electronic spin coherence times at room temperature are in the ms timescale \citep{Gaebel2006,Zhao2012a} and can reach 1~s
with dynamical decoupling at about 77~K \citep{Bar-Gill2013}. Also, entanglement between between qubits separated by three metres has been
demonstrated \citep{Bernien2013}.
A similar system which is gaining much interest and can also be operated at room temperatures is that of defects in silicon carbide \citep{Koehl2011,Yang2014}.}
Another is using the nuclear spins for quantum memory \citep{Ladd2005,Robledo2011,Akhtar2012,Pla2014,Guichard2014,Wolfowicz2015b}.
There is even a proposal for an all-silicon quantum computer using \sitn\ spins \citep{Ladd2002}.
The coherence time of a single \sitn\ nuclear spin was measured at about 6~ms \citep{Pla2014},
in good agreement with measurements in ensembles \citep{Dementyev2003}.

Recently, quantum registers were demonstrated in diamond by combining the central electronic qubit with
proximate nuclear spins \citep{Cappellaro2009,Waldherr2014,Taminiau2014}.
The decoherence mechanisms of nuclear spins proximate to a donor in silicon was studied in \citet{Guichard2014},
yielding coherence times in excellent agreement with the measured timecale of 1~s in \citet{Wolfowicz2015b}.

\section{Extending Coherence Lifetimes}

There are two distinct techniques of proven effectiveness for extending the
coherence lifetime of spin qubits without having to eliminate the nuclear spin impurities.
One is dynamical decoupling, whereby the qubit is subjected to a 
carefully timed sequence of control pulses;
the other is tuning the qubit towards OWPs,
which are sweet-spots for reduced decoherence in magnetic fields.
It is also of interest to combine the two techniques
in order to achieve the longest coherence time.

\subsection{Optimal Working Points}

In 2002, a ground-breaking study of superconducting qubits established the
usefulness of OWPs \citep{Vion2002} which were then studied as
parameter regimes where the system becomes -- to first order -- insensitive
to fluctuations of external classical magnetic fields \citep{Vion2002,Martinis2003,Makhlin2004a,Makhlin2004b,Falci2005,Ithier2005,Steger2011,Cywinski2014}.
More recently, OWPs were studied for coupled InGaAs quantum dots \citep{Weiss2012,Weiss2013}
and in systems with substantial electron-nuclear spin mixing such as the bismuth donor system
\citep{Mohammady2010,Mohammady2012,Balian2012,Wolfowicz2013,Balian2014,Balian2015}.
OWPs for the bismuth donor were investigated theoretically in
\citet{Mohammady2010,Mohammady2012,Balian2012,Balian2014} and \citet{Balian2015} and
also experimentally \citep{Wolfowicz2013,Balian2014}, extending ensemble electronic spin
$T_2$ times in natural silicon from $0.5$~ms \citep{George2010,Morley2010}
to $100$~ms \citep{Wolfowicz2013}.
OWPs have also been investigated in isotopically enriched silicon \citep{Wolfowicz2013}.
Measured electronic spin coherence times near and far from OWPs are summarized in
in \Table{owp}.

\begin{table}[b]
\centering
    \begin{tabular}{ccc}
    \hline \hline
    Sample & $T_2$ far from OWP (ms) & $T_2$ near OWP (ms)  \\ \hline
    Natural (with \sitn) & 0.5  &  100  \\
    Enriched $^{28}$Si       & 20  & 2000 \\ \hline\hline
    \end{tabular}
\caption{Measured electronic spin coherence times $T_2$, illustrating the enhancement of coherence by operating near OWPs.
The values shown are for the bismuth donor in silicon. In natural silicon and far from OWPs, coherence times were measured in
\citet{George2010} and \citet{Morley2010}. Away from OWPs in enriched samples, coherence times were measured in \citet{Wolfowicz2012} and \citet{Tyryshkin2012a},
and coherence times near OWPs were measured in \citet{Wolfowicz2013} and \citet{Balian2014}.}
\label{table:owp}
\end{table}

It is useful to note that a wide variety of important defects in the solid state possess central spin state-mixing.
These include donors in silicon \citep{Morley2015}, NV centres in diamond \citep{Zhao2012a}, transition metals
in II-VI materials \citep{George2013} and rare-earth dopants in silicates \citep{Fraval2005,Wolfowicz2015a}.
The OWPs for spin bath decoherence of donors in silicon are a direct result of this mixing.

Earlier studies of other systems which have sweet-spots for insensitivity to classical field noise \citep{Vion2002}
led to theoretical analyses of the dependence of $T_2$ on field noise \citep{Ithier2005,Martinis2003}, both at and far from the sweet-spots.
In contrast, it was only recently that a general analytical expression for $T_2$ (near and far from OWPs) was obtained
for spin systems decohered by spin baths \citep{Balian2014}.

In \citet{Mohammady2010} and \citet{Mohammady2012}, a set of minima and maxima
were found in the transition frequency-field parameter space of dipole-allowed transitions of the bismuth donor.
These $df/dB = 0$ points, later dubbed `clock transitions' \citep{Wolfowicz2013}, were first identified as OWPs: line narrowing
and reduced sensitivity to temporal and spatial noise in magnetic field $B$ over a broad
region of fields (closely related to $df/dB=0$ extrema) were found.\footnote{
Note that $df/dB=0$ points (CTs) also exist for nuclear transitions of the phosphorus donor \citep{Steger2011}.}
They were also investigated experimentally \citep{Wolfowicz2013}.
However, it was found later that the suppression of spin bath decoherence cannot be reliably explained in
terms of the classical analysis involving $df/dB$ \citep{Balian2012,Balian2014}.
In contrast, the insensitivity to classical field noise such as instrumental noise can in fact be adequately accounted for
using $df/dB$ arguments \citep{Mohammady2012,Wolfowicz2013}.\footnote{
Suppression of nuclear spin bath fluctuations can also be achieved
in self-assembled quantum dots by induced inhomogenous strain \citep{Chekhovich2015}.}

The OWPs represent a potentially complementary technique, effective for both
natural silicon and partially enriched samples. In addition, our work suggests that
OWPs may also be effective in suppressing residual effects such as
donor-donor interactions, which are the limiting decoherence mechanism in
samples with low concentrations of nuclear impurities
\citep{Mohammady2010,Witzel2010,Wolfowicz2012,Witzel2012,Wolfowicz2013}.
Finally, we note that to date, all single-atom donor experiments have used phosphorus donors,
and experiments measuring OWP coherence times have been for ensembles of bismuth donors.

\subsection{Dynamical Decoupling}

Dynamical decoupling is one of the most established methods for extending coherence times.\footnote{See \citep{Carr1954,Meiboom1958,Viola1998,Viola1999,Morton2006,Uhrig2007,Witzel2007a,Witzel2007c,Lee2008,Yang2008c,Morton2008,Biercuk2009,Preskill2011,Witzel2014b,Ma2014,Ma2015,Balian2015}.}
It involves subjecting the qubit spin
to a sequence of microwave or radio frequency pulses.
A wide variety of solid state spin qubits have been studied under dynamical decoupling control; these include
Group V donors in silicon,\footnote{See \citep{Tyryshkin2006,Tyryshkin2010,Wang2011,Pla2012,Pla2013,Wang2012a,Steger2012,Saeedi2013,Ma2014,Muhonen2014,Witzel2014b,Ma2014,Ma2015,Balian2015}.}
nitrogen vacancy centres in diamond \citep{DeLange2010,Zhao2012a,Pham2012,Wang2012b,Bar-Gill2013},
GaAs quantum dots \citep{Zhang2008}, rare-earth dopants in silicates \citep{Fraval2005,Zhong2015}, malonic acid crystals \citep{Du2009}
and adamantane \citep{Peng2011}.

The record coherence time for any spin system in a solid was measured at 6 hours in a rare-earth dopant
using dynamical decoupling \citep{Zhong2015}.
In silicon, the longest coherence time at room temperature exceeds 30 minutes with dynamical decoupling on ensembles of ionized donors in an
isotopically enriched sample \citep{Saeedi2013}.
At cryogenic temperatures, this coherence time is 3 hours.
Coherence times enhanced by dynamical decoupling in ensemble donor experiments
are summarized in \Table{ddpureensemble} (enriched silicon) and \Table{ddnatensemble} (natural silicon).
For single donor devices, the extension of $T_2$ by dynamical decoupling for enriched and natural silicon are shown
in \Table{ddpure} and \Table{ddnat} respectively.
The coherence time of the \sitn\ impurity has been extended to 25~s using dynamical decoupling \citep{Ladd2005}.
As for a \sitn\ spin in proximity to a donor, dynamical decoupling was recently applied to
extend $T_2$ from 1 to 4~s \citep{Wolfowicz2015b}.

\begin{table}[h!!]
\begin{minipage}[c]{\textwidth}
\centering
    \begin{tabular}{ccc}
    \hline\hline
    Coherence time $T_2$ & Hahn spin echo (ms) & Dynamical decoupling (ms) \\ \hline
    Electronic    & 20  & 500   \\
    Nuclear (neutral donor)  & 42,000 & 180,000  \\
    Nuclear (ionized donor)  & 27,000 & 10,800,000 (3 hours)  \\ \hline \hline
    \end{tabular}
\captionof{table}{Measured ensemble coherence times without (Hahn spin echo) and with dynamical decoupling in isotopically enriched $^{28}$Si.
Coherence times for the electron spin were measured in \citet{Tyryshkin2012a}, \citet{Wolfowicz2012} (Hahn), and \citet{Tyryshkin2012b} (dynamical decoupling).
Nuclear spin coherence times for the neutral and ionized donor were measured in \citet{Steger2012} and \citet{Saeedi2013} respectively.}
\label{table:ddpureensemble}
\vspace{0.2in}
    \begin{tabular}{ccc}
    \hline\hline
    Coherence time $T_2$ & Hahn spin echo (ms) & Dynamical decoupling (ms) \\ \hline
    Electronic    & 0.5   & 4    \\
    Nuclear (neutral donor)  & 1000 & --  \\ \hline\hline
    \end{tabular}
\captionof{table}{Measured ensemble coherence times without (Hahn spin echo) and with dynamical decoupling in natural silicon.
Nuclear spin coherence times were measured in \citet{Petersen2013}, \citet{Balian2014} and \citet{Wolfowicz2015b}.
Electronic spin coherence times were measured in \citet{Tyryshkin2006}, \citet{George2010}, \citet{Morley2010} (Hahn), and \citet{Ma2014} (dynamical decoupling).}
\label{table:ddnatensemble}
\vspace{0.2in}
    \begin{tabular}{ccc}
    \hline\hline
    Coherence time $T_2$ & Hahn spin echo (ms) & Dynamical decoupling (ms) \\ \hline
    Electronic    & 1 & 550  \\
    Nuclear (neutral donor) & 20 & 20  \\
    Nuclear (ionized donor) & 1800 & 3560  \\ \hline\hline
    \end{tabular}
\captionof{table}{Measured single-donor device coherence times without (Hahn spin echo) and with dynamical decoupling in isotopically enriched $^{28}$Si.
The coherence times were measured in \citet{Muhonen2014}. Note that for the neutral donor, the limiting decoherence mechanism was unknown,
and is likely to not be of magnetic origin.}
\label{table:ddpure}
\vspace{0.2in}
    \begin{tabular}{ccc}
    \hline\hline
    Coherence time $T_2$ & Hahn spin echo (ms) & Dynamical decoupling (ms) \\ \hline
    Electronic    & 0.2 & 0.5  \\
    Nuclear (neutral donor) & 3.5 & 7  \\
    Nuclear (ionized donor) & 60 & 132  \\ \hline\hline
    \end{tabular}
\captionof{table}{Measured single-donor device coherence times without (Hahn spin echo) and with dynamical decoupling in natural silicon.
The electronic spin coherence times were measured in \citet{Pla2012} and the nuclear ones in \citet{Pla2013}.
The 7 ms value was obtained by private communication with the lead author of \citet{Pla2013}.}
\label{table:ddnat}
\end{minipage}
\end{table}

It is also of practical importance to understand whether 
dynamical decoupling and OWP techniques may be advantageously combined for a quantum bath
of nuclear spins.
In \citet{Cywinski2014}, the two techniques were investigated for insensitivity to classical field noise.
For donor electronic qubits in silicon, it is known that due to inhomogeneous broadening from naturally-occurring $^{29}$Si spin isotopes,
there was a significant gap between the $T_2 \sim 100$~ms
in natural silicon near an OWP \citep{Wolfowicz2013,Balian2014}, and the $T_2\sim 2$~s in isotopically enriched $^{28}$Si with a low donor concentration
at the same OWP \citep{Wolfowicz2013}. Also, dynamical decoupling may be useful when it is convenient
to operate with the magnetic field close to but not exactly at the OWP.
Recently, dynamical decoupling was used to extend $T_2$ near OWPs from $100$~ms to about 1~s \citep{Balian2015,Ma2015}.

\subsection{Summary of Coherence Times}

\begin{figure}[h]
\centering\includegraphics[width=5.8in]{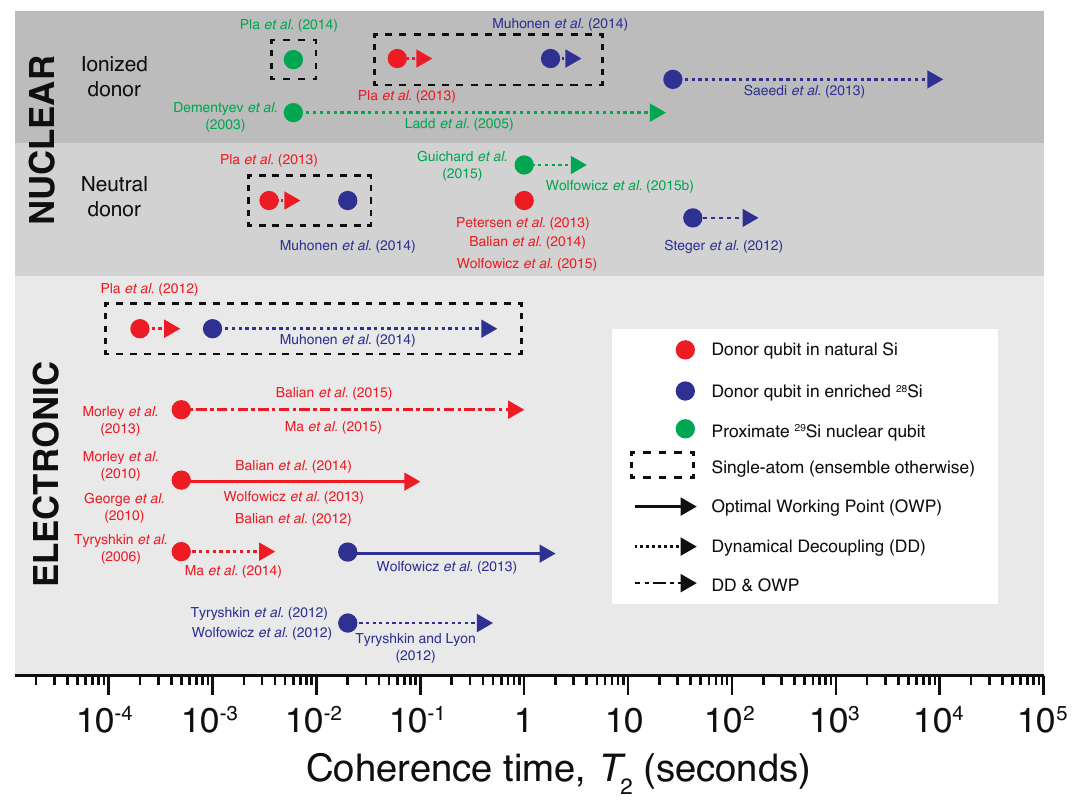}
\caption{
Cryogenic ($T \lesssim 15$~K) coherence times $T_2$ of donor qubits in natural or isotopically enriched silicon, extended by dynamical decoupling, optimal working points or by combining the two methods.
Initial (unenhanced) times are for the Hahn spin echo.
Electronic and nuclear spin coherence times correspond to microwave and radio frequency transitions respectively.
Both single-atom and ensemble measurements are shown.
Coherence times of proximate nuclear qubits are also shown for comparison.
}
\label{fig:Other_t2fig}
\end{figure}

Coherence times for hybrid donor qubits as well as proximate nuclear qubits in silicon are summarized in
\Fig{Other_t2fig}.
It is clear that the best method of enhancing coherence is by combining dynamical decoupling with operation at OWPs.
It can also be seen that ensemble measurements of electronic spin coherence times are in good agreement with measurements in bulk.
Nuclear coherence times are expected to exceed those for the electron due to the smaller gyromagnetic ratio of nuclei.
Finally, by operating near OWPs in natural silicon (even without dynamical decoupling),
coherence times can reach timescales measured in isotopically enriched silicon. We note that dynamical decoupling and operation near OWPs
have not yet been investigated in enriched silicon.

\section{Quantum Theories of Spin Decoherence}

It goes without saying that solving for the joint system-bath dynamics as a closed system is a
practically impossible task due to the large number of bath spins involved and the exponential
complexity of numerical diagonalisation of the Hamiltonian.
The framework of open quantum systems \citep{Breuer2002} offers
good approximations in many systems; however, treating cases with strong system-back action
and environment-memory remains extremely challenging within this framework.
Another inconvenience is that the usual form of Wick's theorem is not available for spin degrees of freedom,
thus preventing the use of Feynman diagrams in many-body spin dynamics \citep{Witzel2006}.
 
For a long time, theories of spin decoherence were based on stochastic models
which were phenomenological in that the noise spectrum of the environment had to be chosen.
See, for example, \citet{Klauder1962}.
The `cluster expansion' was the first ``no-free-parameter'' quantum theory of spin decoherence and was developed
much later in 2006 \citep{Witzel2005,Witzel2006}, following a study considering the
individual intra-bath interaction rates of independent pairs of bath spins \citep{DeSousa2003b}.
The `pair-correlation approximation' immediately followed \citep{Yao2006},\footnote{An early example of the role of entanglement in
decoherence can be found in \citet{Schliemann2002}.}
which coincides with the cluster expansion to second order; i.e.,
involving contributions from independent pairs of bath spins.
The `linked-cluster expansion' \citep{Saikin2007} and `disjoint cluster' \citep{Maze2008} methods followed,
and also accounted for many-body effects beyond the pair correlations in \citet{Yao2006}.
\citet{Saikin2007} also provided a simple diagrammatic representation.

The most general many-body theory is the `cluster correlation expansion' (CCE) \citep{Yang2008a,Yang2008b,Yang2009}
which we use for our numerical calculations.
At its level concerning only correlations from pairs of bath spins,
the CCE corresponds to the pair-correlation approximation \citep{Yao2006}.
The CCE is equivalent to the original cluster expansion for sufficiently large baths \citep{Witzel2006},
and is closely related to the linked cluster expansion \citep{Saikin2007}.
The theory has also been developed for calculations of ensembles of central spins \citep{Yang2009}
and also modified for the case of the central spin system in a spin bath of the same species \citep{Witzel2012}.

It remains an active area of research to identify situations
where the quantum theory of decoherence can be adequately
explained in terms of classical or semiclassical noise models \citep{Balian2014,Witzel2014a,Ma2015}.
The role of $n > 2$-body correlations has also been actively studied.
It is often the case that such many-body results offer corrections
over decoherence driven by the lowest-order contributions \citep{Witzel2010,Witzel2012,Zhao2012a,Ma2014}.
However, near an OWP, independent pair correlations are almost completely suppressed
for low to moderate orders of dynamical decoupling and clusters involving three bath spins
dominate the decoherence dynamics \citep{Balian2015}.

\section{Outline of Thesis}

The thesis is structured as follows.
In \Chap{background}, we summarize the basics of magnetic resonance
for quantum information processing and
describe in detail the theory of spin bath decoherence.
In \Chap{hybridqubit}, the hybrid qubit is introduced as the central spin system,
with emphasis on its state mixing and fast quantum control, using bismuth donors in silicon as an example.
\Chap{interaction} contains experimental measurements characterizing
the hybrid qubit-silicon spin bath interaction for Si:Bi and a theoretical spectral identification of OWPs.
Numerically calculated coherence times of the hybrid qubit in all regimes, including forbidden transitions and OWPs,
are presented and compared with experiment in \Chap{decoherence}. Also included in \Chap{decoherence} are comparisons between
quantum-bath and classical-field decoherence, the suppression of pair correlations, and many-body CCE results.
The analytical formula for coherence times of the hybrid qubit
in a nuclear spin bath is derived in \Chap{formula} and its predictions compared with experiment and numerical calculations.
In \Chap{dynamicaldecoupling}, dynamical decoupling and operation at OWPs are combined in order to maximise coherence times
of the hybrid qubit. \Chap{nucleardecoherence} comprises our study of nuclear impurity qubits proximate to the hybrid qubit in the high-field limit (phosphorus-doped silicon).
Finally, we conclude and present ideas for future work in \Chap{conc}.

\chapter[Spin Decoherence]{Spin Decoherence}
\label{chap:background}

This is the first of two consecutive chapters which primarily serve as
background for the original work presented in the thesis.
We first describe the basic principles behind the experiments
with which we compare our theories.
We proceed with the basic theory of decoherence driven by quantum spin baths.
The discussion also covers the pure dephasing approximation which is used extensively in our work.
The next section reviews the particular decoherence mechanism known as spin diffusion,
together with all the terms in the spin Hamiltonians involved.
Finally, we describe the cluster correlation expansion
which is used to solve for the many-body dynamics and hence calculate coherence times
of a central spin system under pulse control and
interacting with a spin bath with non-zero intra-bath couplings.

\section[Magnetic Resonance for Quantum Information Processing]{Magnetic Resonance for \\ Quantum Information Processing}
\label{sec:magneticresonance}

Spins in solids can be manipulated using magnetic resonance.
The basic principles of magnetic resonance and more advanced experimental techniques
are give in \citet{Schweiger2001}. In this section, we introduce the basic principles
and describe the experiments with which we motivate and compare our theories.

The energies of a spin system are quantized in a static and uniform magnetic field
of strength $B$. By applying a second, time-dependent oscillating field perpendicular to the first and
with frequency matching the energy difference between any two of the quantized energy levels
$\ket{u}$ and $\ket{l}$ ($u\equiv$`upper', $l\equiv$`lower'),
a transition $\ket{u} \to \ket{l}$ is induced between the two levels.
This is valid for any complex Hamiltonian,
provided the excitation frequency is chosen to match the frequency difference between
the desired pair of eigenstates.

If the oscillating field is applied continuously in time, the experiment
is classified as `continuous wave' (CW), otherwise the term `pulsed' is used.
We are mainly concerned with pulsed magnetic resonance for controlling the general quantum state
of a qubit in the basis $\{\ket{0} \equiv \ket{u} , \ket{1} \equiv \ket{l} \}$.
In pulsed magnetic resonance experiments, instead of driving the spin between
its upper and lower states continuously,
sequences of magnetic pulses with specific pulse durations are applied
to navigate the quantum state anywhere on the surface of the Bloch sphere (\Fig{Other_bloch}),
or equivalently, to create arbitrary superpositions of the upper and lower states
as shown in \Eq{qubit}.

Choosing the uniform magnetic field $B$ along the $z$-axis,
transition amplitudes are proportional to the matrix element $\bra{u} \hat{\sigma}^x \ket{l}$ (or $\bra{u} \hat{\sigma}^y \ket{l}$),
where $\hat{\sigma}^x$ ($\hat{\sigma}^y$) is the Pauli-$X$ (-$Y$) operator
and the $x$($y$)-axis is along the excitation field.\footnote{See \App{pauli} for the Pauli operators.}
The transition probability is proportional to the modulus squared of this amplitude.

\subsection{Electron Spin Resonance}

Magnetic resonance experiments in which the excitation frequency is in the microwave range
(i.e.\ corresponding to GHz frequencies)
are termed electron spin resonance (ESR) experiments.
This is because, typically, the spin system being addressed is
an electron spin, with an energy splitting of order GHz in a uniform magnetic field.
The so-called Zeeman interaction of a spin with a uniform magnetic field is discussed in detail in \Sec{zeeman}.
For a magnetic field of magnitude $B$ quantized along the $z$-axis, the good quantum number is the magnetic quantum number,
which for an electron spin takes one of two values $m_S = \pm \tfrac{1}{2}$ corresponding to energies proportional to $\pm \tfrac{B}{2}$ (eigenvalues of the $z$-projection of spin $\hat{S}^z$).
High fidelity single-qubit operations are possible using pulsed ESR \citep{Morton2005}.

The usual ESR selection rule is $|\Delta m_S| = 1$, implying a single flip of the electron spin.
The transition amplitude is proportional to the matrix element $\bra{u} \hat{S}^x \ket{l}$ involving only the electronic spin.
Hence, the intensity of an ESR spectral line is proportional to $|\bra{u} \hat{S}^x \ket{l}|^2$.
Finally, it is important to note that because of control under a microwave field, the time taken to manipulate
electronic spins by pulsed ESR is often on the order of nanoseconds.

\subsection{Nuclear Magnetic Resonance}

Nuclear spin energy splitting are typically of order MHz. Hence, nuclear magnetic resonance (NMR)
requires radio frequencies for resonance. The selection rule involves a single nuclear spin flip and
control is much slower than in ESR, typically on microsecond timescales.
The matrix element required for calculating transition amplitudes and thus probabilities
involves only nuclear $\hat{I}^x$ terms.

\subsection{Electron-Nuclear Double Resonance}

Electron-nuclear double resonance (ENDOR) measures
radio frequency splittings of ESR transitions.
To obtain an ENDOR spectrum, an ESR experiment is performed as a function of a
radio frequency excitation. When the radio frequency radiation is resonant with an NMR
transition, changes are seen in the ESR signal if the populations of the
relevant energy levels change. Thus, associated with an ENDOR spectrum is an ESR transition,
with each of the two levels split. If the latter splittings are of order MHz, they
are observed in the ENDOR spectrum.

\subsection{Rabi Oscillations}

\begin{figure}[h]
\centering\includegraphics[width=6in]{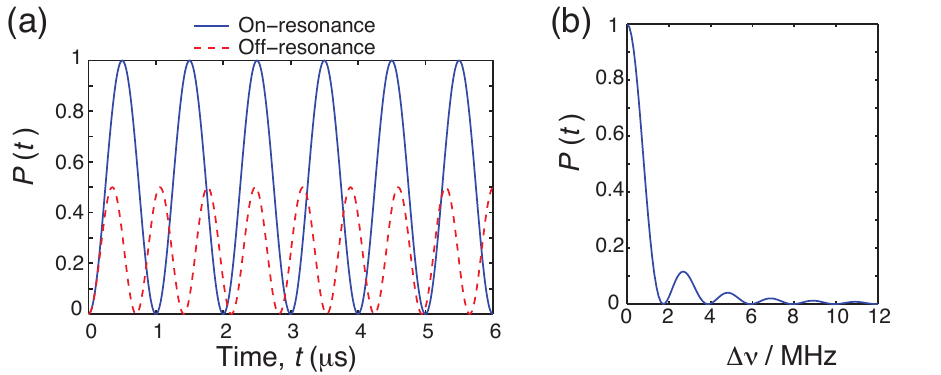}
\caption{
(a) Rabi oscillations on- and off-resonance. The vertical axis is the probability $P(t)$ for the electron spin to occupy the higher
in energy of its two energy levels and is given by \Eq{rabi}.
The blue line is for the on-resonance case when the excitation and splitting frequencies match ($\nu = \nu_{B} =9~\mathrm{GHz}$). The red line is for off-resonance:
$v_{B}$ is increased to $9.001$~GHz.
(b) The probability $P(t)$ decays rapidly as a function of the detuning $\Delta\nu \equiv |\nu-\nu_{B}|$.
Here, $t$ is fixed to half the time period of on-resonance Rabi osciallations ($0.5 ~\mu$s).
}
\label{fig:Other_rabi}
\end{figure}

Coherent quantum control is often demonstrated using a basic CW experiment
whereby the qubit is driven between the upper and lower states by continuous excitation
and in which so-called {\em Rabi oscillations} are observed.
For simplicity, we consider an electron initially in the $m_S = -1/2$ state.
When a sinusoidally oscillating excitation field of frequency $\nu$ is applied,
the probability $P(t)$ as a function of time $t$ for the electron to occupy the higher energy state labelled by $m_{S} = +1/2$
is given by
\begin{equation}
P(t) = \left( \frac{\nu_{1}}{\nu_{r}} \right)^{2} \sin^{2}\{\pi \nu_{r} \left(t-t_{0} \right) \},
\label{eq:rabi}
\end{equation}
where $\nu_{r}$ is the {\em Rabi frequency}:
\begin{equation}
\nu_{r}^{2} = \nu_{1}^2 + \left(\nu-\nu_{B} \right)^{2},
\end{equation}
$\nu_{1}$ is the amplitude of the excitation field in frequency units and
$\nu_B$ is the frequency difference between the two electronic spin states.

The probability $P(t)$ is plotted in \Fig{Other_rabi} for an excitation
field of amplitude $\nu_{1}=1~\mathrm{MHz}$. \Fig{Other_rabi}(a) compares the probability 
for the on-resonance case for which $\nu=\nu_{B}=9~\mathrm{GHz}$ and the case for off-resonance
with a finite frequency difference or detuning $\Delta\nu \equiv |\nu-\nu_{B}| = 1$~MHz.
The probability for the on-resonance case always reaches unity.
As the frequency difference $\Delta\nu \equiv |\nu-\nu_{B}|$ is increased, moving away from resonance, 
the maximum probability drops and the Rabi frequency increases.
The sharp drop in the maximum $P(t)$ as we move away from resonance is illustrated in \Fig{Other_rabi}(b).
Whether on- or off-resonance, decoherence damps Rabi oscillations.

\section{Measuring Coherence Times}

The magnetic resonance experiments in which coherence times are measured
involve special pulse sequences which we now describe. The simplest
of these is the free induction decay (FID), in which a pulse is applied to
flip spins in an initially polarised sample to create superpositions of two of the eigenstates.
The corresponding polar flip angle from either pole of the Bloch sphere to the equator gives the pulse
its name: $\pi/2$-pulse.
After the system evolves in time, the $xy$-plane or in-plane magnetisation of the sample
is measured and is proportional to the coherence.
The Hahn spin echo involves a sequence with one refocusing or $\pi$-pulse and can be classified as
the lowest order dynamical decoupling sequence which is applied to extend
coherence times before making the measurement to determine $T_2$.
Higher-order dynamical decoupling sequences apply a train
of more than one such refocusing pulses.

\subsection{Free Induction Decay}

The simplest way of measuring the spin coherence time $T_2$
is to prepare the desired state of the qubit using an excitation
pulse of the correct duration, then leave it to evolve freely in
its environment. If the qubit is initially polarised in state $\ket{u}$ or
$\ket{l}$, the normalised state after the $\pi/2$-pulse will be the superposition
\begin{equation}
\ket{\psi} = \frac{1}{\sqrt{2}}\left( \ket{u} + e^{i\phi}\ket{l} \right).
\end{equation}

After a period of free evolution of duration $t$ in the qubit's environment,
the off-diagonal of the (reduced) qubit density matrix is proportional to
\begin{equation}
\avg{\hat{\sigma}^{+}} \equiv \avg{\hat{\sigma}^x} + i\avg{\hat{\sigma}^y},
\end{equation}
in which the expectation values are evaluated in the final state immediately before measurement.
The signal in the FID experiment of a single qubit is proportional to
this quantity.
For measurements on an ensemble of $N$ qubits,
the in-plane macroscopic magnetisation vector is the measured quantity:
\begin{equation}
{\bf M}_{xy} \propto \sum_{n=1}^N  \left( \avg{\hat{\sigma}^x}_n \hat{\bf x}  + \avg{\hat{\sigma}^y}_n \hat{\bf y}  \right)
\end{equation}
for the uniform magnetic field along $\hat{\bf z}$ as usual.
Experimentally, it is possible to distinguish between the $\hat{\bf x}$ and $\hat{\bf y}$
components, but the coherence is often quoted as the magnitude of ${\bf M}_{xy}$.
Note that the polarisation of the sample, for example when making a measurement to determine $T_1$,
is related to ${\bf M}_{z}$.
Finally, even for a single qubit, experiments are often repeated
and a time average over initial states of the bath is reported.

We are mainly concerned with the single-spin FID which is the intrinsic coherence time of a single central spin system.
In measurements of ensembles of such systems, the FID $T_2$ time is usually dominated by static inhomogeneous magnetic field broadening
from multiple qubits, and is often quoted as $T_2^*$. The latter coherence time is far shorter than the intrinsic
coherence time $T_2$.

\subsection{Hahn Spin Echo}
\label{sec:hahnecho}

\begin{figure}[h]
\centering\includegraphics[width=2.0in]{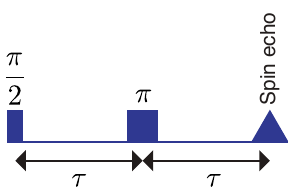}
\caption{
The Hahn echo sequence applies a $\pi/2$-pulse followed by free evolution of time duration $\tau$.
The refocusing $\pi$-pulse follows and a spin echo is observed after a further time period of $\tau$.
To measure the coherence time $T_2$, the sequence is repeatedly performed by varying $\tau$ and the
in-plane magnetisation observed (at the echo time) as a function of $t=2\tau$. 
}
\label{fig:Other_hahn}
\end{figure}

The Hahn spin echo \citep{Hahn1950} sequence removes qubit noise originating from static magnetic fields.
This includes the inhomogeneous field broadening responsible for the short $T_2^*$ ensemble coherence time described above.
Following a $\pi/2$-pulse, the qubit is allowed to evolve for some time period $\tau$ after which
a $\pi$- or refocusing pulse is applied to rotate the state by $180^{\circ}$ about an axis perpendicular to the
Bloch vector on the equator. After a further period $\tau$ of free evolution,
a {\em spin echo} is observed with intensity proportional to the coherence.
The pulse sequence is illustrated in \Fig{Other_hahn}.
To measure the coherence time, the sequence is performed for a range of increasing $\tau$,
and coherence decay is obtained as a function of $t=2\tau$.
The time taken to apply the refocusing times is much shorter than $\tau$,
and in most theoretical analyses, the refocusing pulse is assumed to be
instantaneous.

\subsection{Carr-Purcell-Meiboom-Gill Sequence}
\label{sec:cpmg}

\begin{figure}[h]
\centering\includegraphics[width=2.0in]{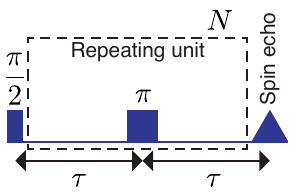}
\caption{
The CPMG dynamical decoupling sequence consists of
the initial $\pi/2$ pulse, followed by the $-\tau-\pi-\tau-$ sequence
repeated $N$ times, after which an echo is observed.
}
\label{fig:Other_cpmg}
\end{figure}

The dynamical decoupling sequence we study is the
Carr-Purcell-Meiboom-Gill (CPMG) \citep{Carr1954,Meiboom1958,Witzel2007a} sequence,
which applies a set of $N$ periodically spaced near-instantaneous refocusing pulses (CPMG$N$) as illustrated
in \Fig{Other_cpmg}. The Hahn spin echo sequence corresponds to CPMG1. The CPMG sequence is capable of
removing noise from time-fluctuating magnetic fields. The frequency of noise removed depends on $N$ or the interval between refocusing pulses $2\tau$.
The experiment to measure $T_2$ is repeated by varying $\tau$ and decoherence is observed as a function of $t = 2\tau N$.\\

Measured coherence times we compare our theories with are either for the Hahn spin echo
or higher-order CPMG sequences on an ensemble of central spin systems.
Nevertheless, we also analyse the simpler single-spin FID which is relevant
for single-spin experiments (not $T_2^*$) and compare it with the Hahn echo.
We also derive our analytical $T_2$ formula for nuclear spin diffusion for
the case of the single-spin FID, and numerically account for the effect of the Hahn spin echo.

\section{Spin Bath Decoherence}

\begin{figure}[h]
\centering\includegraphics[width=3.0in]{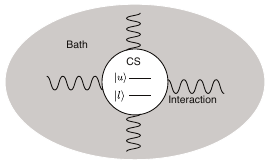}
\caption{
Central spin decoherence of a system interacting with a far larger environment or bath.
The central system (CS) need not be a spin-1/2 with two energy levels and in general
is formed out of two eigenstates of a complex spin Hamiltonian (i.e.\ a transition between upper and lower levels $\ket{u} \to \ket{l}$).
}
\label{fig:Other_csd}
\end{figure}

The decay in coherence of a central spin system interacting with a bath of other spins
can be related to its entanglement with the bath \citep{Breuer2002,Witzel2005,Yao2006,Liu2007,Yang2008a}.
In this section, we describe the problem of {\em central spin decoherence},
as illustrated in \Fig{Other_csd}.

Consider closed system-bath dynamics governed by total Hamiltonian
\begin{equation}
\hat{H}_{\text{tot}} =
          \hat{H}_{\text{CS}} + \hat{H}_{\text{int}} + \hat{H}_{\text{bath}}.
\label{eq:totalH}
\end{equation}
Here, $\hat{H}_{\text{CS}}$ denotes the central spin (or qubit) Hamiltonian completely isolated from the environment.
All system-bath interaction terms are included in $\hat{H}_{\text{int}}$,
while the bath degrees of freedom, including intra-bath couplings (essential for decoherence)
are contained in $\hat{H}_{\text{bath}}$ (\Fig{Other_csd}).

Suppose that at some initial time $t_0$ the central system's state is prepared in a coherent superposition of a pair of its energy eigenstates ($\ket{u}$ and $\ket{l}$).
For example, this is the case after applying a $\pi/2$-pulse in a FID or Hahn spin echo experiment.
Immediately after preparing the state, we assume that the qubit and bath are in a product state (i.e.\ unentangled).
The combined initial system-bath state is thus
\begin{equation}
\ket{\psi(t_0)} = \frac{1}{\sqrt{2}}
\left( \ket{u} + \ket{l} \right) \otimes
\ket{\mathcal{B}(t_0)},
\end{equation}
where the initial bath state is $\ket{\mathcal{B}(t_0)}$.

Now suppose that the system evolves under $\hat{H}_\text{tot}$ (\Eq{totalH}) until time $t$ according to the time-dependent Schr\"{o}dinger equation
\begin{equation}
i \frac{d }{dt} \ket{\psi(t-t_0)} = \hat{H}_{\text{tot}} \ket{\psi(t-t_0)} 
\end{equation}
with $\hbar = 1$.
The formal solution is $\ket{\psi(t-t_0)} = \hat{U}(t-t_0) \ket{\psi(t_0)}$ with the unitary free evolution operator given by
\begin{equation}
\hat{U}(t-t_0)= 
e^{-i\hat{H}_\text{tot}(t-t_0)},
\end{equation}
for time-independent Hamiltonians.
The unitary evolution operator $\hat{U}$ may also represent a dynamical decoupling sequence. For example, for the Hahn spin echo we have
\begin{equation}
\hat{U}(t)= 
e^{-i\hat{H}_\text{tot}t/2}
\hat{\Pi}
e^{-i\hat{H}_\text{tot}t/2}.
\end{equation}
where for simplicity we have chosen $t_0 = 0$.
The free evolution, can be written as follows:
\begin{equation}
e^{-i\hat{H}_\text{tot}t/2} = \sum_{n} \ket{\phi_n} e^{-i E_{n} t/2} \bra{ \phi_n },
\end{equation}
after performing the eigendecomposition of the Hamiltonian $\hat{H}_\text{tot}$ to obtain the energy eigenbasis $\{\ket{\phi_n}\}$ with eigenvalues $\{E_{n}\}$.
Assuming the time taken for the $\pi$-pulse is much shorter than $t$,
the $\pi$-pulse operator is given by
\begin{equation}
\hat{\Pi} = \left(\hat{\sigma}^x + \sum_{n \neq u,l} \outerprod{n}{n} \right)  \otimes \hat{\mathds{1}}_{\mathcal{B}},
\end{equation}
\noindent where $\hat{\mathds{1}}_{\mathcal{B}}$ denotes the bath identity and $\hat{\sigma}^x$ is the Pauli-$X$ gate $\outerprod{u}{l} + \outerprod{l}{u}$.

After evolution to time $t$, the central system and bath states are in general entangled.
Writing the combined system-bath density operator $\hat{\rho}(t) = \ket{\psi(t)} \bra{\psi(t)}$,
the coherence of the system is characterised by the off-diagonal of its reduced density matrix:
\begin{equation}
\rho_\text{CS}^{+-}(t) = \bra{u} \text{Tr}_\mathcal{B} \left[ \hat{\rho}(t) \right] \ket{l} = \bra{u} \left( \sum_k \bra{k} \hat{\rho}(t) \ket{k} \right)  \ket{l},
\end{equation}
which is obtained by tracing out the bath degrees of freedom; the $\{\ket{k}\}$ here form an orthonormal basis for the bath.
The quantity of interest is the off-diagonal
\begin{equation}
\mathcal{L}(t) = \rho_\text{CS}^{+-}(t) / \rho_\text{CS}^{+-}(0),
\end{equation}
normalised such that $\mathcal{L}(t = t_0 = 0) \equiv 1$. For this initial time, the phase information contained in the initial state of the system is fully known.
The normalization to unity is important for the formulation of the cluster correlation expansion as described in \Sec{cce}.
The coherence $\mathcal{L}(t)$ is proportional to $\avg{\hat{\sigma}^{\pm}} \equiv \avg{\hat{\sigma}^x} \pm i\avg{\hat{\sigma}^y}$.
The density operator is Hermitian, so it does not matter which off-diagonal
($\rho_\text{CS}^{+-} \propto \avg{\sigma^+}$ or $\rho_\text{CS}^{-+} \propto \avg{\sigma^-}$) we consider.
Importantly, $|\mathcal{L}(t)|$ is proportional to the signal in an experiment probing the transverse magnetisation.

\subsection{Initial Bath State}
\label{sec:bathstate}

Since nuclear bath energies in a magnetic field $B$ typically exceed intra-bath interaction strengths,
we assume a thermal initial state of the bath (unentangled) \citep{WitzelThesis}:
\begin{equation}
\hat{\rho}_\mathcal{B}(t_0) = \sum_{n} P_n \ket{\mathcal{B}_n(t_0)} \bra{\mathcal{B}_n(t_0)}  \simeq
\bigotimes_n \left( \sum_{m} p_{nm} \ket{b_{nm}} \bra{b_{nm}} \right),
\end{equation}
where $\ket{b_{nm}}$ are eigenstates of the bath Hamiltonian excluding intra-bath interaction terms $\hat{H}^0_{\text{bath}}$.
For thermal equilibrium,
\begin{equation}
\hat{\rho}_\mathcal{B}(t_0) \approx \exp{ \left[ - \frac{\hat{H}^0_{\text{bath}}}{k_B T} \right] },
\end{equation}
where $k_B$ is the Boltzmann constant. Assuming the high-$T$ limit, which is valid for the energies of $\hat{H}^0_{\text{bath}}$ and the temperatures
we consider, the initial bath density matrix reduces to the identity; i.e.\ for a given $n$, the states $\ket{b_{nm}}$ occur with equal probability $p_{nm}$.

We note that for small baths, whether for ensemble measurements or single spins,
the coherence can be sensitive to sampling from the initial ensemble \citep{Yang2009}
and the use of a randomly chosen pure state of the bath is not valid.
However, the baths we consider in general consists of a very large number of spins ($\gg 10^4$) and for such
sufficiently large baths, it is valid to consider a pure initial bath states
chosen at random with equal probability amongst the energy eigenstates of $\hat{H}^0_{\text{bath}}$.
Nevertheless, we consider the case of averaging the complex coherence over
such random initial pure states, both for time-averaged measurements and measurements on ensembles of qubits. In the latter case,
not only the bath states vary for a single realisation, but also for bath spin positions.

\subsection{Pure Dephasing}
\label{sec:puredephasing}

If we assume that during the combined system-bath free evolution the
states of the CS remain unchanged, the final state can be written as
\begin{equation}
\ket{\psi(t)}= \frac{1}{\sqrt{2}}
\left( e^{-iE_ut} \ket{u} \otimes \ket{\mathcal{B}_u(t)} + e^{-iE_lt} \ket{l} \otimes \ket{\mathcal{B}_l(t)} \right).
\label{eq:tangle}
\end{equation}
Here, it is clear that the central system and bath are in general entangled and that the bath evolves differently
$\ket{\mathcal{B}_{u/l}(t)}$ depending on the state of the system $\ket{u}/\ket{l}$. The phases $e^{-iE_{u/l}t}$
are physically not important as they disappear when we take the modulus of $\mathcal{L}(t)$.

It is easy to show that tracing over the bath and taking the off-diagonal of the resulting reduced density
matrix is equivalent to evaluating the overlap between the bath states correlated with the upper and lower
system states:
\begin{equation}
\mathcal{L}(t) \propto \langle\mathcal{B}_u(t)|\mathcal{B}_l(t)\rangle = \langle\mathcal{B}(0) | \hat{T}_u^\dagger \hat{T}_l | \mathcal{B}(0)\rangle.
\label{eq:bathoverlap}
\end{equation}
The measured temporal coherence decays can be simulated if one can 
accurately calculate this overlap. Even for extremely large baths, the initial bath states
are the usual thermal states. Thus, the challenge is to evaluate the unitaries $\hat{T}_u$ and $\hat{T}_l$.

For our systems of interest, the pure dephasing model (i.e.\ keeping only interaction and bath terms which don't depolarise the states of the central system)
is justified since the energies of the system dominate over typical system-bath and intra-bath couplings.
Note that in contrast to the case of an electronic spin-$1/2$ qubit, for the mixed spin qubits which we describe in \Chap{hybridqubit},
if the coherence is evaluated by directly evolving the total Hamiltonian in \Eq{totalH},
the depolarising terms are not just those involving $\hat{S}^x$ and $\hat{S}^y$,
but also $\hat{S}^z$.

\section{Spin Diffusion}

\begin{figure}[h]
\centering\includegraphics[width=3.0in]{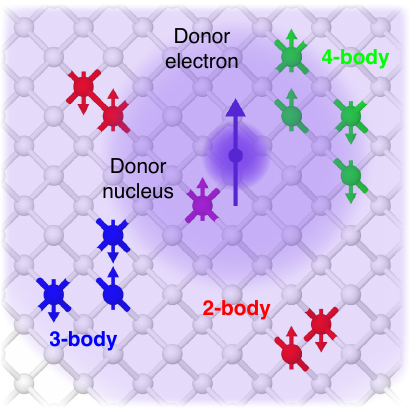}
\caption{Coherences of the central electronic spins are dephased primarily by a surrounding 
quantum bath of clusters of 2, 3, 4 or more nuclear spin impurities (for natural silicon, pictured) or other donors (for isotopically enriched silicon).
Figure adapted from \citet{Balian2015}.
}
\label{fig:Balian2015_correlations}
\end{figure}

We now introduce the mechanism which dominates decoherence in silicon and also in diamond at cryogenic temperatures (i.e. assuming $T$ is small enough
such that $T_1 \gg T_2$, which is satisfied when $T < 15$~K or so).
{\em Nuclear spin diffusion} is the process by which a central electronic spin in a solid decoheres due to a nuclear spin bath \citep{DeSousa2003b,Witzel2005,Yao2006,Witzel2006}.
The term spectral diffusion is also used to describe the same process. The problem can be adapted to cases when the central spin and the bath are of the same species and the
underlying physics of the process is the same \citep{Witzel2010,Witzel2012}. The following discussion concerns nuclear spin diffusion which is often the case encountered in our decoherence studies,
and we use the term {\em spin diffusion} to refer to the general problem regardless of the nature of the central system or spin bath.

In nuclear spin diffusion, bath spins are coupled via the magnetic dipole interaction, for example, between \sitn\ nuclei in silicon or
$^{13}$C nuclei in diamond, both spin-1/2 species.
The scenario of a donor qubit in silicon is illustrated in \Fig{Balian2015_correlations}.
In natural silicon, the fractional abundance of \sitn\ is $f_\text{natSi} = 0.0467$ \citep{DeSousa2003b}.
The bath Hamiltonian is a sum over nuclear Zeeman and dipolar Hamiltonians
\begin{eqnarray}
\hat{H}_{\text{bath}} &=& \hat{H}_\text{D} + \hat{H}_\text{NZ}, \nonumber\\
\hat{H}_\text{NZ}     &=& \sum_{n} \hat{H}_\text{NZ}^{(n)}, \quad \hat{H}_\text{NZ}^{(n)} = \gamma_n B \hat{I}_n^z, \nonumber\\
\hat{H}_\text{D}      &=& \sum_{n < m} \hat{H}_\text{D}^{(nm)}, \quad \hat{H}_\text{D}^{(nm)} = \hat{{\bf I}}_{n} \cdot \mathcal{D}({\bf r}_{nm}) \cdot \hat{{\bf I}}_{m},
\label{eq:nsdbathH}
\end{eqnarray}
where the bath spins $\hat{{\bf I}}_{n}$ have nuclear gyromagnetic ratios $\gamma_n$, $\mathcal{D}$ is the dipolar tensor
and ${\bf r}_{nm}$ is the separation vector between localized nuclear spins labelled $n$ and $m$.
The Zeeman and dipolar interactions are discussed in \Sec{zeeman} and \Sec{dipolar} respectively.
The central spin system interacts with the bath spins primarily through the electron-nuclear hyperfine interaction:
\begin{equation}
\hat{H}_{\text{int}} = \hat{H}_{\text{HF}} = \sum_n \hat{H}_{\text{HF}}^{(n)}, \quad \hat{H}_{\text{HF}}^{(n)}
= \hat{{\bf S}} \cdot \mathcal{J}({\bf r}_{en}) \cdot \hat{{\bf I}}_{n},
\end{equation}
where $\hat{{\bf S}}$ represents the central electron, $\mathcal{J}$ is the hyperfine tensor described in \Sec{hyperfine},
and ${\bf r}_{en}$ is the electron-nuclear separation.
Although anisotropic terms in the interaction Hamiltonian modulate the coherence \citep{Witzel2007b}, they have little effect on the $T_2$ timescale.
Therefore, the isotropic hyperfine interaction can be assumed:
\begin{equation}
\hat{H}_{\text{int}} \simeq  \sum_n J_{\text{F}} ( {\bf r}_n ) \hat{{\bf S}} \cdot  \hat{{\bf I}}_{n} = 
							 \sum_n J_{\text{F}} ( {\bf r}_n ) \left[\hat{S}^z \hat{I}_n^z + \frac{1}{2} (\hat{S}^+ \hat{I}_n^- + \hat{S}^- \hat{I}_n^+)\right],
\label{eq:nsdintH}
\end{equation}
where $J_{\text{F}}$ is the strength of the Fermi contact interaction which depends on the nuclear position ${\bf r}_n$,\footnote{The origin of the coordinate system is taken as the point when the electron-nuclear separation ${\bf r}_{en}$ is zero.}
and is described in \Sec{hyperfine}.
We refer to any terms involving a product of two $z$-spin projections as an `Ising term', such as the first term in the square brackets
on the R.H.S. of \Eq{nsdintH}.

Due to the disparity between the nuclear Zeeman energies and typical dipolar couplings,
the dipolar interaction is usually assumed to be secular as described in \Sec{dipolar}.
The secular dipolar interaction includes only terms containing $\hat{I}^z_n \hat{I}^z_m$ and $\hat{I}^+_n \hat{I}^-_m + \hat{I}^-_n \hat{I}^+_m$.
The latter term is why the phrase `flip-flopping' spins is used to describe such bath dynamics.
Nuclear spin diffusion with an Ising-only hyperfine interaction is further referred to as `indirect flip-flops', to distinguish it from
the $T_1$-like process of `direct flip-flops' which involves the flip-flop of a bath spin with the central spin \citep{Tyryshkin2012a}.

Most of our results are presented for indirect flip-flops in a nuclear spin bath.
However, these results are easily generalizable, especially in the context of mitigating decoherence
driven by indirect flip-flops in a bath which has the same spin species as the central spin system;
for example, in isotopically enriched samples where the abundance of \sitn\ is reduced \citep{Witzel2010,Witzel2012,Tyryshkin2012a}.

\subsection{Zeeman Interaction}
\label{sec:zeeman}

For simplicity, we begin by describing the Zeeman interaction of a magnetic field with a single electron spin in vacuum \citep{Schweiger2001,Weil2007}.
Consider an electron in a static and uniform magnetic field $\boldsymbol{B}$ which we choose along the $z$-axis.
Associated with the electron is the intrinsic angular momentum $\hbar\boldsymbol{S}$ called spin. Due to spin and
the non-zero electronic charge $e$, the electron possesses a non-zero magnetic dipole moment $\boldsymbol{\mu}$ given by
\begin{equation}
\boldsymbol{\mu} = \frac{e}{2m_{e}}\hbar\boldsymbol{S},
\end{equation}
where $m_{e}$ is the electronic mass and $\hbar\boldsymbol{S}$ is the spin
angular momentum vector.
The component of $\hbar\boldsymbol{S}$ along $z$ is quantized: it can take either one of
the values $m_{S}\hbar = \pm\frac{1}{2}\hbar$. Thus, the component of $\boldsymbol{\mu}$ along $z$ is
\begin{equation}
\mu_{z} = \gamma_{e} \hbar m_{S},
\label{eq:muz}
\end{equation}
where the constant of proportionality $\gamma_{e} = e / 2 m_{e}$ is the electron {\em gyromagnetic ratio}. Defining the
Bohr magneton as $\beta_{e}\equiv|e|\hbar / 2m_{e}$ and including the $g$-factor for the free electron
$g_{e}$ needed to relate its magnetic moment to an angular momentum in quantum theory, \Eq{muz} becomes
\begin{equation}
\mu_{z} = - g_{e} \beta_{e} m_{S}
\end{equation}
\noindent where the free electron $g$-factor is measured to be $g_{e} = 2.0023193043617(15)$ and is well predicted
by quantum electrodynamics. Note that this value is for the electron in vacuum and in a solid $g_e$ in general is different.

The energy $U$ of a magnetic dipole moment $\boldsymbol{\mu}$ in a magnetic field $\boldsymbol{B}$ is given by,
\begin{equation}
U = - \boldsymbol{\mu} \cdot \boldsymbol{B}
\end{equation}
and for a single electron, this becomes
\begin{equation}
U = - \mu_{z} B = g_{e} \beta_{e} B m_{S}.
\end{equation}
The two levels, labelled by $m_{S} = \pm1/2$, are referred to as the electronic Zeeman energies, and the energy splitting field $\boldsymbol{B}$ is sometimes
called the Zeeman field.
For a transition between the two states, the frequency $\nu$ of an excitation field $\boldsymbol{B}_{1}$ inducing the transition must match the
energy difference $\Delta U$ between the two states (i.e.\ $h \nu = \Delta U = g_{e} \beta_{e} B$).
By treating the electron as a classical magnetic dipole moment in a static magnetic field, it can be shown that
the electron precesses about the field with frequency $\nu_{B}$, a process known as Larmor precession:
\begin{equation}
\nu_{B} = g_{e} \beta_{e} B / h.
\end{equation}

The Zeeman Hamiltonian describing the response of a general spin $\hat{{\bf S}}$ in ${\bf B}$ is written
\begin{equation}
\hat{H}_\text{Z} = \gamma {\bf B} \cdot \hat{{\bf S}},
\end{equation}
with gyromagnetic ratio $\gamma$ and we have set $\hbar = 1$.
At this point, we note that all our energies are in angular frequency units of rad~s$^{-1}$ ($\hat{H}\rightarrow\frac{\hat{H}}{\hbar}$)
unless otherwise indicated. In some cases, angular frequency units are scaled by $1/2\pi$,
and this is indicated using frequency units Hz.

Choosing ${\bf B}$ along the $z$-axis, $\hat{H}_\text{Z} = \gamma B  \hat{S}^z$.
For spin-1/2 species, $\hat{\bf S} = \hat{\boldsymbol{\sigma}}/2$,
where $\hat{\boldsymbol\sigma} \equiv (\hat{\sigma}^x,\hat{\sigma}^y,\hat{\sigma}^z )$ is the three-vector of Pauli operators (\App{pauli}).
The gyromagnetic ratios for a donor electron in silicon and a \sitn\ impurity, both spin-1/2 species,
are given in \Table{gammas}.
The nuclear gyromagnetic ratios of Group V donors in silicon are given in \Table{donors}.
The sign of $\gamma$ determines whether the classical magnetic moment associated with the spin precesses
in the clockwise or anticlockwise direction about the magnetic field.

\begin{table}[h]
\centering
    \begin{tabular}{cc}
    \hline\hline
    Spin species & $\gamma$ (M~rad~s$^{-1}$~T$^{-1}$) \\ \hline
    Electron in silicon & $+1.7591\times10^{5}$ \\
    \sitn\ nucleus & $+53.1903$ \\ \hline\hline
    \end{tabular}
\caption{Gyromagnetic ratios $\gamma$ for a donor electron in silicon \citep{Feher1959} and a \sitn\ nucleus \citep{Stone2005}.}
\label{table:gammas}
\end{table}

\subsection{Dipolar Interaction}
\label{sec:dipolar}

\begin{figure}[t]
\centering\includegraphics[width=4in]{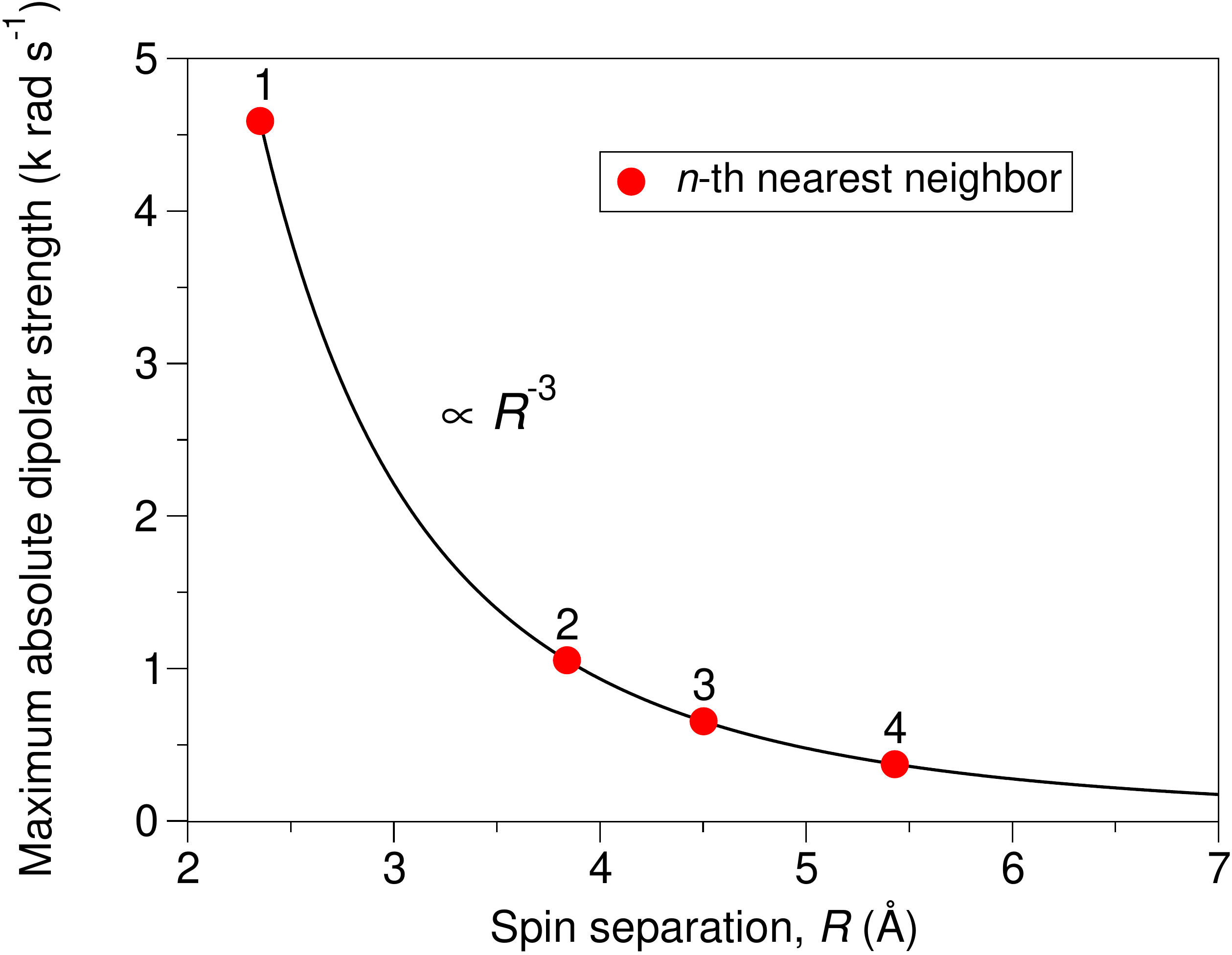}
\caption{
The magnitude of the secular dipolar coupling, shown here between two \sitn\ nuclei with gyromagnetic ratio
$\gamma = 53.1903$~M~rad~s$^{-1}$~T$^{-1}$,
falls as the cube of the separation $R$ between the interacting spins.
Here, the crystal orientation is such that the direction of the magnetic field
is parallel to the line connecting the two spins (i.e.\ $\theta=0$).
The red dots with integer labels mark the strengths for the
1st to 4th nearest neighbor distances in the silicon crystal structure (\App{silicon}).
The nearest neighbor distances are $\tfrac{\sqrt{3}}{4}a_0$,$\tfrac{\sqrt{2}}{2}a_0$,$\tfrac{\sqrt{11}}{4}a_0$ and $a_0$ respectively,
with lattice parameter $a_0 = 5.43$~\AA.
}
\label{fig:Other_dipolar}
\end{figure}

The magnetic dipole interaction \citep{Schweiger2001} between two localized spins
$\hat{\bf I}_n$ and $\hat{\bf I}_m$ with gyromagnetic ratios $\gamma_n$ and $\gamma_m$ is
\begin{equation}
\hat{H}_{\text{D}}^{(nm)}= \hat{{\bf I}}_{n} \cdot  \mathcal{D} ({\bf r}_{nm}) \cdot \hat{{\bf I}}_{m},
\label{eq:dipolar}
\end{equation}
where ${\bf r}_{nm}$ denotes the relative position vector of the two spins
and the components of the dipolar tensor are given by
\begin{equation}
\mathcal{D}_{i j}({\bf r}_{nm}) = \frac{\mu_0}{4\pi r_{nm}^3} \gamma_n \gamma_m \hbar
\left(\delta_{i j} - 3 \frac{r_{nm}^{(i)} r_{nm}^{(j)}}{r_{nm}^2}\right),
\end{equation}
where $\mu_0 = 4\pi \times 10^{-7}$~NA$^{-2}$ is the permeability of free space, $\delta_{ij}$ the Kronecker delta
and $i,j = x,y,z$.

In a sufficiently strong and uniform magnetic field, the dipolar interaction
can be simplified by keeping only secular or energy conserving terms:
\begin{equation}
\hat{H}_{\text{D}}^{(nm)} \simeq C_{nm} \hat{I}^z_n \hat{I}^z_m - \frac{C_{nm}}{4} \left( 
\hat{I}^+_n \hat{I}^-_m + \hat{I}^-_n \hat{I}^+_m \right),
\label{eq:secdipolar}
\end{equation}
with strength $C_{nm}$ given by:
\begin{equation}
C_{nm} = \frac{\mu_0}{4\pi} \gamma_n \gamma_m \hbar \frac{(1 - 3 \cos^2{\theta_{nm}})}{r_{nm}^3}.
\end{equation}
Here, $\theta_{nm}$ is the angle between the line connecting the spins and the $z$-axis.

For coupling among nuclear spins in the silicon spin bath, the dipolar strength is at most
a few k~rad~s$^{-1}$. Since the gyromagnetic ratios of these nuclei are of order tens of M~rad~s$^{-1}$~T$^{-1}$,
the secular approximation is justified for magnetic field strengths as weak as about 100~mT \citep{Witzel2008}.

In order to illustrate the radial dependence of the dipolar interaction,
the absolute of the maximum strength (i.e.\ $|C(\theta=0)|(R)$)
is plotted in \Fig{Other_dipolar} as a function of separation distance $R$ between a pair of \sitn\ nuclei.
For \sitn, the value of $\gamma_n$ is given in \Table{gammas}.

\subsection{Hyperfine Interaction}
\label{sec:hyperfine}

\begin{figure}[h]
\centering\includegraphics[width=4.5in]{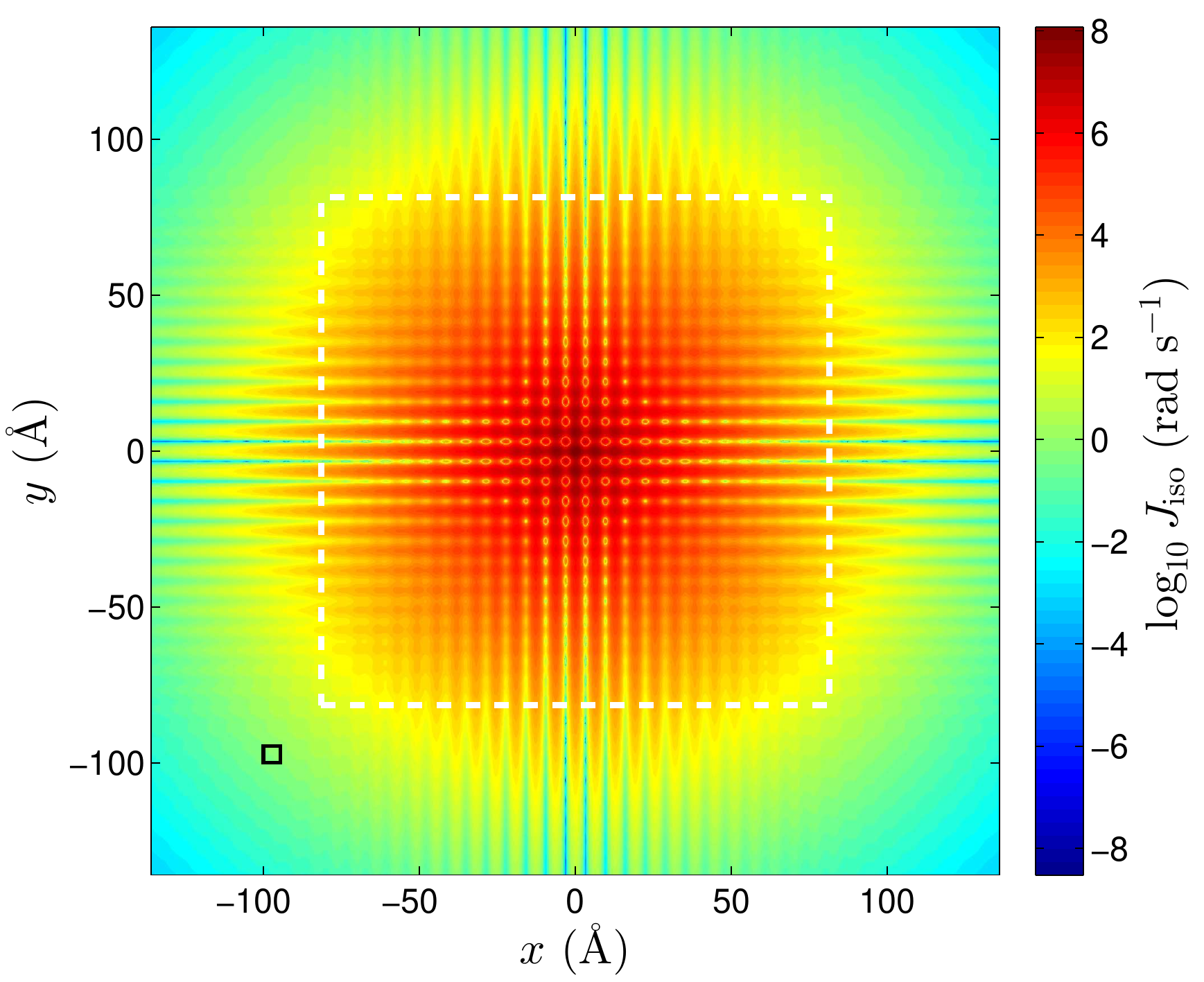}
\caption{
Isotropic hyperfine couplings (Fermi contact only) for a bismuth donor electron
in silicon interacting with a \sitn\ impurity, as a function of distance
between the donor and the impurity. The donor is situated at the origin and the $z=0$
plane is shown. Including spins in the white box is enough for obtaining convergent coherence decays
for nuclear spin diffusion using the cluster correlation expansion.
The black square (of side length $5.43$~\AA) represents the conventional
cubic cell of the diamond cubic crystal structure.
}
\label{fig:Other_hyperfine}
\end{figure}

The magnetic interaction between an electron $\hat{\bf S}$ and localized nuclei $\hat{\bf I}_n$ is essentially
given by \Eq{dipolar}. However, due to the spatial extent of the electron wavefunction in a solid, such as in the case of a donor electron,
we evaluate
\begin{equation}
\hat{H}_{\text{HF}}^{(n)} = \bra{ \Psi( {\bf r}_e  )}  \hat{{\bf I}}_n \cdot  \mathcal{D} ({\bf r}_{en}) \cdot \hat{{\bf S}}   \ket{ \Psi( {\bf r}_e  )}
= \int d^3r_{e} | \Psi({\bf r}_e) |^2 \hat{{\bf I}}_n \cdot  \mathcal{D} ({\bf r}_{en}) \cdot \hat{{\bf S}},
\end{equation}
where ${\bf r}_{en} = {\bf r}_e - {\bf r}_n$ is the electron-nuclear separation \citep{DeSousa2003b}.
In other words, the electron's position is taken into account by evaluating the expectation value
of the interaction in the electron wavefunction $\Psi({\bf r}_e)$ in real space.
This integral has a singularity at ${\bf r}_{en} = {\bf 0}$, or when the electron is at the nuclear site.
The singularity gives rise to the
{\em Fermi contact} interaction:
\begin{equation}
J_\text{F}({\bf r}_n) = \frac{2}{3} \gamma_e \gamma_n \hbar \mu_0  \left| \Psi({\bf r}_n) \right|^2,
\end{equation}
where $\gamma_e$ ($\gamma_n$) is the electronic (nuclear) gyromagnetic ratio.
Importantly, the Fermi contact interaction only contains the nuclear position ${\bf r}_n$ and the origin of the coordinate system is at ${\bf r}_{en} = {\bf 0}$.
The full interaction is expressed using the hyperfine tensor $\mathcal{J}$ which is decomposed into
the Fermi contact and a residual dipolar interaction:
\begin{equation}
\hat{H}_{\text{HF}}^{(n)} = \hat{{\bf I}}_n \cdot  \mathcal{J} ({\bf r}_{en}) \cdot \hat{{\bf S}}
 = J_\text{F}({\bf r}_n) \delta_{ij}
+ \left\langle \Psi({\bf r }_e) \left|  \mathcal{D}_{ij}({\bf r}_{n})  \right| \Psi({\bf r }_e) \right\rangle.
\label{eq:hyperfine}
\end{equation}
The Fermi interaction is isotropic and the anisotropic dipolar part is effective for nuclei at sufficiently large distances from the origin, where
the electron can be assumed localized.
Due to the large mismatch between electronic and 
nuclear gyromagnetic ratios and a sufficiently
strong magnetic field, the hyperfine interaction above can be written in secular form and keeping only an Ising term \citep{DeSousa2003b}:
\begin{equation}
\hat{H}_{\text{HF}}^{(n)} \simeq \left[
J_\text{F}({\bf r}_n) - 
\frac{\mu_0}{4\pi} \gamma_n \gamma_e \hbar \frac{(1 - 3 \cos^2{\theta_{n}})}{r_{n}^3}
\Theta(r_n - r_0) \right]
\hat{S}^z \hat{I}_n^z,
\label{eq:sechyperfine}
\end{equation}
where $r_n \equiv |{\bf r}_n|$ and $\Theta(r_n - r_0)$ is the Heaviside step function; i.e.\
the electron-nuclear residual dipolar interaction is non-zero for $r > r_0$ (for donors in silicon, $r_0 \approx 20$~\AA).

The Kohn-Luttinger donor electronic wavefunction \citep{DeSousa2003b} is
often employed for the silicon donors, to evaluate the probability density at the nuclear site $|\Psi({\bf r }_n)|^2$.
The wavefunction is derived from effective mass theory. It leads to oscillations and near-exponential decay of the hyperfine contact strength according to
\begin{equation}
J_\text{F}({\bf r})= \frac{4}{9} \gamma_e \gamma_n \hbar \mu_0
\left[
F_{1}({\bf r})\cos{(k_{0}x)} +
F_{2}({\bf r})\cos{(k_{0}y)} + 
F_{3}({\bf r})\cos{(k_{0}z)}
\right]^{2}
\end{equation}
where ${\bf r} \equiv {\bf r}_n = (x,y,z)$, $k_0 = (0.85)2\pi/a_0$ and $\gamma_e$ is the electron gyromagnetic ratio in silicon.
The cubic lattice parameter is $a_0$ (see \App{silicon} for the silicon crystal structure) and $\eta$ is the charge density on each crystal site.
The relevant envelope functions are:
\begin{equation}
F_{1}({\mathbf r}) = \frac{\exp{\left[
-\sqrt{\frac{x^{2}}{(nb)^{2}}+\frac{y^{2}+z^{2}}{(na)^{2}}}\right]}}
{\sqrt{\pi(na)^{2}(nb)}},
\end{equation}
\begin{equation}
F_{2}({\mathbf r}) = F_{1}({\mathbf r}) \text{~with~} \{x \to y, y \to z, z \to x\},
\end{equation}
\begin{equation}
F_{3}({\mathbf r}) = F_{1}({\mathbf r}) \text{~with~} \{x \to z, y \to x, z \to y\},
\end{equation}
where $a$ and $b$ are lengths characteristic to the donor
and $n = \sqrt{0.029~\text{eV}/\epsilon_i}$ with the electron ionization energy $\epsilon_i$ in eV.

Numerical values for a Group V donor in silicon interacting with \sitn\ impurities
are given in \Table{hyperfine}. Calculated couplings using the values in
\Table{hyperfine} are plotted in \Fig{Other_hyperfine} as a function of distance
from the donor electron. The electronic and nuclear gyromagnetic ratios are
given in \Table{gammas}, the silicon lattice constant is $a_0 = 5.43$~\AA\ and
the ionization energies of the Group V donors in silicon are in \Table{donors}.

\begin{table}[h]
\centering
    \begin{tabular}{cc}
    \hline\hline
    Parameter & Value \\ \hline
    Charge density $\eta$ & 186 \\
    Length $a$  & $25.09$~\AA \\
    Length $b$  & $14.43$~\AA \\ \hline\hline
    \end{tabular}
\caption{Numerical values for calculating the hyperfine interaction between a donor electron spin in silicon and a \sitn\ spin impurity \citep{DeSousa2003b}.}
\label{table:hyperfine}
\end{table}

We note that for our calculations the residual dipolar interaction in \Eq{hyperfine} is assumed to be secular
and only becomes effective after a distance of $na$ from the origin. The value of $na$ is about 20~\AA\ for silicon donors.

\subsection{Hyperfine-Mediated Interaction}
\label{sec:rkky}

The hyperfine-mediated interaction (also known as the RKKY interaction) \citep{Yao2006,Liu2007},
is a long-range coupling between two nuclear spins mediated
by an electron hyperfine-coupled to each of the two nuclei.
It results from the flip-flop part of $\hat{\bf I} \cdot \hat{\bf S}$
terms in the hyperfine interaction in \Eq{hyperfine}.
For a pair of nuclei, using perturbation theory, it can be approximated as
\begin{equation}
\hat{H}_\text{RKKY}^{(nm)} \simeq -\frac{J_{\text{F},n} J_{\text{F},m} }{ 4\gamma_e B }
\left(  \hat{I}_n^+ \hat{I}_m^- + \hat{I}_n^- \hat{I}_m^+  \right) \hat{S}_z.
\end{equation}
For decoherence of hybrid qubits, the intra-bath dipolar interaction
dominates over the RKKY. Also, for the case of the Hahn echo, it is suppressed.
However, the RKKY becomes important when considering nuclear spin decoherence
in \Chap{nucleardecoherence}.\\

We now proceed to explain the CCE for approximating the many-body dynamics
of a central spin system in a spin bath. The CCE has been extensively applied
for both nuclear spin diffusion (see e.g. \citet{Balian2014}) and spin diffusion due to an
all-dipolar electron spin system (see e.g. \citet{Witzel2012}).

\section{Cluster Correlation Expansion}
\label{sec:cce}

As stated in \Chap{intro}, it is not in practice possible to exactly solve for the
dynamics for tens of thousands of bath spins.
For spin baths with strong back action with the qubit, cluster expansion methods
\citep{Witzel2005,Witzel2006,Yang2008a,Yang2008b,Yang2009,Witzel2012} have enabled realistic numerical
simulations of the joint system-bath dynamics, predicting coherence times in remarkable agreement with experiment
(see e.g. \citet{Balian2014} or \citet{Ma2014}).
Here we describe the most general of these -- the cluster correlation expansion (CCE) \citep{Yang2008a,Yang2008b,Yang2009}.
The CCE also happens to have the simplest formulation which we outline in this section.

In the CCE and analogous formalisms, $\hat{H}_{\text{tot}}$ is diagonalised for sets or `clusters' of bath spins of varying sizes
up to some maximum cut-off size and the coherence decay is obtained from a product over all cluster contributions in the bath.\footnote{We note that the term `cluster' simply refers to a collection of spins and does not imply that these spins must be localized.}
Clusters are illustrated for the case of the hybrid qubit in silicon in \Fig{Balian2015_correlations}.

\subsection{General Formalism}

We first derive the CCE and later discuss its physical motivation.
We wish to calculate the complex coherence function $\mathcal{L}(t)$ as a function of time $t$
(regardless of any dynamical decoupling pulses).
Let the set of all spins in the bath be denoted by $\mathcal{R}$, and write the exact coherence as
$\mathcal{L}_\mathcal{R}(t) \equiv \mathcal{L}(t)$.
This quantity results from exactly solving the closed system-bath dynamics,
then tracing out the entire bath to obtain the off-diagonal of the reduced density matrix of the central system.
Even for tens of spins in the bath, this problem is practically impossible
on a classical supercomputer.

\begin{figure}[h]
\centering\includegraphics[width=2in]{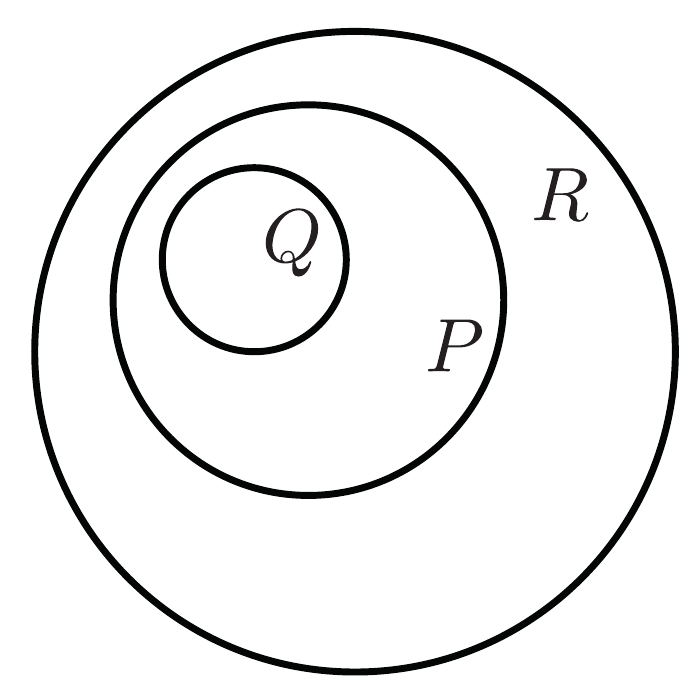}
\caption{
The CCE decomposes the set of all bath spins $\mathcal{R}$ into all its subsets or `clusters' of spins:
$\dots \subseteq \mathcal{Q} \subseteq \mathcal{P} \subseteq \mathcal{R}$.
}
\label{fig:Other_sets}
\end{figure}

We now decompose the bath into all its subsets or clusters $\mathcal{P}$, as illustrated in \Fig{Other_sets},
and note that these subsets include $\mathcal{R}$, the entire bath itself. For a given $\mathcal{P}$,
the coherence $\mathcal{L}_\mathcal{P}(t)$
is evaluated by considering the central system and bath spins contained only in $\mathcal{P}$.
This defines the reduced problem for $\mathcal{P}$.
In other words, the reduced problem for $\mathcal{P}$ has
all bath spins outside of $\mathcal{P}$ completely ``frozen''.

We now expand the reduced problem for $\mathcal{L}_\mathcal{P}(t)$ as a product of cluster correlation terms
$\tilde{\mathcal{L}}_\mathcal{P}(t)$:
\begin{equation}
\mathcal{L}_\mathcal{P}(t) = \prod_{\mathcal{Q}} \tilde{\mathcal{L}}_\mathcal{Q}(t).
\label{eq:cce1}
\end{equation}
\Eq{cce1} essentially defines the CCE. The cluster correlation or `tilde' terms are defined recursively
by re-writing \Eq{cce1},
\begin{eqnarray}
\mathcal{L}_\mathcal{P}(t) &=& \prod_{\mathcal{Q} \subseteq \mathcal{P} } \tilde{\mathcal{L}}_\mathcal{Q}(t) \nonumber \\
&=& \tilde{\mathcal{L}}_\mathcal{P}(t) \prod_{\mathcal{Q} \subset \mathcal{P} } \tilde{\mathcal{L}}_\mathcal{Q}(t),
\label{eq:cce2}
\end{eqnarray}
hence,
\begin{equation}
\tilde{\mathcal{L}}_\mathcal{P}(t)=
\frac{\mathcal{L}_\mathcal{P}(t)}{\prod_{\mathcal{Q} \subset \mathcal{P} } \tilde{\mathcal{L}}_\mathcal{Q}(t)}.
\label{eq:cce3}
\end{equation}
Thus, the cluster correlation term for the cluster of spins $\mathcal{P}$
is given by solving for $\mathcal{L}_\mathcal{P}(t)$ and recursively dividing by lower order
correlations formed by $\mathcal{Q}$, the proper subsets of $\mathcal{P}$.
Once all the cluster correlation terms are obtained,
$\mathcal{L}_\mathcal{R}(t) \equiv \mathcal{L}(t)$
is exactly recovered using the CCE:
\begin{equation}
\mathcal{L}_\mathcal{R}(t) = \prod_{\mathcal{P}} \tilde{\mathcal{L}}_\mathcal{P}(t).
\label{eq:cce4}
\end{equation}

As it stands, the exact CCE (\Eq{cce4}) seems practically useless because the correlation term for $\mathcal{R}$
contains the exact solution $\mathcal{L}_\mathcal{R}$. The strength of
the CCE method becomes evident when the expansion is truncated to include subsets bound
by the number of spins they contain:
\begin{equation}
\mathcal{L}_{[k]}(t) = \prod_{|\mathcal{P}| \leq k } \tilde{\mathcal{L}}_\mathcal{P}(t).
\label{eq:cce5}
\end{equation}

\subsection{Convergence and Heuristics}

\begin{figure}[h]
\centering\includegraphics[width=2.5in]{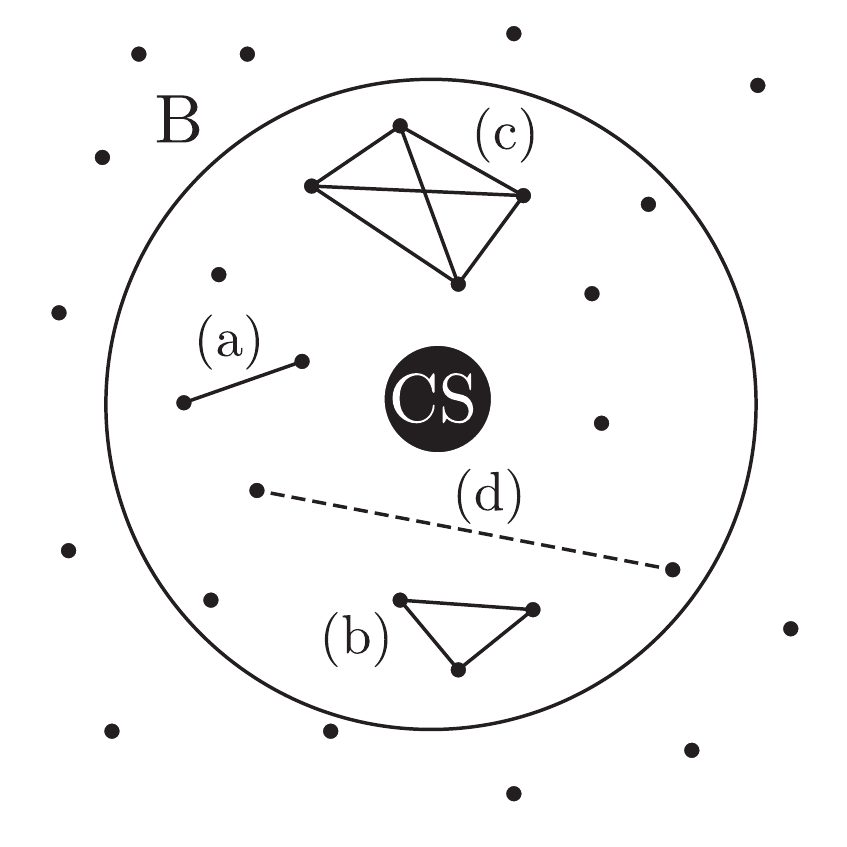}
\caption{
`CS' denotes the central spin system or qubit, in its environment, or bath `B'
of interacting spins. Clusters of 2 (a), 3 (b) and 4 (c) spins are depicted, with all intra-bath interactions shown as lines connecting the spins.
There are no clusters outside the sphere around the qubit, and clusters which include spins far away from one another are excluded (d).
}
\label{fig:Other_cce}
\end{figure}

\Eq{cce5} is the $k$-th order truncation to the CCE (which we denote by CCE$k$) and includes clusters for no more than $k$ spins in each cluster.
The expansion's success is judged by its convergence with respect to $k$, and also how small $k$ is.
The CCE can be said to be converged at $k=k'$ when
$| \mathcal{L}_{[k']}(t) - \mathcal{L}_{[k'+1]}(t) | , | \mathcal{L}_{[k']}(t) - \mathcal{L}_{[k'+2]}(t) |, \dots, \ll 1, \forall t$.
For example, if there is little difference in the coherence decay for CCE2 and the decays for CCE3 and CCE4, then
the CCE is reasonably converged at $k=2$.
Also, the practicality of the CCE is limited by the order of convergence.

Whether or not the CCE converges and the order of convergence depend on the relative interaction
strengths in the bath and interaction Hamiltonians. Practically, heuristic cut-offs are also imposed.
For example, the dipolar interaction decays as the inverse cube of the inter-spin separation. Therefore, only spins
separated by some cut-off distance are allowed to form clusters as the contribution from
farther away spins is relatively negligible. Also, for nuclear spin diffusion,
the extent of the electronic wavefunction imposes a radial cut-off for the superlattice cube or sphere in which clusters are
formed. Heuristic cut-offs are illustrated in \Fig{Other_cce}.

\subsection{Physical Motivation}

The product of coherences in the CCE is motivated by the fact that for two clusters $\mathcal{A}$ and $\mathcal{B}$
that are sufficiently far apart so that the interactions between spins in $\mathcal{A}$ and those in $\mathcal{B}$
are negligible compared to interactions inside the individual clusters,
it can be shown that the combined coherence is well approximated by the product $\mathcal{L}_{\mathcal{AB}} \simeq \mathcal{L}_{\mathcal{A}} \mathcal{L}_{\mathcal{B}}$.
The task is to compute the correction $\mathcal{L}_\text{corr}$ as the interactions between $\mathcal{A}$ and $\mathcal{B}$ are increased, in order to obtain the exact value
of $\mathcal{L}_{\mathcal{AB}}$:
\begin{equation}
\mathcal{L}_{\mathcal{AB}} =  \mathcal{L}_{\mathcal{A}} \mathcal{L}_{\mathcal{B}} \mathcal{L}_\text{corr}.
\end{equation}

For simplicity, consider three spins labelled 1, 2 and 3 and assume that the lowest-order contribution to the coherence is pairwise flip-flops, given by
$\mathcal{L}_{(1,2)} \mathcal{L}_{(1,3)} \mathcal{L}_{(2,3)}$. The correction required to take into account collective many-body flip-flops is simply
\begin{equation}
\mathcal{L}_\text{corr} = \frac{\mathcal{L}_{(1,2,3)} } { \mathcal{L}_{(1,2)} \mathcal{L}_{(1,3)} \mathcal{L}_{(2,3)} } \equiv  \tilde{\mathcal{L}}_{(1,2,3)},
\end{equation}
giving the true or non-factorisable 3-body correlation which we denote using $\tilde{\mathcal{L}}_{(1,2,3)}$.
Since the lowest-order correlations in this example are from two-clusters, the true non-factorizable correlation equals
the in general factorizable correlation: $\tilde{\mathcal{L}}_{(i,j)} = \mathcal{L}_{(i,j)}$.
Thus, for many pairs and triplets in the bath, the coherence is given by
\begin{equation}
\mathcal{L} = \prod_{(i,j)} \tilde{\mathcal{L}}_{(i,j)} \prod_{(i,j,k)} \tilde{\mathcal{L}}_{(i,j,k)}.
\label{eq:ccemot}
\end{equation}
\Eq{ccemot} motivates the systematic expansion \Eq{cce4}, which takes into account $n$-body correlations.

\subsection{Beyond the Standard CCE}

The discussion so far has not made any reference to the initial state of the bath.
For a pure product initial state, the reducible correlation terms $\mathcal{L}(t)$
for each cluster are calculated by simply considering a product state of eigenstates of
the non-interacting bath $\hat{H}^0_{\text{bath}}$.
For relatively small baths, the coherence calculated can be highly sensitive to initial state sampling.
Also, for a non-factorizable or entangled initial bath states,
the CCE described above is not valid. The ensemble-averaged CCE has been developed for these reasons \citep{Yang2009}.
However, these cases are not encountered in our work and the quantities of interest
are largely insensitive to the choice of initial product bath states we use.

Another important modification of the CCE exists \citep{Witzel2010,Witzel2012} in which
Ising interactions with spins external to each cluster are included in addition to the usual interactions within the cluster.
In addition, an efficient method of averaging over initial bath states has been developed \citep{Witzel2012}.
These modifications were required to speed up convergence and to remove numerical instabilities for the all-dipolar problem of a central
electron spin interacting with an electron spin bath.

Finally, we note that the expansion need not formally be a product expansion.
In fact, the CCE has been reformulated using sums and differences of correlation terms \citep{Witzel2014a}.

\subsection{Pair Correlations}
\label{sec:pseudospins}

The lowest non-trivial CCE order for spin diffusion is $k=2$ or CCE2.
This only involves the qubit-bath dynamics involving pairs of spins
in the bath -- pair correlations.

If a pure dephasing model is assumed, as described in \Sec{puredephasing},
the bath dynamics is governed by effective Hamiltonians depending on the state of the
central system. In other words, the total Hamiltonian has a form such that
terms depending on the states of the central system are uncoupled. These state-dependent uncoupled Hamiltonians are:
\begin{equation}
\hat{h}_{i} \equiv \bra{i}(\hat{H}_\text{int} + \hat{H}_\text{bath})\ket{i},
\end{equation}
written for state $\ket{i}$.

An orthonormal basis for two spin-1/2 particles is
$\ket{\uparrow \uparrow}, \ket{\downarrow \downarrow}, \ket{\uparrow \downarrow}, \ket{\downarrow \uparrow}$.
Here, $\ket{ \uparrow \downarrow } \equiv \ket{ \uparrow } \otimes \ket{\downarrow} $, where $\ket{ \uparrow / \downarrow }$ denotes spin up/down.
Therefore, the state-dependent Hamiltonians above act on this basis.
Furthermore, for a secular flip-flop intra-bath interaction, the only matrix elements
involved in the interaction are those involving the states $\ket{\uparrow \downarrow}$ and $\ket{\downarrow \uparrow}$.
The other two polarised states only contribute a phase factor which disappears when the modulus of
the coherence is taken. Note that care must be taken when averaging over initial bath states and coherences of unity in this model must
be accounted for the polarised states. The state-dependent Hamiltonians can be written using $2 \times 2$ matrices in the basis
$\{\ket{\uparrow \downarrow}$ , $\ket{\downarrow \uparrow}\}$:
\begin{equation}
\hat{h}_{i} = a \hat{\mathds{1}} + b \hat{\boldsymbol\sigma}\cdot {\bf H}_{i},
\label{eq:statedependentH}
\end{equation}
where $\hat{\boldsymbol\sigma}$ is the three-vector of Pauli operators 
and ${\bf H}_{i}$ is an effective field about which these so-called `pseudospins' precess.
The length of the state-dependent pseudofields gives the pseudospin precession frequency and is obtained by diagonalising \Eq{statedependentH}.
The identity term in \Eq{statedependentH} is dynamically uninteresting, contributing a constant shift to the pseudospin energies.
This central state-dependent dynamics has been applied for a variety of spin problems including quantum dots and NV centres \citep{Yang2008a,Yang2008b,Yang2009,Yao2006,Yao2007,Liu2007,Zhao2012a},
the latter involving one-spin cluster dynamics at the lowest non-trivial order.

\subsection{Many-Body Correlations}

We use the term `many-body or $n$-body correlations' to refer to qubit-bath dynamics containing non-negligible contributions from clusters including more than two bath spins ($n>2$). For example, in spin diffusion, collective flip-flops of three bath spins are referred to as to 3-body correlations.
It is of fundamental interest to isolate such many-body correlations \citep{Ma2014,Balian2015}. These are rare occurrences,
and most of the decoherence problems in this thesis are described by only considering pair correlations (or CCE2).
However, we shall see that near an OWP and for low to moderate pulsed dynamical decoupling, 3-body clusters dominate the dynamics.

\section{Conclusion}

In conclusion, the current chapter primarily serves as a resource for understanding our theoretical results in subsequent chapters
dealing with decoherence.
The basic principles of magnetic resonance, in particular experiments measuring coherence times were outlined.
The problems of central spin decoherence and spin diffusion were introduced and finally,
we described the CCE for solving for the many-body qubit-bath dynamics.
In summary, here we have described $\hat{H}_\text{bath}$ and $\hat{H}_\text{int}$.
In the next chapter, we discuss $\hat{H}_\text{CS}$ for the hybrid qubit.

\chapter{The Hybrid Qubit}
\label{chap:hybridqubit}

This chapter primarily concerns the central spin Hamiltonian $\hat{H}_\text{CS}$ for our central spin decoherence
problem with total Hamiltonian given by \Eq{totalH}. The central system we consider is in general a
mixed electron-nuclear or `hybrid' qubit.

It is often the case that magnetic resonance transitions are between states which are eigenstates of the magnetic field or Zeeman Hamiltonian described
in \Sec{zeeman}.
The central Hamiltonian may, for example, contain multiple uncoupled spins. 
The basis of eigenstates in this case is the product of states labelled by the magnetic quantum numbers (of $z$-projection spin operators)
of the non-interacting spin species. The term `mixed' here refers to the case when the eigenstates can no longer be approximated using these Zeeman states,
but instead involve their entanglement.\footnote{The term `mixing' in this thesis should not be confused with a probability distribution of pure states. We use
the term to refer to entanglements of high-field eigenstates of a strongly-interacting spin system situated in a magnetic field.}

Electronic spin manipulation times are much shorter compared to those of nuclear spins.
For the hybrid qubit, however, we show that an NMR transition in the high-field limit can be manipulated
on short ESR timescales in the mixing regime. For quantum information applications, the shortest manipulation time
and longest coherence time are highly desirable. The coherence times of the hybrid qubit are the topic of \Chap{decoherence} for the Hahn
spin echo and \Chap{dynamicaldecoupling} for higher orders of dynamical decoupling.

In what follows, we present the spin Hamiltonian of the hybrid qubit and its eigendecomposition.
Examples are the Group V donors in silicon, in particular bismuth donors, due to their exceptionally
strong mixing. We then describe fast quantum control of the hybrid qubit, published in \citet{Morley2013}.
The latter reference also contains numerical simulations of coherence times which we
present in \Chap{decoherence}, and use to confirm the dominant decoherence mechanism in the experiments of \citet{Morley2013}.
This study not only demonstrated fast control, but also relatively long coherence times of the hybrid qubit.

\section{Group V Donors in Silicon}

Our hybrid electron-nuclear qubit can be implemented as one of the Group V hydrogenic impurities in silicon.
These are phosphorus, arsenic, antimony and bismuth in increasing period of the periodic table.
The notation Si:$X$ is used to denote $X$-doped silicon, where $X$
is one of the four donor atoms. Silicon has four valence electrons, thus a Group V
impurity donates a single electron which at low enough temperatures is localized at the substitutional site
of the donor nucleus \citep{Stoneham2001}.
The host impurity nuclear spin is coupled to the donor electron spin via the isotropic Fermi contact hyperfine interaction
(which we describe in detail in \Sec{hyperfine} for the case of a donor electron and a bath nucleus).
Ionization energies of the donors, their nuclear gyromagnetic ratios, nuclear spin quantum numbers and
electron-nuclear hyperfine coupling strengths are given in \Table{donors}.

\begin{table}[h]
\centering
    \begin{tabular}{ccccc}
    \hline\hline
    Donor $X$ & $\epsilon_{i,X}$ (eV) & $\gamma_X$ (M~rad~s$^{-1}$~T$^{-1}$) & $I_X$ & $A_X$ (M~rad~s$^{-1}$) \\ \hline
	Phosphorus $^{31}$P & 0.044 & $-108.41$ & 1/2 & $7.3846\times10^{2}$  \\
	Arsenic $^{75}$As & 0.049 & $-45.95$ & 3/2 &  $1.2467\times10^{3}$  \\
	Antimony $^{121}$Sb & 0.040 & $-64.44$ & 5/2 & $1.174 \times 10^{3}$ \\
	Bismuth $^{209}$Bi & 0.069 & $-43.775$ & 9/2 & $9.2702\times10^{3}$ \\ \hline\hline
	\end{tabular}
\caption{Nuclear gyromagnetic ratios $\gamma_X$, donor electron ionization energies $\epsilon_{i,X}$, nuclear spin total quantum numbers $I_X$
and isotropic hyperfine coupling strengths $A_X$
of the Group V donors in silicon (Si:$X$).
Values for $\epsilon_{i,X}$ and $A_X$ can be found in \citet{Feher1959} and values for $\gamma_X$ and $I_X$ in \citet{Stone2005}.}
\label{table:donors}
\end{table}

\subsection{Spin Hamiltonian and Eigenspectrum}

The effective spin Hamiltonian of a donor in silicon (Si:$X$) in a magnetic field $B$ along the $z$-axis is given by the sum of Zeeman
and hyperfine terms \citep{Mohammady2010,Mohammady2012},
\begin{equation}
\hat{H}_{\text{Si:}X}=
\omega_0 \left( \hat{S}^z  + \delta_X  \hat{I}^z_{X} \right) +
A_X\hat{{\bf I}}_{X} \cdot \hat{{\bf S}},
\label{eq:donorH}
\end{equation}
where the electron Larmor precession frequency is $\omega_0 = \gamma_e B$
and $\gamma_e$ is the gyromagnetic ratio of the electron spin in silicon.
The hyperfine interaction between the host nuclear and donor electron spins
is well-approximated by considering only the isotropic Fermi contact part,
here denoted by $A_X$, to distinguish it from hyperfine coupling to the bath $J_\text{F}$.
The nuclear gyromagnetic ratio is contained in $\delta_X = \gamma_X / \gamma_e$, which is
much smaller than $\gamma_e$ (i.e.\ $|\delta_X| \ll 1$).

For the case of the Zeeman and hyperfine terms having comparable magnitudes ($\omega_0 \sim A_X$),
the matrix representation of the Hamiltonian in \Eq{donorH} is no longer diagonal in the
the Zeeman basis $\ket{ m_S } \otimes \ket{ m_I } \equiv \ket{m_S,m_I}$, where $m_S = \pm 1/2$ and $m_I = -I_X,-I_X+1, \dots , I_X$.
This competition between the two terms is significant for the donor systems at typical magnetic field strengths for ESR
experiments ($B \simeq 0.1 - 0.6$~T), especially for bismuth which has the largest hyperfine strength.

It is easy to show that the sum of spin $z$-projections commute with the Hamiltonian:
\begin{equation}
\left[\hat{H}_{\text{Si}:X}, \hat{S}^z+\hat{I}^z_X\right]=0,
\label{eq:sumzzcommute}
\end{equation}
and thus a good set of quantum numbers for the eigenbasis of \Eq{donorH} are
$-|I_X+S|,-|I_X+S|+1, \dots, I_X+S$. These label the mixed or `adiabatic' eigenstates $\ket{\pm,m}$ which mix the
Zeeman basis to (at most) doublets of constant $m=m_{S}+ m_{I}$ with energies $E^{\pm}_{m} \left( \omega_{0} \right)$.
Due to the commutation relation \Eq{sumzzcommute},
the Hamiltonian \Eq{donorH} can be divided into two one-dimensional Hamiltonians acting on bases
\[\{\ket{1/2,I_X}\} , \{\ket{-1/2,-I_X}\},\] and 2$I$ two-dimensional Hamiltonians acting on
\[\{ \ket{ \pm 1/2 , m \mp 1/2 } , \ket{ \mp 1/2 , m \pm 1/2 } \},\] the latter with $|m| \leq  I_X - 1/2$.
In other words, in its matrix representation in the Zeeman basis, the Hamiltonian is in block form composed of $2 \times 2$
matrices of which there are $2I$ and two $1 \times 1$ entries.
Eigenstates of the one-dimensional Hamiltonians are those which have $m=\pm|I_X+1/2|$ and they remain unmixed at all magnetic field,
with energies
\begin{equation}
E_{m = \pm |I_X + 1/2|} \left( \omega_{0} \right) = \pm \frac{\omega_0}{2}(1 - 2\delta_X I_X) + \frac{A_X I_X}{2}.
\end{equation}
The two-dimensional Hamiltonian for each doublet is
\begin{equation}
\hat{H}_{m \ne \pm |I_X + 1/2|} = \frac{A}{2} \left( R_m \cos{\theta_m} \hat{\sigma}^z + R_m \sin{\theta_m} \hat{\sigma}^x - \epsilon_m \hat{\mathds{1}} \right),
\label{eq:doubletH}
\end{equation}
where
\begin{eqnarray}
\cos \theta_{m} &=& \frac{\Omega_m(\omega_0)}{R_m(\omega_0)}, \\
\sin \theta_{m} &=& \frac{\Delta_m} { R_{m}(\omega_0) },
\end{eqnarray}
and,
\begin{eqnarray}
R_{m}(\omega_0)^{2} &=& \Omega_m^2(\omega_0) + \Delta^2_m, \\
\Omega_m(\omega_0) &=& m + \tilde{\omega}_{0} (1 + \delta_X), \\
\tilde{\omega}_{0} &=& \frac{\omega_{0}}{A_X}, \\
\Delta^2_m &=& (I_X+\tfrac{1}{2})^2 - m^{2}, \\
\epsilon_m &=& \frac{1}{2} \left( 1 + 4\tilde{\omega_0}m\delta_X \right).
\end{eqnarray}
Diagonalising the above Hamiltonian, the eigenstates $\ket{\pm,m}$,
mixed in the product Zeeman basis are given by
\begin{equation}
\ket{\pm,m} = 
a_{m}
\ket{m_{S} =   \pm \frac{1}{2}} \otimes
\ket{m_{I} = m \mp \frac{1}{2}} 
\pm b_{m}
\ket{m_{S} =   \mp \frac{1}{2}} \otimes
\ket{m_{I} = m \pm \frac{1}{2}},
\label{eq:mixedstates}
\end{equation}
with amplitudes
\begin{eqnarray}
a_{m} &=& \cos{\frac{\theta_m}{2}}, \\
b_{m} &=& \sin{\frac{\theta_m}{2}}, 
\end{eqnarray}
and corresponding eigenenergies
\begin{equation}
E^{\pm}_{m} \left( \omega_{0} \right) =
\frac{A_X}{2}\left[ -\frac{1}{2} \left(1+4 \tilde{\omega}_{0} m \delta_X \right) \pm R_{m}(\omega_0)  \right].
\end{equation}
In the notation $\ket{\pm, m}$, the two unmixed eigenstates are thus
$\ket{ + , I_X + 1/2 }$ and $\ket{ - , -|I_X + 1/2| }$.

We often write the adiabatic eigenstates using the notation $\ket{i} = 1,2, \dots , d$, where $d = (2S+1)(2I_X+1) = 4I_X + 2$.
The transformations between the two labels $\ket{i} \leftrightarrow \ket{\pm,m}$ are given by,
\begin{eqnarray}
\ket{i} &=&
  \ket{{ \begin{array}{rl}
				+ &\mbox{for $ 2I_X+2 \le i \le 4I_X + 2$} \\
				- &\mbox{for $ 1 \le i \le 2I_X + 1$}
			 \end{array}},
			 |2I_X+1-i|- S - I_X } \\
\ket{\pm,m} &=&
	\ket{{
					\begin{array}{rl}
					& 3I_X - S + 2 + m \\
					&S+I_X-m
					\end{array}
		}
	}.
\end{eqnarray}

It is clear that $a_m^2 + b_m^2=1$ (eigenstates are normalised).
The difference $(a_m^2 - b_m^2)$ is proportional to the
expectation value of the electron spin $z$-projection,
\begin{equation}
\bra{\pm,m} \hat{S}^z \ket{\pm,m} = \pm\frac{1}{2} (a_m^2 - b_m^2) = \pm\cos{\theta_m} = \frac{\Omega_m(\omega_0)}{R_m(\omega_0)} \equiv \frac{P_i(\omega_0)}{2},
\end{equation}
half the {\em polarisation} $P_i(\omega_0)$ for the $i$-th eigenstate $\ket{i} \equiv \ket{\pm,m}$.
For the unmixed system, we take the limit $\omega_0 \to \infty$ so $R_m \to \Omega_m$ and
the polarisation $\pm1$ for the bare electron is recovered (i.e.\ $ \langle\hat{S}^z\rangle=\pm 1/2 $).

\subsection{Frequency-Field Gradient}

Denoting the excitation frequency for the two levels in a transition as $f$, at $df/dB=0$ points,
decoherence from classical field noise is significantly reduced \citep{Mohammady2012}.
The Hellmann-Feynman theorem \citep{Cohen1977} states that the derivative of the energy with respect to some
parameter in the Hamiltonian, in this case the magnetic field $B$, is obtained by evaluating the derivative of the
expectation value of the Hamiltonian in the corresponding energy eigenstate. Thus,
\begin{eqnarray}
\frac{dE_{i}}{dB} &=& \bra{i} \frac{d\hat{H}_{\text{Si:}X}}{dB} \ket{i} \nonumber \\
&=& \bra{i} \frac{d}{dB} \left(   \gamma_e B \hat{S}_z  + \gamma_X B \hat{I}^z_{X} + A_X \hat{{\bf I}}_{X} \cdot \hat{{\bf S}} \right) \ket{i} \nonumber \\
&=& \gamma_e \langle \hat{S}^z \rangle_{i} + \gamma_X \langle \hat{I}^z_{X} \rangle_{i}.
\end{eqnarray}
The excitation frequency is simply $ f = (E_{i=u} - E_{i=l})/2\pi $ for a magnetic resonance
transition $\ket{i=u} \to \ket{i=l}$ between an upper ($i=u$) and a lower ($i=l$) level.
Therefore, the frequency-field gradient is given by
\begin{equation}
\frac{df}{dB} =  \frac{1}{2\pi} \left[\frac{\gamma_e}{2} (P_u - P_l) + \gamma_X \left(  \langle \hat{I}^z_{X} \rangle_{u} - \langle \hat{I}^z_{X} \rangle_{l}  \right) \right],
\label{eq:dfdB}
\end{equation}
and it can be shown that the $B$ values for which $df/dB=0$, satisfy,
\begin{equation}
0 =  P_u(B) - P_l(B) + \frac{\delta_X (m_l-m_u) }{1+\delta_X},
\label{eq:dfdBzero}
\end{equation}
where $m_{u/l}$ are the $m_S + m_I$ quantum numbers for the two levels.

\subsection{Cancellation Resonances}

When $\Omega_m = 0$, the polarisation $P_i$ vanishes.
Also, the $\hat{\sigma}^z$ term in \Eq{doubletH} is zero,
and thus the eigenstates become those of $\hat{\sigma}^x$:
\begin{equation}
\ket{\Psi^\pm} = \frac{1}{\sqrt{2}}
\left(  \ket{  \frac{1}{2} , m - \frac{1}{2} }
\pm     \ket{ -\frac{1}{2} , m + \frac{1}{2} }     \right).
\end{equation}
Typically, due to the large mismatch between electronic and nuclear gyromagnetic ratios,
$\delta_X \ll 1$ and hence $\Omega_m \simeq m + \tilde{\omega_0}$. Thus,
the Bell-like eigenstates above occur at magnetic fields corresponding
to $\tilde{\omega_0} \simeq -m$. For $-(I_X - 1/2) \leq m \leq 0$, these field
values correspond to Landau-Zener (LZ) crossings.
The LZ points have OWPs (see below) midway between them in $\theta_m$ coordinates.
Fields when the doublet Hamiltonian reduces to a sum of $\hat{\sigma}^z$
and $\hat{\sigma}^x$ with the same coefficients may also
be of special interest.

\subsection{Optimal Working Points}

For coupling of the electronic spin to a nuclear spin bath,
the OWPs considered in this thesis and discussed in \Chap{decoherence}. \Chap{formula} and \Chap{dynamicaldecoupling}
occur when $P_u = P_l$ for the transition $\ket{u} \to \ket{l}$.
Since typically, $\delta_X \ll 1$, some OWP and $df/dB=0$ points (\Eq{dfdBzero}),
can be extremely close in magnetic field.

\section{Bismuth Donor}

\begin{figure}[h]
\centering\includegraphics[width=4in]{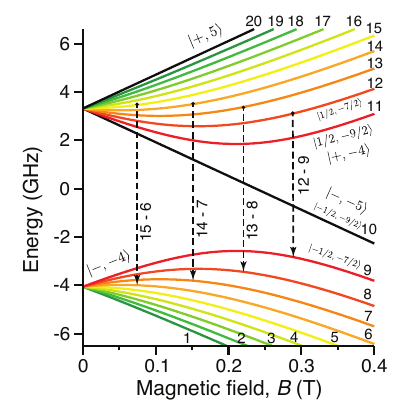}
\caption{
The spectra of donor spin systems such as arsenic, antimony or bismuth (pictured) 
are affected by strong mixing between the electron and host nuclear spin, at magnetic fields $B$
smaller or comparable to the hyperfine coupling $A$, allowing a richer behaviour than 
unmixed electron spins. The plot shows the
eigenspectrum of Si:Bi as a function of magnetic field $B$ labelled in order of increasing energy $\ket{i}$ using integers $1,\dots,20$, in the Zeeman basis
$\ket{m_S} \otimes \ket{m_I} \equiv \ket{m_S,m_I}$ (i.e.\ as $B \to \infty$)
and using the adiabatic basis $\ket{\pm,m}$. Strong mixing of the Zeeman basis is evident in the region $B \simeq 0.1 - 0.3$~T due to competition between the Zeeman
and hyperfine Hamiltonians.
At particular field values termed optimal working points (OWPs),
decoherence can be strongly suppressed.
The arrows indicate the transitions with four of the most significant OWPs. The colours match for the two states in each doublet labelled by $m$.
Figure adapted from \citet{Mohammady2010} and \citet{Balian2015}.
}
\label{fig:Balian2015_sibi}
\end{figure}

The bismuth donor is special among the Group V donors in two aspects. First, the hyperfine coupling is the strongest
of all the Group V donors (\Table{donors}). The large hyperfine value means that there is significant
electron-nuclear mixing for fields in the range `intermediate ESR field regime': $B \simeq 0.1 - 0.6$~T.
Second, the bismuth nucleus has the highest dimension
of Hilbert space since $I_{\text{Bi}} = 9/2$, giving $2(9+1) = 20$ energy levels and thus a large number of states
for possible manipulation in quantum information applications. The strong hyperfine and large $I_\text{Bi}$ lead to
the largest number of OWPs among the silicon donors. We note that for an OWP, $I_X > 1/2$ and hence
the Si:P system ($I_\text{P} = 1/2$) unfortunately does not have any.

\subsection{Energy Levels and $X$-Band Transitions}

The energy spectrum as a function of magnetic field for Si:Bi is shown in \Fig{Balian2015_sibi}.
The eigenstates can be labelled in multiple ways: in order of increasing energy ($\ket{i}$, $i=1,2,\dots,20$),
in the Zeeman basis in the high-field limit ($\ket{m_S,m_I}$, $m_S = \mp\tfrac{1}{2}$, $m_I = -\tfrac{9}{2},-\tfrac{7}{2},\dots,\tfrac{7}{2},\tfrac{9}{2}$),
or the adiabatic basis ($\ket{\pm,m}$, $-5 \leq m \leq 5$).
Since the Zeeman basis is no longer the eigenbasis of the Hamiltonian,
the usual ESR and NMR selection rules presented in \Chap{background} do not apply.
Nevertheless, we refer to $\ket{\pm,m} \leftrightarrow \ket{\mp,m-1}$ and $\ket{\pm,m} \leftrightarrow \ket{\pm,m-1}$ as ESR-type and NMR-type
transitions respectively, noting that $\ket{-,m} \leftrightarrow \ket{+,m-1}$ are dipole forbidden in the high field limit.

\begin{figure}[h]
\centering\includegraphics[width=3in]{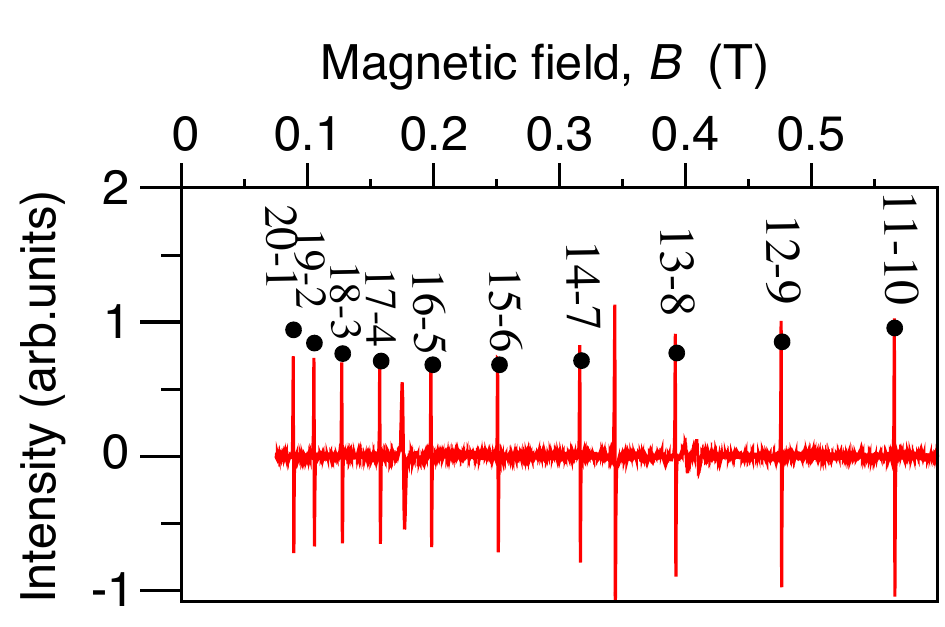}
\caption{
Experimental CW ESR spectrum of bismuth-doped silicon at excitation frequency $f=9.7$~GHz (red lines)
showing good agreement with the resonance positions predicted by theory (black dots). Ten transitions $\ket{i=u} \to \ket{j=l}$ are observed
with $i = 11,\dots,20$ and $j = 10,\dots,1$ (see horizontal dashed line in \Fig{Other_dfdB}). Figure adapted from \citet{Mohammady2010}.
}
\label{fig:Other_xband}
\end{figure}

\begin{figure}[h]
\centering\includegraphics[width=3in]{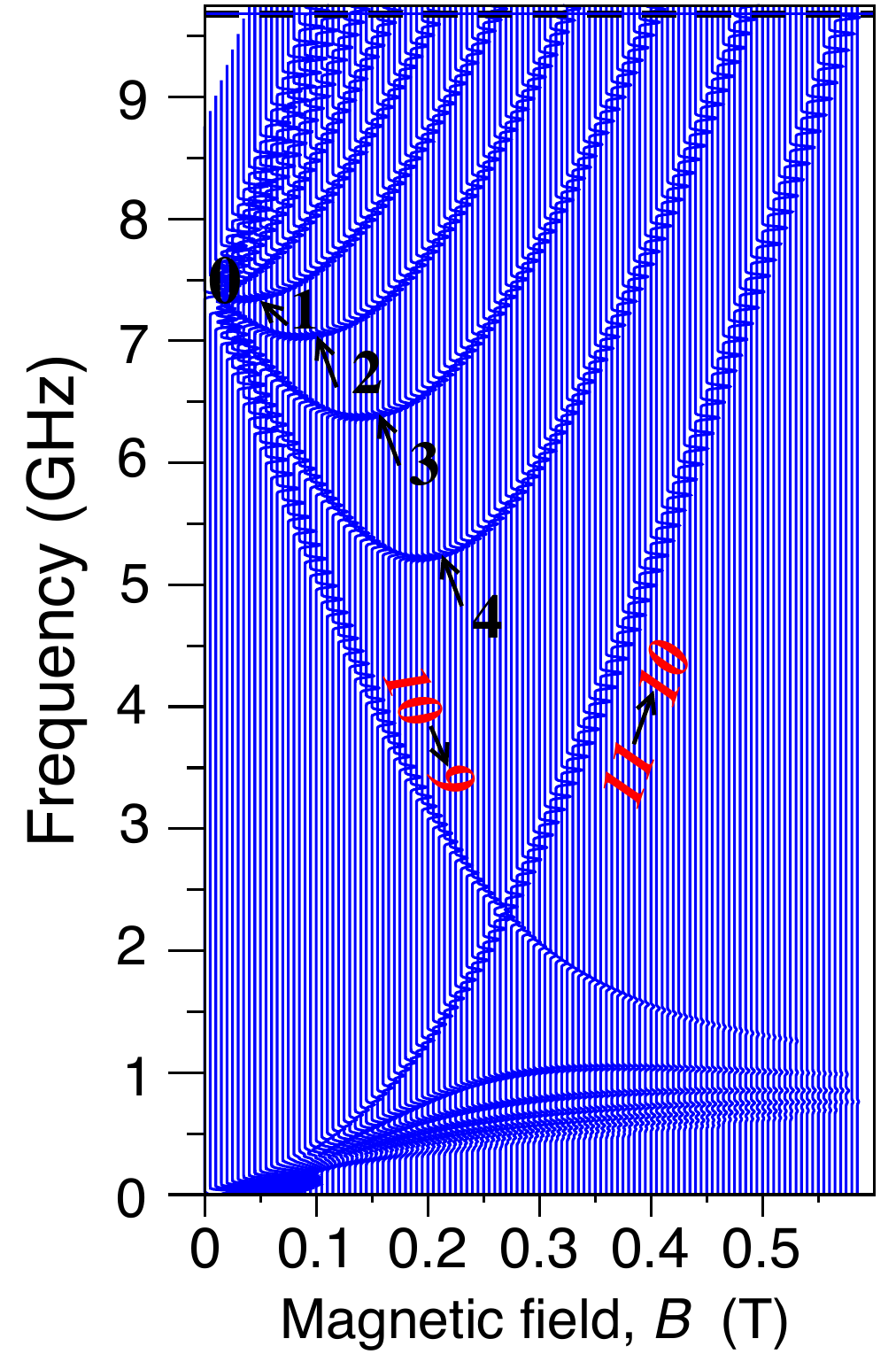}
\caption{
ESR excitation frequency $f$ as a function of magnetic field $B$ for bismuth-doped silicon.
The $df/dB=0$ extrema correspond to magnetic fields at which classical noise decoherence
is significantly reduced. The dashed horizontal line shows the $X$-band excitation frequency at $9.7$~GHz, corresponding to the measured spectrum in \Fig{Other_xband}.
The labelled transitions are those at $S$-band (4~GHz) and the black integers label the cancellation resonances,
where the electronic Zeeman and hyperfine terms cancel in the donor Hamiltonian (\Eq{donorH}).
Figure adapted from \citet{Mohammady2010}.
}
\label{fig:Other_dfdB}
\end{figure}

The CW ESR spectrum for Si:Bi at an excitation frequency of $f=9.7$~GHz ($X$-band) is shown in \Fig{Other_xband} showing 10 spectral lines.
This is a region of weak mixing.
The resonance positions are in excellent agreement with the analytical expressions for the eigenspectrum described above.

\subsection{OWPs, CTs and Other Special Fields}

The OWPs we consider correspond to suppression of decoherence in quantum spin environments.
Si:Bi has four ESR-type and four-NMR type OWPs. These OWPs are all doublets, so there are in fact 16 separate OWP transitions.
The OWPs we primarily consider are the ESR-type ones corresponding to states $\ket{12} \to \ket{9}$
and $\ket{14} \to \ket{7}$ occurring at $B \simeq 0.19$~T and $B \simeq 0.08$~T respectively.
The transitions corresponding to four ESR-type OWPs are labelled in \Fig{Balian2015_sibi},
while the other four correspond to forbidden transitions close by.
It is important to note that all of these couple {\em two neighbouring} avoided crossings.
Selection rules are detailed in \citep{Mohammady2012}, but OWP transitions have $\Delta m =  \pm 1$
which implies that $\langle u | {\hat S}^z| l\rangle = 0$ meaning that magnetic field fluctuations
do not induce bit flips $\ket{u} \leftrightarrow \ket{l}$ (assuming Ising-like coupling to the bath).

An important consequence of the strong mixing in Si:Bi is the existence of multiple $df/dB=0$ maxima and minima in the
$f-B$ parameter space. These are shown in
\Fig{Other_dfdB}.
There are $df/dB=0$ minima for the $|15\rangle \to |6\rangle$,
$|14\rangle \to |7\rangle$, $|13\rangle \to |8\rangle$,
$|12\rangle \to |9\rangle$, and $|11\rangle \to |8\rangle$
transitions in the frequency range $5-7.5$~GHz and two maxima for
$|12\rangle \to |11\rangle$ and $|9\rangle \to |8\rangle$ close to $1$~GHz.
The $df/dB=0$ points are referred to as `clock transitions' (CTs) \citep{Wolfowicz2013} and correspond to
where decoherence arising from classical field noise is reduced \citep{Mohammady2012}.
They are distinct from the OWP points, albeit in close proximity to them in magnetic field.

One might also  consider the possibility of creating a superposition of two states 
$|u\rangle$ and $|l\rangle$ at a single avoided crossing; for example, the superposition
$|11\rangle + |9\rangle$ in \Fig{Balian2015_sibi}, at the avoided crossing between these states at $B \simeq 0.21$~T.
Although the $\ket{11}\to \ket{9}$ transition is never allowed, such a superposition
might be created by a two pulse excitation from level $\ket{10}$.
Both states are at zero energy gradient ($dE_{u,l}/dB=0$) so coherences are to first order insensitive
to dephasing noise; however, as shown in \citet{Mohammady2012},
in that case $\langle u | {\hat S}^z| l\rangle \neq 0$ so magnetic fluctuations couple
the states in the superposition and thus coherence is vulnerable to depolarisation by magnetic noise.

The cancellation resonance points are where
the hyperfine interaction cancels the Zeeman splitting. They occur for Si:Bi
at magnetic fields $\omega_0 \simeq -mA$ and $m  =0, -1, -2, -3, -4$, labelled in \Fig{Other_dfdB}.
They correspond to the avoided crossings seen in \Fig{Balian2015_sibi}.
Finally, the field for which $\tilde{\omega_0} \simeq 7$ corresponds to $df/dB=0$ maxima CTs and NMR-type OWPs for the $\ket{12} \to \ket{11}$
and $\ket{9} \to \ket{8}$ transitions, and is where the doublet donor Hamiltonian is proportional to $(\hat{\sigma}^x + \hat{\sigma}^z)$.

\section{Fast Quantum Control}

The current section concerns pulsed ESR control of
the hybrid qubit on the nanosecond timescale as demonstrated
by experiments in \citet{Morley2013}.
The coherence times for the hybrid qubit for the transitions
and magnetic fields at which the experiments
were performed are presented in \Chap{decoherence}.
The experiments presented herein were carried out by
Dr. Petra Lueders at ETH Zurich with assistance from Dr. Gavin Morley
and Dr. Hamed Mohammady, who are all co-authors of \citet{Morley2013},
which is published work co-authored by S.J.B., contributing to the theme of the thesis.\\

Magnetic resonance involves transitions
between doublets adjacent in $m$ (i.e.\ $m \to m \pm 1$), using the notation of
energy states of the hybrid qubit as described above.
The usual ESR and NMR selection rules correspond to single electronic and nuclear spin flips respectively
as discussed in \Sec{magneticresonance}.
These selection rules are in the usual `unmixed regime' or equivalently in the limit of $B \to \infty$.
The time taken to manipulate electronic spins
by pulsed ESR is often on the order of nanoseconds; much faster than pulsed NMR
manipulation of nuclei which typically takes microseconds.
This is because the gyromagnetic ratio of the electron is of order GHz~T$^{-1}$ whereas
for nuclei it is on the MHz~T$^{-1}$ scale.
If electrons and nuclei were to be used for quantum information processing,
the shortest possible total manipulation time is highly desirable.
Importantly, the manipulation time must also be shorter
than the coherence time in order to implement quantum error correction protocols.

\subsection{Forbidden Transitions}

An important consequence of the strong mixing for the hybrid qubit (implemented as Si:Bi)
is the existence of ESR transitions of its eigenstates
which are ESR-forbidden at high fields but can be manipulated
using fast ESR pulses in the strong mixing or hybrid regime.
Furthermore, as we shall see in \Chap{decoherence}, the qubit coherence times in the hybrid regime can be up to five orders
of magnitude longer than the manipulation times and are limited by
nuclear spin diffusion by a \sitn\ spin bath.
We use the term hybrid because in this regime
the electron and nucleus are hybridized (\Eq{mixedstates}) or near 50:50
superpositions of bare electronic and nuclear spin Zeeman states.

\subsection{$S$-Band Transitions}

\begin{figure}[h]
\centering\includegraphics[width=4.5in]{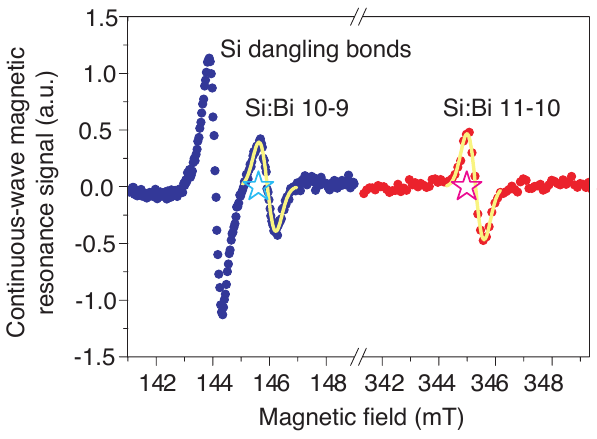}
\caption{
Magnetic resonance spectrum of bismuth-doped silicon (Si:Bi) at 4~K
with 4.044 GHz CW excitation ($S$-band). The
predicted positions of both Si:Bi resonances are shown as stars.
The bismuth dopant concentration is about $3 \times 10^{15}$~cm$^{-3}$
and the magnetic field was perpendicular to the $[111]$ direction of the doped silicon crystal.
Figure adapted from \citet{Morley2013}.
}
\label{fig:Morley2013_sband}
\end{figure}

To access the hybrid regime in Si:Bi, a $4.044$~GHz ESR excitation frequency
was used ($S$-band). As can be seen in \Fig{Other_dfdB},
this corresponds to two transitions: $\ket{10} \to \ket{9}$ and 
$\ket{11} \to \ket{10}$ at $B = 145.6$~mT and $B = 345.0$~mT
respectively.
The CW ESR spectrum at this frequency is shown in \Fig{Morley2013_sband},
in which the resonance positions are well predicted by analytical diagonalisation (\Eq{mixedstates})
of the donor Hamiltonian (\Eq{donorH}).

In the high-field limit at $f \simeq 4$~GHz, the observed transitions are:
\begin{eqnarray}
(\ket{11} \to \ket{10})_{B \to \infty} &\equiv& \ket{m_S = \frac{1}{2} , m_I = -\frac{9}{2} } \to \ket{-\frac{1}{2},-\frac{9}{2}}, \\
(\ket{10} \to \ket{ 9})_{B \to \infty} &\equiv& \ket{m_S = -\frac{1}{2} , m_I = -\frac{9}{2} } \to \ket{-\frac{1}{2},-\frac{7}{2}}
\end{eqnarray} 
and respect the usual selection rules, the first being an ESR transition and the second an NMR one.
Clearly, the second transition violates the ESR selection rule
and is not expected to be seen in ESR spectra.
This is indeed the case in the high field limit.
However, in the hybrid regime, this ESR forbidden
transition is observed as shown in \Fig{Morley2013_sband}.

\begin{figure}[h]
\centering\includegraphics[width=5.5in]{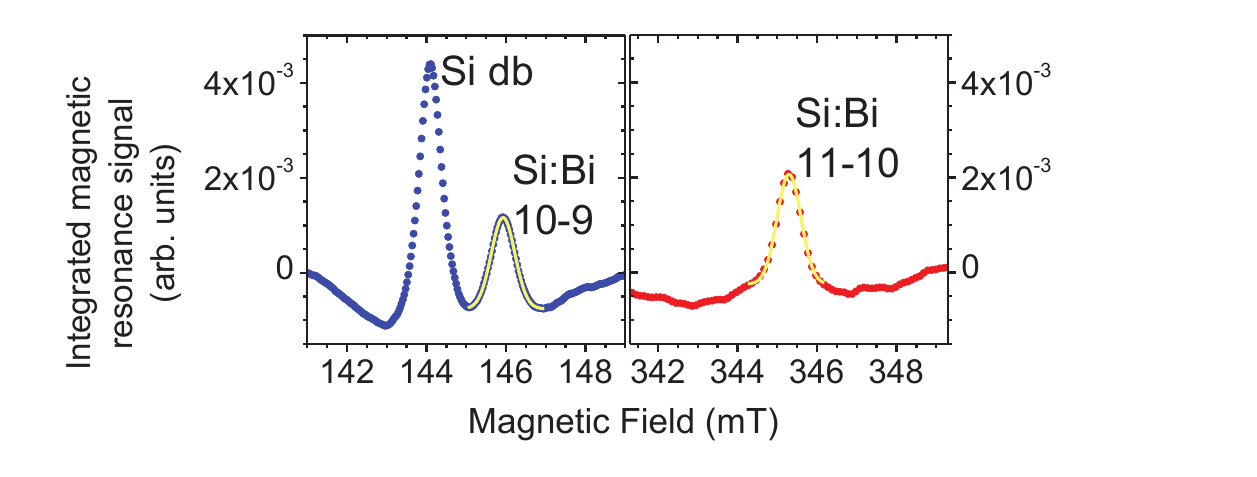}
\caption{
The areas under the Gaussians (obtained by integrating the spectra in \Fig{Morley2013_sband}) are in the
ratio area$_{11-10}$/area$_{10-9}$ = 1.2, in agreement with calculated values. Figure adapted from \citet{Morley2013}. 
}
\label{fig:Morley2013_sbandintensity}
\end{figure}

The ESR transition amplitudes are given by matrix elements of the electron $x$-spin projection:
\begin{eqnarray}
\bra{10} \hat{S}^x \ket{9}  &\propto& \sin{\left(\frac{\theta_{-4}}{2}\right)}, \\
\bra{11} \hat{S}^x \ket{10} &\propto& \cos{\left(\frac{\theta_{-4}}{2}\right)}.
\end{eqnarray}
The states $\ket{11}$ and $\ket{9}$ form the $m=-4$ doublet.
For $B=0.15$~T or the $\ket{11} \to \ket{10}$ transition,
$\theta_{-4} = 0.62\pi$ and for the other transition $\theta_{-4} = 0.28\pi$.
The ratio of the modulus of the transition amplitudes is therefore
\begin{equation}
\frac{|\bra{11} \hat{S}^x \ket{10}|} {|\bra{10} \hat{S}^x \ket{9}|} \simeq 1.1.
\label{eq:ampratio}
\end{equation}
The Rabi oscillation frequency is proportional to the transition amplitude.
Thus, this ratio (in the hybrid regime) predicts the same order of magnitude
manipulation for an ESR transition and a (high-field) NMR one.
Usually, the Rabi oscillations are 103 times slower for NMR relative
to ESR.
The transition rate is proportional to the square of the amplitude,
hence, this ratio predicts a ratio of magnetic resonance intensities
of about $1.1^2 \approx 1.2$.
This calculated value is in agreement with the measured intensities
as shown in \Fig{Morley2013_sbandintensity}.

\subsection{Rabi oscillations}

\begin{figure}[h!]
\centering\includegraphics[width=5.0in]{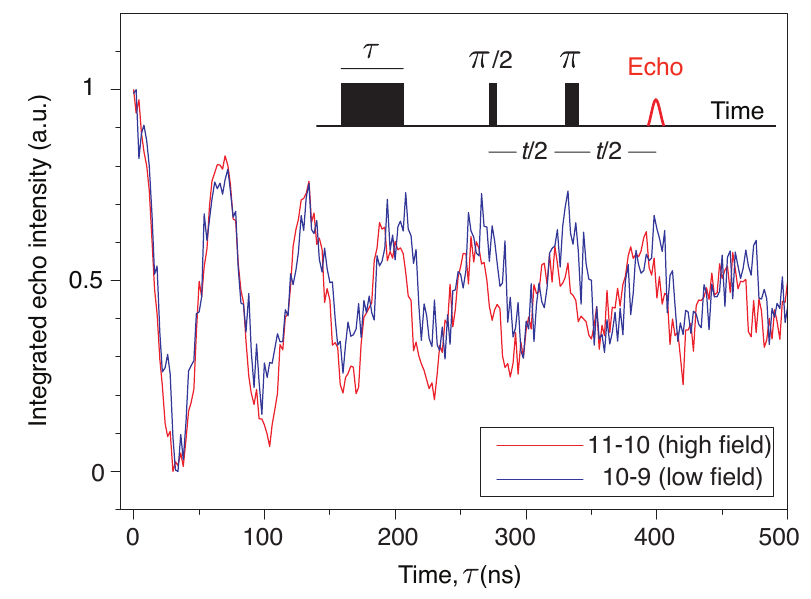}
\caption{Rabi oscillations demonstrate coherent control of both of the 4~GHz
hybrid electron-nuclear transitions. At higher magnetic fields, the 11-10
resonance becomes an ESR transition, whereas the 10-9 resonance
becomes an NMR transition. Controlling this NMR transition in the past
has required $\pi$ pulses of $\geq4~\mu$s, two orders of magnitude longer than the
32~ns $\pi$ pulses we use here. Figure adapted from \citet{Morley2013}.
}
\label{fig:Morley2013_rabi}
\end{figure}

\begin{figure}[h!]
\centering\includegraphics[width=5.0in]{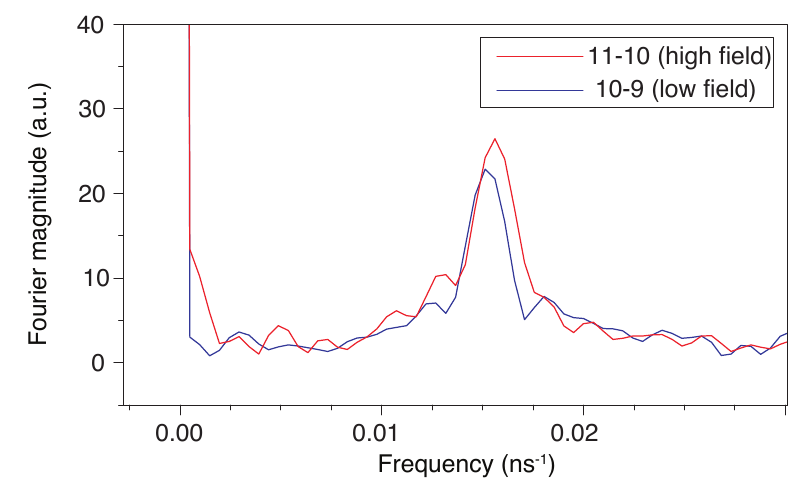}
\caption{.
Fourier transforming the Rabi oscillations in \Fig{Morley2013_rabi}
reveals that the 11-10 transition experiences 10\% faster nutation, as
expected. Pulsed measurements used
16~ns $\pi/2$ pulses and 32~ns $\pi$ pulses with two-step phase cycling. Figure adapted from \citet{Morley2013}. 
}
\label{fig:Morley2013_rabifourier}
\end{figure}

Rabi oscillations were used to demonstrate coherent control of the hybrid
electron-nuclear qubit. The pulse sequence of the experiment is as follows:
$\tau-\pi/2-t-\pi-t-\text{echo}$.
The experiment varies the first pulse $\tau$, rotating the qubit
from the upper towards the lower transition by some polar angle on the Bloch sphere.
The lower eigenstate is reached if the angle is $\pi$ and the time taken is the duration of the $\pi$-pulse as described in \Chap{background}.
The usual Hahn echo experiment follows the first pulse in order
to obtain a good signal. The period of Rabi oscillations is simply
twice the $\pi$-pulse time. The time taken for the $\pi$-pulse is quoted as
the qubit manipulation time.

The Rabi oscillations for the two transitions at 4~GHz are shown in
\Fig{Morley2013_rabi}. The Fourier transformation
of the Rabi signal (\Fig{Morley2013_rabifourier}) reveals that the inverse of the Rabi frequency is
about $1/66$~ns$^{-1}$, corresponding to a qubit manipulation
time of 32~ns. The ratio of the two Rabi frequencies extracted from the experimental data is about 1.1
and is in agreement with the calculation in \Eq{ampratio}.
In the high-field limit, the $\ket{10} \to \ket{9}$ transition
would be a pure NMR transition and thus require manipulation times
of order microseconds.

\section{Conclusion}

In summary, we introduced the spin Hamiltonian of the hybrid electron-nuclear qubit
and analytical expressions for its energies and eigenstates.
We then discussed the bismuth donor as a hybrid qubit.
The calculated eigendecomposition of the hybrid qubit leads to resonance positions in excellent
agreement with experiment at both $X$-band and $S$-band excitation frequencies, commonly used for ESR studies.
By accessing ESR-forbidden transitions by operating in the hybrid regime,
a factor of $125$ speed-up can be achieved in the qubit manipulation time, 
from 4~$\mu$s in the high-field limit to 32~ns in the hybrid regime.
However, we note that this speed-up is limited
by the power of the microwave field driving the transition.
The fundamental speed-up, on the other hand, is limited by the ratio of the electronic
to host nuclear gyromagnetic ratios, and is a factor of $1/\delta_\text{Bi} \approx 4,000$.

\chapter[Interaction of Hybrid Qubit with a Nuclear Spin Bath]{Interaction of Hybrid Qubit \\ with a Nuclear Spin Bath}
\label{chap:interaction}

The content of this chapter was published in \citet{Balian2012}.
Here, we present pulsed ENDOR experiments using bismuth-doped silicon
which enable us to characterise the coupling between the hybrid qubit
and the surrounding spin bath of \sitn\ impurities. This spin bath provides the 
dominant decoherence mechanism (nuclear spin diffusion) at low temperatures
($< 16$~K) for the hybrid qubit. At magnetic fields corresponding to OWPs,
we demonstrate a collapse in the strength of the qubit-bath interaction.
This serves as a clear spectroscopic signature of OWPs at which decoherence
is suppressed as shown in \Chap{decoherence}.

Relevant for decoherence of the hybrid qubit, the experiments suggest that anisotropic hyperfine contributions are comparatively weak,
and isotropic couplings dominate; hence anisotropic couplings can be safely neglected
in calculating hyperfine couplings for our decoherence simulations. Importantly
this means that the form of the suppression of decoherence at OWPs is largely independent of crystal orientation.

The central hybrid qubit is coupled to the \sitn\ spin bath via the electron-nuclear
hyperfine interaction introduced in \Sec{hyperfine}.
In general, the hyperfine interaction is in tensor form and includes
the effects of both isotropic and anisotropic couplings. Anisotropic couplings
depend on the direction of the magnetic field relative to the crystal (or equivalently, the `sample orientation').
The crystal structure of silicon is described in \App{silicon}.

The pulsed ENDOR measurements (as opposed to continuous-wave) also motivate the possibility of
addressing nuclear impurity spins for quantum information applications.
We present decoherence mechanisms for such spins in \Chap{nucleardecoherence}.

\section{Pulsed ENDOR Measurements}

The nature of the qubit-bath interaction is investigated
by means of pulsed ENDOR \citep{Schweiger2001}.
Previous ENDOR studies of Si:Bi used radio frequencies
of at least several hundreds of MHz \citep{Morley2010,George2010},
and thus could not probe the weak couplings to the nuclear spin bath. In contrast,
radio frequencies of a few MHz were used in the work we present.
With this approach, we successfully measured the hyperfine couplings of the bismuth donor
to \sitn\ impurities and determined their anisotropy.

The experiments spectroscopically resolve and characterise a
set of distinct $\mathcal{J}_n$ as defined in \Eq{hyperfine},
corresponding to occupancy of inequivalent lattice sites by \sitn\ impurities,
and whether or not these have any anisotropic character.

\subsection{Experimental Method}

Pulsed ENDOR experiments were carried out by Dr. Micha Kunze and Professor Chris Kay
at UCL who are co-authors of \citet{Balian2012}.
The Davies ENDOR pulse sequence was used \citep{Schweiger2001}, which applies the sequence
$\pi_{\text{mw}}-\tau_1-\pi_{\text{rf}}-\tau_2-\frac{\pi}{2}{\text{\scriptsize mw}}-\tau_3-\pi_{\text{mw}}-\tau_3-\text{echo}$,
where the microwave frequency (mw) is chosen to excite one ESR
transition and the radio frequency (rf) is stochastically varied between
$2-12$~MHz or $2-7$~MHz to excite all nuclear spin transitions
in this region. $256$~ns long $\pi_{\text{mw}}$-pulses and a $128$~ns
long $\frac{\pi}{2}{\text{\scriptsize mw}}$-pulse were used. For optimal signal-to-noise
ratio and resolution, a $\pi_{\text{rf}}$-pulse of $10~\mu\text{s}$ was used.
Pulse delays were set to $\tau_1=1~\mu\text{s}$, $\tau_2=3~\mu\text{s}$, and
$\tau_3=1.5~\mu\text{s}$ and a shot repetition time of 1.3~ms was
employed to give a good signal-to-noise ratio. All experiments were carried out
at 15~K on an E580 pulsed EPR ($\equiv$ESR) spectrometer (Bruker Biospin) equipped with
pulsed ENDOR accessory (E560D-P), a dielectric ring ENDOR resonator
(EN4118X-MD4), a liquid helium flow cryostat (Oxford CF935), and a radio frequency amplifier
(ENI A-500W). The donor concentration of the sample was $3 \times 10^{15}$~cm$^{-3}$ and
the magnetic field was directed perpendicular to the $(111)$ crystal plane.

\begin{figure}[h!]
\centering\includegraphics[width=4in]{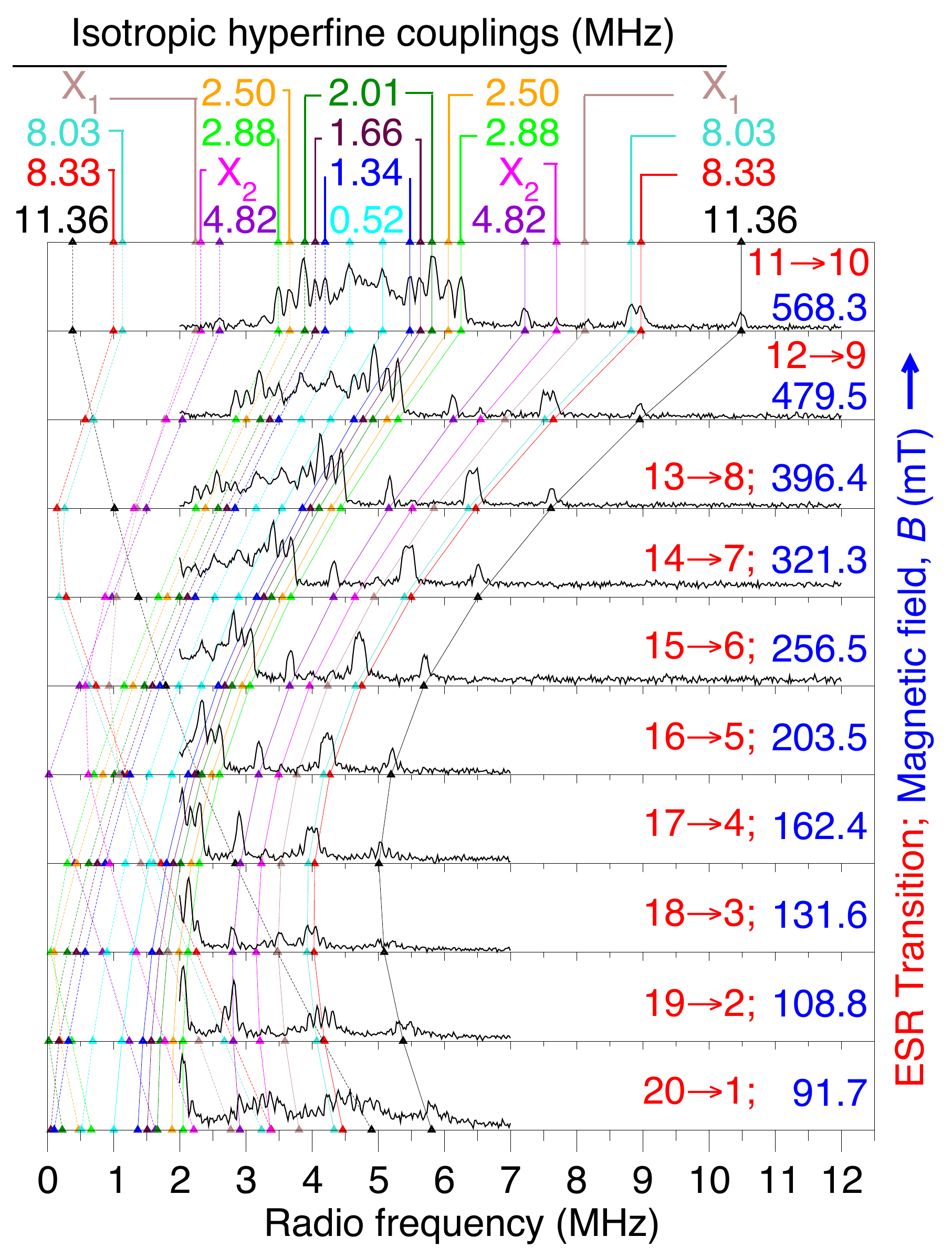}
\caption{
Pulsed ENDOR measured for bismuth-doped silicon with
frequency 9.8~GHz at which ten ESR lines are observed,
the resonance peaks due to interactions of the donor with \sitn\
nuclei at inequivalent lattice sites. The isotropic hyperfine couplings
were extracted from the spectrum at the highest magnetic field.
As the field is varied, the smooth lines follow the resonance positions
according to \Eq{endor}. Solid and dotted lines distinguish
between the two peaks observed for each coupling, each corresponding to one of
the two donor levels involved in the ESR transition.
Only the peaks labelled $X_1$ and $X_2$, in addition to a third pair
not resolved here, were found to show anisotropy from performing ENDOR as a
function of crystal orientation. Figure adapted from \citet{Balian2012}.
}
\label{fig:Balian2012_endor}
\end{figure}

\subsection{Experimental Results}

While not offering the higher frequency resolution attainable with
CW ENDOR \citep{Feher1959,Hale1969a,Hale1969b}, the pulsed ENDOR measurements
permit us to adequately constrain calculated hyperfine couplings and thus
demonstrate the reliability of our numerical simulations of $T_2$.
In particular, we established that isotropic couplings
to the spin bath dominate over anisotropic ones. As mentioned above, a further motivation for {\em pulsed} experiments
in contrast to CW is to investigate the feasibility of an alternative
possibility for QIP: to simultaneously manipulate the \sitn\ spins as
spin-1/2 qubits, along with the donors \citep{Akhtar2012}.
Measured ENDOR spectra at ESR excitation frequency $f\simeq9.755$~GHz are presented in \Fig{Balian2012_endor},
together with a list of the extracted qubit-bath hyperfine couplings.

\section{Calculating Resonance Positions}

For the magnetic field range $B \simeq 0.1-0.6$~T in \Fig{Balian2012_endor},
there is significant mixing of the high-field Si:Bi energy eigenstates (\Fig{Balian2015_sibi}).
We now proceed to derive an expression for the ENDOR resonance positions,
taking into account the effect of the central-state mixing.
We note that in \Eq{hyperfine} describing the hyperfine interaction in \Sec{hyperfine},
the Fermi contact part is always isotropic. The remainder of the interaction
has both anisotropic and isotropic components.
Here, since we are interested in measurements of the interaction,
we make no reference to the Fermi contact part of the interaction
and instead divide the tensor into isotropic and anisotropic components.
Also, since we are measuring the interaction, there is no need to write down the spatial
electronic wavefunction.

\subsection{Hyperfine Tensor}

To investigate the qubit-bath interaction, we add to the central spin Hamiltonian
the interaction Hamiltonian and a single spin-$1/2$ \sitn\ nuclear Zeeman term:
\begin{equation}
\hat{H}_\text{ENDOR} = \hat{H}_\text{Si:Bi} + \gamma_n B \hat{I}^z +  \hat{ {\bf I} } \cdot \mathcal{J} \cdot \hat { {\bf S} },
\end{equation}
where $\mathcal{J}$ here is the hyperfine tensor for coupling to a single \sitn\ nucleus and $\gamma_n$ is the \sitn\ gyromagnetic ratio given in \Table{gammas}.
The donor Hamiltonian $\hat{H}_\text{Si:Bi}$ is given by \Eq{donorH}.
Interactions of \sitn\ nuclei at different lattice sites with the same donor electron can be treated by summing independent such hyperfine terms:
\begin{equation}
\hat{H}_\text{int} = \hat{H}_\text{Si:Bi} + \sum_l \left(
\gamma_n B \hat{I}_l^z +  \hat{ {\bf I} }_l  \cdot \mathcal{J}_l \cdot \hat { {\bf S} }
\right).
\end{equation}
Nuclear \sitn--\sitn\ and $^{209}$Bi--\sitn\ dipolar interactions are much weaker and thus negligible compared to the electron-nuclear hyperfine interaction.
This is because the latter involves the product of electronic and nuclear gyromagnetic ratios
which is $\sim 1000$ stronger than a product of two nuclear gyromagnetic ratios as seen from \Table{gammas} and \Table{donors}.

The hyperfine tensor is diagonal in the molecular frame (MF); this coordinate system is one in which the external field direction
is collinear with the line connecting the central bismuth and \sitn\ sites
\citep{Schweiger2001}:
\begin{equation}
\mathcal{J}^{\text{MF}} =
\left( \begin{array}{ccc}
 				a_\text{iso} - T & 0 & 0  \\
				0 & a_\text{iso} - T & 0 \\
				0 & 0 & a_\text{iso} + 2T \\
				\end{array} \right),
\end{equation}
where $T$ and $a_\text{iso}$ are scalars.
Rotating the operator $\left(\hat{ {\bf I} } \cdot \mathcal{J} \cdot \hat { {\bf S} }\right)^\text{MF}$
by angle $\theta$ towards the laboratory $z$-axis (along $B$), where $\theta$ is the angle between the $z$-axis and the line connecting the \sitn\ spin and the donor site
in the molecular frame,
gives $\left(\hat{ {\bf I} } \cdot \mathcal{J} \cdot \hat { {\bf S} }\right)$
in the laboratory frame in terms of $T$, $a_\text{iso}$ and the rotation angle $\theta$:
\begin{equation}
\hat{H}_\text{ENDOR} =
\hat{H}_\text{Si:Bi} + \gamma_n B \hat{I}^z
+ \alpha \hat{I}^z \hat{S}^z
+ \beta \hat{I}^x \hat{S}^z,
\end{equation}
where
\begin{eqnarray}
\alpha &=& \left[ \left(a_\text{iso} - T\right) + 3T^{2}\cos^{2}\theta \right], \nonumber \\
\beta &=& 3T\sin\theta\cos\theta,
\end{eqnarray}
and we have ignored non-secular terms involving $\hat{S}^x$ and $\hat{S}^y$.
This secular approximation which reduces the hyperfine interaction to the simpler form above
is motivated by the disparity between the central spin (electronic Zeeman and host hyperfine)
and donor-\sitn\ hyperfine energy scales.
Non-secular terms lead to transitions between ESR levels.
To simulate the ENDOR resonance positions,
we wish to describe nuclear transitions (MHz) for each of the two ESR levels (GHz) independently.
The Hamiltonian above appears in \citet{Schweiger2001} and
the term $\hat{S}^z \hat{I}^y$ vanishes if we choose the \sitn\ nucleus to lie in the $xz$-plane.

\subsection{Expression for Resonance Positions}

For a transition between the central spin states
$\ket{u} \to \ket{l}$, we choose the basis formed by the states
\begin{equation}
\begin{array}{rl}
\ket{u}&\otimes~~\ket{\uparrow}, \\
\ket{u}&\otimes~~\ket{\downarrow}, \\
\ket{l}&\otimes~~\ket{\uparrow}, \\
\ket{l}&\otimes~~\ket{\downarrow},
\end{array}
\end{equation}
where $\ket{\uparrow}$ ($\ket{\downarrow}$) is the spin up (down) eigenstate of the non-interacting \sitn\ spin-1/2.
The matrix representation of the $4\times4$ Hamiltonian in this basis is
\begin{equation}
\begin{split}
{\bf H}_\text{ENDOR} &=
\left( \begin{array}{ccc}
 				{\bf h}_u & {\bf 0} \\
				{\bf 0} & {\bf h}_l
				\end{array} \right) \\
{\bf h}_{u/l} &=
\left( \begin{array}{ccc}
E_{u/l} + \frac{\gamma_n B}{2} + \frac{\alpha P_{u/l}}{4} & \frac{\beta P_{u/l}}{4}  \\
\frac{\beta P_{u/l}}{4} & E_{u/l} - \frac{\alpha P_{u/l}}{4} - \frac{\alpha P_{u/l}}{4}
\end{array} \right)
\end{split}
\label{eq:endorH}
\end{equation}
where $E_{u/l}$ are the energies of the central states $\ket{u}/\ket{l}$.
Straightforward diagonalisation of ${\bf h}_{u/l}$
gives the expression for the ENDOR resonance frequency at ESR level $\ket{i}, i= u,l$, written in units of Hz:
\begin{equation}
\Delta f ^{(i)} =
\frac{1}{2\pi} \sqrt{ 
\left( \frac{\gamma_n B}{2}\omega_{0} + \frac{\alpha_{\perp} P_i(B) }{4} \right) \sin^2\theta +
\left( \frac{\gamma_n B}{2}\omega_{0} + \frac{\alpha_{\parallel} P_i(B) }{4} \right) \cos^2\theta
 },
\end{equation}
with $\alpha_{\parallel}$ and $\alpha_{\perp}$ given by
\begin{equation}
\begin{split}
\alpha &= \alpha_{\parallel}\cos^{2}\theta + \alpha_{\perp}\sin^{2}\theta, \\
\alpha_{\parallel} &= a_{\mathrm{iso}} + 2T, \\
\alpha_{\perp} &= a_{\mathrm{iso}} - T.
\end{split}
\label{eq:fullendor}
\end{equation}
Note that in the case when there is no angular dependence of an ENDOR line (i.e.\ $T=0$),
\Eq{fullendor} reduces to
\begin{equation}
\Delta f_{\text{iso},l}^{(i)} (B) =
\frac{1}{2\pi} \left|
\gamma_n B + \frac{a_{\text{iso},l}P_i(B)}{2}
\right|
\label{eq:endor}
\end{equation}
written for coupling to a single \sitn\ at site $l$.
\Eq{endor} above can also be obtained using first order time-independent
perturbation theory with the hyperfine interaction taken as the perturbation Hamiltonian.
It is also in perfect agreement with full numerical
diagonalisation of the Hamiltonian in \Eq{endorH}.

\section{Extracting Qubit-Bath Couplings}

The isotropic couplings in \Fig{Balian2012_endor} were extracted from the
measured spectra by fitting to the data Gaussians of equal width and using the expression
\Eq{endor}.
The same expression and a single set of couplings gave excellent agreement with
data at 10 different magnetic fields (ESR lines). In particular, the observed pattern
of half a dozen highest frequency \sitn\ resonances moving to
a minimum at $B \simeq 0.2$~T, then increasing again, is directly attributable
to mixing of the states of the bismuth donor: i.e., here $|P_i|$ has a minimum.

The following procedures were adopted to extract the central positions of ENDOR peaks.
We first deal with those peaks which showed no angular dependence as a function
of the crystal orientation (isotropic case). We later discuss the
experimental data as a function of crystal orientation.

\subsection{Isotropic Case}

Multiple Gaussians sharing a full width at half maximum (FWHM) of $0.12$~MHz
gave a good fit to the $\ket{11}\rightarrow\ket{10}$ experimental spectrum (ESR line 10, counting from low to high field, which is equivalent
to the $i$ label for the final state).
The central positions of peaks at radio frequencies higher than the \sitn\ nuclear Zeeman frequency ($\gamma_n B \approx 4.8$~MHz)
were used to extract hyperfine couplings employing \Eq{endor} rearranged for $a_\text{iso}$.
We note that this was done only for peaks which showed no angular dependence; the anisotropic case is discussed in the following subsection (\Fig{Other_anisotropic}).
This set of extracted couplings and \Eq{endor} were then used to predict the resonance positions at frequencies lower than
$\gamma_n B \approx 4.8$~MHz for the same ESR line (i.e.\ for the $\ket{11}$ ESR level).
The last step was repeated for each of the two ESR states of the other nine Si:Bi ESR lines observed at $X$-band -- i.e. as
a function of $B$, using \Eq{endor}.

The predicted positions for the nine ESR lines are in excellent agreement with the experimental peak positions as seen in \Fig{Balian2012_endor}.
The appearance and number of multiple sideband peaks at lower fields
(e.g.\ see the lowest-field ESR line) were found to be dependent on the radio
frequency pulse length used in the experiment, and thus such peaks were not attributed to \sitn\ sites.

\subsection{Anisotropic Case}

Ten out of the twelve couplings extracted from data were found to be purely
isotropic. The highest-field spectrum was measured for a range of crystal
orientations and only three weak-intensity ENDOR peaks showed orientation-dependent
frequencies and hence anisotropy. Two are indicated by $X_1$ and $X_2$ in
\Fig{Balian2012_endor}: The corresponding two couplings with non-zero anisotropy
were found to have ($a_{\text{iso,$X_1$}}\simeq2.8$,~$T_{X_1}\simeq2.4$)~MHz and
($a_{\text{iso,$X_2$}}\simeq0.4$,~$T_{X_2}\simeq2.8$)~MHz by fitting the more general form for
resonance positions with non-zero $T$,
\Eq{fullendor}. A previous ESEEM (electron spin echo
envelope modulation) study identified a single anisotropic
coupling \citep{Belli2011}, attributed to $E$-shell (nearest neighbor)
\sitn. The third line we identify is fitted by coupling constants
consistent with the anisotropic coupling in \citet{Belli2011}.
For most crystal orientations, this line is masked by much higher intensity
lines arising from isotropic couplings.

\begin{figure}[h!]
\centering\includegraphics[width=5in]{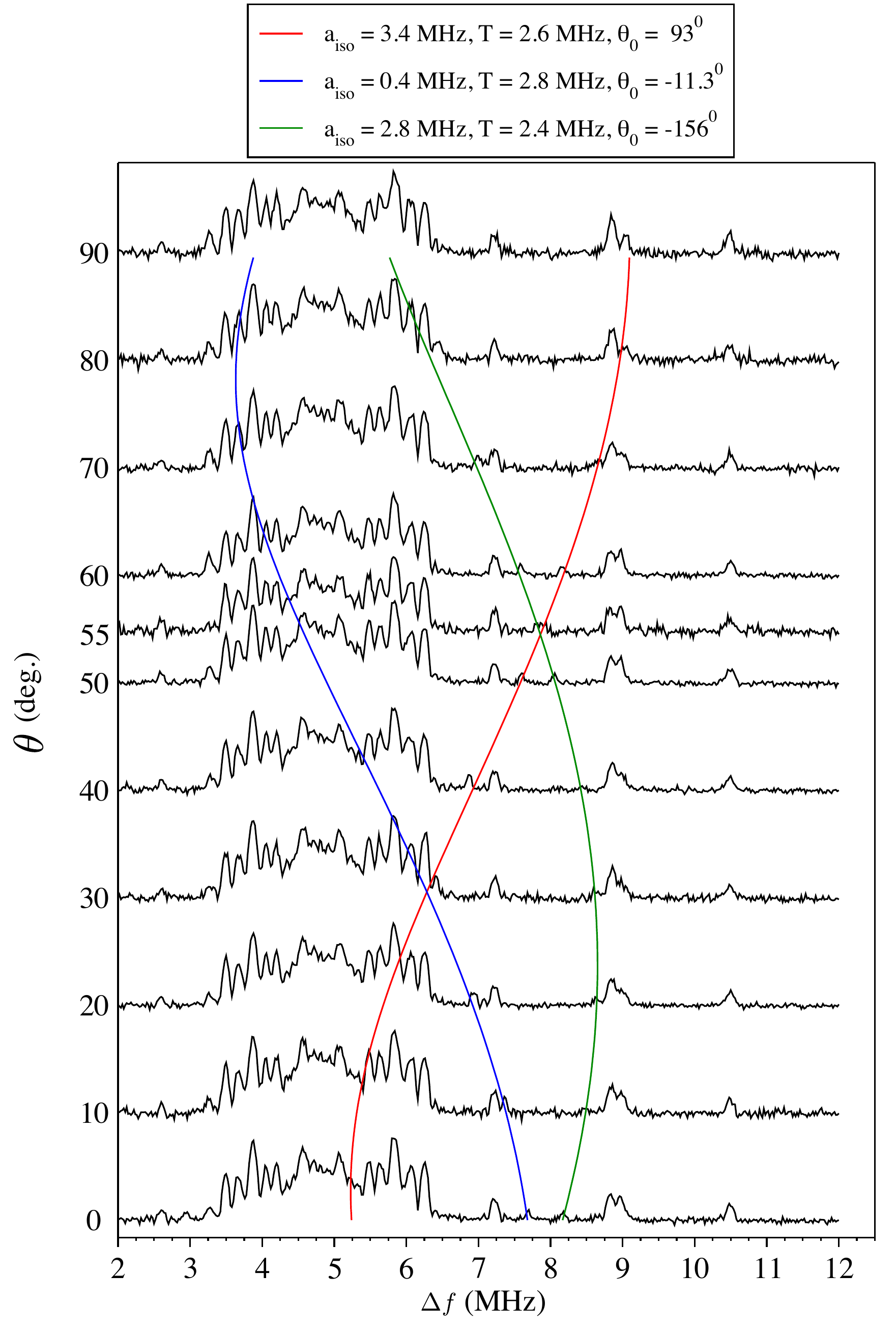}
\caption{
Si:Bi ENDOR spectra for the $\ket{11} \to \ket{10}$ ESR transition at 9.75~GHz microwave excitation obtained as a function of $\theta$,
where $(\theta-\theta_{0})$ is the angle between the external magnetic field and the $[111]$-direction. The three smooth lines are fits of \Eq{fullendor},
and were used to extract values of the isotropic and anisotropic parts of the hyperfine coupling $a_{\mathrm{iso}}$ and $T$ respectively,
as well as the offset angle $\theta_0$ which was not known during the experiments.
}
\label{fig:Other_anisotropic}
\end{figure}

We now describe the procedure of extracting the three peaks which were found to show $\theta$ dependence.
\Fig{Other_anisotropic} shows line 10 spectra obtained for a range of orientation angles
between the external field and the $[111]$-direction of the sample crystal.
However, the other angle required to fully determine the crystal orientation relative to
the magnetic field was unknown in the experiment.
Thus, the functional form $\Delta f^{(11)} \left(\theta - \theta_{0} , a_\text{iso},T\right)$ (\Eq{fullendor})
was used with free parameters for each peak $a_\text{iso}$, $T$ and a constant offset $\theta_{0}$.
We managed to fit three curves with this expression to all the spectra as $\theta$ was varied.
The peak with $\left(a_{\mathrm{iso}},T\right)\approx\left(3.4,2.6\right)~\mathrm{MHz}$ is consistent with the one previously
identified in ESEEM spectra for ESR line 10 [$\left(a_{\mathrm{iso}},T\right)=\left(3.36\pm0.03,2.56\pm0.03\right)~\mathrm{MHz}$ \citep{Belli2011}].
However, this peak is not apparent in \Fig{Balian2012_endor} due to its position and small area at the orientation angle used
to perform the measurements in \Fig{Balian2012_endor}.
The two other couplings were found to have $\left(a_{\mathrm{iso}},T\right)\approx\left(0.4,2.8\right)~\mathrm{MHz}$ and $\left(a_{\mathrm{iso}},T\right)\approx\left(2.8,2.4\right)~\mathrm{MHz}$.
It can be seen that for the peaks with non-zero $T$, the difference between theory and experiment can be up to about $0.1$~MHz at low field
(see, for example, ESR lines 5 to 2 in \Fig{Balian2012_endor}). The deviation is likely due to the experimental uncertainties
in the orientation angle $\theta$. In obtaining the rotation spectra in \Fig{Other_anisotropic}, the experimental uncertainty in $\theta$ was estimated at $\epsilon_{\theta}\approx\pm2^{0}$. This leads to a maximum of $\epsilon_{\Delta f}\approx\pm0.1$~MHz shift in the resonance positions of those
peaks for which values of $a_{\mathrm{iso}}$ and $T$ were extracted from the rotation spectra.

\section{Collapse of Couplings}

The magnitude of the polarisation $|P_i(\omega_0)|$ becomes small
close to OWPs \citep{Balian2014} and $df/dB=0$ minima \citep{Mohammady2010,Mohammady2012}.
Thus, \Eq{endor} tends to the $^{29}\text{Si}$ Zeeman frequency $\gamma_n B$.
This also holds in the anisotropic case (\Eq{fullendor}).
In effect, near these points, the donor might be said to approximately decouple
from the bath. For example, for the ESR transition $|12\rangle \to |9\rangle$
$P_{12}(B)=0$ at $B=157.9$~mT and $P_{9}(B)=0$ at $B=210.5$~mT.
We note that there is however no $B$-field value where both the upper and lower
levels have $|P_i(B)=0|$: As we see later, this is not 
actually essential for complete suppression of spin diffusion.
The actual OWP for suppression of decoherence is at $B=188.0$~mT, where $P_{12}=P_{9}$ \citep{Balian2014}.
For this transition, the $df/dB=0$ point occurs when
\begin{equation}
P_{12} - P_{9} - \frac{2\delta_{\text{Bi}}}{(1+\delta_{\text{Bi}})} = 0.
\end{equation}
The third term on the left is of order $10^{-4}$ and hence the OWP and frequency-field
minimum are very close as discussed earlier.

\begin{figure}[h]
\centering\includegraphics[width=5.2in]{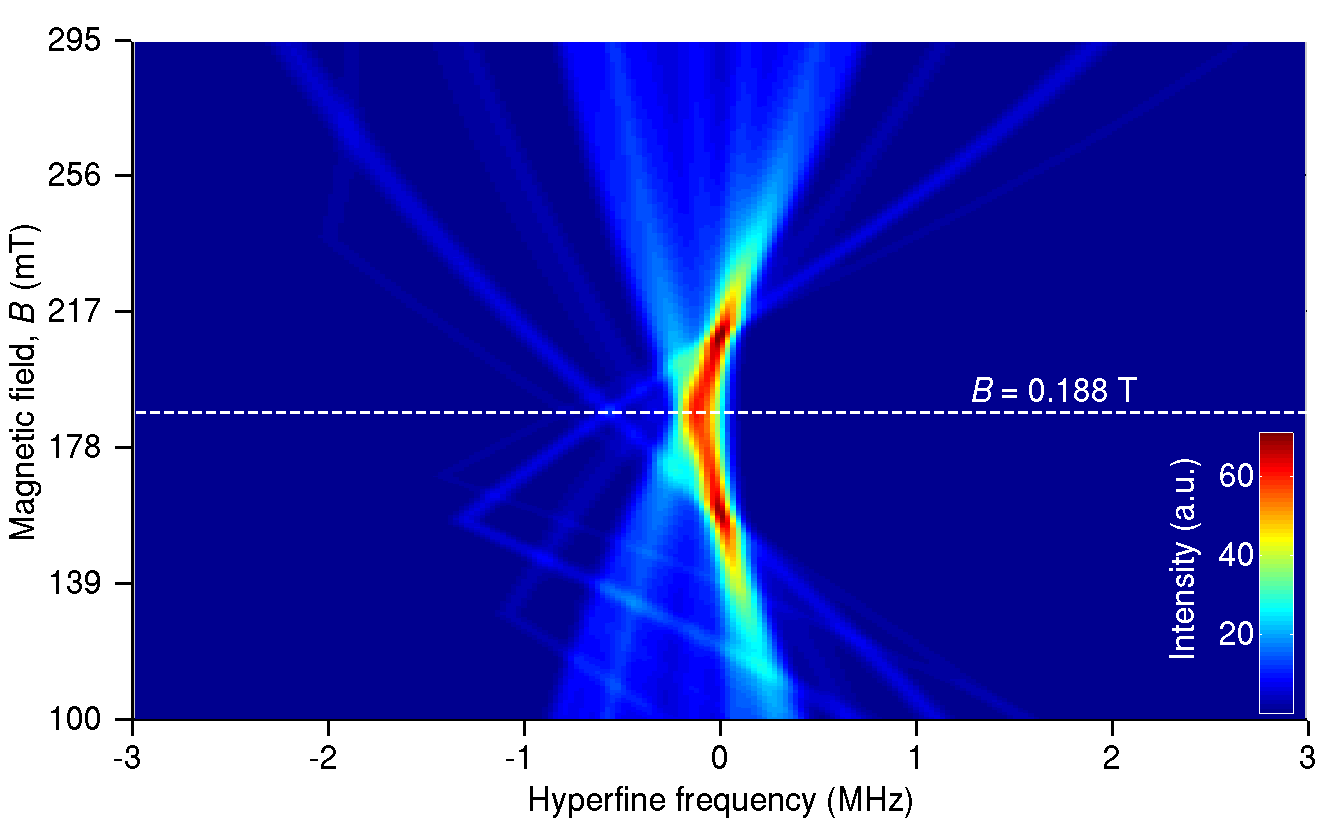}
\caption{
    Simulated ENDOR as a function of magnetic field $B$,
    showing collapse of the hyperfine couplings
    for the $|12\rangle \to |9\rangle$ Si:Bi ESR transition.
    The OWP is at 0.188~T.
    Figure adapted from \citet{Balian2012}.
}
\label{fig:Balian2012_collapse}
\end{figure}

\subsection{Isotropic Case}

\Fig{Balian2012_collapse} shows a colour map of the qubit-bath hyperfine spectrum
for a high density of magnetic fields for the $\ket{12} \to \ket{9}$ Si:Bi ESR line.
The spectra were simulated as a function of $B$,
using \Eq{endor} and centered about the \sitn\ nuclear Zeeman frequency.
Strikingly, as $B$ approaches the OWP at 0.188~mT, the ``comb''
of radio frequency hyperfine lines narrows to little more than the width of a single line.
This suggests a drastic reduction in the value of the hyperfine couplings,
indicating that the bismuth donor has become largely decoupled from the
$^{29}\text{Si}$ spin bath. Note that as we see in \Chap{decoherence}, \Chap{formula} and \Chap{dynamicaldecoupling},
the enhancement of coherence times involves treatment of the qubit-bath entanglement
and the fact that hyperfine couplings are reduced does not fully explain the suppression of decoherence.
Nevertheless, this behaviour of collapse in the couplings does provide a spectroscopic signature
of OWPs or frequency-field extrema.

\begin{figure}[h]
\centering\includegraphics[width=5.2in]{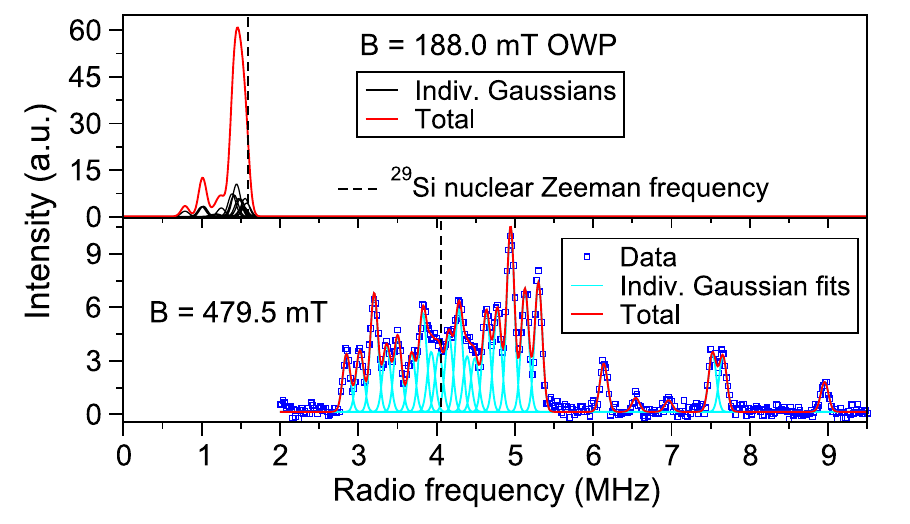}
\caption{
    Simulated ENDOR at the $B=188.0$~mT OWP (upper panel)
    and experimental spectrum at 9.755~GHz (lower panel), for the $\ket{12} \to \ket{9}$ Si:Bi ESR transition.
    Individual Gaussian fits to the data and their sum are also shown in the lower panel.
    Figure adapted from \citet{Balian2012}.
}
\label{fig:Balian2012_line9data}
\end{figure}

\begin{figure}[h]
\centering\includegraphics[width=5in]{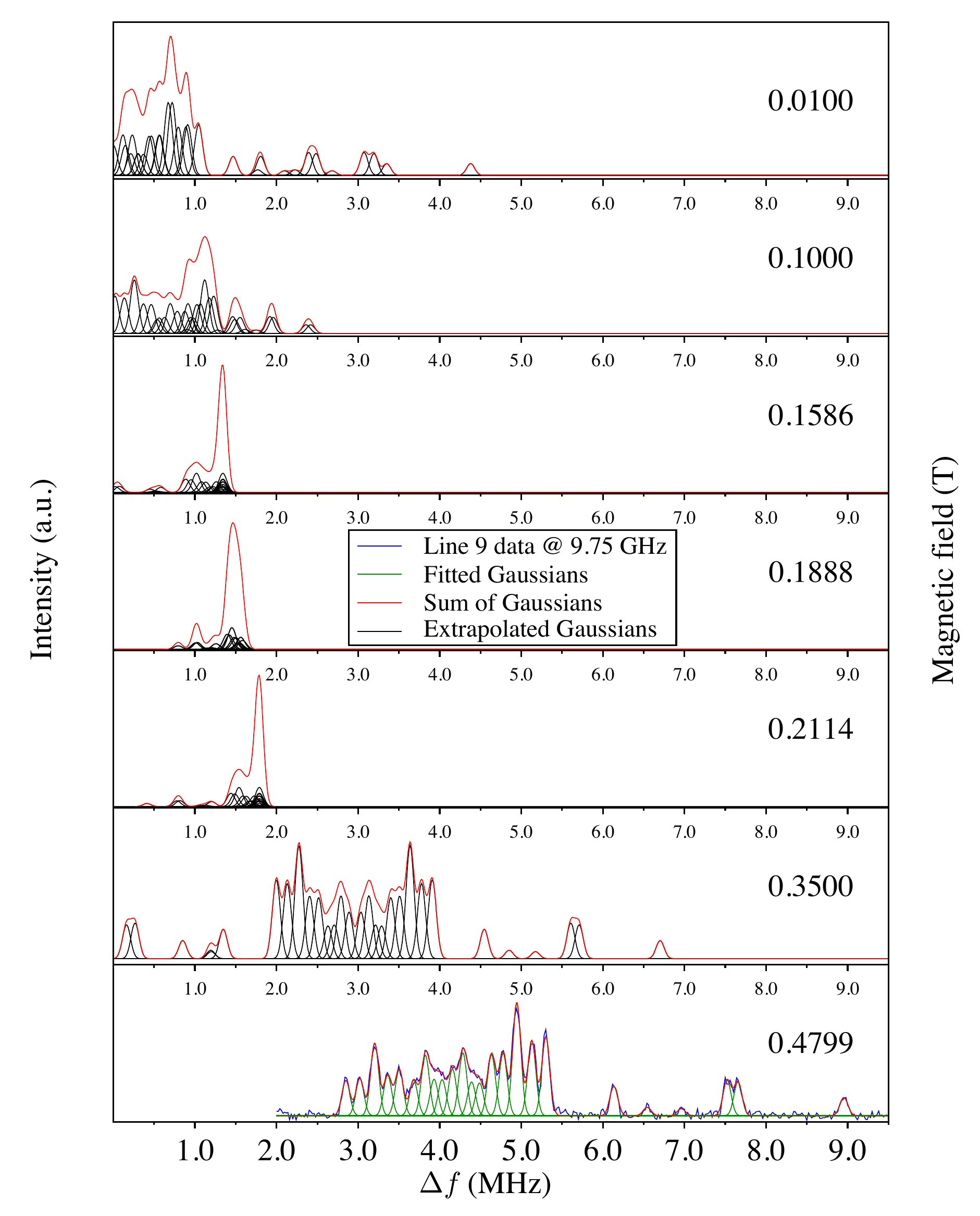}
\caption{
Experimental ($B=0.4799$~T, at ESR excitation frequency 9.75~GHz) and extrapolated (simulated) ENDOR spectra for Si:Bi,
for the $\ket{12} \to \ket{9}$ ESR transition.
Gaussian peaks were fitted to the experimental spectrum.
The spectra for $B=0.2114$~T and $B=0.1586$ correspond to the cancellation resonance for levels $\ket{9}$ and $\ket{12}$ respectively.
The $B=0.189$~T spectrum is near the OWP.
To construct the theoretical spectra, Gaussians peak centres from the experimental spectrum were shifted according to \Eq{endor} as the field $B$ was varied.
}
\label{fig:Other_extrapolate}
\end{figure}

The collapse in the hyperfine couplings is illustrated further in
\Fig{Balian2012_line9data}. The lower panel shows
the measured spectrum at 9.755~GHz. Using our experimentally determined hyperfine couplings,
the corresponding spectrum at the OWP is shown in the upper panel
of \Fig{Balian2012_line9data}, demonstrating clearly the narrowing of the spectrum
[corresponding to the same parameters as \Fig{Balian2012_collapse} but at
the precise field value of the OWP].
The spectra for a range of magnetic fields is also shown in
\Fig{Other_extrapolate}, including at the cancellation resonance points.
The panels in \Fig{Other_extrapolate} can be thought of as horizontal slices of the high-$B$-density plot in \Fig{Balian2012_collapse},
rotated to show the intensity on the vertical axis.
It can be seen clearly that the most significant suppression of the mostly isotropic couplings
occurs near the OWP at $B \simeq 188$~mT.

In constructing the theoretical spectra in \Fig{Balian2012_collapse}, \Fig{Balian2012_line9data} and \Fig{Other_extrapolate},
the various areas under Gaussian peaks and the $\text{FWHM} \approx 0.10$~MHz shared by all peaks were extracted from fitting to the $\ket{12} \to \ket{9}$ experimental spectrum.
For each coupling, the area of the lower frequency $\ket{12}$ peak was set equal to that of the the higher frequency $\ket{9}$ peak.
This was done in order to eliminate to some extent the linear damping in intensity as the radio frequency is lowered in pulsed ENDOR \citep{Schweiger2001}.
Finally, the extracted peak centres were shifted for lower fields according to \Eq{endor}.

It is worth mentioning that near $df/dB=0$ points or OWPs, since $P_u \simeq P_l$,
the ENDOR frequencies for the two ESR levels in the case of isotropic peaks becomes equal (\Eq{endor}),
resulting in a single peak in the ENDOR spectrum for each coupling.

\begin{figure}[h!]
\centering\includegraphics[width=5in]{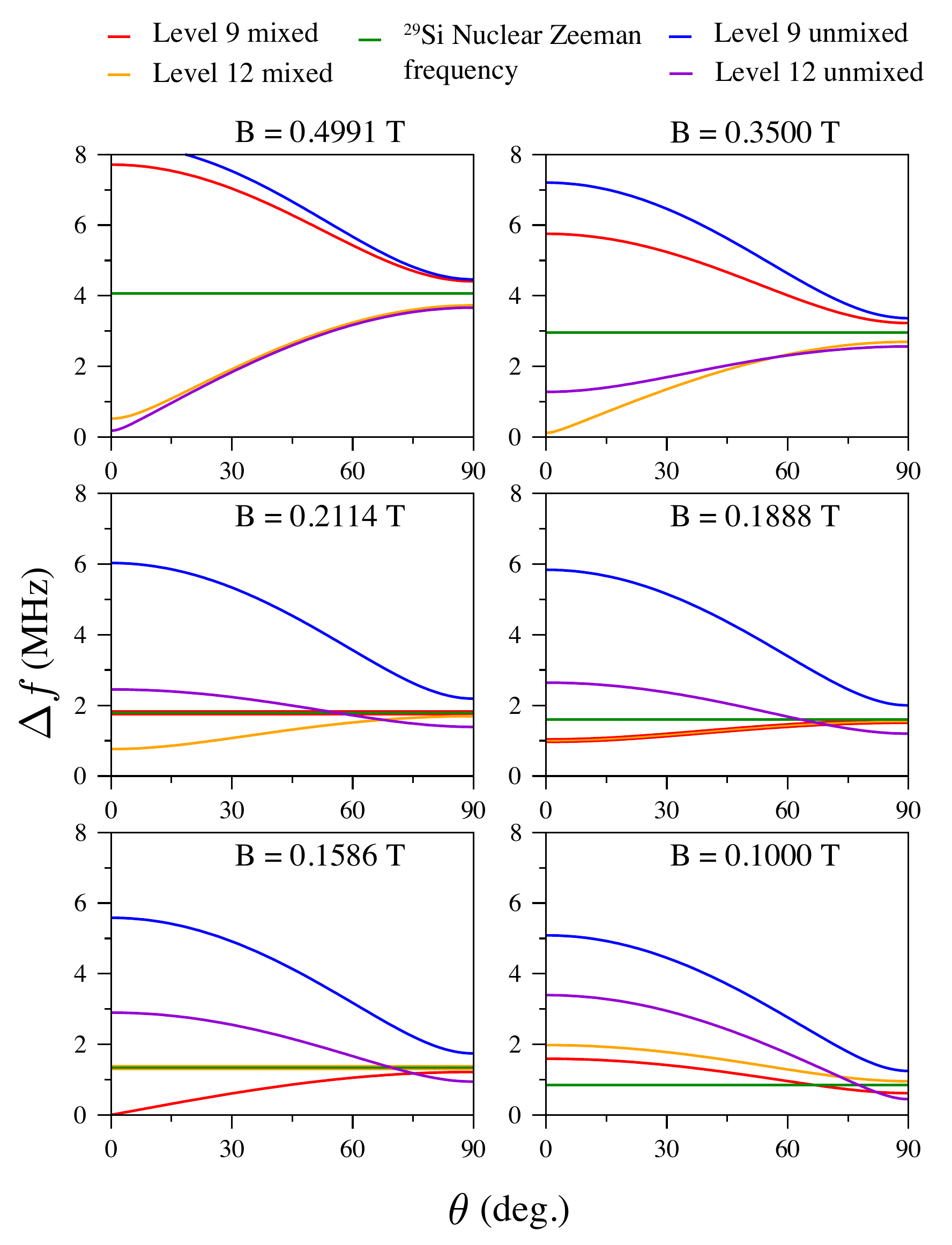}
\caption{
Dependence of the $\left(a_{\mathrm{iso}},T\right)=\left(3.36\pm0.03,2.56\pm0.03\right)~\mathrm{MHz}$ peak \citep{Belli2011} on the crystal orientation angle $\theta$
with and without the mixing polarisation terms for the $\ket{12} \to \ket{9}$ ESR transition in Si:Bi. The field $B=0.4799$~T corresponds to 9.75~GHz. Fields $B=0.2114$~T and $B=0.1586$~T are at the cancellation resonance for levels $\ket{9}$ and $\ket{12}$ respectively, and $B=0.1888$~T is near the OWP. The curves were obtained using
\Eq{fullendor} in the case of mixing included. For curves excluding mixing, the polarisation terms were fixed to $\pm$1 at all fields in \Eq{fullendor}.
}
\label{fig:Other_rotation}
\end{figure}

\subsection{Anisotropic Case}

\Fig{Other_rotation} shows the theoretical $\theta$ dependence of the peak with
$\left(a_{\text{iso}},T\right) = \left(3.36 , 2.56  \right)$ MHz \citep{Belli2011}
as the field is lowered for the $\ket{12} \to \ket{9}$ transition,
for the two cases of allowing $P(B)$ to vary with $B$ (mixing included), and fixing $|P(B)|=1$ (unmixed).
This anisotropic peak was identified for the unmixed level $\ket{10}$ \citep{Belli2011}. The results illustrate the significance of the mixing
term or polarisation $P(B)$ in causing the suppression of $\theta$ dependence, or equivalently the anisotropic part of the hyperfine coupling at fields close to OWP.

\section{Conclusion}

In conclusion, we presented measurements of the hyperfine couplings between a hybrid qubit (bismuth
donor) and a spin bath of $^{29}\text{Si}$ impurities which suggest that isotropic couplings dominate.
In using pulsed ENDOR as opposed to CW, we demonstrated the feasibility of \sitn\ nuclei as qubits. The decoherence of \sitn\ nuclei is the topic of \Chap{nucleardecoherence}.
We further demonstrated the suppression of couplings in both cases of isotropic and anisotropic couplings, serving as a clear signature of OWPs which are discussed
in \Chap{decoherence}, \Chap{formula} and \Chap{dynamicaldecoupling}.
Our results motivate both spectroscopic and decoherence measurements for the Si:Bi system in the excitation frequency range
$5 - 7.5$~GHz for ESR-type OWPs and around 1~GHz for NMR-type ones (\Fig{Other_dfdB}).

We note that by using much higher-resolution CW ENDOR and by performing experiments which adequately sample all crystal directions,
it is possible to map hyperfine couplings to shells of \sitn\ with known positions from the donor site,
as was achieved for the Si:P system in \citet{Hale1969a} and \citet{Hale1969b}.
However, this requires considerable experimental effort; also, as we shall see in later chapters,
hyperfine couplings calculated using effective mass theory (\Sec{hyperfine})
suffice to give coherence decays in excellent agreement with experiment.
Nevertheless, the order of magnitude and range of measured couplings agree 
with the calculated values, and this give us confidence in using the model described in \Sec{hyperfine} for our decoherence studies of the hybrid qubit.
Later in \Chap{nucleardecoherence}, for decoherence of \sitn\ nuclear impurities, we
propose a model which relies heavily on symmetries of the electronic wavefunction,
and thus would benefit much more from experimental `$J \to {\bf r}_n$' mapping offered by CW studies.

\chapter[Coherence Times of Hybrid Qubit]{Coherence Times of \\ Hybrid Qubit}
\label{chap:decoherence}

This chapter concerns the decoherence dynamics of the hybrid qubit.
We investigate the decoherence mechanism and calculate coherence times
using the CCE for the Hahn spin echo and single-spin FID.
We also compare our results to experimental coherence measurements.
Although the Hahn spin echo is the lowest order CPMG dynamical
decoupling sequence, it is often the case that the shortest measured $T_2$ times
for solid-state systems are reported for this sequence (in order to remove
static magnetic field inhomogeneities as discussed in \Sec{hahnecho}).
Therefore, we discuss the Hahn echo in the current chapter, and present higher-order
dynamical decoupling sequences (CPMG$N$, $N>1$) in the next chapter.

We begin by describing our numerical method of calculating coherence times of the hybrid qubit.
We then present calculated coherence times for the hybrid qubit at the forbidden transitions
for which fast quantum control was demonstrated as discussed in \Chap{hybridqubit}.
These were the first CCE calculations taking into account full quantum
state-mixing of the central system for decoherence in a spin bath.
The calculations were performed and published in \citet{Morley2013} by S.J.B. with
supervision from Dr. Wayne Witzel at Sandia National Laboratories, USA.
Prior to these results, only weak state-mixing was investigated for Si:Bi in the unmixed regime by varying an effective gyromagnetic ratio ($\equiv df/dB$)
as a function of $B$ \citep{George2010}.
This treatment limited access to forbidden transitions. Also, as we shall see below and found in \citet{Balian2014} which partly
studies for differences between spin bath and classical noise decoherence,
effective gyromagnetic ratio (or equivalently $df/dB$ \citep{Vion2002}) treatments do not reliable describe decoherence of
the hybrid qubit in all regimes.

In \Chap{hybridqubit}, OWPs were introduced as field values corresponding to
suppression of decoherence in spin baths.
Here, we present the first validation of OWPs as sweet-spots, first published in \citet{Balian2012}.
We proceed to numerically calculate coherence times across orders of magnitude variation as a function of magnetic field $B$,
and find excellent agreement with experiment in nearly all field regimes (published in \citet{Balian2014}).

For the Hahn spin echo near OWPs, we find that in order to simulate coherence times,
one must consider clusters of three interacting spins within the many-body quantum bath,
since independent pairs do not even give finite $T_2$ decay times; i.e.\ OWPs almost completely suppress decoherence driven by pair correlations.
In all regimes except for OWP regions for the Hahn spin echo and moderate CPMG (as we shall see in \Chap{dynamicaldecoupling}),
the usual independent pairs of flip-flopping spins dominate. This finding formed the first part of our
work published in \citet{Balian2015}; the other part concerns $N > 1$ dynamical decoupling and is the topic of \Chap{dynamicaldecoupling}.

\section{Fitting Coherence Decays}

Before presenting coherence calculations and measurements it is important to
briefly discuss how $T_2$ times are extracted from coherence decays.
For high temperatures ($\gtrsim 14$~K), coherence decays for donors in silicon are
exponential and limited by the spin-lattice relaxation $T_1$.
For the lower, cryogenic, temperatures we consider
the coherence times are temperature independent
and are well-fitted to a combination of exponential and stretched exponential
time decays:
\begin{equation}
S = \exp{
\left( -\frac{t}{T'_2} -\left[\frac{t}{T_2}\right]^n \right).
}
\label{eq:t2signalfig}
\end{equation}
For the Hahn spin echo case, $t = 2\tau$, where $\tau$ is the 
time either side of the refocusing pulse.
The stretched exponential part with time constant
$T_2$ in \Eq{t2signalfig} characterises nuclear spin diffusion,
and for this process, the exponent $n$ is known to be about 2 \citep{Witzel2005},
thus resulting in a near-Gaussian decay in coherence.
In \Eq{t2signalfig}, all other decoherence mechanisms, including those arising
from donor-donor interactions from measurements on an ensemble of donors,
are characterised by $T'_2$.
Clearly, the shortest decoherence time
is the limiting one.
Note that for a decay of form $\exp{[-(t/T_2)^n]}$, $T_2$
is the time taken for the coherence to drop to $1/e$
of its initial value and this holds $\forall\ n$.

\section[Numerical Method: CCE with Central State Mixing]{Numerical Method: \\ CCE with Central State Mixing}

The CCE method was described in detail in \Sec{cce}.
It is a well-established method for accurately calculating the coherence decay $\mathcal{L}(t)$
of a central spin system in a quantum spin bath \citep{Yang2008a,Yang2008b,Yang2009}.
The code for our implementation of the CCE is open-source and free to use
under the GNU licence (see \App{spindec} for details and how to cite).
Here, we describe the method used to calculated coherence decays of the hybrid qubit
in a bath of \sitn\ impurities, using our implementation of the CCE. We also present results which
we use to establish numerical convergence of the coherence decays with respect to heuristic cluster cut-offs.

The CCE was applied for nuclear spin diffusion in a dipolar \sitn\ bath coupled
to the qubit via the hyperfine interaction as described in \Chap{background}.
Importantly, the central spin Hamiltonian did not include a bare electron
with an effective gyromagnetic ratio taking account of mixing as in previous studies \citep{George2010}.
Instead, we used the full donor Hamiltonian \Eq{donorH} in \Chap{hybridqubit},
including electron-nuclear mixing.
We did not include phonon-induced relaxation effects in our simulations, as temperatures were
cold enough such that $T_1 \gg T_2$. Also, the donor concentration was much less than that of the \sitn\ nuclei
and hence donor-donor decoherence was expected to be on much longer timescales
and hence neglected in our studies.

\subsection{Initial State}

For our simulations, crystal sites of a cubic silicon superlattice (see \App{silicon} for details)
were uniformly populated with $^{29}$Si nuclei ($I=1/2$) with the natural fractional
abundance of 0.0467 and with equal probability of spin-up ($\ket{\uparrow}$) and spin-down ($\ket{\downarrow}$) forming the pure product initial bath state.
The size of the superlattice giving convergent coherence decays was established as we describe below.

We assumed that the bath is not initially entangled with the bismuth donor. The bismuth donor was then put into an equal superposition of the
two eigenstates being excited:
$\ket{\psi(t_0)} = \tfrac{1}{\sqrt{2}}
\left( \ket{u} + \ket{l} \right) \otimes
\ket{\mathcal{B}(t_0)}$.

\subsection{Dynamics}

For the closed dynamics of each reduced problem (i.e.\ for the bismuth donor and cluster of bath spins),
the total Hamiltonian comprised of nuclear Zeeman and dipolar intra-bath interaction
terms given in \Eq{nsdbathH} (assuming a secular dipolar interaction (\Sec{dipolar})),
as well as the isotropic hyperfine interaction in \Eq{nsdintH}.
Such reduced problem Hamiltonians were decomposed to obtain the free evolution operators $\hat{U}(\tau)$.

Note that in the CCE method, the pure dephasing approximation is not required, and the interaction
Hamiltonian in general includes terms which depolarise the states of the central system.
Our CCE calculations include the $\hat{S}^{-}\hat{I}^+ + \hat{S}^{+} \hat{I}^{-}$
terms in the hyperfine interaction Hamiltonian \Eq{nsdintH}, but we find that these
give small corrections to the case when only $\hat{S}^z \hat{I}^z$ terms are included.
This was expected due to the large mismatch between electronic and nuclear gyromagnetic ratios.

The mixed Zeeman basis of the hybrid qubit was employed as described in \Chap{hybridqubit}.
To save computational time, we note that in our implementation, the full 20-dimensional basis was truncated
to exclude matrix elements which remain uninvolved in the dynamics for all times.
This is because the donor Hamiltonian is in block diagonal form (of at most $2\times2$ matrices) and only those
Zeeman basis states required to fully represent the two energy levels for the transitions are involved in the dynamics.
We note however for levels forming doublets, even for the simple Ising-only hyperfine interaction involving only $\hat{S}^z$-coupling,
since $\bra{\pm,m} \hat{S}^z \ket{\mp,m} \neq 0$, the Zeeman states for the level with which the transition level forms a doublet are also required.
For example, if one of the levels in the transition is $\ket{+,m}$, then the four Zeeman states
$\ket{\pm1/2 , m\mp1/2}, \ket{\mp1/2 , m\pm1/2} $ are in general needed.
Note that for transitions between doublets adjacent in $m$ (and assuming an Ising-only interaction to the bath),
four Zeeman states in total are required.

When including the non-Ising terms in the hyperfine interaction, the 20
basis states are strictly required.
However, our numerical results show that (i) the effect of the non-Ising term is negligible as mentioned above,
and (ii) even if the non-Ising term is included with the truncated basis above, the results
are indistinguishable from the case of including the full 20-dimensional basis
with non-Ising terms. The latter is indicative of the weakening perturbative
effect of the non-Ising terms as an increasing number of spin flips are required to navigate
into states far in $m$ from those involved in the transition.

To calculate the electron-bath hyperfine couplings, we use the Kohn-Luttinger
electronic wavefunction for the bismuth donor in silicon
with an ionization energy of 0.069 eV as described in \Sec{hyperfine}.
Calculated qubit-bath couplings were of the same order as those obtained
from the data in \Chap{interaction}. The experimental data presented in \Chap{interaction} suggested that isotropic couplings to the bath
dominate; hence anisotropic couplings were neglected and the simulations were largely insensitive
to orientation (apart from the orientation dependence in the intra-bath dipolar interaction).

In obtaining the \sitn--\sitn\ dipolar terms it was assumed that the external magnetic field was large
enough to conserve the total \sitn\ Zeeman energy, hence assuming the secular dipolar approximation.
The strength of the dipolar interaction was calculated as shown in \Sec{dipolar},
with the magnetic field direction chosen to match that in experiments with which we compare our calculated decays.

As for dynamical decoupling control (including the 1-pulse Hahn echo),
we assumed negligible duration of the $\pi$-pulse compared to the total evolution time and
applied instantaneous spin-flip pulse(s).
We note that for a spin system with more than a pair of energy eigenstates,
the non-resonant states (i.e.\ $\ket{i \neq u,l}$) must remain unchanged when applying refocusing pulses.
Therefore, the $\pi$-pulse operator was taken as
$(\outerprod{u}{l} + \outerprod{l}{u} + \sum_{n \neq u,l} \outerprod{n}{n}) \otimes \hat{\mathds{1}}_{\mathcal{B}}$,
in which the sum includes the other non-resonant eigenstates which can be formed
in the Zeeman representation of $\ket{u}$ and $\ket{l}$.

Finally, the final state is written as a density matrix
and the partial trace over the bath computed to recover the reduced density matrix
of the central system. The orthonormal basis used to perform
the trace operation is formed of all possible product states
of the cluster bath. The off-diagonal of the reduced density
matrix corresponds to the coherence as described in \Chap{intro} and \Chap{background}.
The resulting reducible coherences are then used in the CCE formalism as
described in \Sec{cce}.

\subsection{Cluster Heuristics and CCE Convergence}

\begin{figure}[h]
\centering\includegraphics[width=3.8in]{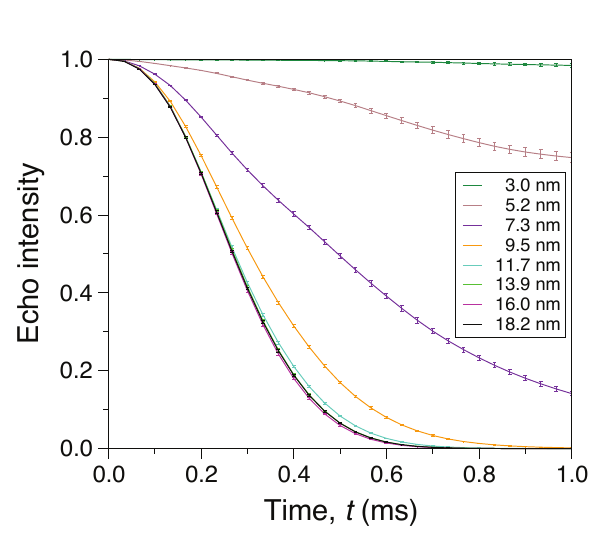}
\caption{
Convergence of the two-cluster correlation expansion for spin echo decays in Si:Bi at 4 GHz with respect to the superlattice size.
Pairs of \sitn\ nuclei with separations up to the 3rd nearest neighbour distance in the silicon lattice were included in the calculation.
The error bars are the standard deviation of the mean intensity after 100 random
spatial and initial state configurations of \sitn\ nuclei, and the external magnetic field was chosen to be $B=0.3446$~T so the
$\ket{11}\to\ket{10}$ Si:Bi transition was excited.
Figure adapted from \citet{Morley2013}.
}
\label{fig:Morley2013_cutoffsA}
\end{figure}

The total size of the spin bath is dictated
by the spatial extent of the wavefunction which decays exponentially
with distance from the donor site, and a superlattice of side length 160~\AA\ (with $10^{4}$ impurities)
gives convergent coherence decays.
\Fig{Morley2013_cutoffsA} shows convergence of the CCE as we increase the lattice space diagonal.
The donor was situated at the centre of a cubic superlattice of side length that was
varied from 30~\AA\ to 182~\AA.
Including spins outside the 160~\AA\ cube had minor effect on the coherence decays.

\begin{figure}[h]
\centering\includegraphics[width=3.8in]{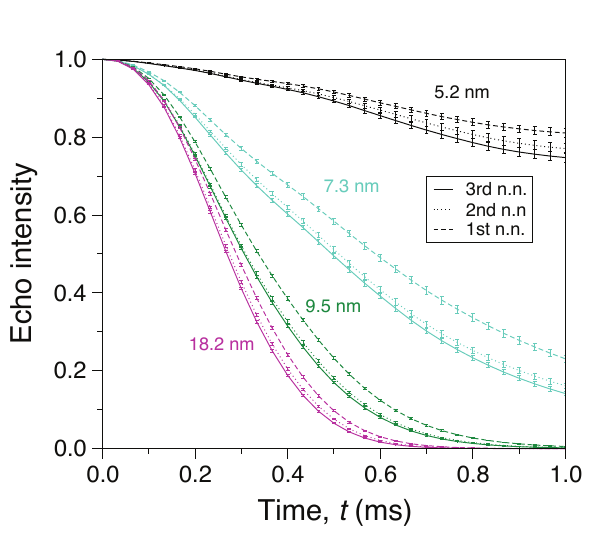}
\caption{
Convergence of the two-cluster correlation expansion for spin echo decays in Si:Bi at 4 GHz with respect to the pair-separation cut-off.
The maximum distance between paired \sitn\ nuclei is increased by pairing 1st, 2nd and
3rd nearest neighbours. Convergence is achieved for the 3rd nearest neighbors. The 1st, 2nd and 3rd nearest neighbour separations in the silicon lattice are
$ \tfrac{\sqrt{3}}{4}a_0$,  $ \tfrac{\sqrt{2}}{2}a_0$, and $ \tfrac{\sqrt{11}}{4}a_0$
respectively, with $a_0 = 5.43$~\AA. The results are compared for a range of lattice sizes.
The error bars are the standard deviation of the mean intensity after 100 random
spatial and initial state configurations of \sitn\ nuclei, and the external magnetic field was chosen to be $B=0.3446$~T so the
$\ket{11}\to\ket{10}$ Si:Bi transition was excited.
Figure adapted from \citet{Morley2013}.
}
\label{fig:Morley2013_cutoffsB}
\end{figure}

Due to cubic decay of the dipolar interaction as the
distance between a pair of $^{29}$Si spins is increased,
it is not necessary to include all spin clusters in the
calculation. At the lowest non-trivial CCE order (CCE2),\footnote{One-cluster contributions (CCE1) have a minor modulation effect on the spin echo decay and were not included in our CCE simulations.}
spins separated by at most the 4-th nearest neighbor distance in silicon ($\sqrt{11} a_0 / 4$, where $a_0 = 5.43$~\AA)
are enough to give convergent decays (provided the CCE converges at the two-cluster level with respect to cluster size).
Convergence in the separation cut-off as it is
increased from the nearest neighbour to the 3rd nearest neighbour distance are shown in
\Fig{Morley2013_cutoffsB}.

To choose three-clusters (i.e.\ including clusters of three bath spins), we loop over all sites in
the crystal and add to each two-cluster only those spins that are
at most separated by $\sqrt{11} a_0 / 4$ from any of the two
spins in the two-cluster. The same procedure was applied to choose higher-order
clusters, by adding spins to clusters one order down.
The numbers of 2, 3, 4, and 5-clusters found with these heuristic cut-offs are each of order $\approx 10^4$.
The total computational time taken to diagonalise all these reduced problems is at most about a day on a desktop
machine for a single initial bath state, with the most intensive (CCE5) of our calculations.\footnote{We note that calculations using all 20 levels of Si:Bi take much longer.}

\subsection{Initial State Averaging}

Finally, in order to simulate multiple experimental runs on the same spatial bath configuration (time-average),
the coherence calculated for each reduced problem can be solved for all the non-interacting bath eigenstates in the cluster
and the results averaged.
This effect can also be simulated by applying the CCE to 
different random configurations of initial bath state (i.e.\ random sampling from the thermal ensemble) and obtaining the mean over these results.
However, this method
takes longer computational time.
Also, for measurements on an ensemble of qubits, the results should be averaged over
different spatial as well as initial state configurations of the bath.
However, we note that the differences in coherence between
the different averages are not significant for the large spin
baths we consider.

Note that all these averages are over the complex coherence (coherent averaging); the real and imaginary parts
of the coherence correspond to orthogonal axes of the in-plane magnetisation.
The modulus of the coherence should strictly only be taken at the end of the calculation.
Nevertheless, applying the modulus to coherences before averaging (incoherent averaging) does not significantly
affect the decays.\\

Before discussing coherence times near OWPs,
we begin by presenting our simulations of $T_2$
for the hybrid qubit at $S$-band frequencies,
where fast quantum control was experimentally demonstrated in \Chap{hybridqubit}.

\section{Coherence Times at Forbidden Transitions}

In this section, we present calculated coherence times of the hybrid qubit in natural silicon
at $S$-band ESR excitation frequency ($f \approx 4$~GHz)
where fast Rabi oscillations were demonstrated as described in \Chap{hybridqubit}
for the forbidden transition of Si:Bi.

Importantly, our CCE calculations enabled us to
attribute the dominant decoherence mechanism for the hybrid qubit at $S$-band
to nuclear spin diffusion from the spin bath of \sitn\ nuclear impurities.
We compare our simulations with experimental data
and establish a coherence time which can be up to five orders
of magnitude longer than the qubit manipulation time in this regime.

\begin{figure}[h!]
\centering\includegraphics[width=5.0in]{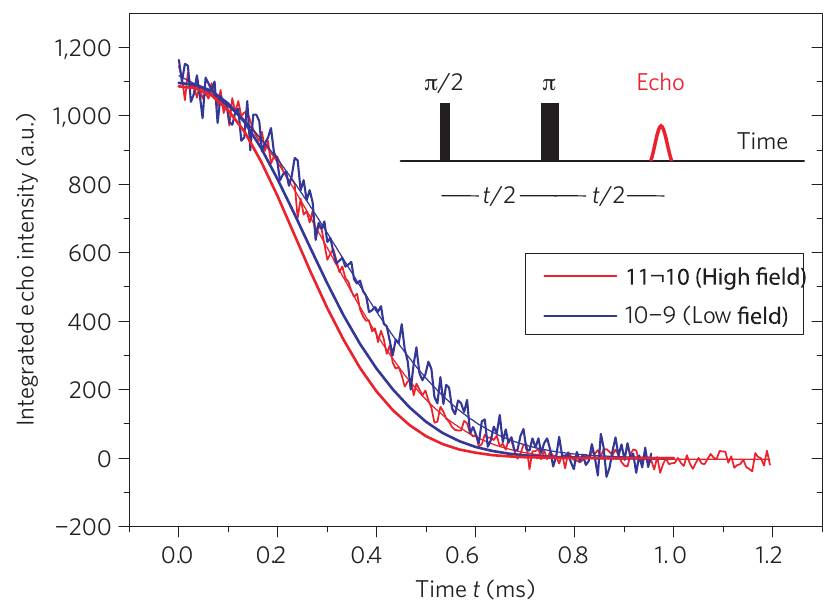}
\caption{
Example spin-echo coherence decays measured for both Si:Bi transitions at 4~GHz,
with a temperature of 10~K. The echo coherence decay is limited by
\sitn\ nuclear spins, as parameterized by $T_2$ in the fitting function
$\exp{( -t/T'_2 - t^n/T_2^n )}$, where $T'_2$ is the spin coherence time expected after isotopic enrichment.
The exponent $n$ was used as a fitting parameter. The smooth thin lines are these
fits whereas the smooth thick lines show a simulation with no free parameters using the
cluster correlation expansion. This shows that \sitn\ impurities dominate the spin
decoherence. The thickness of the line is of the order of the standard
deviation of the mean intensities after 100 random spatial configurations of
\sitn\ nuclei. As discussed in \Chap{background}, the refocusing $\pi$-pulse removes
static magnetic field noise from \sitn\ couplings to the qubit.
The magnetic field direction was perpendicular to the $[111]$ crystal direction.
Figure adapted from \citet{Morley2013}.
}
\label{fig:Morley2013_t2exptheory}
\end{figure}

\begin{figure}[h!]
\centering\includegraphics[width=3.6in]{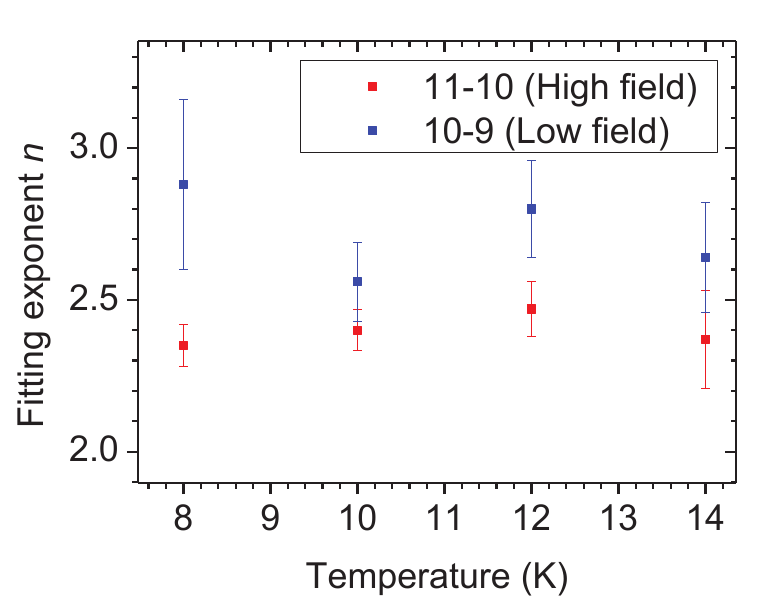}
\caption{
Dependence of the fitting exponent $n$ on temperature, showing the expected range of $n \approx 2$
expected for nuclear spin diffusion. Figure adapted from \citet{Morley2013}.
}
\label{fig:Morley2013_exponent}
\end{figure}

\subsection{Numerical Results}

The calculated decays are shown in \Fig{Morley2013_t2exptheory} as solid lines.
We fitted the decays with the same function as used for fitting to the experimental data (\Eq{t2signalfig}),
and an effectively infinite $T'_2$ (over 1018~s) was obtained,
indicating the lack of an exponential component (this was expected as donor-donor processes were ignored in our calculations).
The values for the exponent $n$ were obtained as 
2.25 for the $\ket{10}\to\ket{9}$ transition and 2.28 for the $\ket{11}\to\ket{10}$
transition with standard errors on the fits of $\pm 0.01$; the values of the exponent
are in agreement with those obtained for nuclear spin diffusion in previous studies \citep{Witzel2005}.
As for the $T_2$ values, these were
$0.314 \pm 0.0005$~ms for $\ket{11}\to\ket{10}$ and $ 0.340 \pm 0.0007$~ms for $\ket{10}\to\ket{9}$(using the
standard errors on the fits).
The value of $T_2$ for the $\ket{11}\to\ket{10}$ transition is slightly shorter than for the $\ket{10}\to\ket{9}$ 
and we attribute this trend to the smaller gradient $df/dB$ for the $\ket{10}\to\ket{9}$ transition.

\subsection{Comparison with Experiment}

Our CCE decays are shown in \Fig{Morley2013_t2exptheory}
and compared to experimental Hahn spin echo decays at $S$-band.
The CCE simulations had no free parameters extracted from the experiment, thus
demonstrating that \sitn\ impurities dominate the spin echo decay
at low temperatures.
Values for the exponent from fits to experimental data are shown in \Fig{Morley2013_exponent}
and are in agreement with the expected nuclear spin diffusion $n$ of near-Gaussian ($n \simeq 2$).
The small discrepancies between theory and experiment are expected to be mainly due to ignoring possibly
undiscovered sources of decoherence in the simulation and limited knowledge of the donor electron
wavefunction. The experimental coherence times also show the expected trend due to $df/dB$.

\begin{figure}[h]
\centering\includegraphics[width=5.0in]{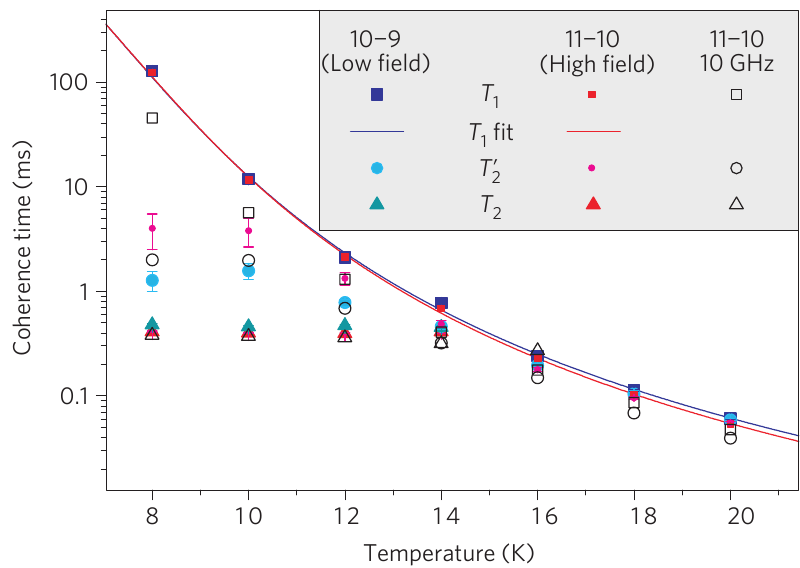}
\caption{
Coherence times of hybrid electron-nuclear qubits as a function of
temperature for both resonances at 4 GHz with previously published data
at 10 GHz \citep{Morley2010} for comparison. The 4 GHz spin-lattice relaxation rates are $1/T_1$.
The error bars show the standard errors, which are in many cases smaller than the symbol.
Figure adapted from \citet{Morley2013}.
}
\label{fig:Morley2013_t2exp}
\end{figure}

\subsection{Discussion}

Measured coherence times in natural silicon, including those extracted from
the experimental decay curves in \Fig{Morley2013_t2exptheory}, are shown in \Fig{Morley2013_t2exp}
as a function of temperature.
It can be seen that $T_2$ can reach 0.4~ms, in good agreement with the predicted theoretical values above, while
$T'_2$ can be about 4~ms. There are no OWPs in this region for the hybrid qubit,
and hence no significant enhancement of $T_2$.
Therefore, the best strategy possible is to use isotopic enrichment
to reach the 4~ms coherence time.

It is important to note that if the hybrid qubit was approximated as
a simple spin-1/2 with an effective field-dependent gyromagnetic ratio as
in \citet{George2010}, it would not be possible to obtain coherence decays for the forbidden transition $\ket{10}\to\ket{9}$.

The $S$-band coherence measurements were performed by Dr. Petra Lueders, Dr. Gavin Morley and Dr. Hamed Mohammady at ETH Zurich.

\section{Coherence Times at Optimal Working Points}

As introduced in \Chap{intro}, the OWPs of the hybrid qubit we consider
are those near where there is suppression of decoherence
from spin bath environments.
OWPs can be understood by considering the qubit-bath entanglement
and the back-action of the qubit on the environment.

\subsection{Loss of Which-Way Information}

OWPs correspond to $B$-field values where the unitaries associated with the upper and lower
central qubit states equalise: $\hat{T}_l  \simeq \hat{T}_u$, occurring when $P_u \simeq P_l$ as found in \citet{Balian2014}.
This means that the combined qubit-bath state after evolution following a $\pi/2$-pulse (\Eq{tangle}) can now be written as
\begin{equation}
\ket{\psi(t)}= \frac{1}{\sqrt{2}}
\left( e^{-iE_ut} \ket{u} + e^{-iE_lt} \ket{l}\right) \otimes \hat{T}_u(t) \ket{\mathcal{B}(0)},
\end{equation}
with the product form preserved (assuming pure dephasing). Crucially, the state is no longer entangled and the `which-way' information in the environment
or the back-action of the central qubit on the environment is minimal. Therefore,
entanglement-induced (quantum bath) decoherence is suppressed:
\begin{equation}
|\mathcal{L}(t)| = |\langle\mathcal{B}(0) | \hat{T}_u^\dagger(t) \hat{T}_u(t) | \mathcal{B}(0)\rangle|
= |\langle\mathcal{B}(0) | \hat{\mathds{1}} | \mathcal{B}(0)\rangle| = 1.
\end{equation}

In practical realisations, OWPs have become closely associated with field values where the frequency-field gradient $df/dB=0$ \citep{Wolfowicz2013}
or `clock transitions'.
However, for donor spin systems, the OWP is close to but not exactly at the $df/dB=0$
point, and not all $df/dB=0$ points are OWPs as shown in \citet{Balian2014}.
In the latter, it was also shown that it is not possible to fit functional
forms of $df/dB$ to describe the coherence $T_2(B)$ for all $B$.

\begin{figure}[h!]
\centering\includegraphics[width=5.0in]{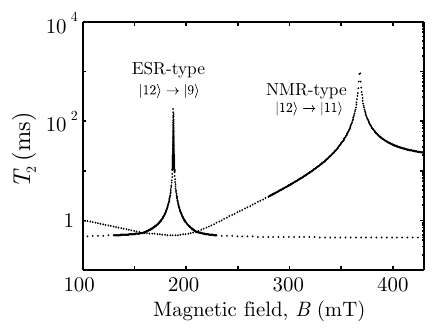}
\caption{
    Suppression of spin bath decoherence near OWPs of the hybrid qubit in a nuclear spin bath (\tsc{nat}Si:Bi).
        Figure was adapted from \citet{Balian2014}.
    Each dot represents the $T_2$ extracted from coherence decays obtained using the cluster correlation expansion (CCE).
    The coherence times are the Hahn spin echo $T_2$, and near the OWPs, these were extracted from the short-time behaviour of CCE2 decays,
    since pair correlations
    are strongly suppressed on the actual timescale of $T_2$ near OWPs.
    Such short-time $T_2$ times are in agreement with those obtained from full decays of the converged CCE3 near OWPs established in \citet{Balian2015} (see \Sec{threeclusterowp} for details).
}
\label{fig:Balian2014_cce}
\end{figure}

\begin{figure}[h!]
\centering\includegraphics[width=5.4in]{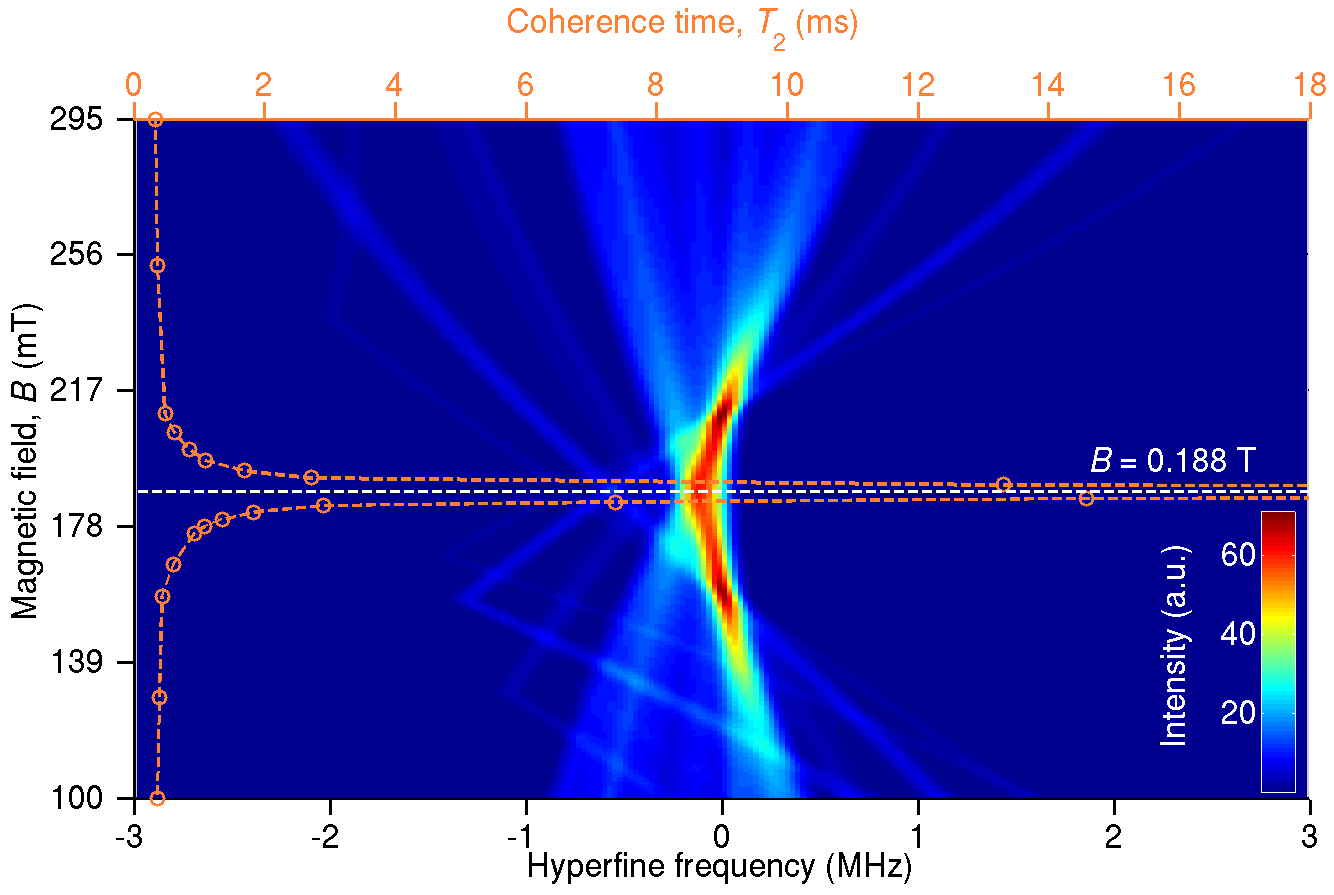}
\caption{
    Suppression of Bi-$^{29}\text{Si}$ spin bath decoherence
    for the $|12\rangle \to |9\rangle$ ESR transition.
    Simulated ENDOR and nuclear spin diffusion
    coherence times $T_2$ (Hahn echo) as a function of magnetic field $B$,
    showing collapse of the hyperfine couplings and a sharp increase
    in $T_2$ as the field approaches the $B=188.0$~mT optimal working
    point (OWP). The dashed line is a fit. Figure adapted from \citet{Balian2012}.
    Coherence times in the OWP region were extracted from the short-time behaviour of CCE2 (Details are given in \Sec{threeclusterowp}).
}
\label{fig:Balian2012_owpt2}
\end{figure}

\begin{figure}[h!]
\centering\includegraphics[width=5.2in]{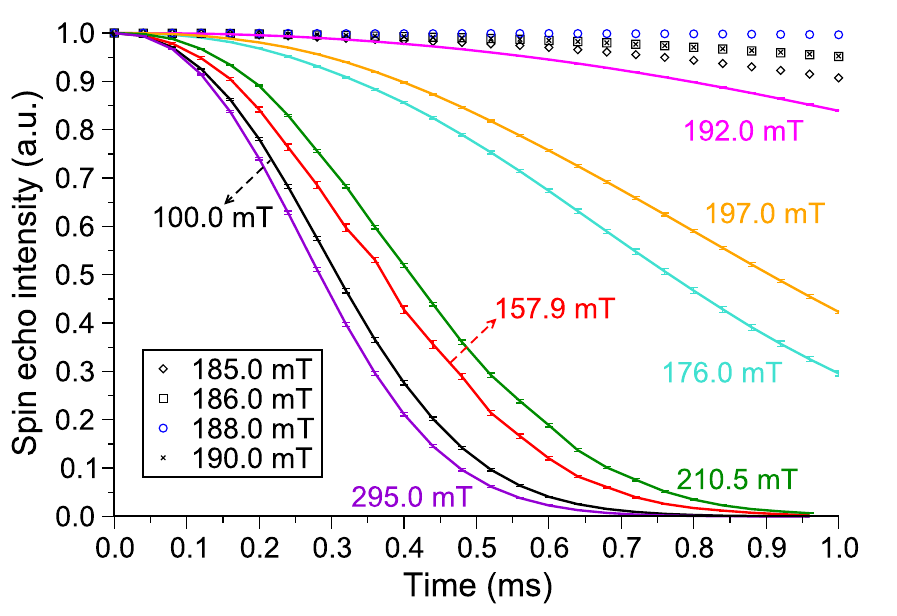}
\caption{
Calculated donor Hahn spin echo decays from which coherence
times in \Fig{Balian2012_owpt2} were extracted. Figure adapted from \citet{Balian2012}.
Decays are for CCE2 (Hahn echo), shown here for short times.
For converged decays near the OWP see \Sec{threeclusterowp}.
}
\label{fig:Balian2012_decays}
\end{figure}

\subsection{Suppression of Nuclear Spin Diffusion}

In this section, we present numerical CCE calculations
showing enhancement of the spin diffusion coherence time $T_2$ at an OWP \citep{Balian2012,Balian2014}.
The orders of magnitude enhancement for coherence times at two
OWPs of the hybrid qubit is illustrated in \Fig{Balian2014_cce}.

The CCE simulations were performed for the $\ket{12} \to \ket{9}$ and $\ket{12} \to \ket{11}$
Si:Bi transitions. Once again, low temperatures were assumed and hence
phonon-induced relaxation effects ignored. We also assumed
that the donor concentration was low such that the decoherence
was dominated by nuclear spin diffusion from the \sitn\ spin-1/2 bath
and not by donor-donor processes.
 
\Fig{Balian2012_owpt2} which shows the behaviour
around the $B=188.0$~mT OWP associated with the 
$|12\rangle \to |9\rangle$ transition,
superposed on the colour map of the qubit-bath hyperfine spectrum (\Fig{Balian2012_collapse}).
The behaviour of $T_2$ was striking and unexpected:
The coherence time predicted by CCE simulations increases asymptotically
at the OWP. Away from the OWP, the results
agree well with experimentally measured values of approximately 0.7~ms 
\citep{George2010}.
In \citet{George2010}, in a regime of weak state-mixing, simulations
using an effective gyromagnetic ratio indicated that $T_2$ was
slightly reduced (by about 5\%) in a regime corresponding far from the OWP (but with some weak $df/dB$ variation).
These results, on the other hand (which in contrast to
\citet{George2010} employed a full treatment of the quantum
eigenstate mixing) show rather an effect very sharply peaked about the OWP:
Nuclear spin diffusion is predicted to be largely suppressed, but over an extremely 
narrow magnetic field range.

\Fig{Balian2012_decays} shows a sample
of CCE spin echo decays from which the $T_2$ times in \Fig{Balian2012_owpt2} were extracted,
and also serves to further illustrate the sharp increase in $T_2$.
Similar suppression is present for other OWPs in Si:Bi.
OWPs are also expected to lead to suppression of
decoherence arising from the interaction with a bath of donors
\citep{Mohammady2010}.

The calculated coherence decays in \Fig{Balian2012_owpt2} and \Fig{Balian2012_decays} are the average over 100 initial
spatial configurations of \sitn.

\subsection{Many-Body Correlations}
\label{sec:threeclusterowp}

\begin{figure}[h]
\centering\includegraphics[width=5.8in]{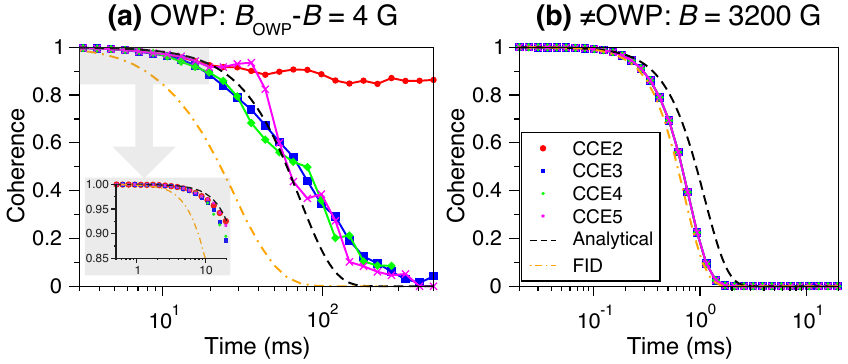}
\caption{Shows quantum many-body calculations of the Hahn spin echo
using the cluster correlation expansion (CCE) method. (a) Near OWPs, calculations using a bath of independent spin pairs only (red, CCE2) 
do not even predict a finite decay time but, surprisingly, calculations with clusters of three spins (blue, CCE3) are already well-converged.
The dashed lines used a closed-form equation derived from the short time behaviour, found in \citet{Balian2014} to yield good agreement with experiments;
this indicates that three-cluster results too give good agreement with measurements. The formula is discussed in \Chap{formula}.
Higher order CCE can encounter numerical divergences (which can be attenuated by ensemble averaging);
this accounts for the discrepancies with CCE5.
(b) Far from the OWP, independent pairs (CCE2) already give results
in good agreement with CCE3-5 as well as experiments.
The single-spin free induction decay (FID) is also shown for comparison.
Note that the analytical formula approximates the decay by a pure Gaussian.
CCE calculations were performed for a bismuth donor in
natural silicon for $B$ along $[100]$ and 
the $\ket{14} \to \ket{7}$ transition for which $B_\text{OWP} = 799$~G.
In (a), $B=795$~G while for (b), $B=3200$~G. Figure adapted from \citet{Balian2015}.}
\label{fig:Balian2015_conv}
\end{figure}

In the high-field regime of the hybrid qubit where
the state-mixing is weak, CCE2 or the pair-correlation approximation is a good
approximation to the coherence when considering only dipolar interactions in the bath
affecting the spin echo, as these are at most a few kHz and hence
perturbative compared to the hyperfine interactions in the MHz range involving the
donor electron. Experimental Hahn echo decays are well predicted by CCE2 in these regimes or for simple spin-1/2
qubits without OWPs \citep{Witzel2006}.
However, this argument does not hold in regions close
to an OWP as seen in \Chap{interaction} where the collapse of hyperfine couplings to the bath was demonstrated.
In the vicinity of the OWPs (where $P_u \simeq P_l$), the CCE2 Hahn echo decay fails to converge at all times and no
decays can be obtained other than initially, for a short time.\footnote{Single-central spin free induction decay (FID), in contrast, gives finite decays at all magnetic fields and is discussed in \Chap{formula}.}

We find that to simulate recently measured Hahn echo decays at OWPs (lowest-order dynamical decoupling), one must consider
clusters of three interacting spins, since independent pairs of bath spins (pair correlations) do not even give finite $T_2$ decay times.
The CCE2 Hahn echo results (for all $t$) are at odds with experiment since they predict infinite coherence times \citep{Balian2014}.
As mentioned above, coherences decay initially, then after a short time, the decays stop.

In order to clarify the origin of the measured coherence decays, we employ
quantum many-body simulations of the system-bath dynamics using the CCE
including contributions from clusters of up to 5 spins (CCE5).
The CCE$3-5$ many-body calculations we undertake are significantly more computationally challenging than CCE2 or pair correlation calculations. 
The converged numerical results are presented in \Fig{Balian2015_conv} both near and far an OWP of Si:Bi for the Hahn spin echo, the latter denoted `$\ne$OWP'.
We show that including three-spin clusters (CCE3) gives converged results while qubit-bath correlations from only spin pairs
(CCE2) give little decay (red line) except at short timescales. The three-spin clusters suffice to give decays in good agreement with experimental results \citep{Wolfowicz2013}.
Importantly, it can also be seen that all orders have similar short time behaviour and that the inclusion of the three-clusters
in effect recovers the short time behaviour of the pair decays.
Hence, the short time behaviour of the pair correlation decay is sufficient
to establish an order-of-magnitude estimate of $T_2$.

It is worth clarifying the physical meaning of the above-mentioned three-cluster result. It is not a matter of enlarging the
quantum bath with additional nuclear spin clusters of the same size. As illustrated in \Fig{Balian2015_correlations},
a three-spin cluster (blue) can be decomposed into three distinct flip-flopping pairs (each nuclear spin 
can contribute to more than one flip-flopping pair). Put simply, if all such three-clusters 
in a given, randomly generated set of impurities in a crystal are decomposed into the constituent flip-flopping pairs,
an infinite decay time is obtained. If, however, the exact same configuration of spin impurities are aggregated into the
`triangle' structures illustrated in \Fig{Balian2015_correlations},
the correct experimental behaviour emerges.
To our knowledge, there is no other example of a central spin system which so
fully eliminates the pair-driven dynamics.

\subsection{Comparison with Experiments}

\begin{figure}[h]
\centering\includegraphics[width=5.0in]{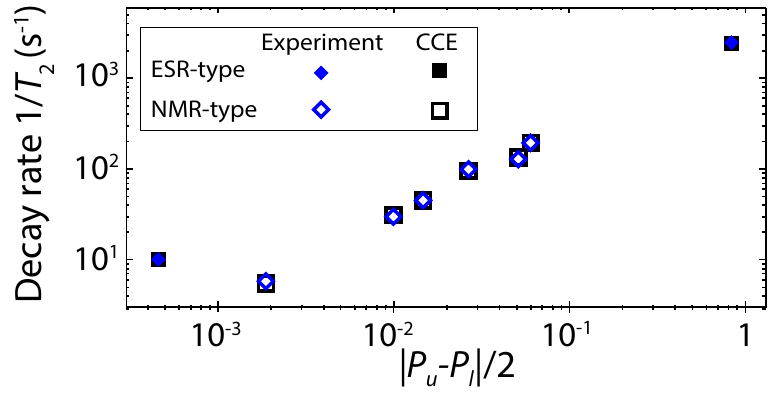}
\caption{
Comparison between theoretically predicted and measured
 $T_2$ in \tsc{nat}Si:Bi for various transitions,
showing remarkable agreement across a wide range of mixing regimes -- magnetic fields and transitions quantified by $|P_u-P_l|$.
The Bi donor concentration was $\leq 10^{16}$~cm$^{-3}$, and decoherence times are limited by \sitn\ spin diffusion.
Figure adapted from \citet{Balian2014}.
}
\label{fig:Balian2014_exp_top_only}
\end{figure}

We compare our CCE simulations (short-time CCE2 Hahn echo) with experiment
across a broad range of magnetic fields and transitions in \Fig{Balian2014_exp_top_only};
for ESR-type transitions in the high-field and OWP regions (filled symbols)
and NMR-type transition (empty symbols) where $T_2$ varies by orders of magnitude.
The measured values are in excellent agreement with the CCE calculations.

Measurements were made at 4.8~K using ESR with a microwave frequency of 9.77 or 7.03~GHz (filled symbols), or electron-nuclear
double resonance (ENDOR) between 200~MHz and 1~GHz using the method described in \citet{Morton2008} (empty symbols),
at magnetic fields between 100 and 450~mT.
The experiments were performed by Dr. Gary Wolfowicz at UCL under supervision from Professor John Morton.

As shown in \Fig{Balian2014_cce},
$T_2$ varies sharply with magnetic field over a small region of $B$; especially for ESR-type OWPs
(corresponding to an order of magnitude over a few G).
Therefore, for direct quantitative comparisons between the calculations and
experimental ensemble measurements, inhomogeneous broadening due to $^{29}$Si,
which has full width at half maximum (FWHM) of about 4~G in natural silicon,
might also have to be considered.
This effect is not only important for predicting the rate of experimental decays but also their shape.
The broadening can be simulated by convolving the calculated decays $\mathcal{L}_B(t)$
with a Gaussian magnetic field distribution with standard deviation $w \simeq 2$~G:
\begin{equation}
D_B(t) = \frac{1}{w \sqrt{2\pi}}
\int{ e^{\frac{-\left(B-B'\right)^2}{2w^2}} \mathcal{L}_B(t) dB'}.
\label{eq:convolution}
\end{equation}

\begin{figure}[h]
\centering\includegraphics[width=3.0in]{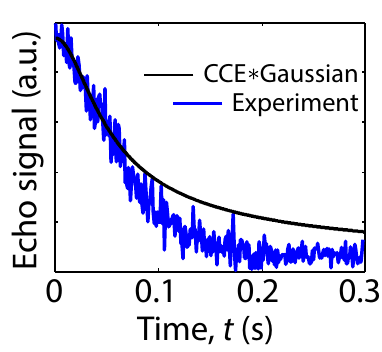}
\caption{
Calculations convolved with Gaussian $B$-field distribution of width 0.42~mT (arising from inhomogeneous broadening from the nuclear spin bath) show an excellent fit with the experimental Hahn echo decay around an ESR-type OWP ($B \sim 80$~mT), with no free fit parameters.
Figure adapted from \citet{Balian2014} and the experimental data was first published in \citet{Wolfowicz2013}.
}
\label{fig:Balian2014_owps_c}
\end{figure}

The ESR-type OWP point in \Fig{Balian2014_exp_top_only} (the first data point where $P_u$ is closest to $P_l$),
is the result of the convolution (with $B = B_\text{OWP}$) shown in \Fig{Balian2014_owps_c},
in good agreement with experiment.
The long-time discrepancy is probably due to the fact that the individual
theoretical decays were obtained using the short-time behaviour of CCE2.

\subsection{Angular Dependence}

\begin{figure}[h]
\centering\includegraphics[width=4.5in]{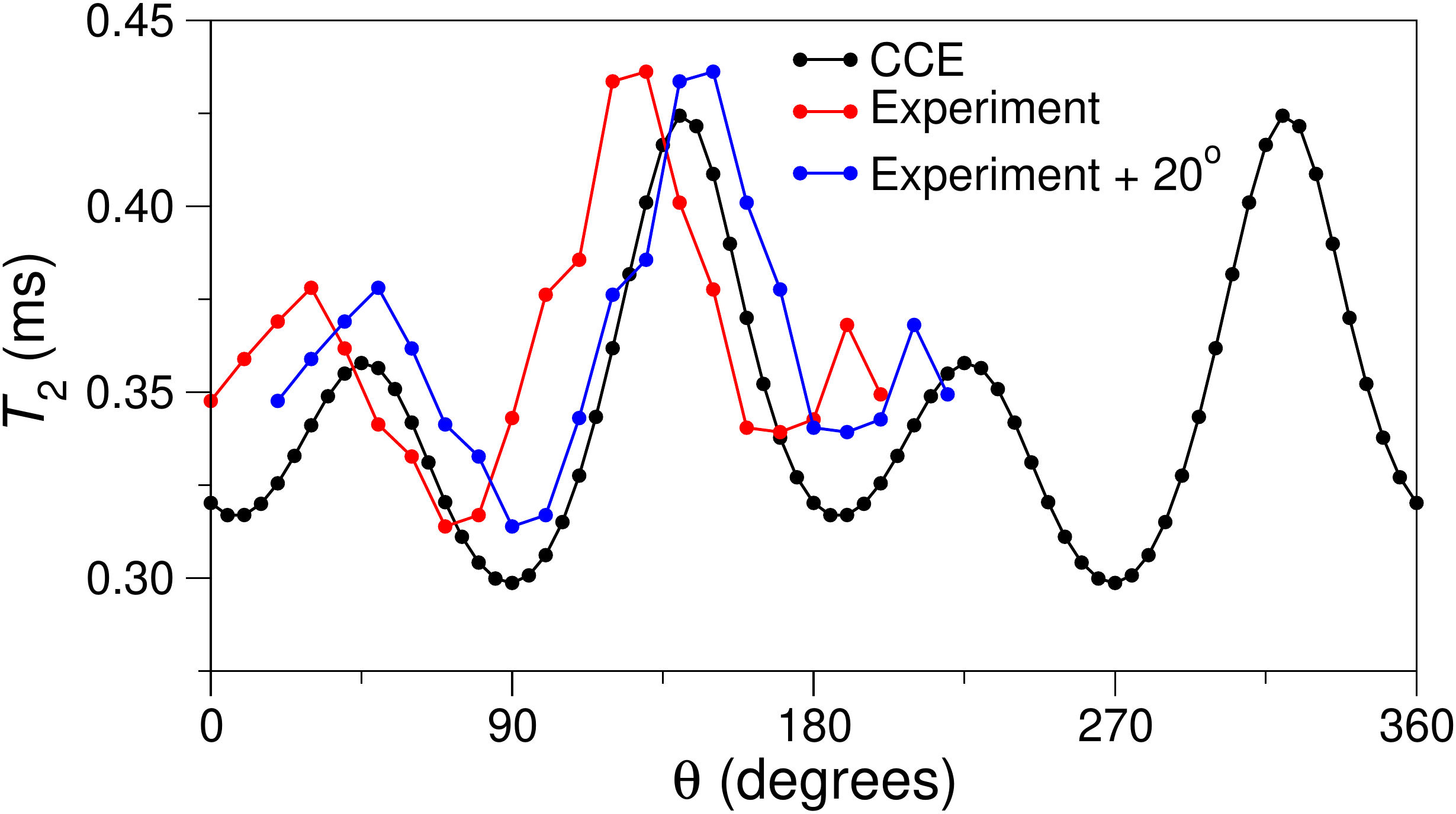}
\caption{
Angular dependence of $T_2$ for an ESR transition of Si:Bi. Rotation was performed about the $[\overline{1} \overline{1} 2]$ axis in the $[\overline{1} 1 0]$ - $[1 1 1]$ plane with $\theta$ from $[\overline{1} 1 0]$. The best match to experiment was obtained for a $5^{\circ}$ tilt in the rotation axis and a zero-offset of $20^{\circ}$.
Figure adapted from \citet{Balian2014}.}
\label{fig:Balian2014_rotation}
\end{figure}

The strength of the dipolar interaction $C_{12}$ depends on the angle between the vector joining the interacting spins and the direction of the magnetic field ${\bf B}$
as discussed in \Sec{dipolar}. As a result, $T_2$ varies with the orientation of the crystal sample relative to ${\bf B}$ \citep{DeSousa2003b,Witzel2006,Tyryshkin2006,George2010}.

$T_2$ was measured as a function of crystal orientation as shown in \Fig{Balian2014_rotation}.
X-ray diffraction using the back-reflection Laue technique showed the rotation axis to be close to $[\overline{1} \overline{1} 2]$.
The external magnetic field is in the rotation plane, defined by the angle $\theta$ such that $\theta=0^{\circ}$ and $\theta=90^{\circ}$ correspond to the field parallel to $[\overline{1} 1 0]$ and $[1 1 1]$ respectively.
However, there are experimental uncertainties in both the initial angle $\theta = 0^{\circ}$ and the position of the rotation axis.
In order to determine the best crystal orientation to use in the simulations,
CCE calculations as a function of orientation were compared with experimental measurements as shown in \Fig{Balian2014_rotation}.
The best match to experiment was obtained for the rotation axis tilted about $[1 1 1]$ by $5^{\circ}$ from $[\overline{1} \overline{1} 2]$,
and a $20^{\circ}$ shift in $\theta$.

All the points in \Fig{Balian2014_exp_top_only} used $\theta = 135^{\circ}$,
except for the first point corresponding to the ESR-type OWP (smallest polarisation difference).
A different sample with the field aligned along $[011]$ was used for the latter point.
The bismuth donor concentration of the ESR-type OWP point was also different ($[\text{Bi}]=10^{16}$~cm$^{-3}$) from the rest of the points
($[\text{Bi}]=3\times10^{15}$~cm$^{-3}$).
The difference in [Bi] is not expected to affect coherence times in the regimes studied, which are not dominated by donor-donor effects.

\subsection{Quantum Bath vs. Classical Noise}

\begin{figure}[h!]
\centering\includegraphics[width=5.0in]{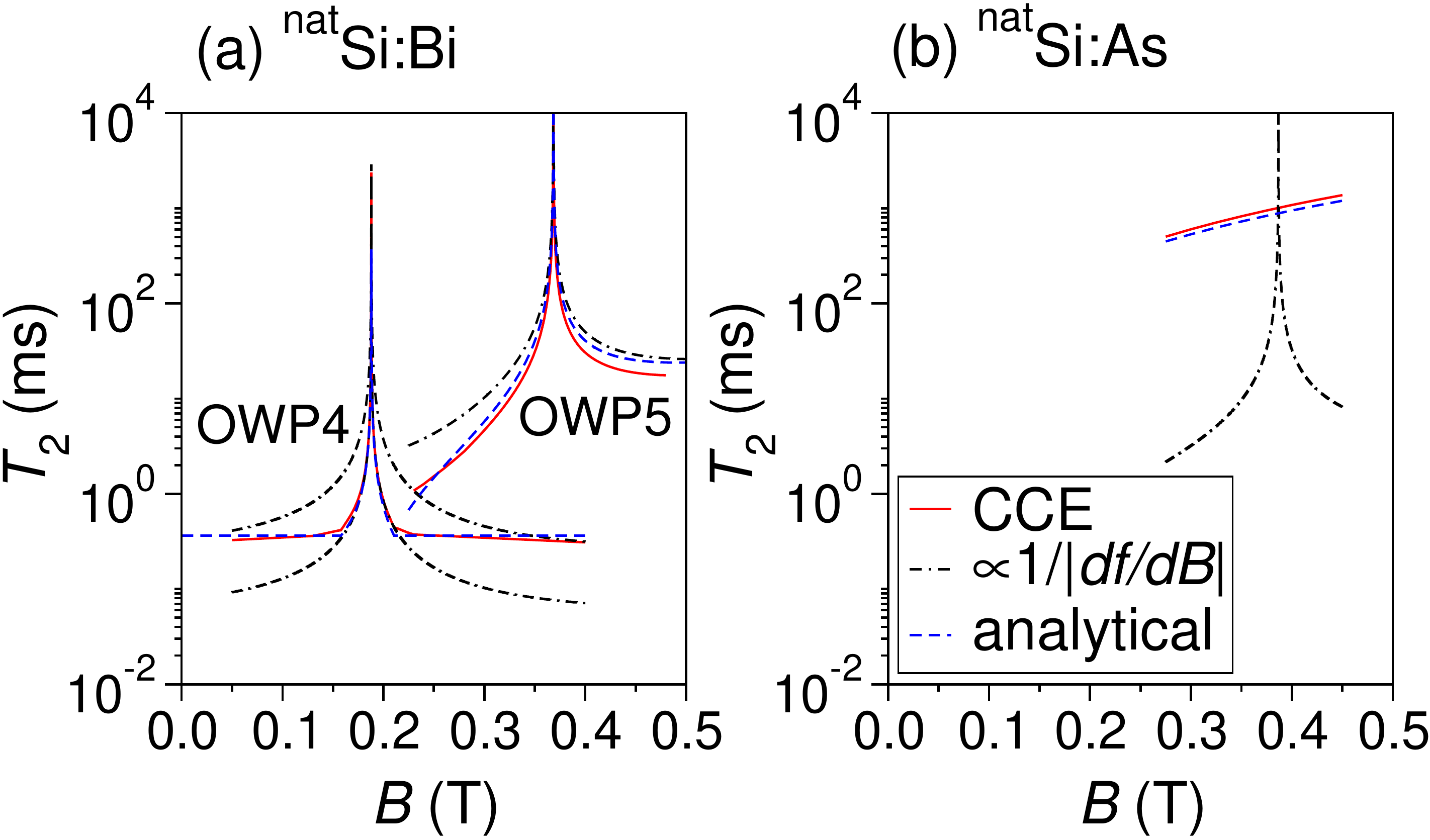}
\caption{Shows that the hybrid qubit coherence time as a function of magnetic field ($T_2(B)$) is not necessarily inversely proportional to the frequency-field gradient $df/dB$. 
Red solid line is $T_2$ calculated using the cluster correlation expansion (CCE); black dotted-dashed line is $T_2 \propto 1/(df/dB)$.
(a) $T_2(B)$ around a typical ESR-type `optimal working point' (OWP) of Si:Bi cannot be fitted by $df/dB$, except locally.
The $df/dB$ lines have been rescaled to fit either the OWP region or the asymptotic regions; they cannot fit both. The blue dashed lines are calculated using the closed-form
formula described in \Chap{formula}. (b) The single NMR-type `clock transition' (CT) of Si:As at $B\simeq 0.39$~T (where $df/dB=0$),
exemplifies a CT which is not an OWP (i.e.\ there is no enhancement in $T_2$). Si:Bi also has such CTs.
Calculations were performed for the natural abundance of $^{29}$Si (4.67\%).
Figure adapted from an earlier version of \citet{Balian2014} (arXiv:1302.1709v3 [cond-mat.mes-hall] (2013) ).}
\label{fig:Balian2014v3_dfdBvsOWP}
\end{figure}

Classical noise models relate coherence times as a function of $B$ to various orders and powers of $df/dB$ \citep{Vion2002,Ithier2005,Martinis2003,Mohammady2012,Wolfowicz2012,Wolfowicz2013}:
\begin{equation}
(T_2(B))^{-1} = C \left\{ \frac{df(B)}{dB}, \left(\frac{df(B)}{dB}\right)^2, \dots, \frac{d^2f(B)}{dB^2}, \left(\frac{d^2f(B)}{dB^2} \right)^2, \dots  \right\},
\end{equation}
where $C$ is a field-independent constant which in general depends on the central spin, bath and interaction Hamiltonians.
The simplest of these is $T_2 \propto 1/(df/dB)$.
This treatment provides an intuitive interpretation of sweet-spots where $df/dB \simeq 0$;
changes in the central-state splitting $\Delta f$ are insensitive to changes in magnetic field fluctuations $\Delta B$
and hence the initial superposition of the qubit has enhanced protection from noise.
This argument is valid for random classical noise such as instrument noise.
However, for the case of $\Delta B$ arising from a quantum spin environment,
we find that $df/dB$ cannot account for the global $B$ dependence of $T_2$.
This is illustrated for Si:Bi in \Fig{Balian2014v3_dfdBvsOWP}(a).
It is only possible to fit $1/(df/dB)$ locally (for some $B$ range, but not all $B$) to CCE $T_2(B)$ curves.

Also shown in \Fig{Balian2014v3_dfdBvsOWP} is that while some of the OWPs are coincident with CTs where $df/dB \rightarrow 0$, 
others (in particular the NMR-type OWPs) are not; such as the CT for Si:As in \Fig{Balian2014v3_dfdBvsOWP}(b).
The reason for this deviation is that $\hat{H}_{\text{int}}$ differs from 
a magnetic field-type term ($\propto (S^z + \delta_X I_X^z)$). 
In other words, while $\hat{H}_{\text{int}}$ determines the form of the interaction between the central spin system
and the bath, it is $\hat{H}_{\text{CS}}$ which determines $df/dB$.
If $\hat{H}_{\text{int}}$ and $\hat{H}_{\text{CS}}$
are of different form, then clock transitions are not OWPs.
In the case of nuclear spin diffusion for Si:Bi systems, for $ B \sim 1$ T, there is
still sufficient mixing between the electronic and nuclear degrees of freedom so that
it is the contact hyperfine interaction ($\propto S^z$ and {\em not} $\propto (S^z + \delta_X I_X^z)$)
which dominates the effect of $\hat{H}_{\text{int}}$,
thus we may neglect the interaction between the bismuth nuclear spin and the bath,
even for NMR-type transitions. However, in this range, the nuclear Zeeman term 
contributes significantly to $df/dB$ for NMR-type transitions (see \Eq{dfdB} for an exact expression for $df/dB$).

\section{Conclusion}

In summary, we presented coherence times of the hybrid qubit in the natural silicon spin bath,
calculated using the CCE, and in numerous parameter regimes;
these are for magnetic fields near and far from OWPs (the latter in the unmixed regime),
both for ESR-type and NMR-type transitions \citep{Balian2014,Balian2015},
and also for ESR-forbidden transitions \citep{Morley2013}.
In all cases, our numerical calculations are in good agreement with experiments.

The coherence times we reported for the hybrid qubit at $S$-band forbidden transitions \citep{Morley2013}
can be longer than those of the pure electron but are shorter than for the case of a nucleus.
The coherence times are five orders of magnitude longer than the timescale for
manipulation (32~ns manipulation time as described in \Chap{hybridqubit}).
Without OWPs, the relevant coherence time for quantum computation with the hybrid qubit is the shortest one;
that of the pure electron spin.
The longest qubit manipulation time is that of the pure nucleus and
dominates the time taken for a quantum computation.
Thus, the hybrid qubit at $S$-band offers the possibility of
preventing the `worst of both worlds': the limiting coherence time is at least as long while the manipulation time
is enhanced by orders of magnitude.

We further presented the first demonstration of suppression of spin bath decoherence at OWPs \citep{Balian2012},
later verified by experiments \citep{Wolfowicz2013,Balian2014}.
Near an ESR-type OWP,
coherence times of the hybrid qubit in natural silicon are increased
from about 0.5~ms to 100~ms. Here, quantum control
can also be achieved with fast ns pulses, as the transition matrix element is primarily
electronic.

An underlying question of physical interest is when decoherence is the result of the magnetic noise 
from independently flip-flopping pairs of spins and when consideration of the many-body nature of the quantum bath is important.
The answer is also of use for practical reasons. For one, if decoherence is due to flip-flopping pairs, there are widely used models (such as the analytical pseudospin
expressions in \Chap{formula}) which can be used to accurately calculate decays. Otherwise, more complex many-body numerics
become essential to simulate and fully understand experimental behaviours.
The clear result is that for the Hahn spin echo (and also for low to moderate pulsed dynamical decoupling as will be seen in \Chap{dynamicaldecoupling}),
the elimination of correlations from independent pairs is so drastic
at OWPs, that many-body numerics (CCE3) is almost indispensable for full understanding and accuracy \citep{Balian2015}.

In summary, for the FID and Hahn spin echo, the latter in regions away from OWPs, pair correlations give converged coherence decays
and hence a reliable $T_2$ can be extracted.
Near OWPs, the Hahn echo only decays for an initial time period much shorter
than the timescale of $T_2$, however, extrapolating the short-time behaviour
for long times gives the correct timescale.
The experiments we compare to are all for the Hahn echo,
and the short-time pair correlations give the correct $T_2$.
We find that up to three spin clusters are needed to achieve converged CCE Hahn echo decays near OWPs.
These higher-order CCE results 
are in agreement with our pair correlation results for the short-time Hahn (and the analytical formula described in \Chap{formula}).

Finally, we illustrated the significant differences between decoherence from classical field noise and quantum bath decoherence
by comparing our CCE calculations to $df/dB$ models \citep{Balian2014}.
The content we present in the next chapter further illustrates such differences and clear signatures of quantum bath decoherence,
by analysing pair correlations and deriving a closed-form $T_2$ formula for the hybrid qubit in a nuclear spin bath.

\chapter[Hybrid Pseudospins and $T_2$ Formula]{Hybrid Pseudospins \\ and $T_2$ Formula}
\label{chap:formula}

In this chapter, we present a derivation for a closed-form $T_2$ formula for nuclear spin diffusion of the hybrid qubit,
first published in \citet{Balian2014}.
The formula also clarifies significant differences between decoherence driven by classical field noise
and quantum bath decoherence.
In order to obtain the formula, we analyse the pair correlations for the single-spin FID case.
We then numerically establish the relationship between the Hahn spin echo and single-spin FID,
scaling the formula to account for the Hahn spin echo case and compare its predictions with experiments
for the Hahn spin echo.

The formula also clearly exposes qualitative differences between decoherence driven by a quantum spin bath
and decoherence due to classical magnetic field fluctuations. We derive the formula by showing that the spin dynamics separate naturally into terms
acting on very different timescales. The formula is valid for the hybrid qubit in both the mixed and unmixed regimes, the latter corresponding to a bare electron spin.
The coherence time is given as a function of the polarisation
(for each level and as described in \Chap{hybridqubit}) which quantifies the mixing as a simple analytical function of $B$.
The formula is
\begin{equation}
T^{u \to l}_2(B,\theta)  \simeq  \overline{C}(\theta) \frac{|P_u(B)|+|P_l(B)|}{\left|P_u(B)-P_l(B)\right|},
\label{eq:t2formula}
\end{equation}
written for a transition $\ket{u} \to \ket{l}$ at magnetic field $B$.
The constant, $\overline{C}(\theta)$, depends only on magnetic field orientation,
the density of nuclear spin impurities and their gyromagnetic ratio. \Eq{t2formula} is shown to give excellent agreement
with CCE numerics and experimental data for both ESR-type and NMR-type transitions near and far from OWP regimes.

\section{Hybrid Pseudospins}

\begin{figure}[h]
\centering\includegraphics[width=3in]{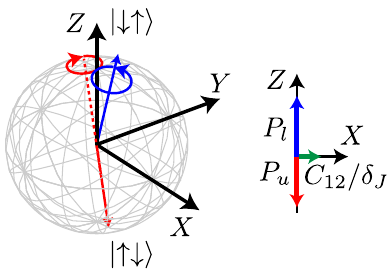}
\caption{
Illustration of the evolution of the bath states in the Hilbert space spanned by $\{\ket{\uparrow \downarrow},\ket{\downarrow \uparrow}\}$ under the influence of their dipole coupling ($C_{12}$) and their mutual detuning $\delta_J$ caused by interaction with the central spin. Figure adapted from \citet{Balian2014}.
}
\label{fig:Balian2014_singlepseudospin}
\end{figure}

In order to investigate the suppression of decoherence at OWPs and also for fields and transitions far from OWPs,
here we analyse only pair correlations which can be treated as independent pseudospins as explained in \Sec{pseudospins}.
For the case of the FID at all magnetic fields and in unmixed regimes far from OWPs for both the FID and Hahn spin echo,
CCE2 gives convergent decays and so an analysis based on pseudospins is fully justified.
Near OWPs however, the Hahn spin echo only decays for some initial time and higher-order numerics
are needed for converged decays as was shown in \Chap{decoherence}.
However, the initial decay of the Hahn echo gives a good indication of the $T_2$ timescale.
All our pseudospin results agree with experimental measurements of $T_2$ for the Hahn echo near OWPs in
this early-decay approximation.

\subsection{Interaction and Bath Hamiltonians}

For donors spin qubits in silicon, one may assume $H_{\text{CS}} \gg H_{\text{bath}}$ and thus ignore non-secular terms in $\hat{H}_{\text{int}}$.
The interaction Hamiltonian \Eq{nsdintH} for the $n$-th pair reduces to Ising form:
\begin{equation}
\hat{H}^{(n)}_{\text{int}} = \sum_{a=1,2}J^{(n)}_a \hat{S}^z \hat{I}^z_i,
\end{equation}
with Fermi contact hyperfine coupling strengths $J^{(n)}_a$.
For the two interacting spin-1/2 bath spins, assuming a large magnetic field and thus keeping only energy conserving terms,
the dipolar interaction in \Eq{nsdbathH}
simplifies to
\begin{equation}
\hat{H}^{(n)}_{\text{bath}} =  C^{(n)}_{12} \hat{I}_1^z \hat{I}_2^z - \tfrac{C^{(n)}_{12}}{4} (\hat{I}_1^+ \hat{I}_2^- + \hat{I}_1^- \hat{I}_2^+)
\end{equation}
where $C^{(n)}_{12}$ is the strength of the dipolar coupling between the two bath spins for pair $n$.
Zeeman terms are also excluded from $\hat{H}^{(n)}_{\text{bath}}$
as these do not contribute to decoherence.
Neglecting the effect of $\hat{H}_{\text{int}}$ on the mixing of the central spin states,
the dynamics is governed by $\hat{h}^{(n)}_{i}$ (conditional on the state of the central spin):
\begin{equation}
\hat{h}^{(n)}_{i} \equiv
\bra{i}(\hat{H}_\text{int}^{(n)} + \hat{H}_\text{bath}^{(n)})\ket{i} = -\tfrac{C_{12}}{4} \hat{\mathds{1}} -\tfrac{1}{4}\hat{\boldsymbol\sigma}\cdot {\bf H}^{(n)}_{i},
\label{eq:pseudospinH}
\end{equation}
where the effective field is ${\bf H}^{(n)}_i=[C^{(n)}_{12}, 0, P_i\delta_J^{(n)}]$.
Here, $\delta_J^{(n)}\equiv (J^{(n)}_1 - J^{(n)}_2)$ is the difference in hyperfine couplings to the bath while
$\hat{\boldsymbol\sigma}$ is the vector of Pauli matrices acting on the bath basis $\{\ket{\downarrow \uparrow} \equiv \ket{\downarrow} \otimes \ket{\uparrow},
\ket{\uparrow \downarrow} \equiv \ket{\uparrow} \otimes \ket{\downarrow} \}$ and $\ket{\uparrow}$ and $\ket{\downarrow}$ denote the nuclear spin-1/2 Zeeman states.
The identity term is dynamically uninteresting; the dynamics can in fact be
considered simply as a precession about ${\bf H}_i^{(n)}$.
Diagonalising the Hamiltonians in \Eq{pseudospinH} gives the pseudospin precession rates
\begin{equation}
\omega_i^{(n)}=\frac{1}{4}\sqrt{(C_{12}^{(n)})^2+ (P_i\delta_J^{(n)})^2},
\end{equation}
while the angle of ${\bf H}_i^{(n)}$ from the $z$-axis is
\begin{equation}
\theta_{i}^{(n)} = \tan^{-1}{[C_{12}^{(n)}/(P_i\delta_J^{(n)})]}.
\end{equation}
The pseudospins are illustrated in \Fig{Balian2014_singlepseudospin}.

For the hybrid qubit, the pseudospin dynamics is in most respects, quite similar to those investigated previously for electron (unmixed)
qubits \citep{Yang2008a,Yang2008b,Yang2009,Yao2006,Yao2007,Liu2007,Zhao2012a}.
However, the main  difference is that for our case, the hybrid pseudospins have the electron $z$-projection $\hat{S}^z$ replaced
by the polarisation term
$P_i \equiv 2 \bra{i} \hat{S}^z \ket{i}$.
While for an electron, $P_i=\pm 1$ is a constant, 
for mixed systems the $P_i(B)$ are strongly field-dependent.
The pseudospin analysis is framed in the pure dephasing approximation and
hence requires that the interaction Hamiltonian has
negligible effect on the mixing of the central spin states themselves, i.e.\ on $P_i$. Since 
$H_\text{CS} \gg H_\text{int}$, this is the case except extremely close to OWPs, where $T_2$ becomes extremely
sensitive to small fluctuations in $P_i$.

\section{Derivation of $T_2$ Formula}

We employ an analysis of the hybrid pseudospins and a range of other approximations described below to derive the formula.
The analysis also provides an intuitive picture of the system-bath dynamics, especially with respect to the magnetic field
approaching an OWP and far from it, and also for NMR-type and ESR-type transitions.

\subsection{Short-Time Behaviour}

The $n$-th cluster decay for a single spin pair $n$ has been investigated analytically for both the FID and
Hahn echo case \citep{Yao2006,Zhao2012a,Witzel2005}.
We emphasize that this is the single-spin FID without inhomogeneous broadening.
In experiment, $T_2$ is normally measured using a Hahn echo pulse sequence, in order to remove strong enhancements in decoherence arising from static inhomogeneities.
Although the Hahn echo can suppress some effects of the dynamics,
the  FID and Hahn $T_2$ times are of the same order, differing by at most 
a factor of $\approx 2$ even at OWPs, so we focus our analysis on the simpler FID expressions.

Although analytical forms for the time decays $\mathcal{L}_{n}^{u \to l}(t)$ from spin pairs are known \citep{Yao2007,Zhao2012a},
a closed form for $T_2$, sufficiently accurate for experimental analysis is more difficult.
Each $\mathcal{L}^{u \to l}_{n}(t)$ is an oscillatory
function, with frequencies given in terms of $\omega_u^{(n)}$ and $\omega_l^{(n)}$
and the full decays combines hundreds or thousands of spin pair contributions.

A usual approach is to expand the decay as a power series $|\mathcal{L}_{n}^{u \to l}(t)|= 1- \sum_{p=1} a^{(n)}_{2p} t^{2p}$ and to infer the order of magnitude of $T_2$ from the early time behaviour.
 However, for important cases like spin diffusion, $a_{2}^{(n)}=0$ while $a^{(n)}_4 \neq 0$, predicting a $\exp[-a^{(n)}_4 t^4]$ decay to leading order \citep{Witzel2005,Yao2006,Yao2007},
in contrast to the observed decays of $\sim \exp[-a^{(n)}_2 t^2]$ for typical spin
systems, where the Taylor coefficient $a^{(n)}_2$ is identified as $1/T_2^2$ . Thus it appears that in that case, one cannot infer the character of the decay
on timescales $t \sim T_2$ from the short time behaviour (i.e.\ on timescales $t \sim \omega_i^{-1}$). 

The observed exponential-quadratic (Gaussian) character of the coherence decay has been demonstrated numerically for both Hahn echo decay and FID
from cluster expansion or linked-cluster expansion simulations \citep{Witzel2006,Saikin2007}.
For the FID case specifically, a crossover from the exponential-quartic to Gaussian behaviour was found on the
microsecond timescale, arising from the combined effect of
many pair cluster contributions \citep{Saikin2007}.
For spin donors in silicon, $T_2$ times are on the millisecond to
second timescales, thus the exponential-quartic regime is not relevant (though it
may be appropriate for GaAs quantum dots, which have shorter $T_2$
times).

Below, we shall see that $T_2$ times sufficiently reliable for experimental
analysis are obtainable analytically if we consider, separately, the different frequency terms involved in
the pair correlation which act on very different timescales.
Thus, we propose a very different explanation for the observed decay form
crossover in the FID which does not require one to combine large
numbers of cluster contributions. We show
that in fact the crossover originates naturally from a
single pair correlation term.

\subsection{Bath State Overlap}

As discussed in \Chap{background}, the decay in coherence of the central spin can be related to its entanglement with the bath.
We assume an initial state such that the qubit and bath are unentangled, and the qubit is prepared
in a coherent superposition of its upper ($\ket{i=u}$) and lower ($\ket{i=l}$) states.
The coherence for the FID in pure dephasing for the $n$-th spin pair is given by
\begin{equation}
|\mathcal{L}^{u\to l}_{\text{FID},n}(|\mathcal{B}^{(n)}(0),t)\rangle| = | \langle \mathcal{B}^{(n)}_u(t) |  \mathcal{B}^{(n)}_l(t) | =
\langle \mathcal{B}^{(n)}_l(0) |   \hat{T}^{(n) \dagger }_u  \hat{T}^{(n)}_l  |\mathcal{B}^{(n)}_u(0) \rangle  |,
\end{equation}
which involves calculating the time-dependent overlap between bath states correlated with the upper and the lower central spin states.

For simplicity, we drop the pair label $n$ in what will follow until summing the contribution from all spin pairs.
The evolution of the bath during the FID of the central spin follows
$\boldsymbol{\mathcal{B}}_i(t)={\bf R}_y(\theta_{i}){\bf R}_z(2\omega_i t) {\bf R}_y^\intercal(\theta_i)\boldsymbol{\mathcal{B}}(0)$
in the matrix representation, where ${\bf R}_y$ and ${\bf R}_z$ represent the usual rotation matrices \citep{Nielsen2010} and $\boldsymbol{\mathcal{B}}(0)$ is the initial bath state in the basis $\{ {(0 \ 1)}^\intercal : \ket{ \uparrow \downarrow} , {(1 \ 0)}^\intercal : \ket{ \downarrow \uparrow} \}$ and in general can be a superposition of $\ket{\uparrow \downarrow}$ and $\ket{\downarrow \uparrow}$.
We can combine the unitaries for the upper and lower state in one matrix and thus the bath overlap can be written as
\begin{eqnarray}
&&\mathcal{L}_{\text{FID}}^{u \to l}(t) =
\boldsymbol{\mathcal{B}}^\intercal(0)
{\bf T}^*_{ul}(\omega^-, \omega^+,t)
\boldsymbol{\mathcal{B}}(0); \nonumber\\
&&{\bf T}^*_{ul}(\omega^-, \omega^+,t) =  \\
&&{\bf R}_y(\theta_{u})\left( \begin{array}{cc}
e^{i \omega^- t} \cos{\theta^-} & e^{i \omega^+ t} \sin{\theta^-}\\
-e^{-i \omega^+ t}\sin{\theta^-}  & e^{-i \omega^- t}\cos{\theta^-} \end{array} \right) {\bf R}_y^\intercal(\theta_l)\nonumber
\label{eq:fidunitary}
\end{eqnarray}
where $\theta^\pm= \frac{1}{2} (\theta_{u}\pm \theta_{l})$ and
$\omega^{\pm}=\omega_u \pm \omega_l$.
We see that expressions for the decays arise naturally in terms of $\omega^{\pm}$ rather than $\omega_u$ and  $\omega_l$ as is usual.\footnote{This is also the case for the Hahn spin echo case.}

For the initial state $\boldsymbol{\mathcal{B}}(0)^\intercal = {(0 \ 1)}$ or ${(1 \ 0)}$, the time decay
 for FID  is given by
\begin{align}
|\mathcal{L}^{u\to l}_{\text{FID}}(\ket{\uparrow \downarrow},t)| &= |\{{\bf T}^*_{ul}(\omega^-, \omega^+,t)\}_{11}| \nonumber \\
&= \left|D^+e^{-i\omega^- t} + D^-e^{+i\omega^- t}
            ~~+ R^+e^{-i\omega^+t} + R^-e^{+i\omega^+t} \right|,
\label{eq:fullfid}
\end{align}
where $R^{\pm} = \frac{1}{2}\sin \theta^-(\sin \theta^- \mp \sin \theta^+)$ while
$D^{\pm} =  \frac{1}{2}\cos \theta^-(\cos \theta^- \pm  \cos \theta^+)$.

\subsection{$T_2$ Weights from Pairs}

\begin{figure}[h]
\centering\includegraphics[width=5in]{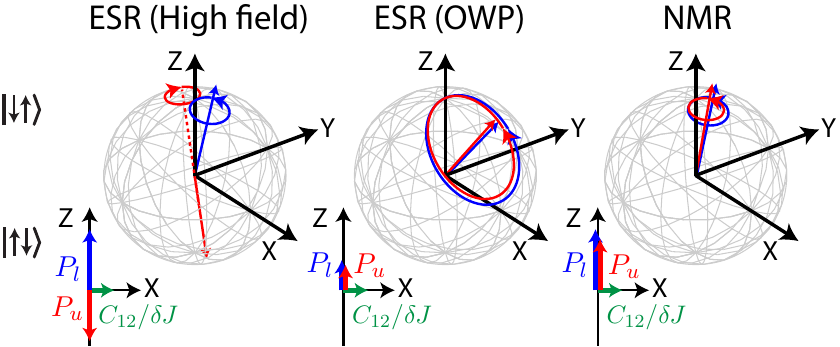}
\caption{
Illustration of the evolution of the bath states in the Hilbert space spanned by $\{\ket{\uparrow \downarrow},\ket{\downarrow \uparrow}\}$ under the influence of their dipole coupling ($C_{12}$) and their mutual detuning caused by interaction with the central spin. At both OWPs and NMR-type transitions, bath trajectories correlated with the upper and lower central spin states follow similar trajectories and hence decoherence is suppressed compared to ESR-type transitions. However, at ESR-type OWPs, $|P_{u,l}| \simeq 0.1$ leads to a larger trajectory and proportionately shorter $T_2$ values relative to NMR-type transitions. Figure adapted from \citet{Balian2014}.
}
\label{fig:Balian2014_pseudospins}
\end{figure}

We now consider the contributions to the coherence which dominate the spin pair correlation
in different regimes and timescales.
We consider \Eq{fullfid} in three principal limits:
\begin{enumerate}[(i).]
  \item For an ESR-type transition in the high-field regime in which the states are not mixed. This corresponds to $P_u \simeq -P_l$.
  \item For an NMR-type transition in the high-field regime, or for any transition near an OWP. Here, $P_u \simeq P_l$.
  \item For an intermediate regime corresponding to a Landau-Zener crossing \citep{Mohammady2010} or cancellation resonance, where one of the $P_i\simeq0$.
\end{enumerate}
The pseudospin evolutions for the first two regimes are illustrated in \Fig{Balian2014_pseudospins}

For either (i) or (ii), since $|P_u| \simeq |P_l|$ then $\omega_u \simeq\omega_l$ and thus $\omega^+/\omega^- \gg 1$.
Hence, we infer that the $R^\pm$ terms act on very
different timescales from the terms proportional to $D^\pm$. We consider the $R^\pm$ and $D^\pm$ terms
separately.
If we set $\omega^-=0$, we obtain the fast oscillating contribution:
\begin{equation}
|\mathcal{L}^{u \to l}_{\text{FID}}(\ket{\uparrow \downarrow},t)|^2 \simeq 1-4\left(D^+ + D^-\right)\left(R^+ + R^-\right) \sin^2 \tfrac{\omega^+t}{2} - 4R^+ R^-  \sin^2 (\omega^+ t).
\label{eq:fast}
\end{equation}
We now extract the contribution of each cluster to the
total decoherence by means of a power expansion; for short times we obtain
\begin{equation}
|\mathcal{L}_\text{FID}^{u \to l}(\ket{\uparrow \downarrow},t)|^2 \approx 1 - \frac{t^2}{T_2^2}  \approx \exp{\left[-\left(\frac{t}{T_2}\right)^2\right]},
\end{equation}
yielding the $n$-th cluster contribution to $T_2$:
\begin{equation}
\left(T^{(n)}_{2}\right)^{-2}  \approx \left[\left(D^+ + D^-\right)\left(R^+ + R^-\right)+ 4R^+ R^-\right]
\left(\omega^+\right)^2.
\label{eq:fidweightsupdown}
\end{equation}
We now perform the incoherent averaging over initial bath states, 
$\left \langle \mathcal{L}^{u \to l}_{\text{FID}}(t)\right\rangle \approx \tfrac{1}{2} + \tfrac{1}{2} |\mathcal{L}^{u \to l}_{\text{FID}}(\ket{\uparrow \downarrow},t)|$
to allow for the fact that approximately half the bath spins are in $\ket{\uparrow \uparrow}$ and $\ket{\downarrow \downarrow}$ states which cannot flip-flop and obtain:
\begin{equation}
\frac{1}{T_2^{(n)}}  \simeq \frac{1}{2}\left|\sin \theta_u -\sin \theta_l\right| \frac{\omega^+}{2},
\label{eq:fidweights}
\end{equation}
noting that the first term is the difference in precession radii of the pseudospins, while the second term denotes the average precession rate.
In terms of the usual flip-flop models, we note that a larger precession radius corresponds to a larger flip-flop amplitude, while a larger precession frequency
corresponds to a higher flip-flop frequency.

\subsection{Separation of Timescales}

We now distinguish between the two regimes (i) and (ii). For (i), for timescales $\ll (\omega^+)^{-1}$, as discussed above, we neglect the slow oscillations
(i.e.\ those in $\omega^-$) in \Eq{fullfid}, which contribute only on very long timescales. We obtain the $n$-th cluster contribution to $T_2$ (\Eq{fidweights})
by Taylor expanding the decay with $\omega^-$ set to zero, i.e.\ using only the fast terms.
For (ii), $\omega^+/\omega^- \gg 1$ is still valid but $|D^{\pm}| \gg | R^{\pm}|$ in \Eq{fullfid},
and the {\em slow} oscillations dominate for timescales
$1/\omega^+ \lesssim t \lesssim 1/\omega^-$. However, expanding these slow oscillations gives precisely the same form
as \Eq{fidweights}.

Using \Eq{fidweights} in all cases, we can estimate a total $T_2$ using
\begin{equation}
\frac{1}{T_2^2} = \sum_{n=1}^{n=N} \left(\frac{1}{T_2^{(n)}}\right)^2,
\label{eq:totalt2}
\end{equation}
where for the converged CCE2 spin bath in natural silicon, $N \simeq 10^4$.

Importantly, including both fast ($\omega^+$) and slow ($\omega^-$) terms the power series (i.e.\ expanding the full decay \Eq{fullfid}),
the quadratic contributions cancel and the pair correlation result simply gives an exponential-quartic dependence (not observed in experiment) at leading order.
Separation of the $\omega^\pm$ timescales is useful not only here, but also potentially in the unmixed ESR regimes of other spin systems.
We proceed to discuss the importance of separating timescales further.

\begin{figure}[h]
\centering\includegraphics[width=3.5in]{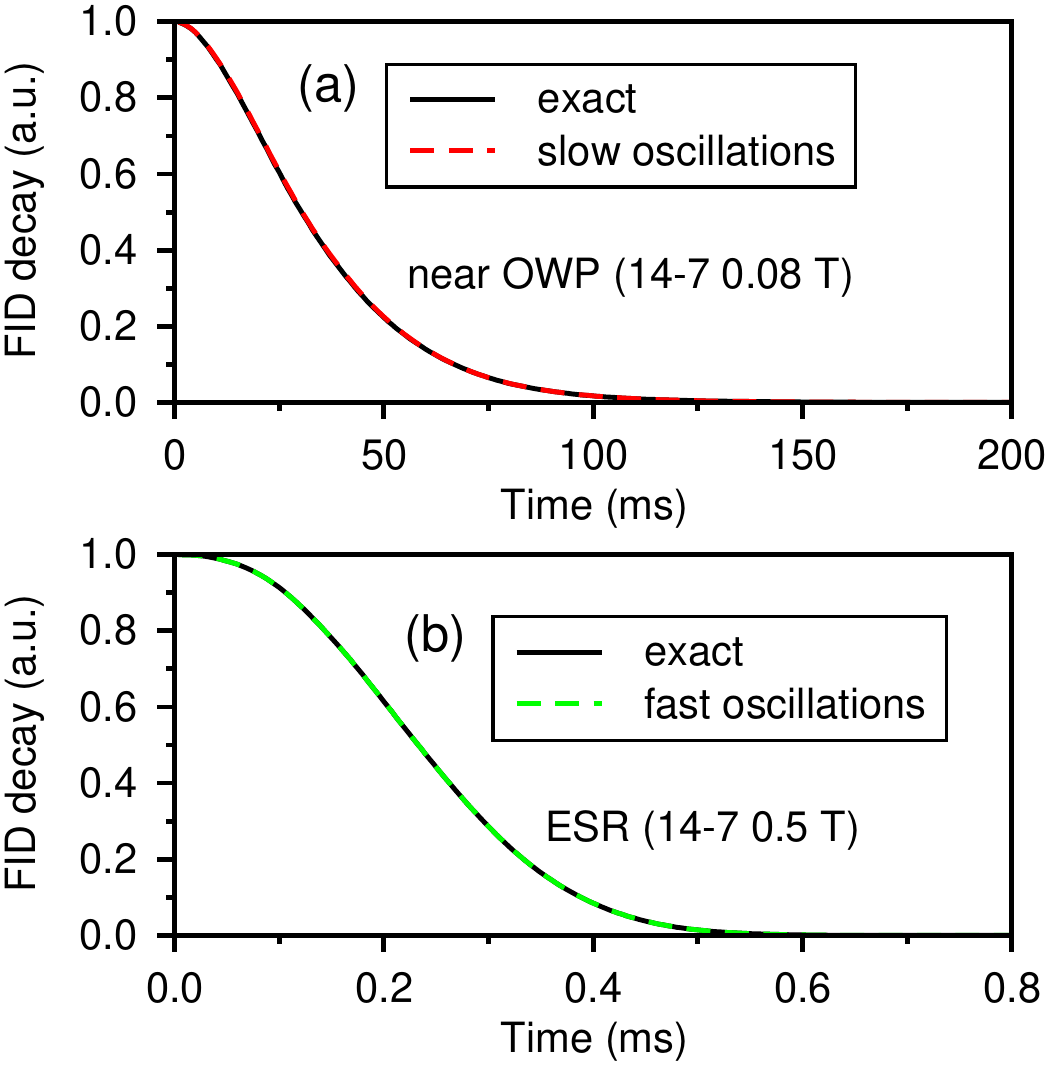}
\caption{
Shows that OWP regimes are dominated by slow oscillating terms while ESR regimes
are dominated by fast oscillating terms in \Eq{fullfid}.
(a) Compares decays obtained from \Eq{fullfid} (exact) with decays obtained from \Eq{slow}
(slow oscillations only). (b) Compares decays obtained from \Eq{fullfid} (exact) with decays obtained from \Eq{fast}
(fast oscillations only). Figure adapted from \citet{Balian2014}.
}
\label{fig:Balian2014_timescales}
\end{figure}

As mentioned above, care is needed when considering the (ii) regimes (OWP and NMR) since here,
$P_u \simeq P_l$ and $\theta_u \simeq \theta_l$
and thus $D^\pm \gg R^\pm$. Here, $D^++D^- \to 1$ while $R^\pm \to 0$.
Decay timescales become long and comparable to $1/\omega^-$ while the $R^\pm$ amplitudes are negligible
and thus the slow oscillating components are important. In that case, we would, in contrast to \Eq{fast},
neglect the fast oscillations. Then we obtain,
\begin{eqnarray}
|\mathcal{L}^{u \to l}_{\text{FID}}(\ket{\uparrow \downarrow},t)|^2 \simeq 1 -  4D^+ D^-  \sin^2 \omega^- t.
\label{eq:slow}
\end{eqnarray}
In this case, $(T_2^{(n)})^{-2}  \approx D^+ D^- \left(\omega^-\right)^2$. 
However, since
 \begin{eqnarray}
\left[\left(D^++ D^-\right)\left(R^++ R^-\right)+ 4R^+ R^-\right]
\left(\omega^+\right)^2 \to  D^+ D^- \left(\omega^-\right)^2
\end{eqnarray}
as  $P_u \to P_l$, the contribution to $1/T_2^2$ from each cluster, in fact, still has the same form
as \Eq{fidweightsupdown}. In other words, the relative weights obtained from the slow, high-amplitude
contributions are quite similar to those obtained by considering the faster, lower oscillations
and thus the $T_2$ expression we derive below is still valid.

In \Fig{Balian2014_timescales}, we show the full temporal decay for all pairs
\begin{equation}
\mathcal{L}_\text{FID}^{u \to l}(\ket{\uparrow \downarrow},t) = \prod_{n}^{N \simeq 10^4} \mathcal{L}^{u \to l}_{\text{FID},n}(\ket{\uparrow \downarrow},t),
\end{equation}
where $\mathcal{L}^{u \to l}_{\text{FID},n}(\ket{\uparrow \downarrow},t)$ are given by \Eq{fullfid}
and compare with the slow terms in an OWP regime (\Fig{Balian2014_timescales}(a)) where
$\mathcal{L}^{u \to l}_{\text{FID},n}(\ket{\uparrow \downarrow},t)$ are given by \Eq{slow}
and the fast terms in the ESR regime (\Fig{Balian2014_timescales}(b))
where $\mathcal{L}^{u \to l}_{\text{FID},n}(\ket{\uparrow \downarrow},t)$ are given by \Eq{fast}.
\Fig{Balian2014_timescales} shows that while the fast terms completely dominate coherence decay in the ESR regime,
the slow terms completely dominate the decays in the OWP/NMR regime yet the form of the weights in the
power expansion is similar: if added, the two contributions thus cancel (albeit briefly) yielding the quartic-exponential decay.
This decay is of course valid on extremely short timescales $t \ll (\omega^+)^{-1}$ but
not on the $T_2$ timescale.

In fact, the fast oscillation behaviour is not entirely straightforward.
For the slow oscillations, \Eq{slow} involves a single 
frequency and an 
approximate $\exp{[-(t/T_2)^2]}$ decay  
is straightforwardly inferred.
For the fast oscillations however, \Eq{fast} may be rewritten as follows for each spin pair:
\begin{eqnarray}
|\mathcal{L}^{u \to l}_{\text{FID}}(\ket{\uparrow\downarrow},t)|^2& \simeq & 1- \sin^2 \theta^- \cos^2 \theta^+ \sin^2(\omega^+ t)  \nonumber \\
 &-&  \sin^2 2\theta^- \sin^2\tfrac{\omega^+t}{2}\nonumber \\
 &-& \frac{1}{4}\sin^2 2\theta^- \sin^2 (\omega^+ t)  \nonumber \\
&=& 1-L_s(t)-\left(L_{l1}(t)-L_{l2}(t)\right).\nonumber \\
\end{eqnarray}
We see that it combines three separate interfering terms, where $L_{l1}$ oscillates at half the frequency of 
the others. In fact, a power expansion of either one of the individual terms $L_s(t)$, $L_{l1}(t)$ and $L_{l2}(t)$
would yield the same 
weights expression \Eq{fidweights}.
It is the ubiquitous nature of this 
$ \left( \sin\theta_u - \sin\theta_l\right)^2 \left(\omega^+\right)^2$ term which underlies the robustness of the field dependence of our $T_2$ expression
derived below.

We note that it is in fact the term
$L_s(t)= \frac{1}{4}\left(\sin \theta_u -\sin \theta_l\right)^2 (\omega^+)^2$
 which yields a quadratic dependence at short times.
However, numerics show that it is the
 $1 -(L_{l1}(t)+L_{l2}(t))$ terms  which 
overwhelmingly determine the decay on longer
 $T_2$ timescales (but actually make little contribution on the $t \ll (\omega^+)^{-1}$ timescale,
where there is once again a brief cancellation of these
near equal amplitude oscillations).

Finally we consider regimes (iii), or the Landau-Zener regimes (there are four such regions for Si:Bi). These do not fit the above analysis,
which assumed $|P_u| \simeq |P_l|$. For the LZ points either $P_u \simeq 0$ or $P_l \simeq 0$.
Thus, assuming $P_u \simeq 0$ we obtain,
\begin{equation}
|\mathcal{L}^{u \to l}_{\text{FID}}(\ket{\uparrow \downarrow},t)|^2 \simeq 1- \sin^2 \theta_u \sin^2 \omega_u t,
\end{equation}
and hence for $ t \ll (\omega_u)^{-1} $, we have simply
\begin{equation}
|\mathcal{L}^{u \to l}_{\text{FID}}(\ket{\uparrow \downarrow}t)|^2 \simeq  1- C_{12}^2t^2.
\label{eq:lzregime}
\end{equation}

\subsection{Strong Coupling Approximation}

\begin{figure}[h]
\centering\includegraphics[width=4.5in]{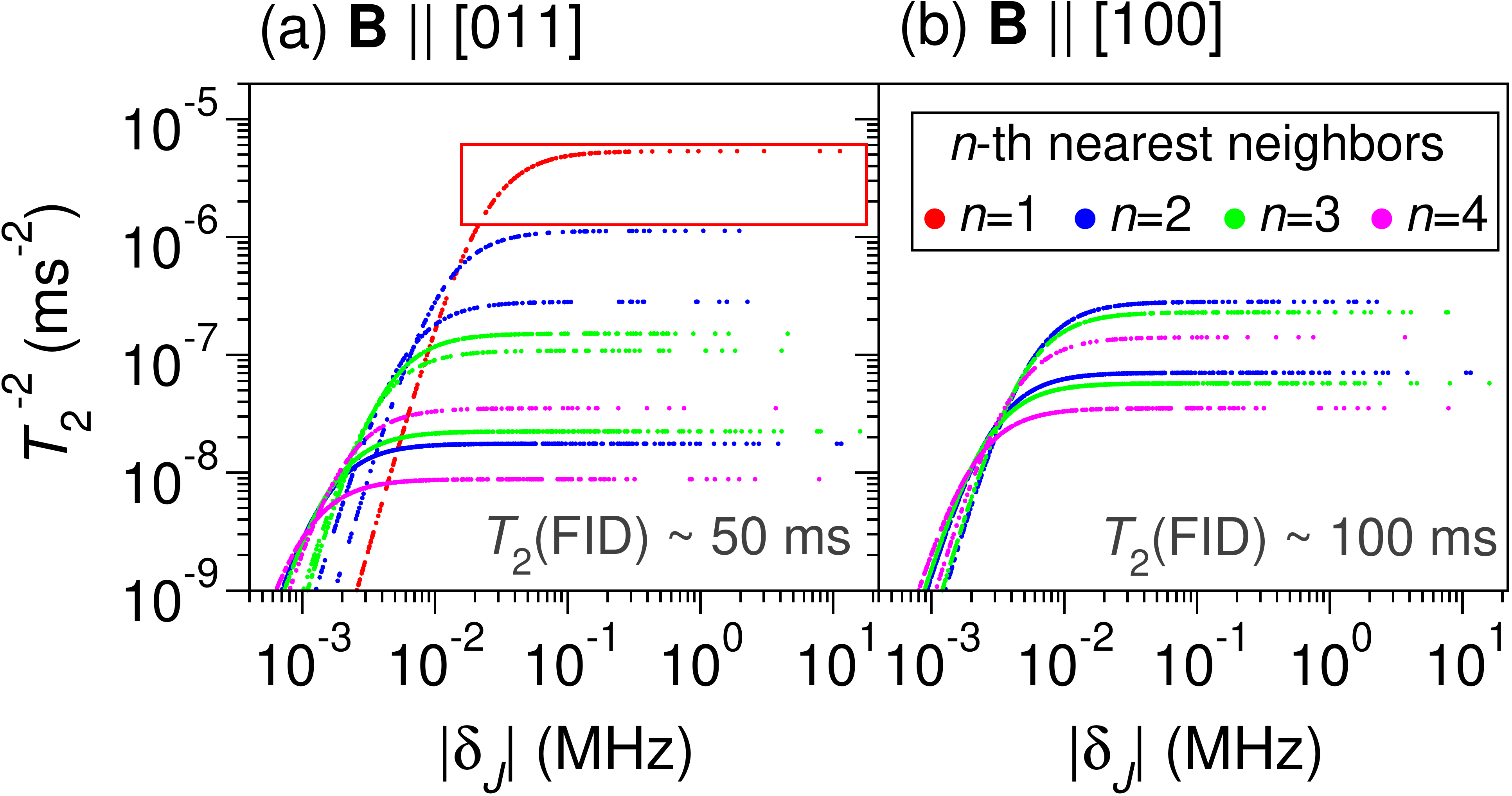}
\caption{The individual contribution of each spin pair in the bath to the total $(1/T_2)^2$ near OWPs, from \Eq{fidweights}. Data are shown for two magnetic field orientations.
For large $|\delta_J|$, coherence times become nearly independent of  $|\delta_J|$. The scale of $T_2$ is set by a comparatively small $N \sim 10^2$ set of strongly-coupled spins ($|P_i\delta_J|\gg |C_{12}|$), illustrated in the red box.
$B=79.8$~mT (about $0.1$~mT offset from the OWP) and $P_i\simeq 0.1$. $\gamma_N = 8.465$~MHz/T for \sitn\
and hyperfine coupling strengths were calculated using the Kohn-Luttinger electronic wavefunction with an ionization energy
of $0.069$~eV for the bismuth electron \citep{DeSousa2003b}. Figure adapted from \citet{Balian2014}.}
\label{fig:Balian2014_weightss}
\end{figure}

In \Fig{Balian2014_weightss}, we use \Eq{fidweights} to evaluate the strength of each $^{29}\text{Si}$ spin pair's individual 
contributions to decoherence of a bismuth donor spin in silicon. We plot $1/(T_2)^2$ for each cluster, as a
function of $|\delta_J|$, in regime (ii) i.e.\ close to OWPs and NMR-type transitions. Strikingly, the spins are grouped into lines of constant $C_{12}$,
corresponding to $n$-th nearest neighbor spins.
Furthermore, for the spin pairs most active in driving decoherence, $1/(T_2)^2$ is only very weakly dependent on $|\delta_J|$.
 The origin of this behaviour is clear from \Eq{fidweights}: for large $|P_i\delta_J|\gg |C_{12}|$, the term $|\sin \theta_u -\sin \theta_l| \propto |\delta_J^{-1}|$  while $\omega^+ \propto |\delta_J|$, eliminating the dependence on the hyperfine coupling between the central spin and bath spins.

 The insensitivity of the decoherence to the coupling between the central spin and the bath 
might at first seem counter-intuitive. However, the physical origin of this effect is
thus:  increasing the
hyperfine detuning $ \propto |\delta_J^{-1}|$ damps the flip-flopping amplitudes;
 however within this model, the decrease in amplitude
is exactly compensated by a corresponding increase in flip-flop frequency.
 We note that without separation of timescales, 
the exponential-quartic decay constants which prevail at times $t \ll \omega_i$ are  dependent
 on $\delta_J^2$  \citep{Yao2007}.
In contrast, our model predicts that a comparatively small number of strongly coupled spins will dominate the 
decoherence, and that their individual contributions to $1/T_2^2$ are approximately equal, although the individual coupling
strengths $ |\delta_J^{-1}|$ vary by orders of magnitude, ranging from $\sim 0.01$ to $10$~MHz. 

To test the validity of this result at $t \sim T_2$ timescales, we ran CCE2 calculations for various field orientations.  
The dipolar coupling, $C_{12}$ is a function of the orientation $\theta$ of the magnetic field and hence the $T_2$ values vary accordingly.
 For $B \parallel \left\langle 011 \right\rangle$, for example, the $N \sim 10^2$ strongest coupled spin pairs suffice to
 set the scale of $T_2$. We have tested our model by running a CCE2 calculation with just 120 nearest-neighbor (NN)
spin pairs (e.g.\ for $B \parallel \left\langle 011 \right\rangle$, $C_{12}^{\text{NN}}=2.4$~kHz) which satisfy $|P_i\delta_J|\gg |C_{12}|$, and confirming the calculated $T_2$ is approximately equal to that considering all $\approx 10^4$ spin pairs.

If we make the strong coupling approximation,
the weights in \Eq{fidweights} can also be written as:
\begin{equation}
\frac{1}{(T_2^{(n)})^2}  \simeq \frac{(\theta_u - \theta_l)^2}{4^2} (\omega^+)^2.
\end{equation}
Then, noting $\theta_i \approx C_{12}/\omega_i$ and $\omega^+ \approx \delta_J (|P_u|+|P_l|)$
we easily obtain $\frac{1}{T_2^{(n)}} \propto  \frac{|P_u- P_l|}{|P_u|-|P_l|}$,
for the cases (i)  and (ii) when $ |P_u| \simeq |P_l|$, which include both 
the unmixed ESR limit as well as the NMR
and OWP limits. Summing the $T_2$ contributions according to \Eq{totalt2}, our final $T_2$ expression is given by \Eq{t2formula}:
$T_2 \simeq  \overline{C}(\theta) \tfrac{|P_u|+|P_l|}{\left|P_u-P_l\right|}.$ 
For most orientations, $\overline{C}(\theta) \approx 4/(C_{12}^{\text{NN}}\sqrt{N})$. However,
as the magnetic field orientation approaches $B\parallel \left\langle100\right\rangle$, the contribution of nearest-neighbor \sitn\ spin pairs vanishes,
while 2nd and 3rd nearest neighbors contribute similarly.
We discuss further details of the orientation dependence of $T_2$ in the next section.

Approaching the high magnetic field limit, ESR-type transitions occur between states where $P_u \simeq - P_l$,
such that $T_2\simeq\overline{C}(\theta)$, while for NMR-type transitions as well as OWPs, $P_u \simeq P_l$, and decoherence by the nuclear spin bath is suppressed.\footnote{In regimes where we can neglect the effect of the central nucleus in $\hat{H}_{\text{int}}$.}
Finally, for the third regime (iii) where one of the $P_i$ is zero, and hence the assumptions made to obtain \Eq{t2formula} are not valid.
Nevertheless, starting from \Eq{fullfid} we obtain \Eq{lzregime} and after the usual bath average and sum over clusters, we find $T_2 \sim \overline{C}(\theta)$
in this regime, and hence \Eq{t2formula} remains a reasonable approximation here.

\section{Angular Dependence}

\begin{figure}[h]
\centering\includegraphics[width=5.8in]{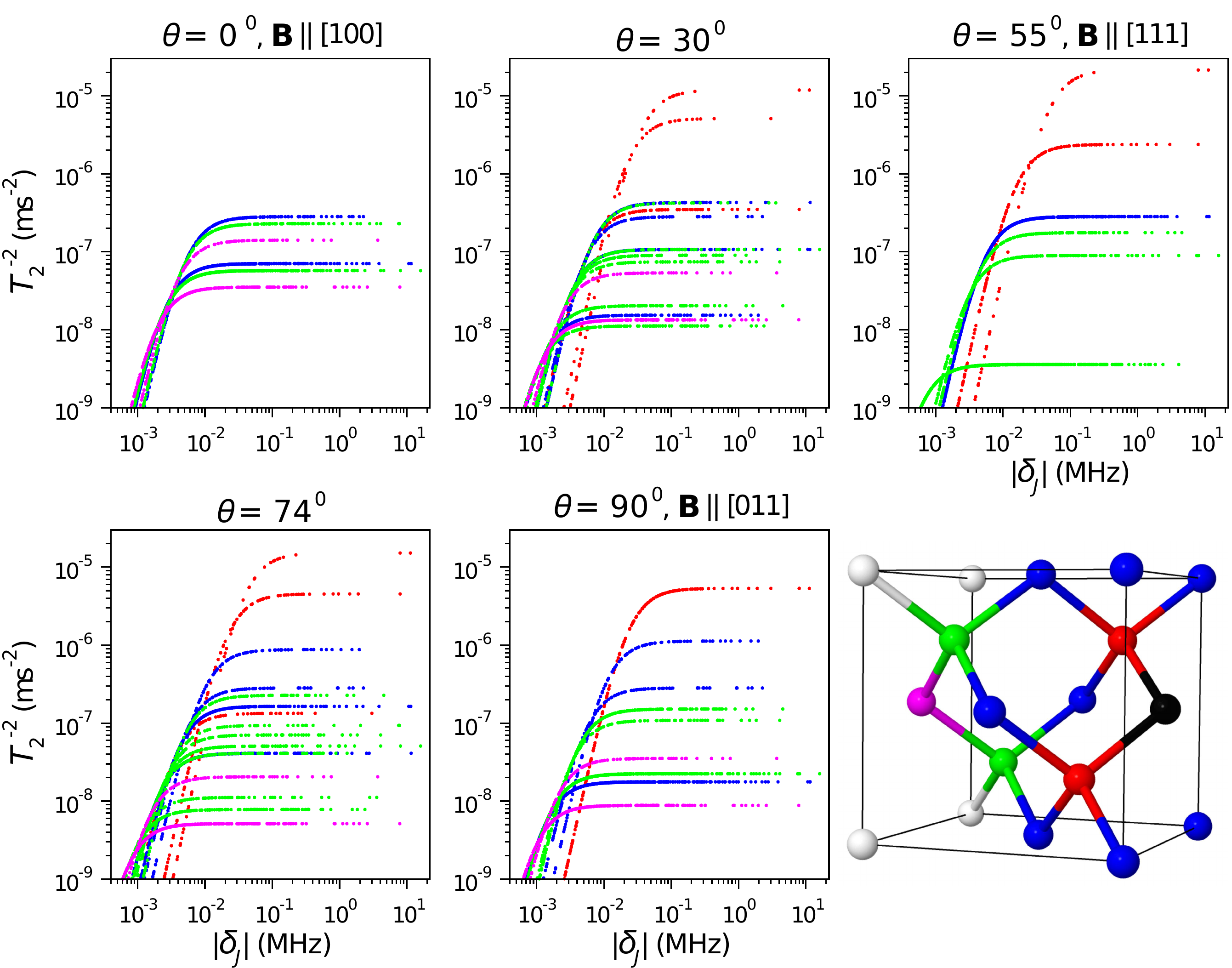}
\caption{
Theoretical contributions of spin pairs to $T_2(\text{Hahn})$, coloured according to $n$-th nearest neighbors relative to the black nucleus as illustrated in the last panel.
First nearest neighbors dominate decoherence for
rotation angles $\theta \gtrapprox 30^{\circ} $.
At $\theta=0^{\circ}$, first nearest neighbor contributions are diminished and second and third nearest neighbors contribute the most to $T_2$.
Rotation is performed about $\left[01\bar{1}\right]$ in the $\left[011\right]- \left[100\right]$ plane,
with $\theta$ from $\left[100\right]$. Figure adapted from \citet{Balian2014}. Details of the silicon crystal structure are given in \App{silicon}.
}
\label{fig:Balian2014_weightsl}
\end{figure}

\begin{table}[!t]
\centering
\begin{tabular}{ccc} \hline \hline
Rotation angle $\theta$ & 1-NN contribution to $\overline{C}(\theta)$ & Numerical $\overline{C}(\theta)$ \\ 
(degrees) & (ms) & (ms)\\ \hline
$90$   &   $0.37$   &  0.40 \\ 
$74$ &   $0.35$   &  0.39 \\ 
$55$   &   $0.32$   &  0.37 \\ 
$30$ &   $0.41$   &  0.45 \\ 
$0$ & None. 2--NN, 3--NN contributions: $0.97$ & $1.1$ \\ \hline\hline
\end{tabular}
\caption{Numerical values of the dipolar prefactor $\overline{C}(\theta)$ compared to $\overline{C}(\theta)$ when including only nearest neighbor spin pairs,
demonstrating that first nearest neighbors set the scale of $T_2$ for rotation angles $\theta \gtrapprox 30^{\circ}$. For $\theta = 0^{\circ}$, 1--NNs do not contribute at all and 2,3--NNs largely determine $T_2$. The total number of strongest spin pairs for each orientation was chosen such that the $T_2$ obtained was about $70 - 80\%$ of the total $T_2$ when including all spin pairs in the bath. Rotation is performed about $\left[01\bar{1}\right]$ in the $\left[011\right]- \left[100\right]$ plane,
with $\theta$ from $\left[100\right]$. Table adapted from \citet{Balian2014}.
}
\label{table:Balian2014_orientation}
\end{table}

Due to angular dependence of the dipolar interaction, $T_2$ varies with the orientation of the crystal sample relative to ${\bf B}$ \citep{DeSousa2003b,Witzel2006,Tyryshkin2006,George2010}.
The dipolar prefactor $\overline{C}(\theta)$ in our analytical $T_2$ formula (\Eq{t2formula}) depends on $C_{12}$ and is thus a function of crystal orientation.
The prefactor is defined as
\begin{equation}
\overline{C}(\theta) = \frac{4}{\sqrt{\sum_s{ N_s \left(C_{12}^{(s)}\right)^2  }}},
\label{eq:prefactor}
\end{equation}
where $s$ labels a unique value of spin pair dipolar strength $C_{12}^{(s)}$, or ``shell'', which occurs $N_s$ times.
We see below that including shells up to $s=3$ gives a good estimate of $\overline{C}(\theta)$, although for most angles $s=1$ suffices.

We now proceed to determine the full angular dependence of $\overline{C}(\theta)$. The various $1/T_{2}^2$ contributions of \sitn\ spin
pairs as a function of crystal rotation angle are shown in \Fig{Balian2014_weightsl}.
The data in \Fig{Balian2014_weightsl} was generated from \Eq{fidweights} near the ESR-type OWP
(for rotation around the $\left[01\bar{1}\right]$ crystal direction)
of Si:Bi in natural silicon, however, our results are independent of $B$ and the central donor species,
up to a scaling factor on 1/$T_2^2$ contributions. 

In \Fig{Balian2014_weightsl}, the different shells are labelled according to whether the interacting spins are first,
second, third or fourth nearest neighbors (1--, 2--, 3--, 4--NNs). The total $T_2$ is obtained by summing $1/T_2^2$ contributions from all spin pairs in the bath.
We pick the strongest $N$ spin pairs (i.e., those with the largest $1/T_2^2$ contribution) such that the sum over $1/T_2^2$ is about $70 - 80\%$ of the total $T_2$,
and find that $N \simeq 270$ for $\theta = 0^{\circ}$ and $N \simeq 100$ for all the other rotations considered.
Contributions from 1--NNs are dominant for $\theta \gtrapprox 30^{\circ}$. In \Table{Balian2014_orientation}, we show that 1--NNs suffice to set the scale of $T_2$ for
$\theta \gtrapprox 30^{\circ}$ by comparing $\overline{C}(\theta)$ obtained from only 1--NNs to $\overline{C}(\theta)$ extracted from numerical CCE2 $T_2$ and using \Eq{t2formula}.
For $\theta = 0^{\circ}$, 2--NNs and 3--NNs contribute the most, without any 1--NNs being involved in setting the scale of $T_2$. Including only the strongest 2--NN and 3-NN
contributions, for $\theta = 0^{\circ}$ we find $\overline{C}(0^{\circ}) \simeq 0.97$~ms, compared to $\overline{C}(0^{\circ}) = 1.1$~ms obtained using the numerical $T_2$.
Thus, using the estimated $\overline{C}(\theta)$ values in the first column of \Table{Balian2014_orientation}
provides a reasonable estimate of the dipolar prefactor $\overline{C}(\theta)$ as a function of crystal rotation.

\section{Relating Hahn Spin Echo to FID}

\begin{figure}[h]
\centering\includegraphics[width=4in]{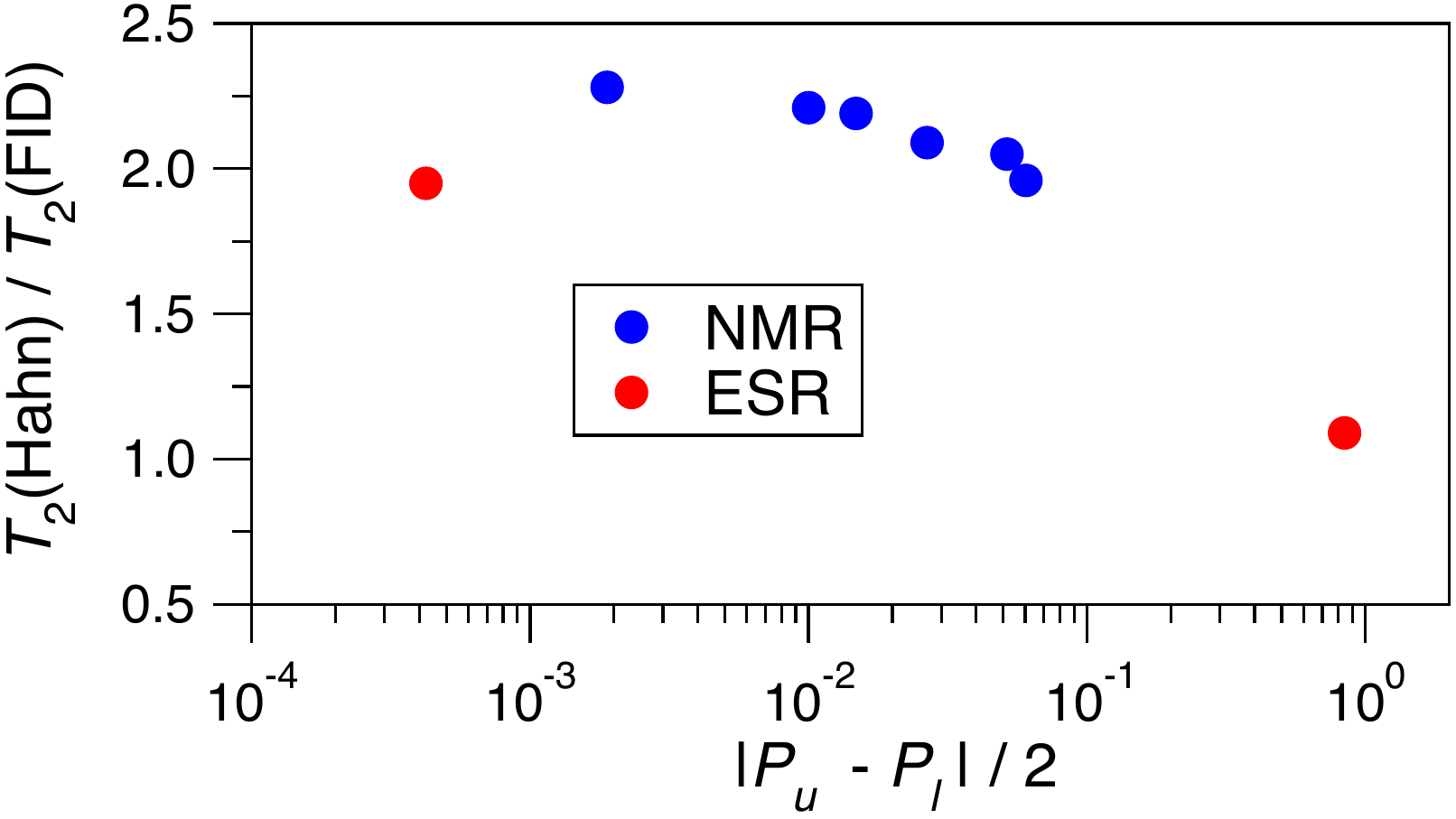}
\caption{
Comparison of calculated $T_{2}(\text{Hahn})$ and $T_{2}(\text{FID})$
for the various ESR-type and NMR-type transitions of Si:Bi for which $T_2$ was measured (\Fig{Balian2014_exp}) covering a wide magnetic field range.
Near OWPs (where $\left|P_u - P_l\right| \ll 1$), $T_{2}(\text{Hahn}) / T_{2}(\text{FID}) \simeq 2$. Figure adapted from \citet{Balian2014}.
}
\label{fig:Balian2014_hahnfid}
\end{figure}

For the Hahn echo case, $\mathcal{L}^{u \to l}_{\text{Hahn}}(2t)=
\boldsymbol{\mathcal{B}}(0)^\intercal
{\bf T}^*_{ul}(\omega^+, \omega^-,t){\bf T}_{ul}(\omega^+, \omega^-,t)\boldsymbol{\mathcal{B}}(0)$,
noting the exchange in order of $\omega^\pm$ relative to the FID
case in \Eq{fidunitary}. The analysis for the Hahn case is less straightforward, but nevertheless for (ii),
we estimate, using numerical CCE2 results at short-times, that near NMR-type transitions and OWPs, $T_2(\text{Hahn}) \approx 2\times T_2(\text{FID})$,
while  $T_2(\text{Hahn}) \approx  T_2(\text{FID})$ elsewhere.

While FID and Hahn echo decays are generally of the same order, within about $5$~mT of 
an OWP, our calculated CCE2 Hahn echo (pair correlations) shows non-decaying
behaviour at timescales beyond a few ms -- as seen in \Chap{decoherence}.
In contrast, the FID shows converged, near-Gaussian decays to zero intensity for all timescales and magnetic
fields. Nevertheless, there is always a period of initial near-Gaussian decay for the Hahn echo near OWPs from which we
extract $T_{2}(\text{Hahn})$. This initial period of convergence is extended to
longer times as higher order cluster contributions are taken into
account as shown in \Chap{decoherence}.
We estimate numerically the
ratio $T_{2}(\text{Hahn}) / T_{2}(\text{FID})$ as shown in \Fig{Balian2014_hahnfid}
and find that
\begin{equation}
T_{2}(\text{Hahn}) / T_{2}(\text{FID}) \approx 2,
\end{equation}
near OWPs (where $\left|P_u - P_l\right| \ll 1$).
Far from OWPs, the coherence times are to within 10\% as can be seen for the last ESR point in \Fig{Balian2014_hahnfid} near $P_u - P_l \simeq 2$.
In the previous chapter, we saw that fully-converged higher-order CCE calculations are in agreement with \Eq{t2formula} (with the factor of 2 scaling)
and the short-time CCE2 results near OWPs.

\section{Comparison with Frequency-Field Gradient}

\begin{figure}[h]
\centering\includegraphics[width=4.5in]{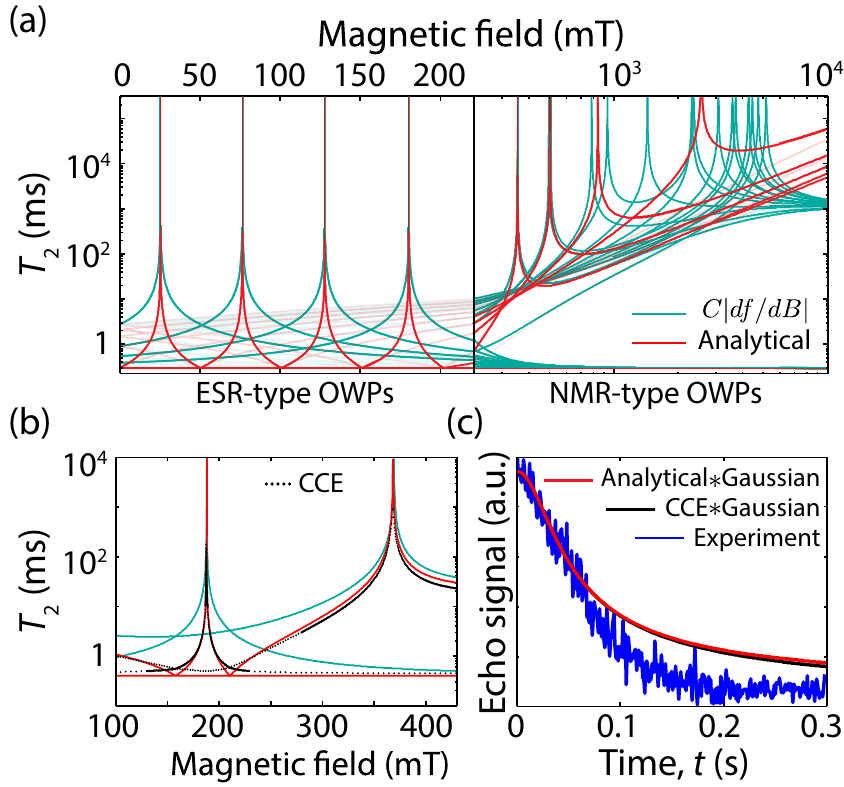}
\caption{
(a) The predicted $T_2$ values as a function of magnetic field for a variety of allowed transitions in Si:Bi, using \Eq{t2formula} derived in the text (labelled `analytical'), show eight OWPs where decoherence is suppressed. We also plot the magnetic field-frequency gradient ($df/dB$); though scaled by an arbitrary constant in order to match the range of estimated $T_2$ values, the discrepancies with \Eq{t2formula} are evident. In the left panel, transitions with no OWP are shown only faintly.
(b) The analytical expression \Eq{t2formula} derived in the text is in good quantitative
agreement with CCE2 numerics, but $df/dB$ is not.
(c) Calculations convolved with Gaussian $B$-field distribution of width 0.42~mT (arising from inhomogeneous broadening from the nuclear spin bath) show an excellent fit with the experimental Hahn echo decay around an ESR-type OWP ($B \sim 80$~mT) \citep{Wolfowicz2013}, with no free fit parameters.
Figure adapted from \citet{Balian2014}.
}
\label{fig:Balian2014_owps}
\end{figure}

We recall that the important mixing parameters $P_i$ in the $T_2$ formula
(\Eq{t2formula}) may be evaluated analytically for an arbitrary donor species,
for all field values. Also, the formula is perfectly valid for the case of a simple electronic spin (unmixed regime).
In this section, we
investigate deviations of our formula from the $T_2 \sim df/dB$ dependence 
that one might expect from classical noise models.
We also compare the predictions of the formula with numerical CCE calculations, and in the next section present comparisons with experiments.
Also, we use the sensitivity of $T_2$ on magnetic field in the vicinity of OWPs \citep{Balian2012}
as a test of \Eq{t2formula}.
The parameters we use are for the bismuth donor in the $^{29}$Si nuclear spin bath (in both mixed and unmixed regimes).

In \Fig{Balian2014_owps}, we plot \Eq{t2formula} for Si:Bi for allowed ESR-type and NMR-type transitions across a range of magnetic fields.
It shows close agreement with numerical CCE calculations including the effect of $\hat{H}_{\text{int}}$ on $P_i$ (i.e.\ with non-secular terms
in $\hat{H}_{\text{int}}$ included).
Both \Eq{t2formula} and CCE have distinctly different signatures from a curve proportional to $df/dB$, which would be expected in
the case of classical field noise; and they cannot be fitted (except locally) by powers of $df/dB$.
A comparison of \Eq{t2formula} with experiment is shown in \Fig{Balian2014_owps}(c).

\Fig{Balian2014_owps}(a) illustrates eight OWPs where $T_2\rightarrow\infty$: four ESR-type and four-NMR type
transitions (these OWPs are all doublets, so there are in fact 16 separate OWP transitions).
The form of \Eq{t2formula} clarifies the origin of these discrepancies. For low fields, ($ B \lesssim 1$~T)
the denominator of \Eq{t2formula} is $|P_u - P_l| \approx  df/dB$. Thus, it is the numerator ($|P_u| + |P_l|$), 
which accounts largely for the deviation from the form expected for analogous classical noise ($T_2 \propto 1/(df/dB)$).

At higher fields (left panel of \Fig{Balian2014_owps}(a)), we 
see that the formula does not coincide with the `false positives' of CTs (where $df/dB \rightarrow 0$).
The reason of such false predictions of suppression of decoherence by CTs was given in \Chap{decoherence}.
In \Fig{Balian2014v3_dfdBvsOWP}(b), it can be seen that the formula is in agreement with CCE (no suppression),
while $df/dB$ predicts suppression of decoherence.

In summary, in \Eq{t2formula}, it is the denominator ($|P_u - P_l|$) which sets the position of the OWPs:
at these points the bath evolution becomes independent of the state ($\ket{u}$ or $\ket{l}$) of the 
central spin, and so the system-bath entanglement is zero (\Fig{Balian2014_pseudospins}). However it is the 
numerator (which can vary by an order of magnitude in the range $0 \leq B \leq 1$~T) which 
provides the most distinct signature of the ``back-action'' between the quantum bath and central spin.

\section{Comparison with Experiments}

In this section the formula is compared with
the experimental data already presented in \Chap{decoherence} for the hybrid qubit near and far from OWPs and for both ESR-type
and NMR-type transitions (\Fig{Balian2014_exp_top_only}), obtained for Si:Bi in natural silicon.
The experimental data was collected by Dr. Gary Wolfowicz and Professor John Morton at UCL.

\begin{figure}[h]
\centering\includegraphics[width=4in]{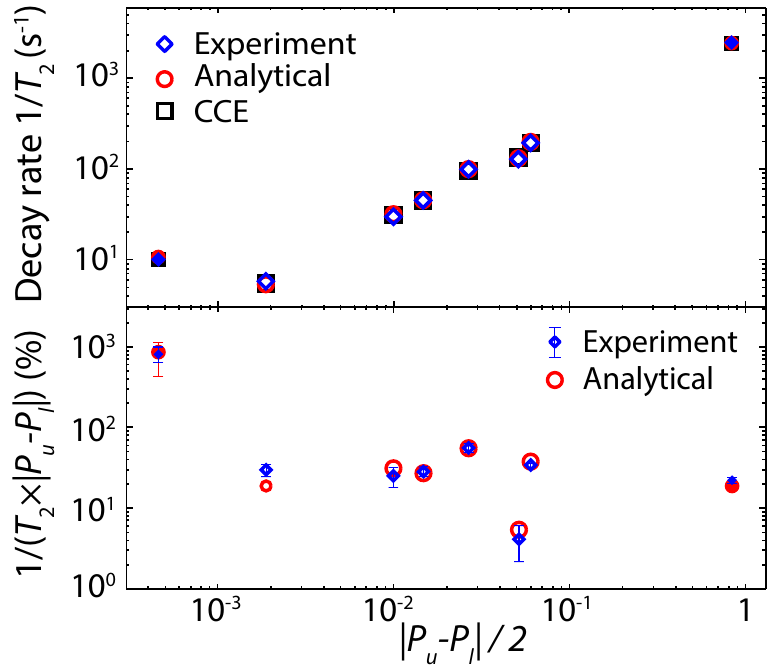}
\caption{
Comparison between theoretically predicted and measured
 $T_2$ in \tsc{nat}Si:Bi for various transitions,
showing remarkable agreement across a wide range of mixing regimes $|P_u-P_l|$. The label `analytical' refers to \Eq{t2formula}. Measurements were made at 4.8~K using ESR with a microwave frequency of 9.77 or 7.03~GHz (filled symbols), or electron-nuclear double resonance (ENDOR) between 200~MHz and 1~GHz using the method described in \citet{Morton2008} (empty symbols), at magnetic fields between 100 and 450~mT. These parameters are all in the regime where $|P_u-P_l|\approx df/dB$. The Bi donor concentration was $\leq 10^{16}$~cm$^{-3}$, and coherence times are limited by \sitn\ spin diffusion. The theoretical points are based on a predicted value for $\overline{C}(\theta) = 0.42$~ms.
In the lower panel, the decay rates are normalised by $|P_u-P_l|$ to highlight the effect of $|P_u|+|P_l|$, and shown relative to the case
when $|P_u|=|P_l|$. Figure adapted from \citet{Balian2014}.
}
\label{fig:Balian2014_exp}
\end{figure}

\Fig{Balian2014_exp} shows $T_2$ measurements of ESR-type transitions towards the high-field regime, where $|P_u-P_l|\simeq2$,
 and $T_2$ for a variety of different NMR-type transitions where $|P_u-P_l|$ varies by two orders of magnitude.
 It can be seen that the formula gives excellent agreement with the measured values.
 The primary variation in $T_2$ is due to the $|P_u-P_l|$ term; this is divided out in the lower panel of \Fig{Balian2014_exp},
 where the additional variations due to $|P_u|+|P_l|$ are apparent in the experiment.

As discussed in \Chap{decoherence},
the donor ESR line is inhomogeneously broadened by unresolved coupling to \sitn, leading to an effective
Gaussian magnetic field variation across the ensemble (FWHM of $0.42$~mT for Si:Bi in natural silicon).
Therefore, to predict the measured $T_2$ at an ESR-type OWP, we convolve \Eq{t2formula} with
the corresponding Gaussian magnetic field profile -- \Eq{convolution} with $\mathcal{L}_B(t)$ replaced by $e^{-\left(t/T_2\right)^2}$,
where $T_2$ is given by \Eq{t2formula} and $B = B_\text{OWP}$.
The convolution $D(t)$ is found to give a non-Gaussian decay, and reaches its $e^{-1}$ value at $100$~ms as shown in \Fig{Balian2014_owps}(c),
in close agreement with the experimental value of $93$~ms for the Si:Bi $\ket{14}\to\ket{7}$ OWP \citep{Wolfowicz2013}.
The convolution sums $T_2(B)$ contributions which vary over orders of magnitude and thus represents a sensitive test of \Eq{t2formula} around an ESR-type OWP.

\Eq{t2formula} gives divergent $T_2$ values at the OWP; comparison with CCE indicates that it becomes unreliable within $\sim 0.01$~mT of the OWP and non-secular
terms cap the maximum $T_2 \lesssim 10$~s. However, the inhomogeneous broadening enables us to use \Eq{t2formula}
to predict the measured (finite) $T_2$ at an ESR-type OWP by the convolution described above.

The experimental data was obtained for $\theta=135^{\circ}$ as described in \Chap{decoherence}, and this
corresponds to $\overline{C}(135^\circ) \simeq 0.4$~ms using \Table{Balian2014_orientation}.\footnote{For the ESR-type OWP point, i.e.\ with the lowest polarisation difference, $B \parallel [011]$ and the value of $\overline{C}(\theta)$ is similar to
the other points which all have $\theta=135^{\circ}$. Thus, the same prefactor was used for all the points in \Fig{Balian2014_exp}.}

We emphasize that the derivation of \Eq{t2formula} involves a range of approximations.
Assumptions have been made regarding the strong coupling approximations, the importance of certain spins
and the numerically estimated FID to Hahn scaling --
$T_2$ times from the FID formula are doubled when comparing with experiments near OWPs.
Thus, while one might expect agreement with experiment within a factor of two,
the agreement with data we obtain over such a large range is remarkable and indicates that
the form of $T_2$ predicted by \Eq{t2formula} persists even for the higher-order CCE calculations presented in \Chap{decoherence}.

\section{Conclusion}

In summary, we have shown that a field dependence given by
$T_2(B) \propto \left(|P_u|+|P_l|\right) \times \left(|P_u-P_l|\right)^{-1}$,
distinctly different from classical field noise which yields $T_2(B) \propto 1/(df/dB)$,
is a generic and robust feature of mixed electron-nuclear spin systems, valid over a broad range of ESR-type and NMR-type transitions both close to and far from OWPs.
The range also includes the {\em unmixed} case in the limit $P_u = -P_l$.

By inspection of the short-time behaviour of the form of single-central spin FID 
decays (which can be given analytically for each pair cluster), the simple closed-form equation gave remarkable and accurate quantitative
agreement with experiment in all regimes. Although only based on pair correlations, the agreement was excellent in regimes spanning orders 
of magnitude changes in $T_2$, whether in the unmixed limit of a spin-$1/2$ or at OWPs.
The universal validity of \Eq{t2formula} is worthy of discussion. 
Farther than about $100$~G from the OWP, and where CCE is converged at the pair correlation level, there is little
difference between single-spin FID and Hahn echo decays;
thus, it is not surprising that an equation obtained by considering the pair contribution to
FID can accurately model the Hahn echo experiments.
Its validity within the OWP regions, however, is not yet fully understood.
In particular, it remains unclear why a single $\overline{C}(\theta)$ prefactor suffices
to accurately estimate experimental $T_2$, whether very far or very close to OWPs; and to describe different
OWP regions (of which there are 16 for Si:Bi, with $P_{u,l}$ values varying by close to an order of magnitude).
This is especially surprising as the underlying cluster dynamics (CCE2 or CCE3) 
is not unchanging. While not providing an explanation, \Fig{Balian2015_conv} further demonstrates
the validity of \Eq{t2formula} by comparing to higher-order CCE in regimes where
the CCE has not converged at the pair correlation level (i.e.\ near OWPs).

The divergence of \Eq{t2formula} at the exact OWP point (where $P_u=P_l$)
is not physically significant. In full quantum results, whether FID or converged Hahn, non-Ising terms suppress the divergence
and in experiments, line broadening due to $^{29}$Si prevents $B = B_\text{OWP}$.
In any case, depending on the donor concentration, for $T_2 \gtrsim 0.2 - 2$~s,
other mechanisms arising from donor-donor flip-flops contribute significantly to decoherence.

In addition to use of an OWP, decoherence by nuclear spin diffusion can be suppressed by enrichment of the host using a spin-zero isotope
(e.g.\ using enriched $^{28}$Si) \citep{Tyryshkin2012a}. The effect of reducing the nuclear spin concentration on $T_2$ is explicit in the $\overline{C}(\theta)$ term,
but it also causes narrowing of the ESR linewidth and hence reduces the effective magnetic field distribution to a narrower range around the OWP.
As the nuclear spin concentration becomes negligible, other decoherence processes become dominant, including couplings to other (e.g.\ donor)
spins which can similarly be analysed for a quantum-correlated bath.

The next chapter investigates the application of dynamical decoupling sequences
by operating near OWPs, partly to determine the best strategy of maximizing $T_2$
for the hybrid qubit.

\chapter[Dynamical Decoupling of Hybrid Qubit]{Dynamical Decoupling of \\ Hybrid Qubit}
\label{chap:dynamicaldecoupling}

In this chapter, by means of quantum many-body calculations,
we investigate the effects of dynamical decoupling pulse sequences far from and near OWPs for
the hybrid qubit subject to decoherence from the silicon nuclear spin bath \citep{Balian2015}.
One of our aims is to clarify where and to what extent,
the independent pair contributions dominate for a quantum bath.
Another aim is to establish the best strategy 
for maximising coherence times of the hybrid qubit.

\section{Maximizing $T_2$}

As mentioned in \Chap{intro}, one way of extending coherence times $T_2$ in silicon is to employ
isotopically enriched silicon (with no nuclear impurities).
However, the nuclear spin bath has technological advantages, which is the topic of \Chap{nucleardecoherence}.
It is thus important to understand whether 
dynamical decoupling and OWP techniques may be advantageously combined for a quantum bath
of nuclear spins, without having to resort to isotopic enrichment.
For donor electronic qubits in silicon, it was shown that due to inhomogeneous broadening from naturally-occurring $^{29}$Si spin isotopes,
there is a significant gap between the $T_2 \sim 100$~ms
in natural silicon near an OWP \citep{Wolfowicz2013,Balian2014} and the $T_2\sim 2$~s in isotopically enriched $^{28}$Si with a low donor concentration
at the same OWP \citep{Wolfowicz2013}.
Also, dynamical decoupling may be useful when it is convenient to operate with the magnetic field close to but not exactly at the OWP.

We employ up to CCE5 and compare coherence decays at an OWP with regimes far from an OWP (denoted by `$\ne$OWP').
We find that while operating near OWPs, dynamical decoupling sequences require
hundreds of pulses for a single order of magnitude enhancement of $T_2$,
in contrast to regimes far from OWPs, where only about ten pulses
are required.

\section{Many-body Correlations}

We also show that for low to moderate numbers of pulses ($N \approx 1 - 16$), not unlike the Hahn spin echo in \Chap{decoherence},
decoherence at OWPs is no longer fully driven by
non-interacting pairs of bath spins, but instead involves the dynamics of clusters
of at least three interacting bath spins coupled to the qubit.
In contrast, for $\neq$OWP regimes, Hahn decays are well described by CCE2 \citep{Witzel2006}.
A recent analysis based on the linked-cluster expansion method indicated that, for even $N$, there is full suppression
of the contribution from independent pairs \citep{Ma2014}.
In fact, in this work we find that the independent pairs, using CCE2, give $T_2$ in the correct experimental timescale
regardless of $N$, although for modest (even) numbers of pulses $N = 2,4,6,8$ there can be a significant discrepancy between CCE2 and CCE4.
For larger $N$, we find that CCE numerics including only independent pairs (CCE2) once again gives converged decays in all regimes whether
in OWP or $\ne$OWP regimes, so many-body calculations become progressively less important as $N \to \infty$.
Hence, the only case of complete suppression of pair correlations occurs near the OWP, for the Hahn spin
echo and $N \lesssim 16$ pulse dynamical decoupling.

\section{Correlation Time vs. Quantum Treatment}

It is well established that for dynamical decoupling to be effective, the pulse spacing
$\tau=t/2N$ for a sequence of $N$ control pulses (where $t$ is the total evolution time)
cannot exceed the correlation time of the bath noise.
But the relevant correlation time, in turn, is an emergent property of the underlying
microscopic quantum bath, comprised of typically $\sim 10^4-10^5$ significant clusters of spins
of different coupling strengths, different sizes and subject to
varying degrees of back-action from the central qubit. Therefore,
to quantitatively simulate the response to dynamical decoupling, a realistic simulation
of the combined system-bath dynamics at the microscopic level is important.

We also present an analysis involving hybrid pseudospins to understand the 
degree of suppression of the usually dominant contribution from independent
pairs of flip-flopping spins within the many-body quantum bath.
Simple analytical expressions for the behaviour of independent bath pairs coupled to the qubit
aid understanding in all the regimes we consider.\\

\begin{figure}[h]
\centering\includegraphics[width=2.5in]{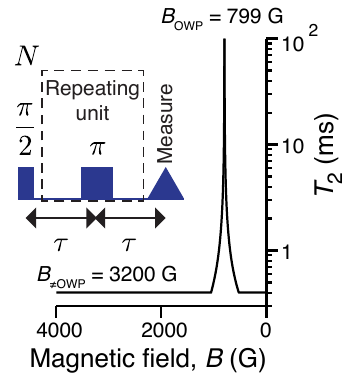}
\caption{
Illustrates coherence enhancement as $B \to B_{\textrm {OWP}}$ (the Hahn spin echo time $T^{(1)}_2$ is plotted). 
The OWP is for a bismuth donor in natural silicon, investigated 
experimentally in \citet{Wolfowicz2013} and \citet{Balian2014}.
The OWP curve was calculated using the analytical formula \Eq{t2formula}.
OWP results are for the $\ket{14} \to \ket{7}$ transition for which $B_{\textrm {OWP}} = 799$~G.
Inset: The CPMG dynamical decoupling sequence consists of
the initial $\pi/2$ pulse, followed by the $-\tau-\pi-\tau-$echo sequence
repeated $N$ times, as described in \Sec{cpmg}.
Figure adapted from \citet{Balian2015}.
}
\label{fig:Balian2015_ddowp}
\end{figure}

For our dynamical decoupling calculations, we have chosen the CPMG sequence which
applies a set of $N$ periodically spaced near-instantaneous pulses (CPMG$N$) as described in \Sec{cpmg}.
The OWP we investigate is for the $\ket{14}\to\ket{7}$ transition of Si:Bi.
The pulse sequence and OWP are illustrated in \Fig{Balian2015_ddowp}.

\section{Low and Moderate Pulsed CPMG}

\begin{figure}[h!]
\centering\includegraphics[width=4.5in]{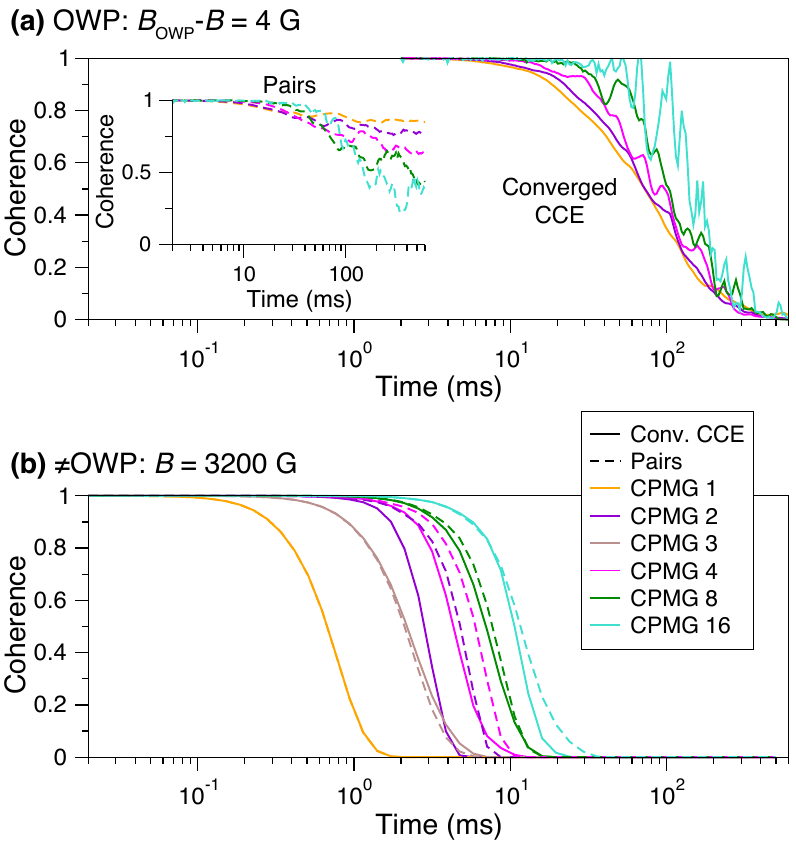}
\caption{
Shows dependence of the coherence on the number of dynamical decoupling pulses $N$,
(a) near an optimal working point (OWP) and (b) far from an OWP, for modest numbers of $N$.
(a) For $B$ close to $B_{\textrm{OWP}}$, the $T_2$ times show comparatively little response to dynamical decoupling. 
Further, even though the initial coherence is extended with increasing $N$, the decays become ever more oscillatory.
For low $N$, the independent pairs contribution is largely eliminated.
Inset of (a): Showing complete suppression of the independent pairs contribution near an OWP;
but showing also its gradual revival as $N$ increases.
(b) In contrast, far from the OWP,
substantial (order of magnitude) enhancement of the $T_2$ time by dynamical decoupling is achieved with a moderate (preferably even)
number of pulses. Decays for independent pair contributions (dashed lines, CCE2) and the converged quantum many-body numerics
(solid lines, CCE4) are also compared, indicating that as $N \gtrsim 10$, once again, the independent pair contribution is sufficient.
CCE calculations were performed for CPMG$N$ on a bismuth donor in
natural silicon for $B$ along $[100]$ and for
the $\ket{14} \to \ket{7}$ transition for which $B_\text{OWP} = 799$~G.
In (a), $B=795$~G while for (b), $B=3200$~G.
The converged CCE in (a) corresponds to CCE3.
Figure adapted from \citet{Balian2015}.
}
\label{fig:Balian2015_decays}
\end{figure}

Details of the CCE simulations are the same as described in \Chap{decoherence},
but for the CPMG sequence. The results shown are for the same single realisation of bath spin positions
and the same initial product bath state. In this section we discuss CPMG$N$ with $N = 1-16$ and present
higher-order CPMG in a later section below.

The coherence decays for the Hahn spin echo ($N = 1$) were presented in \Chap{decoherence}.
In \Fig{Balian2015_decays}, these are plotted together with CPMG for $N$ up to 16,
both near (\Fig{Balian2015_decays}(a)) and far (\Fig{Balian2015_decays}(b)) from OWPs, the decays
corresponding to converged CCE, and also pair correlations for the case close to OWPs (inset of \Fig{Balian2015_decays}(a)).
One notable feature of the comparison near and far from OWPs is the insensitivity of OWP behaviour
to low numbers of pulses, in sharp contrast to the $\ne$OWP regime
where there is a factor of 3 ``jump'' in $T_2$ from CPMG1 to CPMG2;
while for larger $N$, we find $T_2 \sim N$.
However, OWPs are extremely effective at suppressing decoherence:
for the point shown near the OWP, $T_2 \simeq 100$~ms already at
CPMG1, while away from the OWP, to obtain comparable values, $N\simeq 100$ pulses are required as will be shown later.
Previous studies, including the recent study in \citet{Ma2014} of the ESR dynamics of a phosphorus donor at $X$-band frequencies
(a system without OWPs for electron qubit decoherence and which is comparable to our $\neq$OWP regime),
observed a sharp increase  in the coherence time between CPMG1 and CPMG2.
Analysis with the linked-cluster expansion method suggested that spin-pair contributions were fully suppressed \citep{Ma2014}.
But we show CCE2 in $\neq$OWP regimes to still give a reasonable approximation to the
magnitude of the observed $T_2$ time, for both CPMG1 and CPMG2.
In the case away from an OWP, the FID is very similar to CPMG1.
This is in contrast to the OWP, where CCE2 gives no decay at all, while the FID gave 
decay curves comparable to converged CCE3 (and \Eq{t2formula}).
Thus, there is a drastic change from FID to CPMG1 at OWPs; in contrast,
for regimes away from an OWP, there is little change between FID and CPMG1,
but a strong enhancement for CPMG$N$ with $N>1$.

We find that three-spin clusters not only restore the CCE2 short-time decay, but in fact
suffice to give results converged with respect to the many-body dynamics 
(i.e.\ there is little difference between CCE3 and CCE5) for both Hahn echo decays 
and modest $N \lesssim 20$ pulse numbers. For larger $N$, we find that CCE2 once again gives converged decays.

For any kind of spin diffusion, whereby the magnetic noise arises from flip-flopping
(e.g.\ dipolar-coupled) spins in the bath (i.e.\ ``indirect flip-flops''),
the noise from non-interacting pairs of bath spins provides a reasonable estimate for
the $T_2$ timescale of measured echo decays.
For the case of isotopically-enriched samples, where donor-donor dynamics replaces the nuclear bath,
larger spin clusters (CCE3 to CCE6) represent a quantitative correction \citep{Witzel2010,Witzel2012}.
In our case, unlike in \citet{Ma2014}, CPMG$N$ with small $N$ eliminates pair dynamics in the sense that CCE2 (using
only clusters of two bath spins) does not even give a finite $T_2$ time. Such complete suppression,
and also in the absence of any higher order $N>1$ dynamical decoupling is quite exceptional.

The quantum numerics do evidence a clear dependence of the pair contribution on pulse number $N$.
For example, in the inset of \Fig{Balian2015_decays}(a),
we have shown that, for a given field $B$ in the vicinity of the OWP,
as $N$ increases to $N\simeq16$, the pair contribution once again gives significant decay.
To suppress decay for $N=16$ one must choose a value of $B$ even closer to the OWP.
In fact this was one of our main findings: whether at OWPs or far from OWPs,
our comparisons between many-body CCE3-5 and calculations involving only
pairs show that increasing $N$ gradually restores the importance of the pair contribution,
relative to $N=1$ or $N=2$, where many-body effects are seen to make the dominant contribution.

\section{Pseudospin Analysis}

We now proceed to analyse correlations from independent pairs in order to obtain insight on the effect of
dynamical decoupling near and far from OWPs.
We employ the well-established pseudospin model of the system-bath dynamics as was used for the single-spin FID in \Chap{formula}.

After preparing the initial qubit superposition, the CPMG$N$ pulse
sequence can be summarized as $[\hat{T}(\tau) - \pi - \hat{T}(\tau)]^N$, with final evolution time $t=2 N \tau$.
The unitaries $\hat{T}(\tau)$ represents free evolution and $\pi$ denotes the refocusing pulse which flips
between $\ket{u}$ and $\ket{l}$: $\ket{u}\bra{l} + \ket{l}\bra{u}$ but leaves all other central states and the bath unperturbed.
Note that for the case of the numerical CCE calculations of any order (including central state depolarising terms),
$\hat{T}(\tau)$ represents free evolution under the total Hamiltonian in \Eq{totalH}.
For the case of pseudospins, the pure dephasing approximation must hold.

In the pair correlation approximation or CCE2, the coherence decay is simply given by
\begin{equation}
\mathcal{L}(t) = \prod_k |\mathcal{L}_k^{[N]}(t)|,
\label{eq:ddcce2}
\end{equation}
where $|\mathcal{L}_k^{[N]}|$
is the decay contribution from the $k$-th spin pair for CPMG$N$ and the product is over
all spin pairs in the bath.
The analysis below considers the individual pair decay envelopes $|\mathcal{L}_k^{[N]}|$
of which there are $\approx 10^4$ in the bath and
we have dropped the label $k$ for clarity.

In order to evaluate the bath state overlap (\Eq{bathoverlap}) or equivalently the decoherence $\mathcal{L}^{[N]}(t)$ as described in \Chap{background},
we must first evaluate
\begin{equation}
\ket{\mathcal{B}_{u,l}(t)} = \hat{T}^{[N]}_{u,l} \ket{ \mathcal{B}(0) },
\label{eq:ddtime1}
\end{equation}
where the unitaries $\hat{T}^{[N]}_{u,l}$ for CPMG$N$ are given by a
product sequence of $\hat{T}^{[0]}_u$ and $\hat{T}^{[0]}_l$, which correspond to evolutions
under the pseudospin Hamiltonians $\hat{h}_{u,l}$ (\Eq{pseudospinH}):
$\hat{T}^{[0]}_{u,l} = \exp{[-i \hat{h}_{u,l}]}$.
Refocusing pulses simply switch between $u$ and $l$ in applying the unitaries.
For example, for the simple case of the Hahn spin echo (i.e.\ CPMG1), the unitaries are given by
$\hat{T}^{[1]}_{u,l} = \hat{T}_{u,l}^{[0]} \hat{T}_{l,u}^{[0]}$ (note the order of $u$ and $l$).

We first diagonalise $\hat{h}_{u,l}$. We can now write, for the Hahn echo ($N=1$)
\begin{equation}
 \hat{T}^{[1]}_{u,l} = A_0 \hat{\mathds{1}} -i {\bf{A}}_{u,l} \cdot  \hat{\boldsymbol\sigma},
\label{eq:ddunitary2}
\end{equation}
where ${\bf A}_{u}=(A_x,A_y,A_z) $ and $\hat{\boldsymbol\sigma}$ is the vector of Pauli matrices acting on the
bath basis: $\{ \ket{\downarrow \uparrow},\ket{\uparrow \downarrow} \}$.
The ${\bf A}_{u,l} $ components depend on time and can easily
be given explicitly in terms of the pseudospin parameters \citep{Lang2015}:
\begin{align}
A_0(\tau) &= \cos(\omega_u\tau)\cos(\omega_l\tau) - \sin(\omega_u\tau)\sin(\omega_l\tau)\cos(\theta_u - \theta_l),  \nonumber \\
A_x(\tau) &= \sin(\omega_l\tau)\cos(\omega_u\tau)\sin(\theta_l) + \cos(\omega_l\tau)\sin(\omega_u\tau)\sin(\theta_u), \nonumber   \\
A_y(\tau) &= - \sin(\omega_u\tau)\sin(\omega_l\tau)\sin(\theta_u - \theta_l), \nonumber   \\
A_z(\tau) &= \sin(\omega_l\tau)\cos(\omega_u\tau)\cos(\theta_l) + \cos(\omega_l\tau)\sin(\omega_u\tau)\cos(\theta_u),
\end{align}
where $\theta_i$ and $\omega_i$ are the pseudofield angles and pseudospin frequencies respectively (see \Chap{formula} for details and the pseudospin Hamiltonian).
The only term which is not invariant with respect to the exchange $u \leftrightarrow l$ is $A_y$
and thus ${\bf A}_{l}= (A_x,-A_y,A_z) $.

The coherence envelope for each spin pair
$ |\mathcal{L}^{[N]}(t)| \propto
| \langle \mathcal{B}(0) | \hat{T}^{\dagger[N]}_l \hat{T}^{[N]}_u | \mathcal{B}(0)\rangle|$ is obtained simply from
$\hat{{L}}^{[N]}(t) \equiv \hat{T}^{\dagger[N]}_l \hat{T}^{[N]}_u$.
For both CPMG1 and CPMG2, the unitarity of the evolution of
upper relative to lower states is broken by a term proportional to $A_y$. For CPMG1,
\begin{equation}
\hat{L}^{[1]}(t)= \hat{\mathds{1}} -2i A_y \hat{\sigma}_y \hat{T}^{[1]}_u.
\label{eq:ddCPMG1}
\end{equation}
We can consider higher sequences; since $\hat{T}^{[2]}_u=\hat{T}^{[1]}_u\hat{T}^{[1]}_l$ and $\hat{T}^{[2]}_l=\hat{T}^{[1]}_l\hat{T}^{[1]}_u$, we obtain for CPMG2:
\begin{equation}
 \hat{L}^{[2]}(t)= \hat{\mathds{1}} - 4 i A_y (A_z \hat{\sigma}_x-A_x \hat{\sigma}_z)  \hat{T}^{[2]}_u.
\label{eq:ddCPMG2}
\end{equation}
Both the above general expressions apply equally to either OWP or the $\ne$OWP regimes. 
The only important difference between these regimes is that
$ \theta_u \to \theta_l$ for the approach to an OWP
and $ \theta_u = \pi-\theta_l$ for the spin away from the OWP.
Alternatively, from the explicit expressions for the components of
${\bf A}_{u,l} $, we see that the OWP condition is $A_y \to 0$; since $A_y$ is the prefactor to both 
the above expressions, CPMG1 and CPMG2 are equally suppressed at OWPs.

For the thermal initial bath states $ \ket{\downarrow \uparrow}$ or $ \ket{\downarrow \uparrow}$,
the temporal coherence decay for the bath spin pair is
$ |\mathcal{L}^{[N]}(t)| =| \bra{ \downarrow \uparrow } \hat{L}^{[N]}(t) \ket{ \downarrow \uparrow } |
=| \bra{ \uparrow \downarrow } \hat{L}^{[N]}(t) \ket{\uparrow \downarrow} |$ (the states $\ket{ \downarrow \downarrow }$ and $\ket{ \uparrow \uparrow }$
are not involved in pure dephasing decoherence).

We can easily obtain the coherence decay envelopes for CPMG1 in general, assuming pulse interval $\tau$:
 \begin{equation}
|\mathcal{L}^{[1]}(t=2\tau)|^2= 1- 4 A_{y}^2 A_{0}^2,
\label{eq:ddCPMG1EA}
\end{equation}
emphasising that $A_0 \equiv A_0(\tau)$, ${\bf A}_{u,l} \equiv {\bf A}_{u,l}(\tau)$.
For arbitrary even numbers of pulses, CPMG$N$ such that $N/2$ is an integer,
\begin{equation}
\mathcal{L}^{[N]}(t=2 N \tau)= 1 - \frac{2 A_{y}^2}{A_{y}^2 + A_{0}^2} \sin^2 \left[\frac{N\phi(\tau)}{2}\right],
\label{eq:ddCPMG2N}
\end{equation}
where $\cos \phi(\tau)=A_0(2\tau)$. An equivalent expression was obtained in \citet{Zhao2012a}.
Both expressions \Eq{ddCPMG1EA} and \Eq{ddCPMG2N} are equally valid for both regimes (OWP and $\neq$OWP).

\subsection{Near Optimal Working Points}

The only important difference between these regimes is that
$ \theta_u \to \theta_l$ for the approach to an OWP
and $ \theta_u = \pi-\theta_l$ for the spin away from the OWP.
Alternatively, from the explicit expressions for the components of
${\bf A}_{u,l} $, we see that the OWP condition is $|A_y| \to 0$.
Thus, the suppression of qubit-bath correlations from pairs for OWPs is of the same order
for CPMG1, CPMG2 or any other even-pulsed CPMG: for all bath spin pairs equally, the decay due to correlations from each independent pair
uniformly tends to zero as $(A_{y})^2 \to 0$ as $B \to B_{\textrm{OWP}}$.

The dependence on $N$ is entirely contained in the $\sin^2 [N\phi(\tau)/2]$ term. If $N\phi(\tau) \ll 1$
then increasing $N$ has a strong amplifying effect on the signal,
while if $N\phi(\tau) \gg 1$, increasing $N$ simply results in oscillatory behaviour.
 Near OWPs, from the expression for $A_0(2\tau)$, we see that if $\theta_u=\theta_l$,
$\phi(\tau)/2 \simeq (\omega_u+\omega_l)\tau$. Hence we only expect a
 response to dynamical decoupling if $\tau$ is sufficiently small (i.e.\ if $\tau \lesssim (\omega_u+\omega_l)^{-1}$).

\subsection{Far from Optimal Working Points}

In contrast, for CPMG away from an OWP, the $A_y^2$ prefactor is still there, but is not small.
The origin of the suppression of correlations from independent pairs for small numbers of pulses is more subtle
to analyse with the pseudospin model.
For CPMG2 in the $\ne$OWP limit, we obtain
\begin{equation}
|\mathcal{L}^{[2]}(t)|^2 = 1 - 64 A_{y}^2 A_{0}^2 A_{x}^4.
\label{eq:ddCPMG2E}
\end{equation}
The large jump in $T_2$ from CPMG1 to CPMG2 was also analysed in \citet{Ma2014}.
In the notation of \citet{Ma2014}, we see that 
for CPMG1, the decay envelope is of order $n_x^2$, while for
CPMG2 it is of order $n_x^6 n_z^2$, where 
$n_x= \sin{\theta_u} = \sin{\theta_l}$ while
$n_z= \cos{\theta_u} =- \cos{\theta_l}$. Since the bath spans all
angles $|\theta_{u,l}| =[0,\pi/2]$ one cannot {\em a priori} assume  
$\sin{\theta_{u,l}}$ is small.
However, previous numerical 
studies support the idea
that those spin pairs which have $|J_1 -J_2| \gg |C_{12}|$ (i.e.\ are strongly coupled to the central system)
and therefore small pseudospin angles, dominate the Hahn echo
contribution \citep{Balian2014} (see strong coupling approximation \Chap{formula}).
For CPMG2, such strong-coupled spin pairs are  
strongly suppressed, and so $T_2$ becomes dominated by more weakly coupled spin pairs
which are less effective in decohering the qubit.

\begin{figure}[h!]
\centering\includegraphics[width=4.5in]{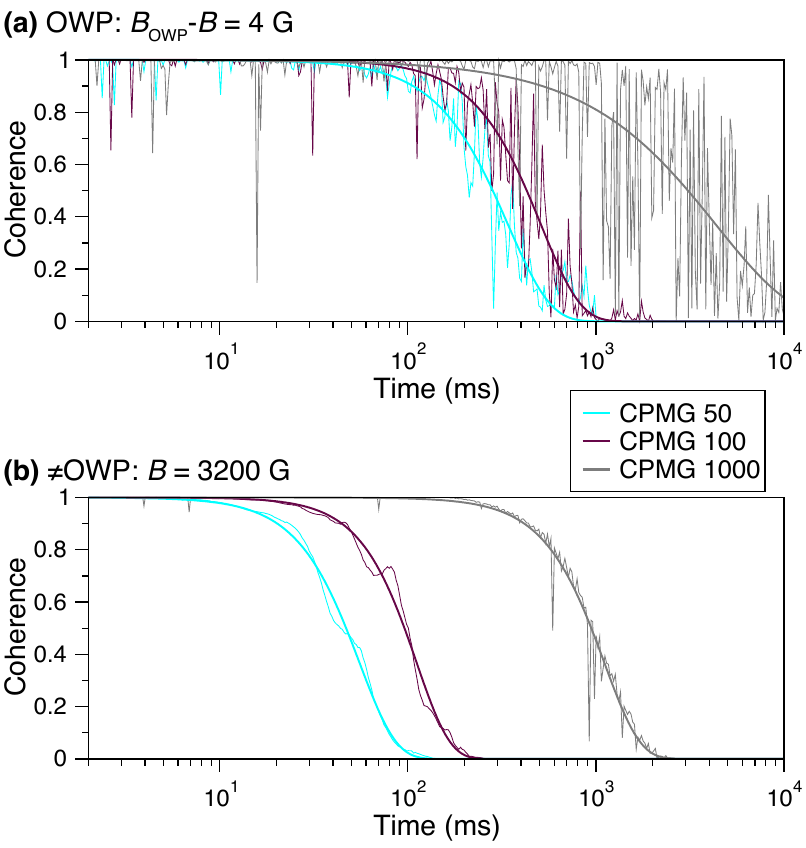}
\caption{
Shows coherence decays for large numbers ($N$) of dynamical decoupling pulses 
(a) near and (b) far from OWPs;
as shown in the inset of \Fig{Balian2015_decays}(a), for such high $N$, correlations from independent pairs once again dominate the decays
in all regimes so CCE2 is converged and plotted.
The behaviour at OWPs is now sensitive to $N$ but
the decays here become increasingly oscillatory 
as $N$ and $T_2$ both become large; we attribute this to large numbers of bath spin-pair frequencies becoming resonant with the pulse spacing.
It indicates the behaviour one might expect in a single-shot single spin study. The smooth
lines are fits to the decays and indicate the expected coherence decay after ensemble averaging. 
CCE calculations were performed for a bismuth donor in
natural silicon for $B$ along $[100]$ and $B_\text{OWP}=799$~G.
Figure adapted from \citet{Balian2015}.
}
\label{fig:Balian2015_hcpmg}
\end{figure}

\section{High Order CPMG}

We now investigate CPMG$N$, with $50 \leq N \leq 1000$ pulses.
For such large $N$, decays from independent pairs only (CCE2) are restored as well as the sensitivity to dynamical decoupling at OWPs.
Even for $N=16$ (\Fig{Balian2015_decays}) we see that the initial period of no decay $\mathcal{L}(t) \sim 1$ is prolonged.
For larger $N$ (\Fig{Balian2015_hcpmg}), the enhancement of coherence even at OWPs is clear, but however,
the decays become extremely noisy.
The noise can be attributed to the timescales of individual nearby spin clusters and the time interval between pulses.
For these long coherence times ($T_2 \sim 1$~s) there are very large numbers of resonances.
The CPMG sequence provides a means of amplifying noise from nearby clusters whenever
 pulse intervals become resonant with the characteristic cluster frequency.
 While this makes CPMG a valuable technique for spin detection \citep{Zhao2012b},
large numbers of such resonances are undesirable if the aim is to protect qubit coherence.
In contrast, far from OWPs, the decays for high $N$ remain relatively smooth.
While the noise at OWPs can be mitigated by ensemble averaging, this is likely to introduce a
considerable disadvantage in terms of single-shot operation of a single hybrid qubit.

\begin{figure}[h!]
\centering\includegraphics[width=2.5in]{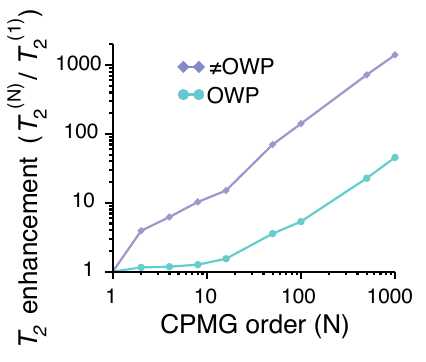}
\caption{
Effect of dynamical decoupling (CPMG with even pulse numbers $N$) as $N \to \infty$.
Plots $T^{[N]}_2/T_2^{[1]}$ showing enhancement of the electron spin coherence time $T_2$ as a function of pulse number $N$,
relative to the $N=1$ Hahn echo value. We find that while dynamical decoupling far from the OWP enhances $T_2$ by an order of
magnitude with about 10 pulses, in contrast, close to an OWP, enhancement is marginal for dynamical decoupling with low $N$.
For high $N$, enhancements near and far from OWPs become comparable.
Even-pulsed CPMG is shown as it is more effective than CPMG with odd numbers of pulses.
The coherence times are when the CPMG decays in \Fig{Balian2015_decays} and the fits to the decays in \Fig{Balian2015_hcpmg} have fallen to $1/e$.
Results are for Si:Bi in natural silicon for the $\ket{14} \to \ket{7}$ transition for which $B_{\textrm {OWP}} = 799$~G.
For the field value near the OWP ($B=795$~G), $T_2^{[1]} \simeq 96$~ms
while $T_2^{[1]}\simeq 0.79$~ms in the $\ne$OWP regime ($B=3200$~G).
Figure adapted from \citet{Balian2015}.
}
\label{fig:Balian2015_t2vsN}
\end{figure}

\section{Summary of Coherence Times}

In sum, we have seen that a key difference between OWP and $\neq$OWP behaviours
 arises from the $A_y^2 \propto \sin^2 ({\theta_u-\theta_l})$
 prefactor which globally suppresses all independent pair contributions on the approach to an OWP, 
and accounts for the drastic effect at OWPs, but which is independent of $N$ 
and has little effect far from OWPs.
 However, to analyse decays
 resulting from dynamical decoupling one must consider the remainder of the expression in
\Eq{ddCPMG2N}, which reflects the dependence on $N$.

The ineffectiveness of dynamical decoupling near OWPs for small $N$ can also be understood with an
intuitive picture considering the relevant timescales of the system.
For dynamical decoupling to be effective, the time interval between pulses ($t/2N$) must be shorter than to the correlation time of the bath $\tau_c$.
Since typical intra-bath interactions are at most a few kHz, $\tau_c \sim 1$~ms.
Near the OWP, $\omega_u \simeq \omega_l$ and $\theta_u \simeq \theta_l$, so the frequency
of the bath noise spectrum ($\sim \omega_{u,l}$) is appreciably higher than $1/\tau_c$
and thus dynamical decoupling becomes ineffective in extending the coherence time $T_2 \gg \tau_c$.
At short times and for high $N$ however ($t/2N < \tau_C$), dynamical decoupling does protect the central system
as evidenced for CPMG16 in \Fig{Balian2015_decays}(a) and higher $N$ in \Fig{Balian2015_hcpmg}.
In contrast, dynamical decoupling is far more effective in extending $T_2$ away from the OWP and for relatively
small $N$ (\Fig{Balian2015_decays}(b)); although the pseudospin frequencies are comparable,
the pseudospin fields are in opposing directions ($\theta_u \simeq \pi - \theta_l$), thus, the frequency
of noise is much slower and becomes comparable to $1/\tau_c \sim 1/T_2$.

The enhancement of coherence times relative to the Hahn spin echo is shown 
for increasing $N$ in \Fig{Balian2015_t2vsN}.
As an OWP is approached, dynamical decoupling gives little enhancement
in $T_2$ with increasing $N$ for the first 100 or so pulses,
in sharp contrast to regimes far from an OWP,
where $T_2$ scales roughly as $N$ and there is a substantial enhancement already 
between $N=1$ and $N=2$.

\section{Inhomogeneous Broadening}

\begin{figure}[h!]
\centering\includegraphics[width=4.5in]{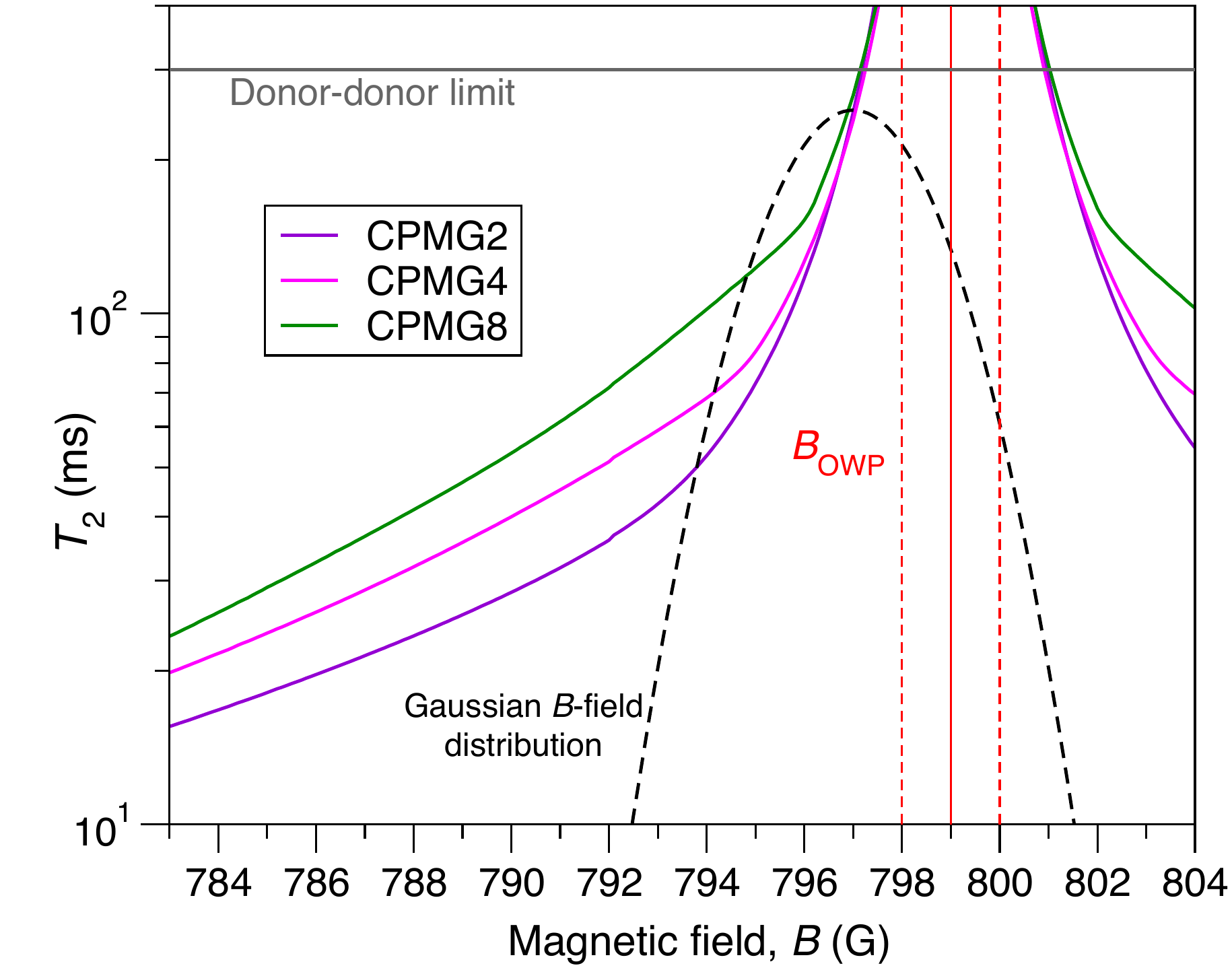}
\caption{
Sharp $B$-field dependence of $T_2$ for various
CPMG orders near an OWP. Inhomogeneous broadening from \sitn\ nuclei
can be incorporated by convolving the decays with a Gaussian $B$-field distribution centred about
$B$ (here centred about $797$~G) and with standard deviation $w \simeq 2$~G (dashed line).
For a donor concentration of $3\times10^{15}$~cm\tsc{-3},
$T_2$ is limited by donor-donor processes at about $300$~ms \citep{Wolfowicz2013}.
The $T_2$ lines were calculated for bismuth donors in natural silicon using the CCE up to 3rd order
and for $B \parallel [\bar{1} 1 0]$. The OWP under investigation is shown in red
at $799$~G. Figure adapted from \citet{Balian2015}.
}
\label{fig:Balian2015_t2estimate}
\end{figure}

Finally, it is important to note that for direct quantitative comparisons between our
dynamical decoupling calculations and experimental ensemble measurements,
inhomogeneous broadening due to $^{29}$Si nuclei might also have to be factored in.
This is because there is a sharp $B$ dependence of $T_2$ near an OWP over a narrow $B$ region.

The sharp variation of $T_2$ with $B$ over a few G near an OWP is shown in
\Fig{Balian2015_t2estimate} for various orders of CPMG.
Inhomogeneous broadening of $B$ due to \sitn\ impurities
has a FWHM of about $4$~G in natural silicon.
As described in \Chap{decoherence} and \Chap{formula},
the broadening can be simulated by convolving the decays 
with a Gaussian $B$-field distribution according to \Eq{convolution}.

Depending on the donor concentration, donor-donor processes
may also need to be included. For example, for a donor concentration
of $3\times10^{15}$~cm\tsc{-3}, $T_2$ near an OWP is limited by direct
flip-flops of the central donor
with other donors in the ensemble \citep{Wolfowicz2013}.
The measured $T_2$ in isotopically enriched samples ranges from $0.2 - 2$~s
($T_2$ for a donor concentration of $3\times10^{15}$~cm\tsc{-3} is 300~ms).
Therefore, care should be taken to include donor-donor processes
very near the OWP (within about 1~G), where nuclear spin diffusion coherence times
are comparable to those of donor-donor processes.

\section{Conclusion}

Understanding the interplay between OWPs and dynamical decoupling 
involves understanding of the quantum behaviour as a function of 
the two limits $B \to B_{\textrm{OWP}}$ and $N \to \infty$ corresponding to
approaching an OWP and simultaneously increasing the number of dynamical decoupling pulses.

For the Hahn spin echo case (\Chap{decoherence}) and for
low to moderate pulsed dynamical decoupling,
pair correlations are drastically suppressed and 3-body CCE is essential for predicting and understanding decays near OWPs.
Even away from OWPs, it was shown that many-body effects make an appreciable contribution for $N \lesssim 10$.
However, once $N \to \infty$ there is little difference between pair correlation and higher-CCE results in all regimes.

For practical applications,
one can hope to identify the best strategy for enhancing the coherence of donor qubits
whilst still keeping the nuclear spin bath of naturally occurring silicon
for its potential technological use.
We note that the magnetic field in \Fig{Balian2015_decays}(a) and \Fig{Balian2015_hcpmg}(a) is about 4 G from the actual OWP.
Although in a theoretical calculation one can obtain a longer $T_2$ at a point closer to the OWP without dynamical decoupling,
the ESR linewidth due to \sitn\ impurities restricts the measured value at the OWP. 

By operating near OWPs without dynamical decoupling, the maximum achievable $T_2$ is 0.1~s due to inhomogeneous
broadening from the environmental nuclei \citep{Wolfowicz2013,Balian2014}.
For isotopically enriched samples in which the nuclear spin bath
is nearly eliminated, $T_2$ at the OWP was measured to be about 1~s
and is limited by decoherence mechanisms involving donor-donor interactions \citep{Wolfowicz2013}.
Therefore, to bridge this single order of magnitude difference in $T_2$ for an ensemble at OWPs without
resorting to isotopic enrichment, dynamical decoupling should be applied with at least a few hundred pulses.
The effect of dynamical decoupling in extending coherence times near an OWP
is marginal with a moderate number of pulses (up to $N\sim16$)
in contrast to the usual regimes far from OWPs.
For high donor concentrations, the timescale of donor-donor decoherence is comparable to the
$T_2$ obtained in a nuclear spin bath, hence one might also want to investigate suppressing
those mechanisms with dynamical decoupling.

However, combining dynamical decoupling with OWPs is not without its drawbacks.
As $T_2$ and $ N \to \infty$, individual few-spin clusters in a silicon bath may become resonant
with the dynamical decoupling pulse spacing, resulting in very noisy decays in single central spin realisations.
Although ensemble measurements are unaffected by this noise, this means that for single-qubit
operations, if OWPs can be exploited, their extraordinary potential for coherence suppression (which has no noisy decay behaviour without dynamical decoupling)
may be sufficient.

At the time of writing, \citet{Ma2015} (with experiments by Dr. Gary Wolfowicz and Professor John Morton) reported $T_2 \simeq 1$~s near an OWP with 128 dynamical decoupling pulses
in natural silicon. This is in broad agreement with our predictions as can be seen in \Fig{Balian2015_hcpmg} where we predict $T_2 \simeq 0.5$~s with 100 pulses with 
$B$ shifted by 4~G from the OWP; the measurements were made 1.5~G from the OWP and for 28 more pulses \citep{Ma2015}.

\chapter[Decoherence of Nuclear Spins Proximate to Hybrid Qubit]{Decoherence of Nuclear Spins \\ Proximate to Hybrid Qubit}
\label{chap:nucleardecoherence}

The content of this chapter differs from the rest in that the decoherence dynamics studied
is for nuclear spins; specifically, for a \sitn\ spin-1/2 \citep{Guichard2014}.
Hyperfine couplings of proximate \sitn\ sites in an ensemble have been spectroscopically resolved by CW ENDOR \citep{Hale1969a,Hale1969b}
and also pulsed ENDOR techniques \citep{Morton2008,Balian2012,Wolfowicz2015b} as described in \Chap{interaction}.
The pulsed experiments in \Chap{interaction} demonstrate the feasibility of quantum control of such \sitn\ spins.
Their coherence may also be investigated by pulsed ENDOR.
As mentioned in \Chap{intro}, the Hahn echo coherence time for a \sitn\ qubit in a spin bath formed of other \sitn\ spins
is about 5~ms \citep{Dementyev2003}. Recently, \citet{Pla2014} measured $T_2$ of a single \sitn\ spin to be 6.4~ms, close to the measured
bulk value reported in \citet{Dementyev2003}.

Here, we investigate the situation of a nucleus in close proximity to the hybrid qubit (or {\em proximate} nuclear spin).
We show that in this case, coherence times can reach the second timescale \citep{Guichard2014}, in agreement with ensemble measurements \citep{Wolfowicz2015b}.
Thus, the hybrid qubit in a sense enhances the coherence of proximate nuclear spins.
In this scenario, one can imagine a long-lived quantum register (memory) implemented
as the nuclear spin, while processing is carried out on the donor qubit.
As of now, there is no $T_2$ measurement of a single proximate nuclear spin in the presence of the donor.
This parameter is of interest to potential future realisations using nuclear spin registers in combination with electronic qubits in silicon,
analogous to the situation involving NV centres and $^{13}$C nuclei in diamond \citep{Cappellaro2009,Waldherr2014,Taminiau2014}.

The theoretical work in this chapter was motivated by experiments measuring
coherence times of proximate nuclear spins by Dr. Gary Wolfowicz, Dr. Pierre Mortemousque and Professor John Morton.
The measurements were performed on Si:P in the high-field limit.
Hence, the hybrid qubit we study here is in the unmixed ESR limit, implemented as
Si:P which is equivalent to Si:Bi in the limit $B \to \infty$ for our purposes; i.e.\ a simple electronic spin-1/2.
The theoretical analysis was published in \citet{Guichard2014} and the experimental
measurements were very recently reported in \citet{Wolfowicz2015b}.

In short, we investigate the decoherence mechanism of a proximate nuclear impurity spin in the quantum bath formed of other impurity spins.
We propose two models of decoherence which give coherence times on the same timescale.
Both are spin diffusion models, analogous to the case of decoherence for the hybrid qubit discussed in
\Chap{decoherence}, \Chap{formula} and \Chap{dynamicaldecoupling}; however, they differ from usual spin diffusion problems in terms
of the properties of clusters which dominate the decoherence dynamics.
The first is a very large nuclear spin bath far from the nuclear qubit,
comprising $\gtrsim 10^8$ weakly contributing spin pairs.
The second involves decoherence driven by pairs of symmetrically sited nuclear spin pairs,
due to symmetries of the donor electron wavefunction. There are only of order $10^2$ such ``equivalent pairs''.
In previous studies (i.e.\ for decoherence of donor spins or non-proximate nuclear spins),
both models produce negligible contributions to coherence decays as will be explained below.
Both models give $T_2$ times of order 1~s in agreement with measured proximate nuclear spin coherence times,
confirming the suitability of proximate nuclei in silicon as very long-lived spin qubits.
We also note that if equivalent pairs represent a measurable source of decoherence,
nuclear coherence decays could provide sensitive probes of the symmetries of electronic wavefunctions.

\section{The Frozen Core}

\begin{figure}[h!]
\centering\includegraphics[width=4.5in]{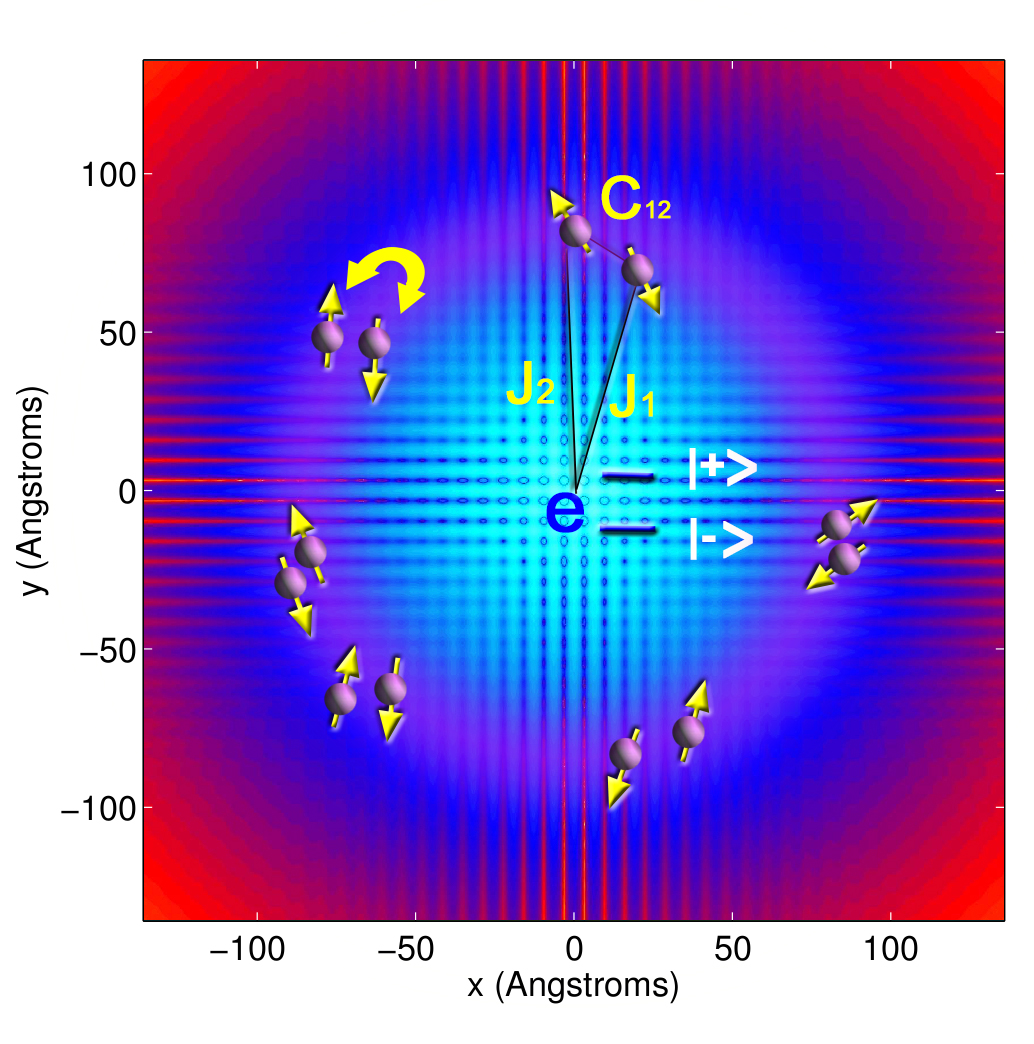}
\caption{Decoherence of electronic spin qubits (or equivalently hybrid qubit in the unmixed limit with the two levels separated by ESR frequencies)
by a flip-flopping nuclear spin bath in natural silicon.
The background plots the spatial electronic wavefunction; blue denotes the strong-detuning
region, where the energy cost of a bath spin flip $\Delta_e^\pm \propto \pm(J_1-J_2)$ exceeds the
strongest intra-bath coupling $C_{12}$; it thus corresponds to the usual definition
of the ``frozen-core'' region.  However, electronic spin decoherence is dominated by an active zone (purple colour)
of pairs of  nuclear spins which are actually {\em within} the blue {\em strongly} detuned region,
with  $|\Delta_e^\pm/C_{12}| =|(J_1-J_2)/C_{12}| \sim 10$ for Si:P (see \Chap{formula} for details).
The reason is that, while for large $|\Delta_e^\pm|$ flip-flop amplitudes are strongly damped,
qubit state-dependence of the quantum bath evolution, essential for the entanglement between the
electronic spin and bath which produces decoherence,
is also proportional to $ \Delta_e^\pm$. Spin pairs for which $J_1= J_2$ (equivalent pairs) have
no effect on {\em electronic} decoherence and were not considered in previous studies.
Figure adapted from \citet{Guichard2014}.
}
\label{fig:Guichard2014_elec}
\end{figure}

Proximate spins lie within a so-called ``frozen core'' region, where the donor electronic hyperfine interaction
strongly suppresses nuclear dynamics. This is partly the reason for the lack of understanding of the decoherence dynamics
of proximate nuclear spins in silicon.
A pair of proximate nuclear spins can interact not only via direct dipolar coupling (\Sec{dipolar}),
but also via the long-range interaction mediated by the central donor electronic spin (\Sec{rkky}).
In both cases, the two spins may flip-flop and this results in decoherence of either an electronic or nuclear qubit, whichever is coupled to the flip-flopping pair.
However, in the case of strong hyperfine coupling between the nuclear impurity pair and the electron spin, the resulting energy detuning on each of the two members of the
pair overwhelms the dipolar coupling, suppressing the flip-flop dynamics and in turn suppressing decoherence
within the frozen core region.

The idea of the frozen core is well-established in the ESR community \citep{Khutsishvili1967,Wald1992,George2010},
but more recently there has been interest in utilizing it as a reservoir of protected qubits \citep{Mildren2013} for the reasons outlined above.
Therefore, it is of interest to determine coherence times of proximate nuclear spins in the frozen core.
The boundary radius $R_\text{FC}$ of the frozen core is commonly set as the distance at which hyperfine coupling strengths
have decreased to values comparable to the dipolar interactions between neighbouring nuclear spins \citep{Mildren2013}.
Representative values of the latter may be inferred from measured linewidths; for example, the $127$~Hz linewidth of $^{29}$Si
in natural silicon \citep{Hayashi2008} corresponds to an estimated $R_\text{FC} \approx 80$~\AA\ for Si:P.
As mentioned above, coherence times of proximate nuclear spins far outside the frozen core, are a few ms.

The argument that large energy detunings in the frozen core drastically suppress nuclear dynamics is not new.
However, spin bath decoherence in terms of entanglement between the qubit (whether electronic or nuclear)
and the environment has not been previously investigated in the frozen core.
Before discussing nuclear-qubit spin decoherence, it is useful to summarise decoherence of the hybrid qubit with reference to the frozen core,
in terms of detuned flip-flop nuclear bath dynamics -- \Fig{Guichard2014_elec}
(see \Chap{decoherence} and \Chap{formula} for a comprehensive account).
The large detunings in the frozen core lead to clear differences between decoherence of hybrid (or electronic) and 
proximate qubits, even when in both cases the same nuclear bath drives decoherence.
Below, we take a more careful look at what is meant by the frozen core and where,
precisely, its boundaries lie.
For example, for the electronic qubits, decoherence is in fact dominated by impurities
which lie within the usual definition of the frozen core as was shown in \citet{Balian2014} (see \Chap{formula} for details),
and as illustrated in \Fig{Guichard2014_elec}, since the detuning fully contributes to qubit-bath entanglement.
Decoherence of proximate qubits, discussed in detail for the rest of this chapter,
is summarised in \Fig{Guichard2014_farbath} and \Fig{Guichard2014_nuc}.

\section{Decoherence Dynamics of Nuclear Qubits}

The two models we consider for proximate qubit decoherence in silicon arise from the usual
pairwise flip-flops of nuclear impurities, but under extreme conditions, not encountered in previous decoherence studies
such as that presented in \Chap{decoherence} and \Chap{formula}.

As described in \Chap{background} and \Chap{decoherence},
it is not necessary to include all combinatorially allowed spin clusters;
within $350$~\AA\ of the donor site, there are in total $\sim 10^{10}$ \sitn\ spin pairs.
The smaller fraction of significantly contributing clusters 
are found by numerical search of each randomly populated lattice realisation by restricting the selection to,
for instance, pairs within a certain distance and hyperfine coupling strength as explained in \Chap{background}.
However, inside the frozen core, applying the normal distance (or coupling strengths) thresholds turned out to be unreliable.
Also, there is a drastic difference between the choice of spin clusters which must be included in the quantum bath for each of our two models;
a few dozen for the equivalent pairs model, $\sim 10^8$ for the far bath model.

Regardless of the choice of spin clusters, the basic decoherence dynamics of proximate qubits is of the same physical origin as for the hybrid qubit
in the unmixed limit -- pairwise flip-flops are predominately responsible for dephasing.
The process is equivalent to the usual CCE2 (pair correlations) but with detuning from the donor electron spin.
Before proceeding, we note that since the central and bath spins are of the same species and
higher correlations arising from larger clusters may be required for high accuracy \citep{Witzel2010,Witzel2012},
but in both our models would represent only a minor quantitative correction, unlike the case encountered for the hybrid qubits near OWPs, where
there was complete suppression of the pair correlation (as seen in \Chap{decoherence}).
We summarise the basic decoherence mechanism of spin diffusion below, but including the suppressive effect
of state-independent detuning provided by the electron spin,
and emphasising the importance of state-dependent detuning which drives decoherence.

For the nuclear qubit ($I=1/2$), we use the notation $\ket{\pm}$ to represent the upper (spin-up; $+$) and lower (spin-down; $-$)
states. The initial state after a $\pi$-pulse is
\begin{equation}
\left| \psi(t=0) \right\rangle =
\frac{1}{\sqrt{2}}
\left(\left|+\right\rangle+\left|-\right\rangle \right) \otimes \left| \mathcal{B}(0) \right\rangle \otimes \left| \phi (0) \right\rangle
\label{eq:product}
\end{equation}
where $\left| \phi (0)\right \rangle$ denotes the initial spin state of
the effective donor electronic spin-1/2 which is not resonant with the external control pulses.

\subsection{Spin Hamiltonian}

The central spin Hamiltonian is simply a sum of Zeeman terms
\begin{equation}
\hat{H}_\text{CS} = \gamma_e B \hat{S}^z + \gamma_n B \hat{I}^z_A,
\end{equation}
where the qubit is labelled $A$ and the high-field limit is assumed for the coupled
electron-nuclear donor at an ESR-type transition (i.e.\ no host donor nuclear spin terms).
Written for a pair of nuclear bath spins labelled by $\hat{\bf I}_l$ ($l=1,2$),
the interaction Hamiltonian is
\begin{equation}
 \hat{H}_{\textrm{int}}= \sum_{l=1,2} \left(
 \hat{\bf S} \cdot \mathcal{J}_l + \hat{\bf I}_{A}
 \cdot \mathcal{D}_{l} \right)
 \cdot \hat{\bf I}_l,
\end{equation}
whereby the qubit is coupled to the electron spin via the hyperfine interaction ($\mathcal{J}$; \Eq{hyperfine})
and to the bath spins via the dipolar interaction ($\mathcal{D}$; \Eq{dipolar}), which we take to be of secular form (\Eq{secdipolar}).
Because of the high-field limit and the large mismatch between electronic and nuclear gyromagnetic ratios,
the hyperfine interaction can be approximated to include
only Ising terms as given in \Eq{sechyperfine},
with the residual electron-nuclear dipolar interaction becoming effective farther than about 20~\AA\ from the donor site for Si:P.
The long-range effect of non-Ising terms
in the hyperfine interaction (RKKY; \Sec{rkky})
are added as a correction to the intra-bath dipolar interaction as described in the following subsection below.
Note that for simplicity, we do not include explicitly the hyperfine term coupling the electron to the resonant nuclear qubit $A$ ($J_A$),
which is only significant in the specific (but minor) contribution from direct flip-flopping processes.

Finally, the bath Hamiltonian is given by a sum of two nuclear Zeeman terms and the secular dipolar interaction (\Eq{secdipolar}):
\begin{equation}
\hat{H}_\text{bath} = \gamma_n B ( \hat{I}^z_1 + \hat{I}^z_2 ) +
 C_{12} \hat{I}^z_1 \hat{I}_2^z - \frac{C_{12}}{4}(\hat{I}^+_1\hat{I}_2^- + \hat{I}^-_1 \hat{I}_2^+).
\end{equation}

\subsection{Nuclear Pseudospins}

Under the action of the total Hamiltonian, the initial product state in \Eq{product} evolves into an entangled state:
\begin{equation}
\ket{\psi(t)}= \frac{1}{\sqrt{2}} \left(
\left[
\ket{+} \otimes \ket{\mathcal{B}_+ (t)}
+
\ket{-} \otimes \ket{\mathcal{B}_- (t)}
\right]
\otimes
\ket{\phi(t)}
\right),
\label{eq:wf}
\end{equation}
where we have omitted the central-state phases as we take the modulus of the coherence below.
As before, the coherence is given by the bath state overlap $|\mathcal{L}(t)| \propto |\innerprod{\mathcal{B}_+(t)}{\mathcal{B}_-(t)}|$
and in the pair correlation approximation $\mathcal{L}(t) = \prod_n \mathcal{L}^{(n)}(t)$,
where $\mathcal{L}^{(n)}(t)$ is the contribution from the $n$-th spin pair.
To apply the pseudospin model as was done in earlier chapters, the pure dephasing
approximation is required (perturbative corrections due to state depolarisation are possible to implement later in the formulation).
We have already assumed an Ising form for the hyperfine interaction.
As for the dipolar interaction of the qubit $A$ to the bath spins,
we numerically find that the non-Ising terms (direct flip-flops)
give negligible contribution to decoherence for both models.
Neglecting such terms,
the pseudpospin Hamiltonians for the nuclear qubit in a bath of two flip-flopping nuclei are written
\begin{equation}
 \hat{h}_\pm= \frac{1}{4}(\Delta^{\pm} \hat{\sigma}_z + C_{12} \hat{\sigma}_x)
\label{eq:equa8}
\end{equation}
where we have omitted the identity term which does not contribute to pseudospin coherence decays.
The Pauli operators act on the basis $\{\ket{\downarrow \uparrow},\ket{\uparrow \downarrow}\}$
of the two-spin-1/2 bath.

Crucially, the detuning is $\Delta^\pm = \Delta_e \pm (C_{1A}-C_{2A})$ for the proximate nuclear qubit,
where $C_{1A}$ and $C_{2A}$ are the dipolar coupling strengths between each bath nucleus and the qubit.
In this case, the electronic detuning $\Delta_e \equiv |J_1 - J_2|$ represents a potentially large contribution which is not sensitive
to the qubit state. For the Hahn echo case the coherence decay is given by:
\begin{equation}
|\mathcal{L}^{(n)}(t)| \simeq \left|1-2\alpha^{(n)}(\alpha^{(n)}+ i \beta^{(n)})\right|,
\label{eq:Decay}
\end{equation}
where
\begin{align}
\alpha &= \sin(\omega_+ t) \sin(\omega_- t) \sin {(\theta_+ - \theta_-)}, \nonumber \\
\beta  &= \sin(\omega_+ t) \cos(\omega_- t) \sin \theta_+ + \sin (\omega_- t) \cos (\omega_{+} t) \sin \theta_-,
\end{align}
while  $\theta_\pm=\tan^{-1}\left(C_{12}/\Delta^\pm\right)$, the eigenvalues $\omega_\pm = \dfrac{1}{4}\sqrt{(\Delta^\pm )^2+(C_{12})^2}$, and we have dropped the pair
label $n$ for convenience.
The larger $\theta_\pm$, the larger the amplitudes of the flip-flopping of nuclear spin-pairs which drives the decoherence.

When considering the contribution of flip-flopping pairs that are within the frozen core, we obtained excellent
agreement between the pseudospin equations above and numerical CCE2 provided that the well-known perturbative correction
for the non-Ising hyperfine terms, i.e.\ the RKKY interaction (\Sec{rkky}), was added to the dipolar coupling when using \Eq{Decay}.
The RKKY interaction was included by adding to $C_{12}$ the term $(J_{1} J_{2})/{\gamma_e B}$.
Note that for a numerical CCE calculation, the long-ranged interaction
emerges naturally if the hyperfine interaction with non-Ising terms is included and the above-mentioned correction should not be applied.

For pairs in the frozen core with different hyperfine couplings to the electron spin, $|\Delta^\pm| \simeq \Delta_e \gg |C_{12}|$, 
thus $\theta_\pm \simeq 0$ and flip-flops become too strongly suppressed.  
The qubit state sensitivity enters in \Eq{Decay} mainly via the $\sin{(\theta_+ - \theta_-)}$ 
prefactor and is also suppressed by $\Delta_e$.
This imposes the further condition  $|\Delta^\pm_n|=|(C_{1}^A-C_{2}^A)| \gtrsim \Delta_e$ for a single individual pair to contribute
appreciably to the decay.
Central to our modelling is the identification of spin clusters within the frozen core which can contribute non-negligibly to the decoherence of a proximate spin.
We now consider two models and apply them to the particular case of natural Si:P.

\section{Far Bath Model}

\begin{figure}[h!]
\centering\includegraphics[width=4.5in]{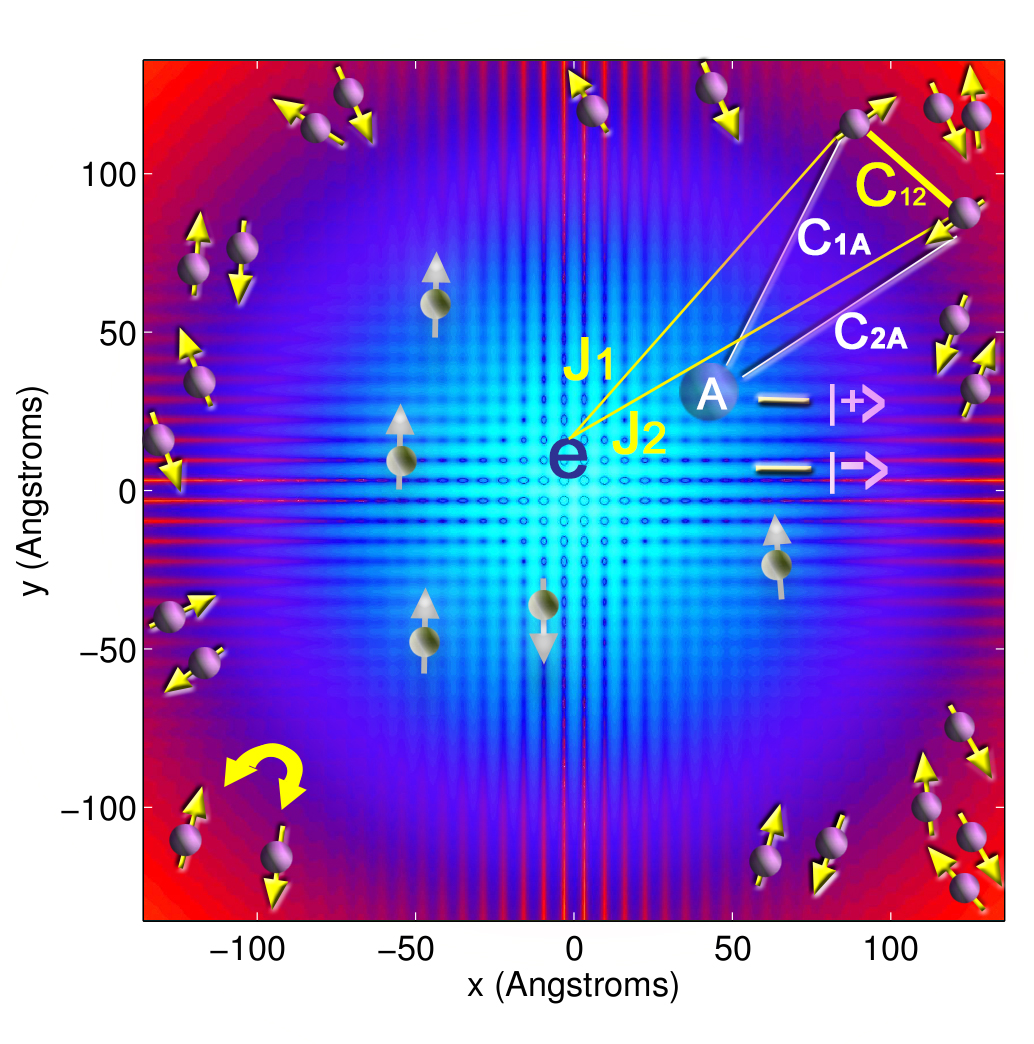}
\caption{Decoherence of a proximate nuclear spin qubit (labelled ``A'') by a quantum bath of nuclear spin pairs outside the frozen core.
In contrast to electron spin decoherence (for which the detuning is fully state-dependent, see \Fig{Guichard2014_elec}), 
the detuning is now $\Delta_e+\Delta_n^{\pm}$: there is now potentially a very large {\em state-independent}
component $\Delta_e \propto (J_1-J_2) $ which simply damps the bath noise, in addition to a {\em state-dependent} component
 $\Delta_n^{\pm} \propto \pm(C_{1A}-C_{2A})$  which leads to qubit-bath entanglement and thus decoherence.  
For large $R$ (distance from donor site), the bath spin interaction with both the electron spin and nuclear 
qubit is dipolar, thus  $|\Delta_n^{\pm}/\Delta_e| \sim 10^{-4}$ so very weak contributions from an extremely large bath
of $10^8$ pairs for $50 \lesssim R \lesssim 350$~\AA\
must be combined to obtain a converged decay.
Figure adapted from \citet{Guichard2014}.}
\label{fig:Guichard2014_farbath}
\end{figure}

In this far bath model, we consider the decoherence from distant nuclear spin pairs, which are outside
the frozen core and thus can flip-flop appreciably. We show that the typical contribution is so weak that we must include of order
$10^8$ flip-flopping pairs outside the frozen core, at distances $R=50-350$~\AA\ from the donor site, in order to obtain results converged with respect to
bath size. In contrast, typical quantum-bath calculations of electronic spin decoherence require  $\sim 10^3- 10^4$ pairs to
obtain convergence.
The far bath model is illustrated in \Fig{Guichard2014_farbath}.

From numerical simulations with a very large spin bath, we find that
distant spin pairs outside the frozen core radius $R_{\textrm{FC}}$ individually make an extremely small
contribution to decoherence: the $ \alpha \propto \sin {(\theta_+ - \theta_-)}$ prefactor
scales the coherence decays in \Eq{Decay}, since $ |\mathcal{L}(t)| \sim 1 - \alpha^2(\dots)$.
We can also show that the approximate weight of the $n$-th pair, is of order
$(1 / T_2^{(n)} )^2 \propto  \sin^2 {(\theta^{(n)}_+ - \theta^{(n)}_-)}$ \citep{Balian2014} (similar to the FID case in \Chap{formula}),
assuming also the temporal character of the associated magnetic noise is relevant: in other words,  
flip-flop frequencies $\omega_\pm$ for the given pair cannot be orders of magnitude different from $ \sim 1/T_2$.
For a non-negligible contribution we would expect that $ N_p \sin^2 {(\theta_+ - \theta_-)} \sim 1$
where $N_p$ is a representative number of contributing spin pairs.

\subsection{Convergence}

Before presenting numerical convergence tests, we first give a heuristic argument in order to establish the size of the convergent far bath.
From the pseudospin model (leaving out the $n$ labels),
\begin{equation}
\sin {(\theta_+ - \theta_-)} \simeq \frac{2 C_{12}}{\omega}  \ \frac{C_{1A}-C_{2A}}{\omega},
\end{equation}
since
$\omega_\pm \simeq \omega = \tfrac{1}{4}\sqrt{\Delta_e^2+C_{12}^2}$.
Above, the factor $2C_{12}/\omega$ determines whether the pair can flip-flop
appreciably and is significant if $|C_{12}/ \Delta_{e}| \sim 1$. The second factor, $(C_{1A}-C_{2A})/\omega$ determines state
distinguishability.
For the far spins, the hyperfine mediated correction plays little role
since $J_1$ and $J_2$ are small. 
For distances $ R \gtrsim 100$~\AA, where the Fermi contact component of the hyperfine interaction becomes small, the 
residual dipolar hyperfine interaction still makes a contribution to the detuning which is much
larger than $(C_{1A}-C_{2A})$. Here,
\begin{equation}
\frac{(C_{1A}-C_{2A})}{\omega} \sim \frac{(C_{1A}-C_{2A})}{J_1-J_2} \sim \frac{\gamma_n}{\gamma_e} \simeq 10^{-4}. 
\end{equation}
Thus, the contribution of each such far bath spin pair is $\left(\tfrac{\gamma_n}{\gamma_e}\right)^2 \sim 10^{-8}$, so
only a far bath with $N_p \sim 10^{8}$ contributing spin pairs can produce significant decay. At very large $R$, however,
$(C_{1A}-C_{2A})/\omega  \to (C_{1A}-C_{2A})/C_{12}$ .
But there is a minimum value of the interaction $C_{12} \equiv C^\text{min}_{12}$ where $(C^\text{min}_{12})^{-1}$  sets a timescale
below which the bath noise is too slow to contribute.
As $R \to \infty$, then $(C_{1A}-C_{2A})/C^\text{min}_{12} \to 0$,
so there is a maximum radius $R_\text{max}$ beyond which the far bath does not contribute significantly to decoherence.

\begin{figure}[h!]
\centering\includegraphics[width=4.5in]{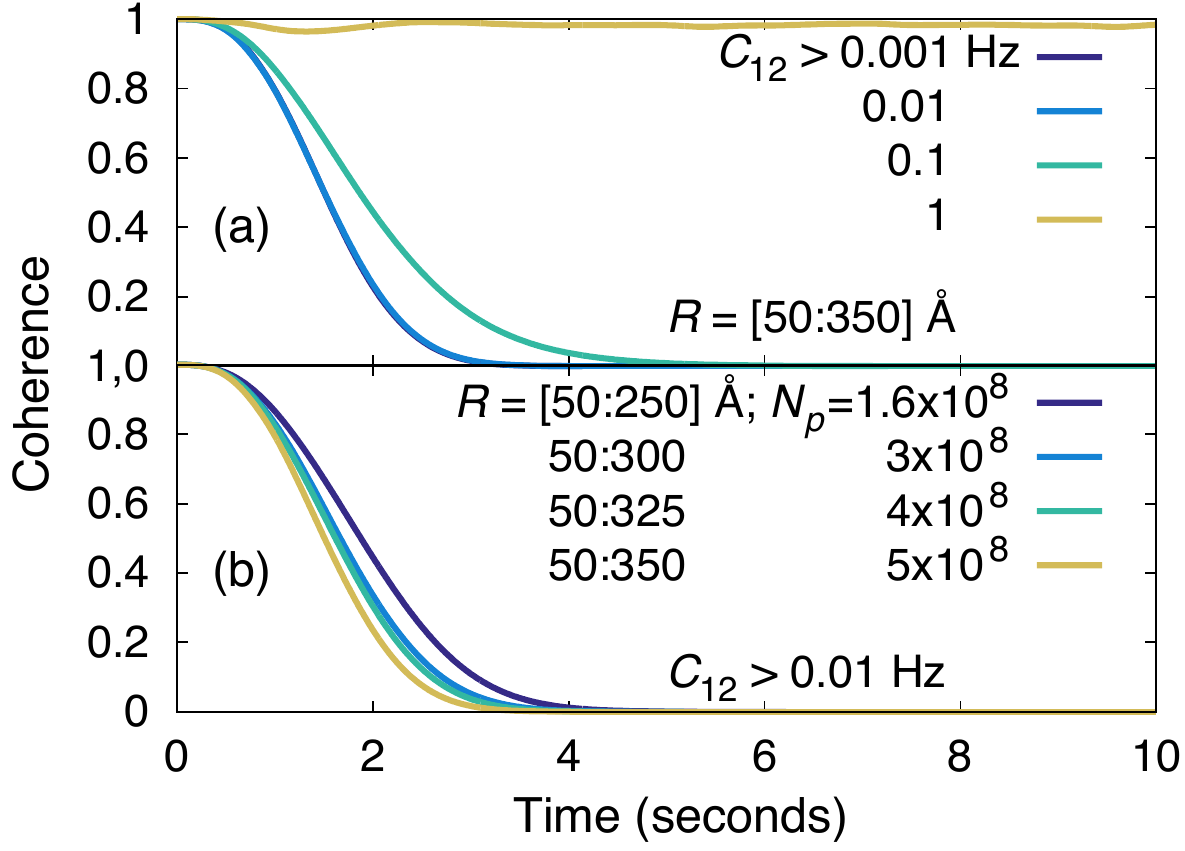}
\caption{Convergence of large bath model with respect to intra-bath dipolar coupling (a) and  with respect 
to bath size (b). The figure indicates that decoherence is dominated by spins with $C_{12} \sim 0.01 -1$~Hz and
a bath of spins within $R \lesssim 350$~\AA~ of the origin, combining the contributions from $5 \times 10^8$ spin pairs.
Calculations were performed for the case of Si:P, for $X$-band and magnetic field orientation $B_0=[1 0 0]$, yielding a $T_{2n}$ of 2 s for a
single nuclear $^{29}$Si spin sited at the origin. This represents an estimate for the upper bound for the coherence time
if the far bath is the dominant process.
Due to the large nuclear spin bath, the coherence decays are insensitive to the choice of random spatial realisation of the bath.
Figure adapted from \citet{Guichard2014}.
}
\label{fig:Guichard2014_conv}
\end{figure}

The analysis above was tested numerically by means of CCE2 calculations using a very large bath of nuclear spin pairs
(excluding contributions from the second model we present below)
and testing the effect on coherence decays of increasing the size of the bath.
\Fig{Guichard2014_conv} shows convergence with respect to bath size for Hahn echo decays, for a nuclear spin at the origin (thus
expected to give an upper bound on the coherence).
The $C_{12} \equiv C^\text{min}_{12} \sim 0.01-0.1$~Hz bound indicates that the pairs are within
$40-50$~\AA\ of each other and the calculation is converged with respect to bath size if we include $5 \times 10^8$ spin pairs
within $R_\text{max} \lesssim 350$~\AA\ of the origin.  
The scale of the bath is remarkable, in comparison with comparable electronic spin or hybrid qubit decoherence calculations with $\sim 10^4$ pairs.

Although it is computationally feasible to solve for a bath of this magnitude by CCE2 or analytical pseudospin methods, the uniformity of the bath means that
it is reliable to evaluate $\mathcal{L}$ in a smaller but geometrically representative sample of the bath.
In addition, no averaging over bath realisations (bath spin positions) was required; the results are insensitive to whether one has a single spin or an ensemble.

\begin{figure}[h!]
\centering\includegraphics[width=4.5in]{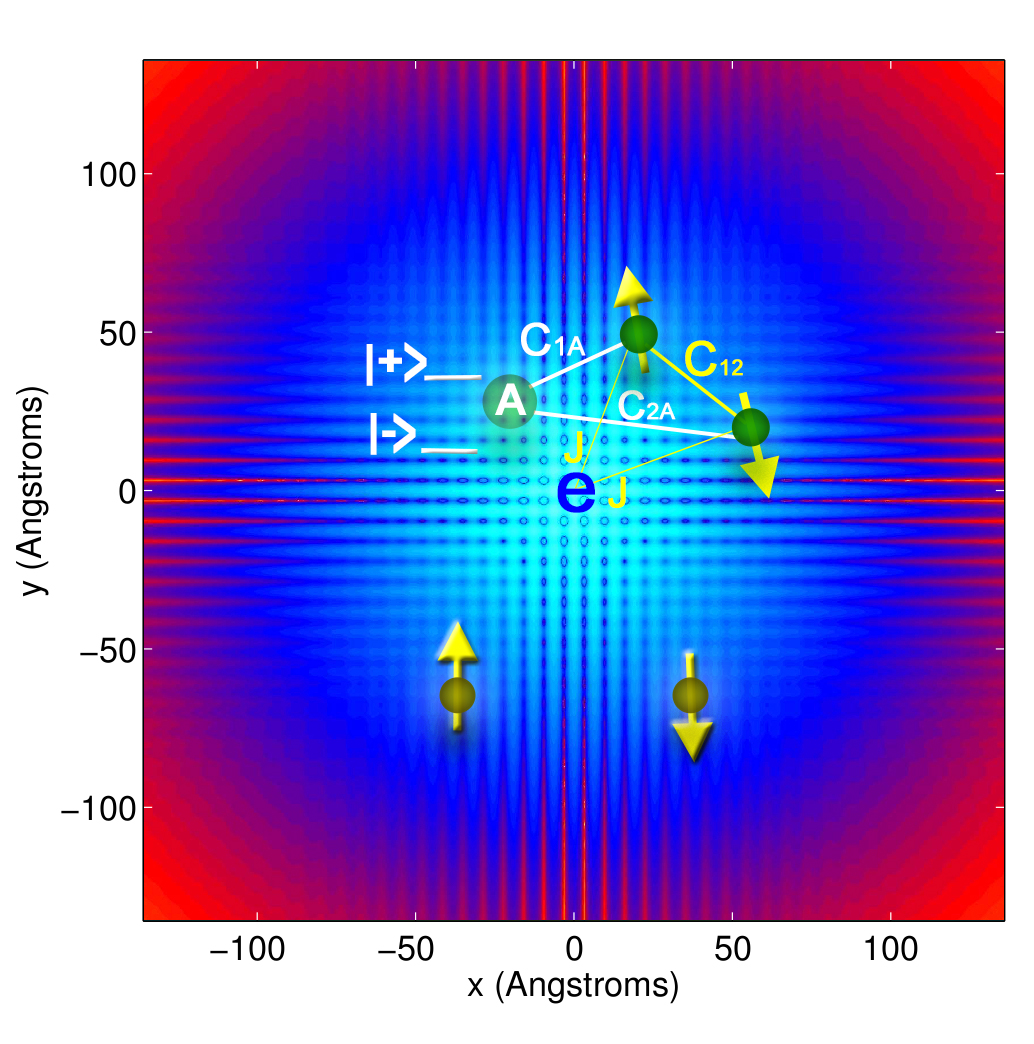}
\caption{Decoherence of a proximate nuclear spin qubit (labelled ``A'') by a quantum bath of nuclear spin pairs inside the frozen core.
See also \Fig{Guichard2014_farbath} for a comparison with decoherence outside the frozen core.
The detuning on flip-flopping bath pairs is $\Delta_e+\Delta_n^{\pm}$; i.e., a sum of a potentially very large
state-independent component $\Delta_e \propto (J_1-J_2)$, which damps decoherence in addition to a state-dependent one
$\Delta_n^{\pm} \propto \pm(C_{1A}-C_{2A})$ which drives decoherence.  
In the frozen core there are comparatively few spin impurities. For equivalent pairs however, $J_1=J_2\equiv J$ so
 $\Delta_e \simeq 0$. Their density is determined by the symmetry of the electronic wavefunction.
The requirement for strong state-selective detuning  implies also that one member of the pair must be
close enough to the qubit to allow appreciable direct dipolar coupling (as opposed to long-range 
coupling between nuclear spins mediated by the electron spin). Pairs which also satisfy this requirement
(exemplified by the upper, but not the lower, equivalent pair) are rare but even a few dozen suffice
to exceed the contribution of the $\sim 10^8$ far-bath spin pairs shown in \Fig{Guichard2014_farbath}.
Figure adapted from \citet{Guichard2014}.}
\label{fig:Guichard2014_nuc}
\end{figure}

\section{Equivalent Pairs Model}

In our second ``equivalent pairs'' (EP) model, the dephasing noise arises from a few dozen nuclear spin pairs, well within the frozen core, for which:
\begin{enumerate}[(i).]
  \item The members of the pair are symmetrically sited relative to the central spin and thus have equivalent values of the hyperfine detuning.
  \item At least one member is sufficiently close to the nuclear qubit to have a significant dipolar interaction, while the other can be remote.
  The nuclear spins interact via the long-ranged hyperfine interaction mediated by the electron. 
\end{enumerate}
The indirect flip-flopping of these EPs is found to be most significant, but we include also the rarer contribution of direct flip-flops between
the nuclear central spin and any equivalent partner it might have. We obtain $T_{2}$ values in the seconds timescale 
both for individual realisations (relevant to single donor experiments) and also for ensemble averages over many realisations.
The EP model is illustrated in \Fig{Guichard2014_nuc}.

\subsection{Counting Equivalent Sites}

The isotropic part of the hyperfine interaction is modeled using the Kohn-Luttinger wavefunction as described in \Sec{hyperfine}.
For phosphorus donors, the ionization energy is $0.044$~eV (\Table{donors}).
We can estimate the local densities of suitable EPs in the isotropic case before considering effects from any anisotropies in the hyperfine coupling
arising from the residual electron-nuclear dipolar coupling.

In our simulations, the full lattice size ranges over $[-N,N]$ cubic cells of diamond cubic for each dimension, resulting in $8N^3$ unit cells and hence $64N^3$
total atomic sites (see \App{silicon} for details of the diamond cubic crystal structure).
Equivalent sites are those with the same hyperfine interaction obtained using the Kohn-Luttinger wavefunction.
We begin by considering the allowed coordinates of the impurities in the crystal.
Owing to the symmetry of the system, each site possesses several potential equivalent partners,
for which positions can be deduced from any allowed permutations of $(\pm n_1,\pm n_2,\pm n_3)$ and which lie on the surface of shells
of radius $R=\frac{a_0}{4}\sqrt{n_1^2+n_2^2+n_3^2}$, where ${\bf n} = (n_1,n_2,n_3)$ is the integer vector
for an atomic site and $a_0$ is the cubic lattice parameter.
By consideration of the symmetries of the wavefunction, we can assign each vector
$\bf{n}$ to a shell $s$ comprising $n_s = $ 48, 24, 12, 8, 6 or 4 partners and we first obtain $\mathcal{N}_{n_s}(N)$,
the number of shells comprising $n_s$ partners within a radius of $R = N a_0$ from the center.

The ranges are adjusted to ensure summation over {\em complete} shells as follows.
First, we divide site vectors into three classes as shown in \Table{Guichard2014_classes}.
Summations must range between $[-N,N]$ for the 2 coordinates and $[-N,N-1]$ for the 0 coordinate of class 1
sites giving $4N^2(2N+1)$ number of sites;
between $[-N,N]$ for class 2 giving $(2N+1)^3$ number of sites; and finally between $[-N,N-1]$ for class 3 giving $8N^3$ number of sites.

\begin{table}[h]
\centering
    \begin{tabular}{ccc}
    \hline\hline
    Class 1 & Class 2 & Class 3  \\ \hline
    (0,2,2) & (0,0,0) & (3,3,3) \\
    (2,0,2) & ~ & (3,1,1) \\
    (2,2,0) & ~ & (1,3,1) \\
    ~       & ~ & (1,1,3) \\ \hline\hline
    \end{tabular}
\caption{The 8-site basis of the diamond cubic crystal structure (\App{silicon}) grouped into three classes. Class $X$ site vectors are obtained by modulo 4 translations of class $X$ basis
vectors.}
\label{table:Guichard2014_classes}
\end{table}

For each class, the contribution to a shell comprising $n_s$ partners within a radius of $R=Na_0$ of the center as a function of $N$ is summarized in
\Table{Guichard2014_tab1}.
Additionally, class 2 contributes as $8N$ to $n_s=8$ and as $6N$ to $n_s=6$ and class 3 contributes as $8N$ to $n_s=4$.

\begin{figure}[h]
\centering\includegraphics[width=4.5in]{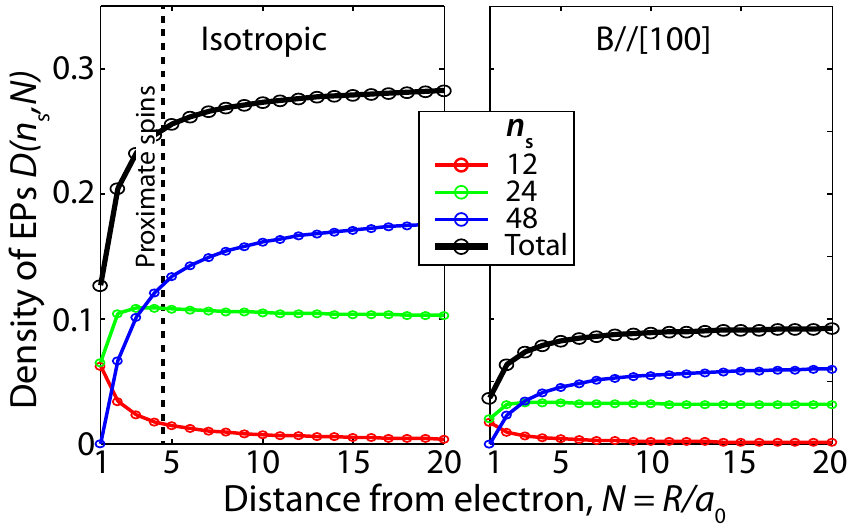}
\caption{Density of equivalent pairs (EPs) as a function of distance from the donor site.
The separate contributions from different types of shells is shown, as well as the total density, assuming a purely isotropic contact interaction (left) or a correction for anisotropic behaviour (right). The density of EPs is approximately constant for $ R \gtrsim 10$~\AA,
but the innermost proximate spins typically interact with fewer EPs. Figure adapted from \citet{Guichard2014}.}
\label{fig:Guichard2014_density}
\end{figure}

\begin{table}[h]
\centering
  \begin{tabular}{cccc}
    \hline\hline
     $n_s$ & 48 & 24 & 12 \\
    \hline
    Class&&& \\
    1 & $24N^2(N-1)$ & $12N(3N-1)$ & $12N$ \\
     &&& \\
    2 & $8N(N-1)(N-2)$ & $36N(N-1)$ & $12N$ \\
    &&& \\
    3 & -- & $16N(N-1)(2N-1)$ & $24N(2N-1)$\\
    \hline\hline
  \end{tabular}
  \caption{Class contribution to the equivalent sites group as a function of $N$. Table adapted from \citet{Guichard2014}.}
  \label{table:Guichard2014_tab1}
\end{table}

We can now obtain estimates for $\mathcal{N}_{n_s}(N)$:
\begin{eqnarray}
\mathcal{N}_{12}(N)&=&4N^2\nonumber\\
\mathcal{N}_{24}(N)&=&\frac{4}{3}N(N^2-1)+N^2,\nonumber\\
\mathcal{N}_{48}(N)&=& \frac{2}{3} N^3-N^2+\frac{N}{3},
\label{eq:sitenumb}
\end{eqnarray}
while $\mathcal{N}_{8}(N)=\mathcal{N}_{6}(N)=N, \mathcal{N}_{4}(N)=2N$. Then assuming a binomial distribution, taking an abundance of $p=0.0467$ for nuclear spin impurities in natural silicon, the estimated average number of significant EP in each shell is:
\begin{eqnarray}
\overline{\zeta}_{n_s} \simeq \sum_k \binom{n_s}{k} p^k (1-p)^{n_s-k} \ \frac{k(k-1)}{2},
\label{eq:avpair}
\end{eqnarray}
where $ \binom{n}{k}=\frac{n!}{k!(n-k)!} $ is the binomial coefficient. For the two dominant shells $\overline{\zeta}_{48} \sim 2.3$ and $\overline{\zeta}_{24} \simeq 0.6$. In these cases, it is quite likely that any impurity spin has an equivalent partner {\em somewhere}, albeit remotely located. Nevertheless, due to the long-range electron-mediated coupling, a $C_{12}$ of about tens of Hz is present. Within a sphere of radius $N$ cubic cells, we expect the total number of EP to be simply:
 \begin{eqnarray}
 \mathcal{N}_{\textrm{EP}} \simeq \sum_{s} \overline{\zeta}_{n_s} \mathcal{N}_{n_s}(N).
\label{eq:avpair2}
\end{eqnarray}
For instance, within a radius of $R=100$~\AA, we find $\mathcal{N}_{\textrm{EP}} \simeq 19,000$. For a proximate nucleus however, one member of the pair must be dipolar-coupled to the resonant spin (see caption of \Fig{Guichard2014_nuc}) which is relevant within about $m \sim 3$ cubic cells.
Each nuclear qubit thus interacts with other nuclei in the neighbouring $(2m)^3 \sim 200$ cells. We can define a density of spin pairs:
\begin{eqnarray}
D(n_s,N=R/a_0)= \frac{\overline{\zeta}_{n_s}\mathcal{N}_{n_s}(N)}{(2N)^3},
\label{eq:loc}
\end{eqnarray}
which gives the mean number of EPs in each cubic cell as a function of distance, $R=Na_0$ from the electron. We see in \Fig{Guichard2014_density} (left panel)
that the mean number for large $R$ is about $0.2-0.3$ pairs per cubic cell, thus each nuclear qubit interacts with $\sim 50$ potential EPs if anisotropy
(right panel of \Fig{Guichard2014_density}) is neglected.

\subsection{Effect of Anisotropy}

For the numerical calculations of the echo decays, we carried out a careful search, retaining about 500 equivalent spin pairs and
averaging over 100 realisations of a randomly generated lattice population with $4.67\%$ 
of sites occupied by \sitn\ spins. Two sets of calculations of the Hahn echo decays were carried out.
The first employed only the isotropic contact interaction and neglected anisotropic components of the hyperfine interaction.
These calculations provide a lower bound for the $T_{2}$ and predicted decay rates $T_{2} \sim 0.2-0.3$~s for different values of the
proximate qubit-electron hyperfine coupling $J_A$.

\begin{figure}[h!]
\centering\includegraphics[width=4.5in]{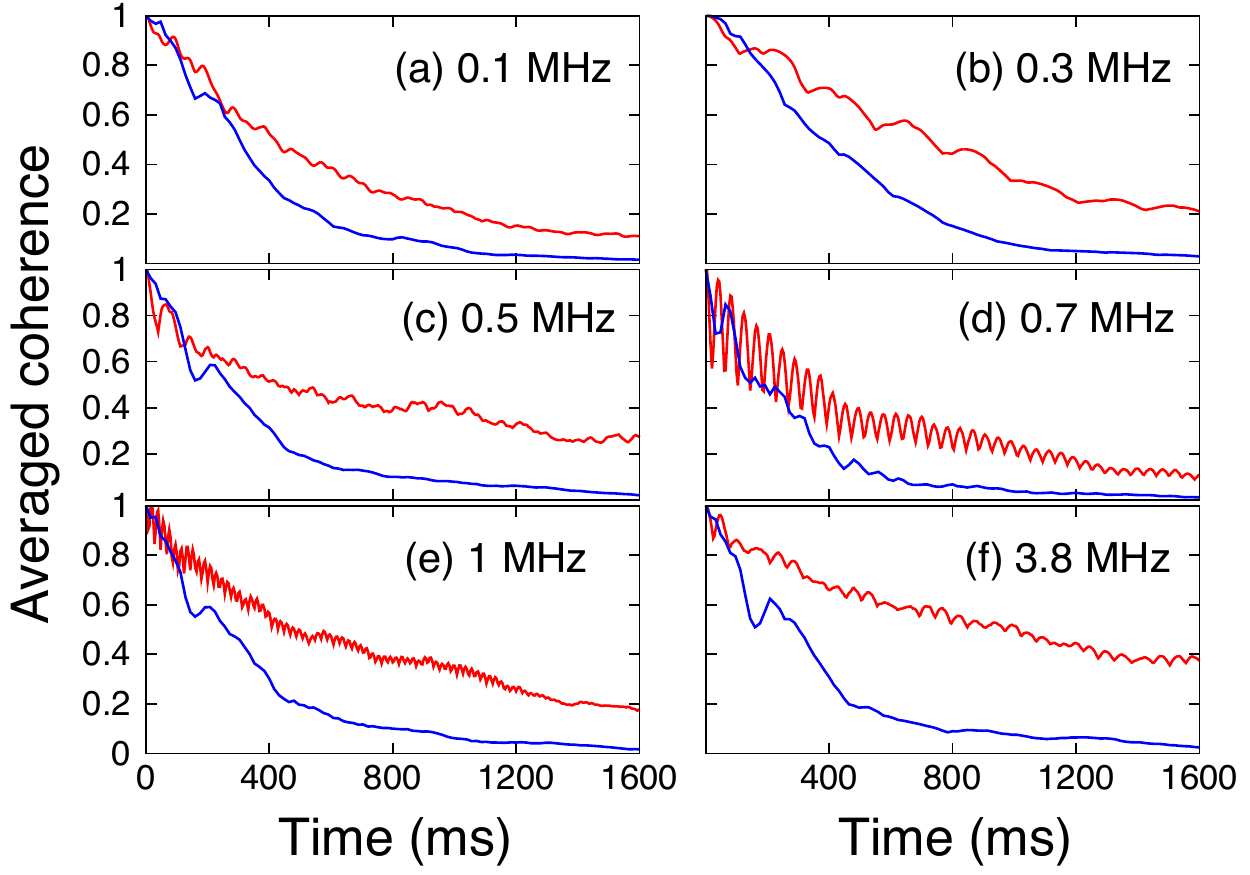}
\caption{Simulations of coherence decays of a set of proximate nuclear qubits
corresponding to a range of electron-qubit hyperfine couplings $J_A$ in MHz.
The blue lines correspond to isotropic electron-bath coupling only and yield $T_{2} \approx 0.2-0.3$~s;
red lines show the effect of symmetry reduction due to the anisotropy of couplings: we compare the effect of
desymmetrisation if we constrain EPs to have in addition the same
orientation condition (i.e.\ same $(\hat{\bf{n}}_B \cdot {\bf n})^2$).
The effect is to produce $T_{2}$ in the seconds timescale. Figure adapted from \citet{Guichard2014}.}
\label{fig:Guichard2014_aniso}
\end{figure}

A second set of calculations attempts to account for the anisotropy, which is less easy to calculate reliably. We assumed that any degree of
anisotropy detunes spin pairs so much that their contribution became negligible. In effect, this model provides an upper bound for the expected  $T_{2}$
as not all shells are affected by anisotropy.
Thus to remain an equivalent pair we required that spins have the same $(\hat{\bf{n}}_B \cdot {\bf n})^2$, where $\hat{\bf{n}}_B$ is the direction of the magnetic field. The
effect is to reduce the symmetries but to increase the number of shells, i.e.\ for $\hat{\bf{n}}_B = [1,0,0]$ and the main shells with
$n_s=48, 24, 12$ partners, we have $n_s \to n_s/3$ and therefore $\mathcal{N}_{n_s} \to 3\mathcal{N}_{n_s}$ (see right panel of \Fig{Guichard2014_density}).
The two sets of calculations are compared in \Fig{Guichard2014_aniso}.

\section{Coherence Times}

\begin{figure}[h!]
\centering\includegraphics[width=4.5in]{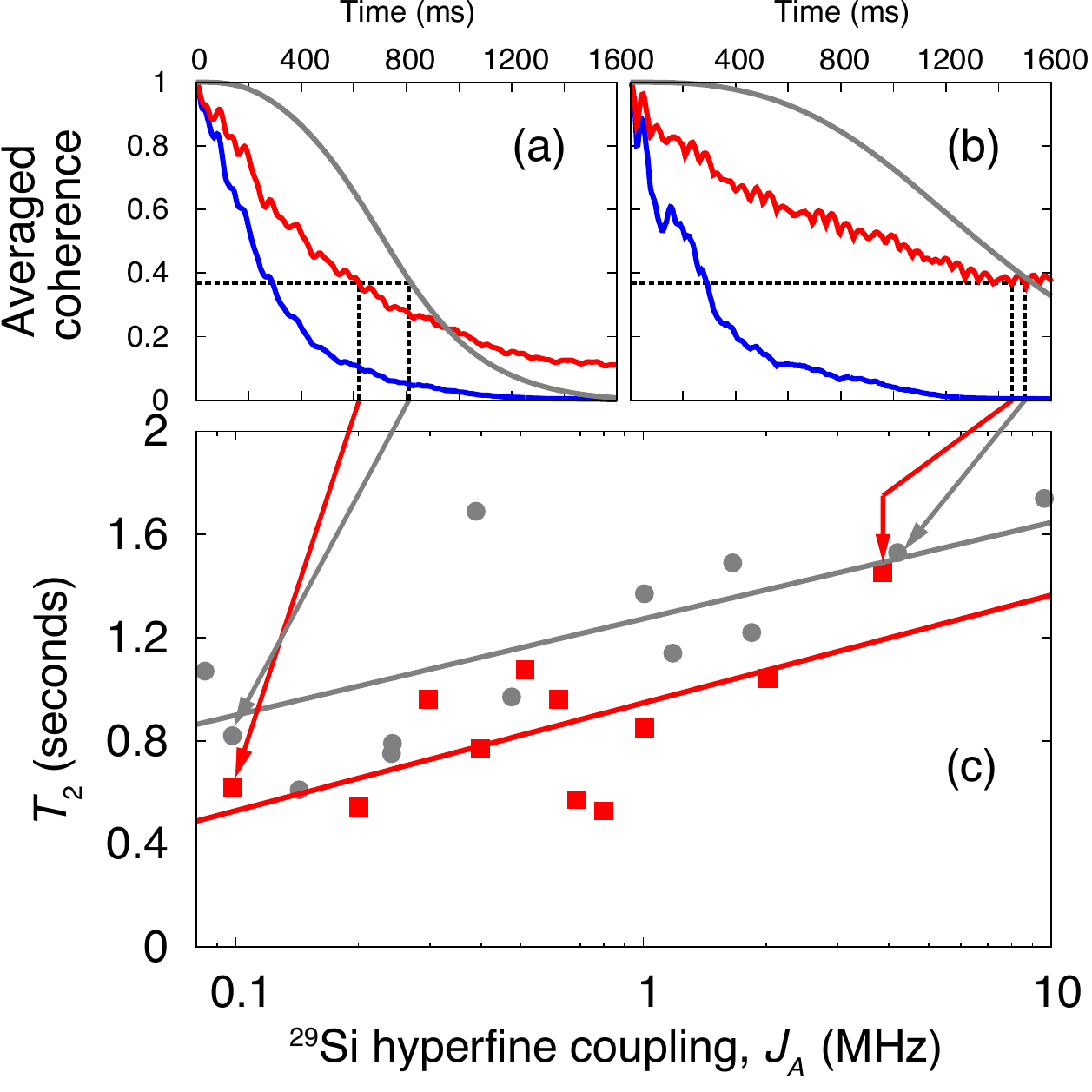}
\caption{Top panels: Calculated Hahn echo decays for proximate spins 
in a Si:P system in natural silicon for (a) $J_A = 0.1$~MHz and (b) $J_A = 3.8$~MHz, where $J_A$ is the hyperfine coupling between the proximate spin and the donor electron.
Red or blue correspond to decoherence driven by equivalent pairs (EP) while grey corresponds to far bath decoherence.
The blue lines include only the isotropic part of the electron-bath hyperfine interaction, while the red lines include both isotropic and anisotropic contributions.
Bottom panel (c): Calculated $T_{2}$ values from both models (red dots for EP model and grey dots for far bath).
There is a weak trend for $T_{2}$ to increase as the hyperfine coupling increases (red line is a fit),
possibly indicative of the decreasing density of EPs as the distance from the donor site $R \to 0$.
In the far bath model, the slight increase in decoherence with lower $J_A$ (grey line) reflects the fact that the lower $J_A$
proximate spins are slightly closer to the far bath.
Coherence times were obtained from decays averaged over 100 spatial realisations of the bath,
but typical single realisations gave the same timescale of decoherence.
Figure adapted from \citet{Guichard2014}.
}
\label{fig:Guichard2014_t2}
\end{figure}

Coherence times from calculated coherence decays employing each model and for a representative set of proximate nuclear spins (quantified by $J_A$)
are shown in \Fig{Guichard2014_t2}. Example coherence decays are also shown in \Fig{Guichard2014_t2}.
In both cases, $T_{2}$ is of order 1~s with a weak dependence on $J_A$ -- the coherence times tend to increase with larger $J_A$,
a trend also seen in the experiments in \citet{Wolfowicz2015b}.

For the EP model, we treat the anisotropic
correction simply as a symmetry lowering effect (see top panel of \Fig{Guichard2014_t2}).
This is plausible as the resultant detuning would be extremely large.
Presently, it is not possible to fully include anisotropy using the Kohn-Luttinger wavefunction.
The dipolar correction within this framework was included with a Heaviside function (\Sec{hyperfine}),
and is thus neglected for $R \lesssim 20$~\AA. Given other uncertainties,
the two EP calculations (with and without the anisotropy correction) in the top panels of \Fig{Guichard2014_t2} provide an upper and a lower bound to $T_{2}$.
As both results are on the seconds timescales, they suffice
for the practical aim of establishing the proximate nuclear spins as useful qubits. 

If the actual Si:P wavefunction exhibits a degree of spatial 
symmetry comparable with the Kohn-Luttinger wavefunction,
then the EPs could be the dominant mechanism, albeit only slightly.
However, it is likely that such symmetries are at least partly broken; in that case, the far 
bath would limit $T_{2}$. Given the uncertainties in the wavefunction model, at present
it is not possible to determine accurately the contributions of EPs relative to the far bath,
but fortuitously as the timescales are comparable, one can still conclude that the resulting $T_{2}$ is about
1~s.

To facilitate comparison with ensemble experiments, the EP results are averaged over many realisations
(the far bath model coherence decays are almost completely insensitive to ensemble averaging).
In the EP model, decoherence is primarily due to the indirect flip-flop process and this arises from only several dozen such EPs.
Thus, although results from single donors fluctuate between realisations, the corresponding order of magnitude for $T_2$
remains on the 1~s timescale, whether ensemble averaging is carried out or not.
The exception is the atypical realisation where the central spin happens to have an equivalent site it can directly flip-flop with.
For proximate central spins, which occur usually in inner shells with $n_s=4,8,12$, this is unlikely.
We find that the small subset of such realisations decohere rapidly. They contribute little to
the ensemble averaged $T_{2} \sim 1$~s values but would clearly be unsuitable as qubit registers unless
some strategy to exploit the degeneracy is envisaged. 
Finally and surprisingly, although the decay curves have a different shape, the $T_{2}$ values from both models are comparable.

\section{Conclusion}

Here we investigated the coherence of nuclear spins lying within the so-called ``frozen core'' surrounding the hybrid qubit in silicon,
within a quantum bath framework. We also calculated the coherence using a very large far bath of spins lying outside the frozen core.  
We introduced a previously unstudied model, based on equivalent pairs (EPs) deep within the frozen core, which we argue
would limit the coherence of proximate nuclear spins -- provided the electronic wavefunction has the symmetries of the Kohn-Luttinger wavefunction
(or even an alternative model with comparable levels of symmetry).
Within the EP model, decoherence is primarily due to an indirect flip-flop process arising from a few dozen such EPs.
Given the importance of carefully locating the few dozen or so most important EPs which can be quite widely separated,
distance or coupling strength thresholds which are usually applied for spin bath decoherence are not reliable.
The EP model of dephasing is quite generic: equivalent sites may play a role in any solid state
qubit system with a sufficiently dense nuclear spin bath.
Our quantitative results, however, are at best indicative. A more refined investigation could consider improved wavefunctions \citep{Pica2014} and more detailed inclusion
of the effect of anisotropy. Experimental investigations including dependence on symmetry-breaking mechanisms (such as crystal orientation and strain) will be useful
to test our proposal of the EP model.

In summary, we considered two decoherence models for proximate nuclear qubits, either of which, given certain assumptions, might contribute.
Both models predict \sitn\ nuclear spin coherence times of order 1~s (using only a Hahn spin echo and no higher dynamical decoupling),
which is consistent with recent experimental measurements \citep{Wolfowicz2015b} showing such spins could be useful as potential qubits.
If electronic symmetries are important, then a strategy for breaking such symmetries with external fields might be considered to obtain an even
longer $T_{2}$; if the far bath is dominant, partial isotopic enrichment might be more useful (consideration of \sitn\ nuclear spin registers in the
present study naturally precludes full enrichment).
We note that below 5~K, the electronic relaxation time $T_1$ is also above 1~s and would not limit the nuclear spin coherence.

\chapter{Conclusions}
\label{chap:conc}

The work presented in this thesis is the first theoretical study of decoherence of
hybrid electron-nuclear spin qubits driven by a quantum spin bath, and also of nuclear qubits proximate to the hybrid qubit.
By considering the full quantum state-mixing of the hybrid qubit in many-body calculations of coherence decays using the cluster
correlation expansion (CCE), we obtain coherence times in perfect agreement with
experiment across orders of magnitude variation with applied magnetic field and for an extensive set of resonant
transitions. We also propose two decoherence mechanisms for proximate nuclear qubits: the equivalent pair and
the far bath models, but expect the far bath to dominate experimental coherence decays because equivalent pair decoherence
relies heavily on the symmetries of an approximate theoretical model for the donor wavefunction.\footnote{We note that in a previous study of Overhauser field decay for quantum dots on much longer timescales (minutes), dynamics is driven by flip-flopping nuclear spins of similar energy interacting via the hyperfine-mediated interaction \citep{Latta2011}. However, in our equivalent pairs model knowledge of symmetries of the wavefunction in silicon is essential in locating the spins with
equivalent hyperfine coupling. Also, in our model at least one of the bath spins has to be within a certain distance from the nuclear qubit
for appreciable decay due to the hyperfine-mediated interaction.}

The physical system we use to implement the hybrid qubit is the mixed donor in silicon,
focusing in particular on bismuth donors due to their atypically strong state-mixing.
As for the quantum bath, we consider the natural spin-1/2 bath of silicon.
However, we expect our theoretical methods to be applicable to other
solid-state systems such as quantum dots or nitrogen vacancy centres in diamond.
Our numerical CCE implementation is capable of handling any complex multi-spin central and bath Hamiltonians (see \App{spindec} for details of the code).

Of interest to the experimental quantum information processing community,
we provide the theoretical means for reliable calculations of coherence times of hybrid qubits
in all magnetic field regimes, including the case of the simple unmixed qubit.
We also demonstrate how to enhance coherence times using
operation at optimal working points (OWPs), by employing dynamical decoupling pulse sequences,
or by combining the two methods.
We derive a closed-form formula for coherence times which gives remarkably accurate predictions
and is simple to use despite capturing the qubit-bath back-action without the need for a detailed
numerical many-body calculation.
The first pulsed magnetic resonance (ENDOR) experiments were presented for
bismuth donors in the silicon spin bath, using which we find
clear spectroscopic signatures of OWPs and characterise the hybrid qubit-nuclear spin bath interaction Hamiltonian.
The pulsed experiments also demonstrate the feasibility of controlling proximate nuclear impurity spins.

When it comes to addressing more fundamental issues such as the extent to which
classical magnetic field fluctuations are valid for describing decoherence driven by quantum baths or
the role of many-body correlations in qubit-bath dynamics,
we demonstrate that the hybrid qubit in a nuclear spin bath presents a good test bed for investigating such issues.
In particular, we demonstrate that near OWP regions, a classical noise model
based on frequency-field gradients does not reliably reproduce coherence times.
The formula for coherence times also identifies clear differences between
the classical and quantum bath models.
In addition, we present the only case where
the usual qubit-bath correlations from pairs of bath spins
are almost entirely diminished and clusters of three spins
in the many-body bath are required for converged coherence decays.
This occurs near OWPs and for low to moderate orders of dynamical decoupling.

Throughout our presentation, we have compared most of our theoretical results with experimental measurements,
finding excellent agreement in nearly all cases.
In particular, theory-experiment comparisons were made for coherence calculations of the hybrid qubit for forbidden transitions,
ESR-type and NMR-type transitions, near and far from OWPs and by combining OWPs with dynamical decoupling.
Our predicted timescale of decoherence for proximate nuclear qubits also agrees with experiment.

A brief summary of the main outcomes of the work was also given in \Sec{outcomes}. For the rest of this chapter, we focus
on the potential of the hybrid qubit (or a mixed electron-nuclear spin system) for quantum information applications,
and discuss other future work motivated by our results, including a
proposal for isolating increasing orders of many-body correlations.

It is clear that the hybrid qubit offers significant advantages for quantum computing and memory
applications compared to the case of uncoupled electronic or nuclear spins.
We focus on three aspects: fast quantum control at forbidden transitions, operation at OWPs, and
coherence enhancement of proximate nuclear spins.

First, operating at ESR-forbidden transitions in the hybrid regime
gives a factor of 125 speed-up of quantum control relative to the high-field regime albeit
at the expense of shorter coherence times;
if longer (pure nuclear) coherence times are required (i.e.\ for memory rather than processing),
the magnetic field may be ramped up to the high-field regime.
Nevertheless, qubit manipulation times are five orders of magnitude
shorter than coherence times in the hybrid regime. We established that in this forbidden transition regime of proven fast control,
coherence times are limited by nuclear spin diffusion.
Unfortunately, this 4~GHz excitation frequency region does not correspond to an OWP region.
Thus, coherence times here could be extended by isotopic enrichment or dynamical decoupling.
Second, for transitions with OWPs, two orders of magnitude enhancement in coherence times are achievable
without the need for any isotopic enrichment or multi-pulse dynamical decoupling control.
Fast microwave-pulsed quantum control is also achievable for ESR-type OWPs.
Further enhancement of coherence at OWPs in natural silicon is possible, but requires a large number
of dynamical decoupling pulses -- of order hundreds for a single order of magnitude enhancement.
Third, the hybrid qubit (in the unmixed regime) features a frozen core, protecting proximate
nuclear spins from quantum bath noise and leading to almost three orders of magnitude nuclear coherence time enhancement,
thus making proximate nuclei ideal for use as quantum registers, while the hybrid qubit is employed for processing.
Finally, we note that a universal set of quantum gates using the hybrid qubit are detailed in \citet{Mohammady2012}.

\section{Future Work}

As mentioned above, methods similar to those presented herein
may be applicable to other spin systems. For OWPs in donors, a host nuclear spin quantum number exceeding $I=1/2$ is required.
Hence, all silicon donors except for phosphorus can implement the hybrid qubit.
To date, all single-donor measurements have been made on phosphorus donors and all OWP experiments
on ensembles of bismuth donors.
Systems of interest in quantum information processing which warrant modification or extension of our methods
include quantum dots \citep{Webster2014}, especially coupled dots \citep{Weiss2012,Weiss2013} and also nitrogen vacancy centres in diamond \citep{Zhao2012a}.
Also, some materials have a rich variety of nuclear impurities; hence, heteronuclear spin baths become relevant.

Better knowledge of the donor electron wavefunction is expected to improve our results for both the hybrid qubit and proximate qubits.
More accurate models than the standard Kohn-Luttinger model exist, such as the model presented in \citet{Pica2014}.
First, it is expected that limitations of the Kohn-Luttinger wavefunction contribute to
the small deviations between calculated and measured coherence decays, whether for proximate nuclear or hybrid qubit decoherence.
More importantly, the equivalent pair model relies heavily on symmetries of the wavefunction.
One way of gaining more information about the wavefunction is to better characterise the nuclear spin bath by performing
continuous wave magnetic resonance which offers higher resolution than pulsed methods and which
can be used to assign hyperfine couplings of the donor electron spin to nuclear spin impurity positions or shells \citep{Hale1969a,Hale1969b}.
We note that if the equivalent pair model is confirmed by some other means,
it could be used to test the validity of symmetries in wavefunction models.
Also, in characterising the interaction to the spin bath, a future improvement is
thorough assessment of the errors in obtaining the anisotropic hyperfine couplings from rotation spectra,
for which data with higher angular resolution and knowledge of all relevant crystal directions are needed.
Finally, understanding the dependence of sidebands in ENDOR spectra on the radio frequency pulse length should lead to an
improved comparison between calculated and measured peak positions at low field.

The validity of the analytical formula for coherence times near OWPs remains partially understood.
The field-independent prefactor in the formula was derived from pair correlations.
The short time approximation was also used.
However, it is not entirely clear how the formula is successful for longer times.
In fact, the convolution of decays predicted by the formula near OWPs
does show significant discrepancy with experiment at longer times (beyond $T_2$) as seen in \Fig{Balian2014_owps_c}.
Three-clusters are essential for coherence decays as a function of time reaching zero-coherence near OWPs.
More remarkable is that the formula gives accurate predictions for a range of OWPs (where the pair of polarisations
vary by about an order of magnitude) and with large variations in coherence time
regardless of the fact that the derivation is based on an analysis of only pair correlations.
One thing that is clear is that the field-dependent component of the formula is a robust feature in all regimes. 
Also, more work is needed to understand the factor of about 2 to 3 difference between
the Hahn spin echo and FID coherence times near OWPs, and the reduction of this factor to about 1.1 as
the field is set far from the OWP.

As for comparing classical and quantum bath decoherence,
very recently it was reported
that the nuclear spin bath is of ``semi-classical'' nature near OWPs \citep{Ma2015}.
This seems to contradict a main result in this thesis.
However, a closer look at the study reveals that
a quantum many-body calculation (CCE) is in fact carried out to determine
the correlation function of the bath which includes a back-action term ($|P_u| + |P_l|$).
Nevertheless, at the start of the thesis, we state that qubit-bath back-action (or central state dependence of the bath evolution)
is a defining characteristic of what we mean by a quantum bath, and we compare
to previous classical models involving only a property of the qubit (frequency-field gradient).
Hence, the conflict between our work and the study in \citet{Ma2015} is simply resolved by the fact
that we use a different definition of classicality than the one in \citet{Ma2015}.

For isotopically enriched samples, where the nuclear spin bath is effectively eliminated,
decoherence is driven by donor-donor processes.
The same spin system, a donor spin, constitutes both the central system and bath.
Decoherence is driven by either spin diffusion from a flip-flopping bath of donors coupled to
the central donor (indirect flip-flops) or flip-flops between the central donor and a bath donor 
(a relaxation-type process involving single-spin bath clusters and known as direct flip-flops).
Indirect flip-flops are similar to the nuclear spin diffusion process and
are expected to be suppressed near OWPs.
However, it is not clear if there are any practically accessible sweet-spots for suppressing direct flip-flops (see for example, \citet{Monteiro2014}),
which limit coherence times in enriched silicon at OWPs.
If refocusing pulses are applied on the all-dipolar donor system,
decoherence is introduced as a process known as instantaneous diffusion \citep{Tyryshkin2012a}
which reduces coherence times from seconds to the millisecond timescale.
Again, it is not entirely understood whether or not special magnetic field
regions are experimentally feasible for the hybrid qubit for suppressing this process \citep{Monteiro2014}.
In order to obtain accurate coherence decays for donor-donor decoherence,
a modified CCE with external cluster awareness is needed for good accuracy \citep{Witzel2012}.
Note that near OWPs, coherence times from donor-donor processes can be comparable to nuclear bath
coherence times in natural silicon, provided the donor concentration is high enough.
Also, dynamical decoupling for donor-donor processes should be investigated,
as well as in combination with OWPs.
Finally, it has been previously demonstrated for phosphorus-doped silicon that line-broadening effects caused
by nuclear impurity spins in the bath suppress donor-donor flip-flops \citep{Witzel2010},
thus future studies must consider partial isotopic enrichment and a mixture of
donor-donor and impurity-related decoherence mechanisms; i.e.,
what is the ideal donor and impurity concentration for maximizing coherence times?

The width of spectral lines in a spin impurity environment presents a limit to how closely the magnetic field can be
adjusted towards the OWP. By investigating partial enrichment, the best compromise between
coherence times of the hybrid qubit and the density of proximate nuclear spins could be established.
As for proximate nuclear qubits, we note that there is no $T_2$ measurement of a single proximate spin.
Coherence decays from single realisations may show oscillations due to the few equivalent pairs.
Also, studies on wavefunction symmetry-breaking mechanisms could provide a good test of our
equivalent pairs model.

It should be mentioned that as an OWP is approached,
the likelihood of numerical divergences for high-order CCE increases. This is especially true for four-clusters and surprisingly less so for five-clusters.
The reason is not fully understood but the divergences depend on the choice of
bath realisations, are mitigated by averaging over many realisations and most likely result numerically from divisions by near-zeros.

Finally, additional work is needed to establish the many-body nature of the silicon spin bath at OWPs.
The 3-body correlations can have their origin in three-clusters containing three flip-flopping spins or
renormalised flip-flopping pairs; i.e.\ two-clusters with detuning from frozen external spins.
The possibility of isolating many-body correlations of increasing order (i.e.\ 3-body, 4-body, \dots)
as the field approaches closer to an OWP should also be investigated.
Since the linewidth in natural silicon limits how close to an OWP the magnetic field can be set,
many-body correlations should be studied for increasingly sparse spin baths with reduced impurity concentrations.

\appendix

\chapter{SpinDec Library}
\label{app:spindec}

\mtcaddchapter
\mtcaddchapter
\mtcaddchapter
\mtcaddchapter
\mtcaddchapter
\mtcaddchapter

\minitoc

\section{Overview}

SpinDec is a C++ library for spin decoherence calculations written by S.J.B..\footnote{\url{seto.balian@gmail.com}} It is free to use and can be downloaded from
Bitbucket or CCPForge:

\url{https://bitbucket.org/sbalian/spindec}

\url{https://ccpforge.cse.rl.ac.uk/gf/project/spindec/}

\noindent It should be cited as it appears in the thesis bibliography (see reference \citet{SpinDec} and \Sec{spindeclicense} for its BibTeX entry).

The code solves for the many-body dynamics of a central spin system coupled to
an interacting spin bath.
It supports any complex multi-spin Hamiltonian for both the central and bath Hamiltonians.
Spin Hamiltonians can be constructed by creating abstract spin interaction graphs and defining
edges for interactions among single spin vertices.

The cluster correlation expansion is implemented, special methods for donors in silicon are available and
methods are included for CPMG control.
The repository also includes an executable for calculating CPMG dephasing\footnote{This includes FID and Hahn spin echo as well as higher pulsed CPMG.} of donors in silicon
interacting with a \sitn\ nuclear spin bath.
Special methods for quantum dots and heteronuclear spin baths are under development.

This appendix includes installation and usage instructions for SpinDec.

\section{Installation}
\label{sec:spindecinstall}

The code has been tested on Linux and Mac OS X.
Installation requirements are as follows:

\begin{itemize}
\item Eigen (free): For linear algebra.

\url{http://eigen.tuxfamily.org}

\item CMake (free): For building and installing.

\url{http://www.cmake.org/}

\item boost (free): For program options. Only required for executable.

\url{http://www.boost.org/}

\item {\em Optional}: Intel MKL optimization for Eigen.

\url{https://software.intel.com/en-us/intel-mkl/}

\end{itemize}

The following installation instructions are for a Unix-like environment
without root privileges. It also assumes you have Mercurial (\texttt{hg})
installed.\\

To get the code, type in a terminal with \texttt{bash},

\lstset{style=custombash}
\singlespacing
\begin{lstlisting}
mkdir spindec
cd spindec
hg clone http://www.bitbucket.org/sbalian/spindec .
\end{lstlisting}
\doublespacing

Now let's build it.

\lstset{style=custombash}
\singlespacing
\begin{lstlisting}
mkdir build
cd build
cmake .. -DCMAKE_INSTALL_PREFIX=/home/myusrname
make -j4 all
make install
\end{lstlisting}
\doublespacing

The argument to \texttt{cmake} defines the installation location
(here set to `myusrname', which you should change to your user name).
If you invoke \texttt{cmake} without specifying the location, it will
install to your default system prefix (requires root).
The \texttt{-j4} option to \texttt{make} parallelises the build process using 4 cores.
Make sure this number does not exceed the number of cores for your machine.

If all went well, you should now have the static library and
executables. The library, headers and executables should be
located in the following directories respectively: \texttt{/home/myusrname/lib},
\texttt{/home/myusrname/include}, \texttt{/home/myusrname/bin}.

Optionally, to enable MKL optimization in Eigen,
pass \texttt{-DSPINDEC\_USE\_MKL=ON} to \texttt{cmake}. Obviously, you will need the
proprietary MKL libraries for this step. For further customization,
see the included file \texttt{CMakeLists.txt} (you will need to understand \texttt{cmake}).

\section{Usage}

Documentation generated with \texttt{doxygen} is available, detailing the structure of the code and explaining its use.
To use SpinDec, just include the \texttt{SpinDec/base.h} header in your source and link with
\texttt{libspindec}. Also, the executable for nuclear spin diffusion \texttt{spindec-dsnsd} has a \texttt{---help} option.

Help can also be found directly in header files (\texttt{.h} extension in \texttt{include/SpinDec}).
Source files (\texttt{.cpp} extension in \texttt{src/}) may contain more information,
usually geared more towards implementation.
See \Sec{spindecexample} for examples of usage.

\section{Examples}
\label{sec:spindecexample}

First, follow the instructions under ``Installation'' (\Sec{spindecinstall}).
Then, open the file \texttt{spindec/tests/spindec-test-cce.cpp}
to see the source with
comments on the example problem. Using the cluster correlation
expansion up to 3rd order, it calculates the Hahn spin echo decay of
a central donor spin in silicon for a nuclear spin bath
(spin-1/2 \sitn\ with a natural abundance of 4.7\%).
To see the decay, run \texttt{spindec-test-cce}.
The test executable is not installed, so run it in the build
directory.
The source for \texttt{spindec-dsnsd} (\texttt{src/spindec-dsnsd.cpp}) should also serve
as a good example.

\section{License and How to Cite}
\label{sec:spindeclicense}

SpinDec is free to use under the GNU General Public License.
See \texttt{LICENSE} file for more details.
If you use any part of the code in a publication,
please cite it as follows:

S. J. Balian, Spindec: C++ Library for Spin Decoherence,
\url{http://www.bitbucket.org/sbalian/spindec} (2011-2015).

Below is the BibTeX entry:

\lstset{style=customtex}
\singlespacing
\begin{lstlisting}
@misc{SpinDec,
  title = {SpinDec: C++ Library for Spin Decoherence},
  author = {Balian, Setrak J.},
  howpublished = {http://www.bitbucket.org/sbalian/spindec},
  year = {2011--2015}
}
\end{lstlisting}
\doublespacing

\section{Version and History}

As of June 15, 2015 the version number is {\bf 0.9} (late-stage beta - debugged, tested, profiled, optimized and checked for memory leaks).
SpinDec is based on bits and pieces of code for certain spin baths
and central spin systems. These were written by S.J.B. starting
in 2011. Methods used to efficiently fill Hamiltonian matrix elements
were originally written in Fortran by Professor Tania Monteiro.
Development for SpinDec as an open-source project started in May 2013.

\chapter{Pauli Operators}
\label{app:pauli}

Any two-level observable can be described
by linear combinations of the {\em Pauli operators} $\hat{\sigma}_k$, $k=1,2,3$ \citep{Audretsch2007}
and the $2\times2$ identity.
The operators act on the 2-dimensional Hilbert space and satisfy
\begin{equation}
\hat{\sigma}^i \hat{\sigma}^j = \delta_{ij} \hat{\mathds{1}} + i \sum_{k=1}^{3} \epsilon_{ijk} \hat{\sigma}^k,
\label{eq:pauli1}
\end{equation}
where $i,j=1,2,3$, $\delta_{ij}$ is the Kronecker delta ($\delta_{ij} = 1$ for $i=j$, $\delta_{ij} = 0$ for $i \neq j$),
$\hat{\mathds{1}}$ is the identity operator and the tensor $\epsilon_{ijk}$ is totally antisymmetric in all indices with $\epsilon_{123} = 1$.
Equivalently, the condition in \Eq{pauli1} can be written in terms of the commutators and anti-commutators
\begin{equation}
\left[ \hat{\sigma}^i, \hat{\sigma}^j \right]  = 2i  \sum_{k=1}^{3} \epsilon_{ijk} \hat{\sigma}^k,
\end{equation}
\begin{equation}
\left\{ \hat{\sigma}^i, \hat{\sigma}^j \right\} = 2 \delta_{ij} \hat{\mathds{1}}.
\end{equation}
The Pauli vector in the Cartesian basis $\hat{\boldsymbol{\sigma}} = (\hat{\sigma}_x , \hat{\sigma}_y, \hat{\sigma}_z )$
is often used to refer to the operators.
The operators are Hermitian ($(\hat{\sigma}^{k})^\dagger = \hat{\sigma}^k$), unitary ($(\hat{\sigma}^{k})^\dagger = (\hat{\sigma}^{k})^{-1}$) and
traceless ($\text{Tr}[\hat{\sigma}^k] = 0$), and have eigenvalues $+1$ and $-1$.
Taking the orthonormal basis of eigenvectors of $\hat{\sigma}_z$ as $\{\ket{0},\ket{1}\}$,
the matrix representations of the operators in this basis are the {\em Pauli matrices}:
\begin{align}
\boldsymbol{\sigma}^x = 
\left(
\begin{array}{cc}
0 & 1 \\
1 & 0
\end{array}
\right),&&
\boldsymbol{\sigma}^y = 
\left(
\begin{array}{cc}
0 & -i \\
i & 0
\end{array}
\right),&&
\boldsymbol{\sigma}^z = 
\left(
\begin{array}{cc}
1 & 0 \\
0 & -1
\end{array}
\right).
\end{align}

The electron has a total spin quantum number $S = 1/2$ and its spin angular momentum operators
are related to the Pauli operators according to
$\hat{\bf S} = (\hat{S}^x , \hat{S}^y, \hat{S}^z ) = \hat{\boldsymbol{\sigma}}/2$.
For a nuclear spin-1/2, we write $I = 1/2$ and $\hat{\bf I} = \hat{\boldsymbol{\sigma}}/2$.

In quantum computing, the Pauli operators represent simple single-qubit gates.
For example, the Pauli-$X$ gate ($\ket{0}\bra{1} + \ket{1}\ket{0}$) flips $\ket{0}$ ($\ket{1}$) to $\ket{1}$ ($\ket{0}$) \citep{Nielsen2010}.

\chapter{Silicon Crystal Structure}
\label{app:silicon}

\begin{figure}[h!]
\centering\includegraphics[width=3in]{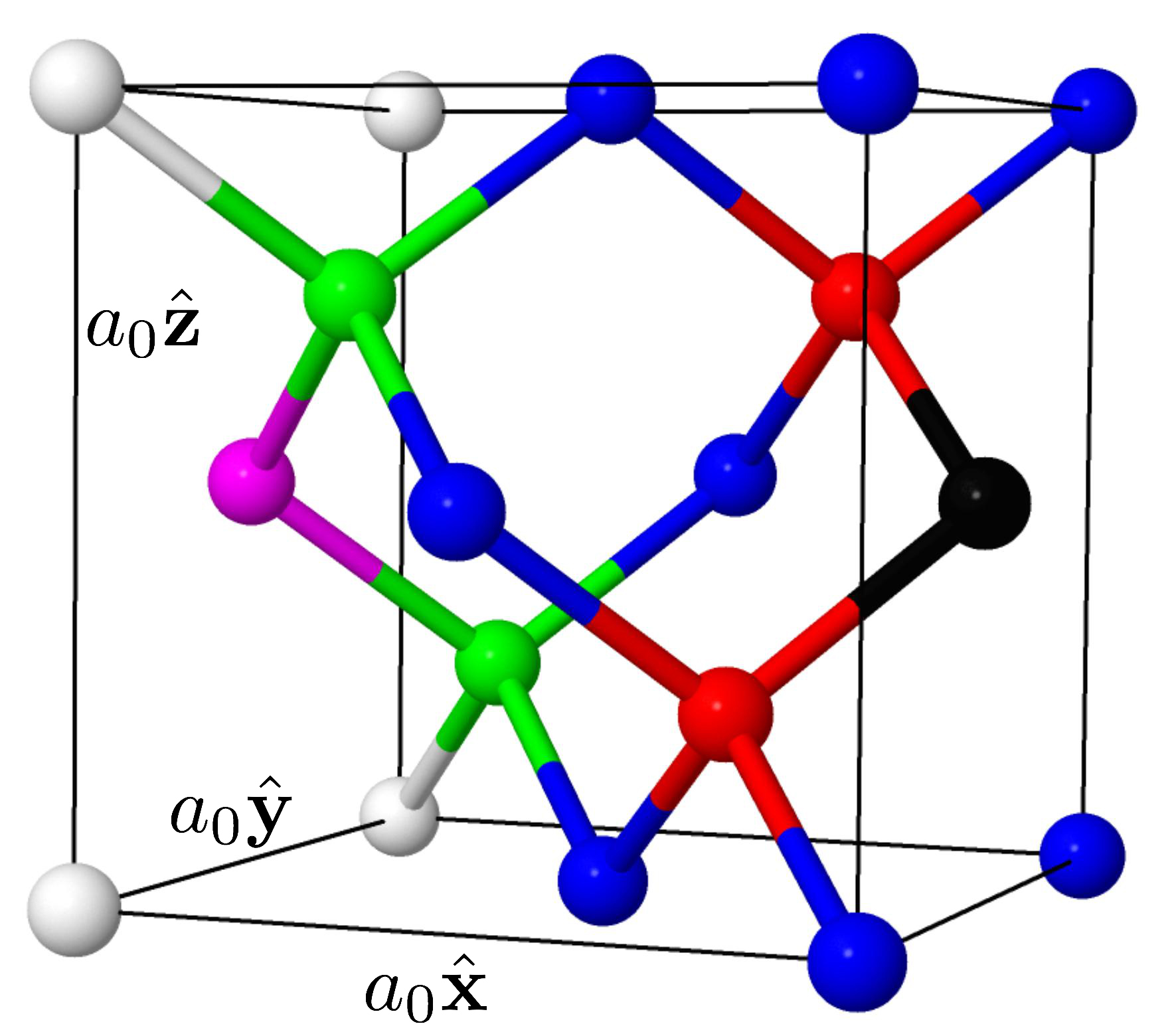}
\caption{
Conventional cubic cell of the diamond cubic crystal structure.
The nearest neighbors distances (illustrated relative to the black atom)
are $\tfrac{\sqrt{3}}{4} a_0$ (red), $\tfrac{\sqrt{2}}{2}a_0$ (blue), $\tfrac{\sqrt{11}}{4}a_0$ (green) and $a_0$ (purple), where
$a_0$ is the lattice parameter.
In Miller index notation \citep{Kittel1996}, [100], [010] and [001] correspond to the directions along $\hat{\bf x}$, $\hat{\bf y}$ and $\hat{\bf z}$ respectively.
For example, [111] is along the body diagonal.
Figure adapted from \citet{Balian2014}.
}
\label{fig:Balian2014_diamondcubic}
\end{figure}

Silicon forms the diamond cubic crystal structure \citep{Kittel1996}.
The diamond cubic is two face-centred cubic (FCC) lattices,
one displaced from the other by a distance of $\frac{1}{4} a_0$ along the body diagonals, where
$a_0$ is the conventional cubic cell lattice parameter (i.e.\ side length of the cubic cell).
For silicon, $a_0 = 5.43$~\AA.
The conventional cubic cell of the diamond cubic is illustrated in \Fig{Balian2014_diamondcubic}.

The crystal structure can also be described by a simple cubic lattice
and an 8-site basis. We represent all atomic sites (the crystal structure)
by an integer vector ${\bf n} = (n_1,n_2,n_3)$, the integers obtained
from modulo 4 translations in all directions of the 8 basis vectors.
The 8 basis vectors are:
(0,0,0),
(0,2,2),
(2,0,2),
(2,2,0),
(1,1,3),
(1,3,1),
(3,1,1),
and
(3,3,3).
The transformation to convert from integer to real space ${\bf R}$ is
\begin{equation}
{\bf R}_{(n_1,n_2,n_3)} = \frac{a_0}{4}\left( n_1 \hat{\bf x} + n_2 \hat{\bf y} + n_3 \hat{\bf z} \right),
\end{equation}
in Cartesian coordinates, where the axes are parallel to those of the conventional cubic cell.


\bibliographystyle{apsrmp4-1}

\singlespacing

\bibliography{refs}

\doublespacing

\end{document}

%% file: user_settings.tex
\title{Quantum-Bath Decoherence of Hybrid Electron-Nuclear Spin Qubits}
\author{Setrak Jean Balian}
\newdate{thesis_date}{28}{08}{2015}

%% file: user_preamble.tex
\usepackage{amsmath,amssymb}
\usepackage{bm}
\usepackage[usenames,dvipsnames]{xcolor}
\usepackage{dsfont}
\usepackage{amsbsy}
\usepackage{mathtools}

\usepackage{hyperref}

\usepackage{enumerate}
\usepackage{fancyhdr}
\usepackage{setspace}
\usepackage{caption}
\usepackage{minitoc}
\usepackage[intoc]{nomencl}
\usepackage[nottoc]{tocbibind}

\usepackage{natbib}

\newcommand{\ket}[1]{\left|{#1}\right\rangle}

\newcommand{\bra}[1]{\left\langle{#1}\right|}

\newcommand{\avg}[1]{\left\langle{#1}\right\rangle}

\newcommand{\innerprod}[2]{\left\langle{#1}|{#2}\right\rangle}
\newcommand{\outerprod}[2]{\left|{#1}\right\rangle\left\langle{#2}\right|}

\newcommand{\Eq}[1]{Equation~(\ref{eq:#1})}
\newcommand{\Fig}[1]{Figure~\ref{fig:#1}}
\newcommand{\Table}[1]{Table~\ref{table:#1}}
\newcommand{\Sec}[1]{Section~\ref{sec:#1}}
\newcommand{\App}[1]{Appendix~\ref{app:#1}}
\newcommand{\Chap}[1]{Chapter~\ref{chap:#1}}

\newcommand{\sitn}{$^{29}$Si}

\newcommand{\tsc}[1]{\textsuperscript{#1}}

\usepackage{listings}
\usepackage{textcomp}
\input{./user_codestyle.tex}

\newcommand{\nthn}[1]{}

%% file: user_codestyle.tex
\definecolor{nicegreen}{rgb}{0,0.6,0}
\definecolor{purplish}{rgb}{0.58,0,0.82}
\definecolor{light-gray}{gray}{0.95}

\lstdefinestyle{custombash}{
  backgroundcolor=\color{light-gray},   
  basicstyle=\ttfamily\small,        
  breakatwhitespace=false,         
  breaklines=true,                 
  captionpos=b,                    
  commentstyle=\color{nicegreen},    
  extendedchars=true,              
  keepspaces=true,                 
  keywordstyle=\color{blue},       
  language=bash,                 
  otherkeywords={mkdir},            
  showspaces=false,                
  showstringspaces=false,          
  showtabs=false,                  
  stepnumber=1,                    
  stringstyle=\color{purplish},     
  tabsize=2,                       
  title=\lstname                   
}

\lstdefinestyle{customcpp}{
  backgroundcolor=\color{light-gray},   
  basicstyle=\ttfamily\small,        
  breakatwhitespace=false,         
  breaklines=true,                 
  captionpos=b,                    
  commentstyle=\color{nicegreen},    
  extendedchars=true,              
  keepspaces=true,                 
  keywordstyle=\color{blue},       
  language=C++,                 
  showspaces=false,                
  showstringspaces=false,          
  showtabs=false,                  
  stepnumber=1,                    
  stringstyle=\color{purplish},     
  tabsize=2,                       
  title=\lstname                   
}

\lstdefinestyle{customtex}{
  backgroundcolor=\color{light-gray},   
  basicstyle=\ttfamily\small,        
  breakatwhitespace=false,         
  breaklines=true,                 
  captionpos=b,                    
  commentstyle=\color{nicegreen},    
  extendedchars=true,              
  keepspaces=true,                 
  keywordstyle=\color{blue},       
  language=TeX,                 
  showspaces=false,                
  showstringspaces=false,          
  showtabs=false,                  
  stepnumber=1,                    
  stringstyle=\color{purplish},     
  tabsize=2,                       
  title=\lstname                   
}

%% file: user_publications.tex
%
%

\begin{itemize}
\item \textbf{Chapters 3 \& 5}: G. W. Morley, P. Lueders, M. Hamed Mohammady, {\bf S.J.B.}, G. Aeppli, C. W. M. Kay, W. M. Witzel, G. Jeschke,
and T. S. Monteiro, Nature Materials {\bf 12}, 103 (2013).

\item \textbf{Chapters 4 \& 5}: {\bf S.J.B.}, M. B. A. Kunze, M. H. Mohammady, G. W. Morley, W. M. Witzel, C. W. M. Kay, and T. S. Monteiro,
Physical Review B {\bf 86}, 104428 (2012). 

\item \textbf{Chapters 5 \& 6}: {\bf S.J.B.}, G. Wolfowicz, J. J. L. Morton, and T. S. Monteiro, Physical Review B {\bf 89}, 045403 (2014).

\item \textbf{Chapters 5 \& 7}: {\bf S.J.B.}, R.-B. Liu, and T. S. Monteiro, Physical Review B {\bf 91}, 245416 (2015).

\item \textbf{Chapter 8}: R. Guichard, {\bf S.J.B.}, G. Wolfowicz, P. A. Mortemousque, and T. S. Monteiro, Physical Review B {\bf 91}, 214303 (2015).

\end{itemize}

%% file: user_abstract.tex
%

\setstretch{1.5}
 
A major problem facing the realisation of scalable solid-state quantum computing
is that of overcoming decoherence -- the process whereby phase information encoded in a quantum bit (`qubit') is
lost as the qubit interacts with its environment. Due to the vast number of environmental
degrees of freedom, it is challenging to accurately calculate decoherence times $T_2$,
especially when the qubit and environment are highly correlated.

Hybrid or mixed electron-nuclear spin qubits, such as donors in silicon,
are amenable to fast quantum control with pulsed magnetic resonance.
They also possess `optimal working points' (OWPs) which are sweet-spots for reduced decoherence in magnetic
fields. Analysis of sharp variations of $T_2$ near OWPs was previously based on insensitivity
to classical noise, even though hybrid qubits are situated in highly correlated quantum environments,
such as the nuclear spin bath environment of \sitn\ impurities.
This presented limited understanding of the underlying decoherence mechanism
and gave unreliable predictions for $T_2$.

In this thesis, I present quantum many-body calculations of the qubit-bath dynamics,
which (i) yield $T_2$ for hybrid qubits in excellent agreement with experiments
in multiple regimes, (ii) elucidate the many-body nature of the nuclear spin bath and (iii)
expose significant differences between quantum-bath and classical-field decoherence.
To achieve these results, the cluster correlation expansion was adapted to include
electron-nuclear state mixing.
In addition, an analysis supported by experiment was carried out to characterise the
nuclear spin bath for a bismuth donor as the hybrid qubit, a simple analytical formula for $T_2$ was derived
with predictions in agreement with experiment, and the established method of dynamical decoupling was
combined with operating near OWPs in order to maximise $T_2$.
Finally, the decoherence of a $^{29}$Si spin in proximity to the hybrid qubit was studied,
in order to establish the feasibility for its use as a quantum register.

\doublespacing

%% file: user_acknowledgements.tex
%

\setstretch{1.5}

First and foremost, I'd like to acknowledge my parents, Alin and Ohan, and my brother, Vartan,
for constantly encouraging me to pursue a career in science.
I am infinitely indebted to my parents for their unconditional support
both moral and financial. Without them none of this work would have been possible.

I express my sincere gratitude to Dr. Mischa Stocklin for generously
co-funding the EPSRC studentship via the Stocklin-Selmoni Impact Studentship.
I acknowledge with much respect my supervisor Professor Tania Monteiro for her guidance and endless patience.
Special thanks to Dr. Gavin Morley, who as secondary supervisor
also provided much support. I'd like to thank Dr. Wayne Witzel
for agreeing to host and supervise my visit at Sandia.
I also acknowledge Professors Ren-Bao Liu, Chris Kay
and John Morton for many useful discussions.

I thank Dr. Gary Wolfowicz, Dr. Hamed Mohammady, Dr. Roland Guichard,
Dr. Wen-Long Ma, Dr. Micha Kunze and Jacob Lang for many illuminating
discussions about our shared work in physics. I am lucky to have had
Dr. Martin Uhrin both as a friend and a walking (open source) library for
programming and computational issues. I'd also like to thank Dr. Bobby Antonio,
Dr. Leonardo Banchi, Dr. Georg Schusteritsch and Dr. Costas Lazarou
for always being open to discussion about physics or programming.

Thanks to all my friends and colleagues at UCL, without whom my time
in London would not have been the same -- Michela Venturelli, Dr. Duncan Little, Na\"iri Usher,
Kostas Konstandinos, Enrico Compagno, Raffaele Nolli, Roberta Guilizzoni, Dr. Alexandros Gerakis, Dr. Roberto Lo Nardo and
Chris Perry.

Finally, thank you Radhika Patel for patiently enduring my occasional selfishness
at the time of writing and for showing me that a positive outlook goes a long way.

\doublespacing

%% file: user_abbreviations.tex
\nomenclature{OWP}{Optimal working point ($P_u=P_l$).}
\nomenclature{CCE}{Cluster correlation expansion.}
\nomenclature{EPs}{Equivalent pairs.}
\nomenclature{CT}{Clock transition ($df/dB=0$).}
\nomenclature{ESR}{Electron spin resonance.}
\nomenclature{NMR}{Nuclear magnetic resonance.}
\nomenclature{ENDOR}{Electron-nuclear double resonance.}
\nomenclature{Natural silicon}{Silicon with 4.67\% \sitn\ spin impurities.}
\nomenclature{Enriched silicon}{Silicon with reduced \sitn\ spin impurities.}
\nomenclature{Hybrid qubit}{Multi-spin system with eigenstates which mix the {\em Zeeman basis}.}
\nomenclature{Zeeman basis}{Basis of product eigenstates of a sum of uncoupled Zeeman (high magnetic field) spin Hamiltonians.}
\nomenclature{Hybrid regime}{Magnetic field region of {\em hybrid qubit} where there is strong mixing of the {\em Zeeman basis}.}
\nomenclature{Unhybridized regime}{Magnetic field region of {\em hybrid qubit} where there is no longer strong mixing of the {\em Zeeman basis}.}

\nomenclature{$\mathcal{L}(t)$}{Coherence as a function of time $t$.}
\nomenclature{$\tau$}{Dynamical decoupling pulse spacing in time.}
\nomenclature{$N$}{Number of dynamical decoupling pulses.}
\nomenclature{$T_2$}{Coherence time.}
\nomenclature{$T_1$}{Relaxation time.}
\nomenclature{$B$}{Zeeman magnetic field strength.}
\nomenclature{$A_X$}{Strength of hyperfine interaction between the donor electron spin and host nuclear spin of Si:$X$.}
\nomenclature{$J$}{Strength of hyperfine interaction between an electron spin and a bath nuclear spin.}
\nomenclature{Si:$X$}{$X$-doped silicon.}
\nomenclature{$\hat{\bf S}$}{Electron spin operator, $\hat{\bf S} = (\hat{S}^x,\hat{S}^y,\hat{S}^z)$.}
\nomenclature{$\hat{\bf I}$}{Nuclear spin operator, $\hat{\bf I} = (\hat{I}^x,\hat{I}^y,\hat{I}^z)$.}
\nomenclature{$\mathcal{J}$}{Hyperfine tensor.}
\nomenclature{$J_\text{F}$}{Fermi contact strength.}
\nomenclature{$\mathcal{D}$}{Dipolar tensor.}
\nomenclature{$C$}{Secular dipolar interaction strength.}
\nomenclature{NV}{Nitrogen vacancy.}
\nomenclature{$df/dB$}{Transition frequency-magnetic field gradient.}
\nomenclature{Cryogenic temperatures}{Temperature $T \lesssim 15$~K.}
\nomenclature{FID}{Free induction decay.}
\nomenclature{CPMG}{Carr-Purcell-Meiboom-Gill dynamical decoupling pulse sequence.}
\nomenclature{LZ}{Landau-Zener.}
\nomenclature{Bloch sphere}{Points on a continuous spherical surface which have a one-to-one correspondence with quantum states of a {\em qubit}.}
\nomenclature{Qubit}{Quantum bit or two-level system.}

\nomenclature{ESR-type transitions}{$\ket{\pm,m} \leftrightarrow \ket{\mp,m-1}$ transitions.}
\nomenclature{NMR-type transitions}{$\ket{\pm,m} \leftrightarrow \ket{\pm,m-1}$ transitions.}
\nomenclature{$\ket{u} \to \ket{l}$}{Magnetic resonance transition from an upper ($\ket{u}$) to a lower ($\ket{l}$) level in energy.}
\nomenclature{$\hat{\boldsymbol{\sigma}}$}{Vector of Pauli operators, $\hat{\boldsymbol{\sigma}} = (\hat{\sigma}^x,\hat{\sigma}^y,\hat{\sigma}^z)$.}
\nomenclature{$\gamma_e$}{Electronic spin gyromagnetic ratio.}
\nomenclature{$\gamma_n$}{Nuclear spin gyromagnetic ratio.}
\nomenclature{$m_S$}{Electronic spin magnetic quantum number.}
\nomenclature{$m_I$}{Nuclear spin magnetic quantum number.}
\nomenclature{$S$}{Electron spin quantum number ($S = \tfrac{1}{2}$).}
\nomenclature{$I$}{Nuclear spin quantum number.}
\nomenclature{$I_X$}{Silicon donor (Si:$X$) host nuclear spin quantum number.}
\nomenclature{$\omega_0$}{Electron spin Larmor frequency, $\omega_0 = \gamma_e B$.}
\nomenclature{$\ket{\pm,m}$}{Adiabatic eigenstate of mixed electron-nuclear spin system ({\em hybrid qubit} implementation).}
\nomenclature{$\ket{i}$}{Eigenstates of {\em central spin system} labelled in order of increasing energy.}
\nomenclature{$\hat{H}_\text{bath}$}{Bath spin Hamiltonian.}
\nomenclature{$\hat{H}_\text{int}$}{Interaction spin Hamiltonian.}
\nomenclature{$\hat{H}_\text{CS}$}{{\em Central spin system} or {\em qubit} spin Hamiltonian.}
\nomenclature{$\hat{H}_\text{tot}$}{Total spin Hamiltonian, $\hat{H}_\text{tot} = \hat{H}_\text{CS} + \hat{H}_\text{int} + \hat{H}_\text{bath}$.}
\nomenclature{$\hat{h}_{i}$}{Pseudospin Hamiltonian for central level $\ket{i}$.}
\nomenclature{$\hat{U}(t)$}{$e^{i \hat{H} t}$, for Hamiltonian $\hat{H}$.}
\nomenclature{$\hat{T}_i(t)$}{$e^{i \hat{h}_i t}$, for pseudospin Hamiltonian $\hat{h}_i$.}
\nomenclature{$X$-band}{9.7~GHz ESR excitation frequency.}
\nomenclature{$S$-band}{4~GHz ESR excitation frequency.}
\nomenclature{ESR-forbidden transition}{Magnetic resonance transition ESR-forbidden at high fields (i.e., a pure NMR transition at high fields.)}
\nomenclature{$\ket{\uparrow}$}{Spin-up.}
\nomenclature{$\ket{\downarrow}$}{Spin-down.}
\nomenclature{$P_i$}{Polarisation for level $\ket{i}$, $P_i = \bra{i} \hat{S}^z \ket{i}$.}
\nomenclature{FWHM}{Full width at half maximum.}
\nomenclature{$\overline{C}(\theta)$}{Dipolar prefactor of $T_2$ {\em formula}, with crystal orientation angle $\theta$.}
\nomenclature{$T_2$ formula or expression}{$T_2 = \overline{C}(\theta) \tfrac{|P_u(B)|+|P_l(B)|}{|P_u(B) - P_l(B)|}$.}

\nomenclature{$\omega^{\pm}$}{$\omega^{\pm} = \omega_u \pm \omega_l$.}
\nomenclature{$\theta^{\pm}$}{$\theta^{\pm} = \theta_u \pm \theta_l$.}
\nomenclature{$\theta_i$}{{\em Pseudofield} angle from $z$-axis of {\em Bloch sphere}.}
\nomenclature{$\omega_i$}{{\em Pseudospin} frequency.}
\nomenclature{$\delta J$}{$J_1 - J_2$.}
\nomenclature{$\neq$OWP}{Far from OWPs.}
\nomenclature{RKKY}{Ruderman-Kittel-Kasuya-Yosida or hyperfine-mediated interaction.}
\nomenclature{$\ket{u}$}{Upper central system level.}
\nomenclature{$\ket{l}$}{Lower central system level.}

\nomenclature{$\epsilon_i$}{Donor electron ionization energy.}
\nomenclature{$a_0$}{Cubic lattice parameter.}
\nomenclature{$\hbar$}{Planck constant divided by $2\pi$.}
\nomenclature{$T$}{Temperature.}
\nomenclature{$k_B$}{Boltzmann constant.}
\nomenclature{Diamond cubic}{Crystal structure of silicon or diamond.}
\nomenclature{$\mu_0$}{Permeability of free space.}
\nomenclature{Coherent or equal superposition}{$\tfrac{1}{\sqrt{2}}\left( \ket{u} + \ket{l} \right)$}
\nomenclature{$\ket{\mathcal{B}(0)}$}{Initial bath state.}
\nomenclature{$\ket{\mathcal{B}_i}$}{Bath state correlated with central system state $\ket{i}$.}
\nomenclature{$T_2^*$}{Ensemble FID coherence time.}
\nomenclature{$T_2'$}{Coherence time sometimes used to distinguish donor-donor decoherence from {\em nuclear spin diffusion} decoherence characterized by $T_2$.}
\nomenclature{Direct flip-flops}{{\em Flip-flops} between {\em central spin system} and bath spins.}
\nomenclature{Indirect flip-flops}{{\em Flip-flops} between bath spins coupled to a {\em central spin system}.}
\nomenclature{Flip-flops}{Spin Hamiltonian terms of the form $ \hat{S}_1^- \hat{S}_2^+ + \hat{S}_1^+ \hat{S}_2^-$.}
\nomenclature{Central spin system}{In general, a multi-spin system resonant with control pulses and distinguished from bath or environmental spins.}
\nomenclature{Spin bath}{Environment surrounding a {\em central spin system} and made up of spin species.}
\nomenclature{Quantum bath}{Environment with strong system-environment back-action (involving entanglement). A {\em spin bath} is an example.}
\nomenclature{Instantaneous diffusion}{Decoherence caused by flipping bath spins as well as the {\em central spin system} with the application of $\pi$-pulses.}
\nomenclature{$\pi$-pulse}{Refocusing pulse as used in the Hahn spin echo sequence for example.}
\nomenclature{$\pi/2$-pulse}{Pulse to create a {\em qubit} superposition.}
\nomenclature{Spin diffusion}{Any {\em indirect flip-flop}-type decoherence mechanism.}
\nomenclature{Nuclear spin diffusion}{Spin diffusion of an electronic spin or {\em hybrid qubit} in a nuclear {\em spin bath}.}
\nomenclature{Proximate nuclear spins}{Nuclear spins nearby an electronic spin or {\em hybrid qubit} in the {\em frozen core} region.}
\nomenclature{Frozen core}{A region of strong detuning caused by an electronic spin or {\em hybrid qubit} where nuclear {\em flip-flops} are suppressed.}
\nomenclature{State-dependent detuning}{Part of the total detuning on bath {\em flip-flop} dynamics which is important for driving decoherence.}
\nomenclature{State-independent detuning}{Part of the total detuning on bath {\em flip-flop} dynamics which always suppresses decoherence.}
\nomenclature{SpinDec}{C++ spin decoherence library written by S.J.B..}
\nomenclature{Pseudospin}{Two-level system described by the space spanned by the non-polarized eigenstates of a two-spin-1/2 system
($\{\ket{\downarrow \uparrow},\ket{\uparrow \downarrow}\}$).}
\nomenclature{Ising terms}{Spin Hamiltonian terms involving products of only $z$-projection operators.}
\nomenclature{non-Ising terms}{Spin Hamiltonian terms not involving products of $z$-projection operators, and often involving {\em flip-flop} terms.}
\nomenclature{Pure dephasing}{Decoherence involving no central state depolarisation.}
\nomenclature{$\hat{\sigma}^{\pm}$}{$\hat{\sigma}^x \pm i\hat{\sigma}^y$, and similarly for $\hat{S}^{\pm}$ and $\hat{I}^{\pm}$.}
\nomenclature{CCE$k$}{CCE truncated at, or calculated up to, $k$-th order.}
\nomenclature{$n$-body correlations}{Non-trivial {\em qubit}-bath dynamics arising from $n$ bath spins and involving entanglement between the {\em qubit} and bath.}
\nomenclature{Pair correlations}{{\em $2$-body correlations}.}
\nomenclature{Pseudofield}{{\em Pseudospin} precession axis.}